# Introduction to High-Resolution Inelastic X-Ray Scattering

Alfred Q.R. Baron

*Materials Dynamics Laboratory, RIKEN SPring-8 Center, RIKEN, 1-1-1 Kouto, Sayo, Hyogo 679-5148 Japan*
baron@spring8.or.jp

**Abstract**

This paper reviews non-resonant, meV-resolution inelastic x-ray scattering (IXS), as applied to the measurement of atomic dynamics of crystalline materials. It is designed to be an introductory, though in-depth, look at the field for those who be may interested in performing IXS experiments, or in understanding the operation of IXS spectrometers, or those desiring a practical introduction to harmonic phonons in crystals at finite momentum transfers. The treatment of most topics begins from ground-level, with a strong emphasis on practical issues, as they have occurred to the author in two decades spent introducing meV-resolved IXS in Japan, including designing and building two IXS beamlines, spectrometers and associated instrumentation, performing experiments, and helping and teaching other scientists. After an introduction that compares IXS to other methods of investigating atomic dynamics, some of the basic principles of scattering theory are described with the aim of introducing useful and relevant concepts for the experimentalist. That section includes a fairly detailed discussion of harmonic phonons. The theory section is followed by a brief discussion of calculations and then a longer section on spectrometer design, concepts, and implementation, including a brief introduction to dynamical diffraction and a survey of presently available instruments. Finally, there is discussion of the types of experiments that have been carried out, with a focused discussion on measurements of superconductors, and brief discussion (but with many references) of other types of crystalline samples.

## Table of Contents

## Table Of Abbreviations

| | |
|---|---|
| CDW | Charge Density Wave |
| DAC | Diamond Anvil Cell |
| DHO | Damped Harmonic Oscillator |
| HRM | High Resolution Monochromator |
| INS | Inelastic Neutron Scattering |
| IFC | Interatomic Force Constant |
| IUVS | Inelastic Ultraviolet Spectroscopy |
| IXS | Inelastic X-Ray Scattering<br>(In the present paper: specifically meV resolved non-resonant investigations) |
| NIS | Nuclear Inelastic Scattering<br>(also, sometimes, NRVS, or, occasionally, NRIXS, is used instead) |
| NRIXS | Non-Resonant Inelastic X-Ray Scattering<br>(Generally with resolution on the 0.1 eV scale for electronic interactions) |
| NRS | Nuclear Resonant Scattering |
| PSC | Post Sample Collimation |
| RIXS | Resonant Inelastic X-Ray Scattering |
| SA | Spherical Analyzer |
| SIXS | Soft (X-ray) Inelastic X-Ray Scattering |
| TDS | Thermal Diffuse Scattering |
| XAFS | X-Ray Absorption Fine Structure |
| XPCS | X-Ray Photon Correlation Spectroscopy |

# 0. Preface to the 2018 and 2020 Revisions

This paper is a revised version of a review prepared in 2014 and disseminated in 2015 and 2016 (Baron), and then revised in 2018 and 2020. The 2018 version updated references, and generally tried to make the paper more readable and precise. It also extended the discussion of spectrometers with a more detailed comparison of spherical analyzers (SA) and post-sample collimation (PSC) optics (section 4.8). The 2020 revision includes further updates in references and changes to the discussion about multi-phonon scattering, and detailed balance.

# 1. Preliminaries

## 1.1 Introduction

Inelastic x-ray scattering (IXS), broadly interpreted, refers to a large collection of photon-in -> photon-out measurement techniques that are used to investigate many disparate aspects of materials dynamics*. The breadth of the science involved, and the sophistication of the instrumentation, can make the field somewhat daunting for newcomers. Table 1 lists some of the techniques that fall under the broad interpretation of the heading, while figure 1 schematically shows some of the excitations that may be probed. Methods range from Compton scattering that may be used to probe electron momentum densities, to x-ray Raman scattering that is, to a first approximation, similar to XAFS, probing atomic bonding, to resonant inelastic scattering which can probe collective and localized electronic states, as well as magnetism, to non-resonant techniques that, with sufficiently high flux (and, usually, poorer resolution), can be used to probe band-structure and electronic excited state structure, and, with higher resolution, to probe atomic dynamics.

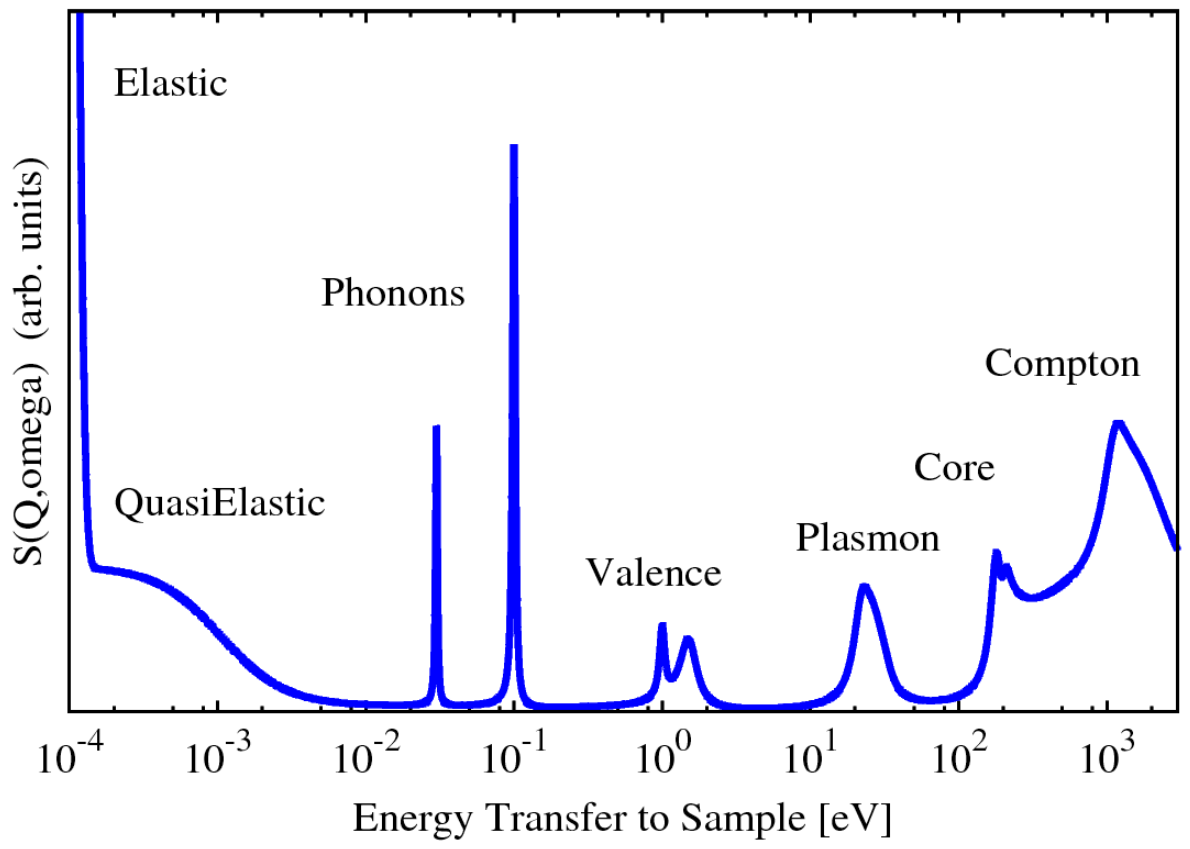

**Figure 1. Conceptual diagram of excitations that may be probed using non-resonant IXS.** Energy scales and intensities are approximate. Magnons (say 0.01-1 eV) are not shown but may be accessed via resonant inelastic x-ray scattering. The energy scale of the core excitations will depend significantly on the material and that for Compton scattering on the x-ray energy and momentum transfer. Valence excitations include both band structure and localized and non-localized excitations. Momentum transfer is not considered, but can have a huge impact on intensity.

The present paper will focus on high-, meV-resolution, non-resonant IXS as applied to the investigation of atomic dynamics, especially phonons in crystalline materials. It will provide an introduction to x-ray scattering appropriate for the consideration of these measurements, a reasonably detailed description of the instrumentation and optical issues relevant to spectrometer performance, and a selection of examples drawn from recent literature. The goal is to provide an introduction that is appropriate for a motivated but uninitiated reader - we will try to clearly define most of the concepts used, and provide a selection of references for more information. Further, while discussing the background for the scattering theory, constant effort is made to make contact with practical issues and examples.

The driving impetus for this paper is the *present opportunity* afforded by meV-resolved inelastic x-ray scattering. The concepts for understanding IXS largely date back to the 50s and 60s when neutron sources and spectrometers began to allow measurement of dynamical properties of materials over atomic-scale correlation lengths. However, with the development of synchrotron sources, x-rays have undergone an increase in source brilliance of more than 10 orders of magnitude in the last 50 years, so, at least in some respects, the x-ray spectrometers have now out-stripped the neutron spectrometers. In particular, with the increased specialization of synchrotron beamlines over the last 2 decades, there are now multiple facilities where one, as a user, can show up with a small piece of crystal - say 1 mm, or 0.1 mm, or

* A comprehensive treatment of many aspects of IXS, but not the meV-resolved work discussed here, can be found in (Schülke, 2007).

even 0.01 mm in diameter - and walk away with extremely clean, meV-resolved measurements of the dynamic structure factor, $S(\mathbf{Q},\omega)$. One can investigate phonon dispersion, interactions, and line shapes, doping dependencies, etc., with relative ease: in IXS, the backgrounds tend to be small, there are no "spurions", there is no complex resolution coupling, and (for better or worse) no contribution from magnetic excitations. Thus there is an opportunity to investigate dynamics in qualitatively new way with x-rays, and perhaps, when combined with modern computer modeling, to begin to reverse the exodus from atomic dynamics of crystals (with its large number of modes) to magnetic dynamics (often simpler, with only one or two magnetic atoms per unit cell), or, even better, to involve new people in investigations of atomic dynamics.

| | Technique | Incident Photon Energy | Energy Transfer | Information Content |
|---|---|---|---|---|
| 1 | Compton | 100 keV | keV | Electron Momentum Density<br>Fermi Surface Shape |
| 2 | Magnetic Compton | 100 keV | keV | Density of Unpaired Spins |
| 3 | RIXS<br>Resonant IXS | 4-15 keV | 0.1-50 eV | Electronic & Magnetic<br>Structure & Excitations |
| 4 | SIXS<br>Soft X-Ray (Resonant) IXS | 0.1 - 2 keV | 0.05 - 5 eV | Electronic & Magnetic<br>Structure & Excitations |
| 5 | NIS (NRVS)<br>Nuclear IXS | 10 - 30 keV | 1-100 meV | Element Specific Phonon<br>Density of States (DOS) |
| 6 | X-Ray Raman | 10 keV | 50 -1000 eV | Edge Structure, Bonding,<br>Valence |
| 7 | NRIXS<br>Non-Resonant IXS | 10 keV | 0.01 - 10 eV | Electronic Structure<br>& Excitations |
| 8 | IXS<br>High-Resolution IXS | 10-25 keV | 1-100 meV | Atomic Dynamics &<br>Phonon Dispersion |

**Table 1: List of techniques broadly grouped under the heading of Inelastic X-Ray Scattering** including common abbreviations and *approximate* photon energies. Techniques 3 and 4 (RIXS and SIXS) use *resonant* scattering so both the incident energy and the analyzers must be tuned near to the relevant electronic resonance/edge. These resonant techniques often have high rates, but can be complex to interpret. Technique 5 (NIS) employs scattering from nuclear (Mossbauer) resonances. Discussion of methods 1-3, 6 and 7 can be found in the book by Schülke (Schülke, 2007), while a review of (3) also relevant to (4) can be found in (Ament, et al., 2011) and a review of NRS methods, (5), can be found in (Gerdau & de Waard, 2000).

Historically, the possibility to apply x-ray scattering to measure atomic dynamics had long been tacitly recognized in the presence of thermal diffuse scattering (TDS) from crystals. However, energy-resolved measurements of this scattering faced considerable challenges due to the required high resolution: Å wavelength x-rays have energies ~10 keV, so resolving the ~meV excitations associated with atomic dynamics requires energy resolution, $\Delta E/E$, at the level of $10^{-7}$. None-the-less with the advent of synchrotron radiation sources, and relying on the existence of highly perfect silicon crystals, such experiments were discussed as early as 1980 (Dorner, et al., 1980). This was followed by additional work (including (Graeff & Materlik, 1982),(Fujii, et al., 1982)) and a published proposal for a spectrometer (Dorner et al.1986), and then construction and demonstration of the first inelastic x-ray scattering (IXS) spectrometer at HASYLAB at DESY in Hamburg, Germany (Dorner, et al., 1987) (Burkel et al.1987) in the 80's. More in-depth discussion of the background and historical development of IXS as a technique through this time may be found in (Burkel, 1991). Then, with the development of 3rd generation sources, and the success of the first IXS beamline at the European Synchrotron Radiation Facility (ESRF) in the mid 90's, there has been sustained effort in the field, and meV-resolved IXS has come into its own as a technique.

This paper will review high-resolution non-resonant IXS measurements from the point of view of scattering theory, instrumentation, and application to current issues in materials science and condensed matter physics. It builds on, and is designed to complement, the existing literature. Other papers that have covered similar topics include (Burkel, 2000) (Sinn, 2001) (Krisch & Sette, 2007) (Baron, 2009) and, of necessity, there will be some overlap. However, this paper will both attempt to emphasize things that have not been discussed, or not been discussed so thoroughly, in the previous papers, and will choose a different selection of scientific examples.

## 1.2 IXS: Concept and Spectral Gallery

It is useful at the outset to introduce the basic concept of an IXS measurement and some examples of spectra. These can be kept in mind during the discussions that follow.

The modern IXS spectrometer is essentially a rather large x-ray diffractometer with an energy-scan option. A schematic and a photograph of (part of) one spectrometer are shown in figure 2. The bandwidth of the incident beam is reduced to the meV-level using a high-resolution monochromator (HRM), and the beam scattered from the sample passes through an analyzer with similar resolution. Such spectrometers operate at momentum transfers between about 0.5 and 100 $nm^{-1}$ and energy transfers, typically, up to ~300 meV, with momentum resolution between ~0.05 and 1 $nm^{-1}$ and energy resolution between about 0.6 and 6 meV, depending on the details of the setup. As will be discussed in section 4, the high resolution required leads to rather large (~10m scale) spectrometers. A measurement usually proceeds by (1) moving the spectrometer two-theta arm (which holds the analyzer - or analyzer array) so the analyzer will intercept the scattered radiation at the momentum transfer of interest[+], and then (2) scanning the energy of the incident beam while holding the analyzer energy (and position) constant. The resulting spectrum, intensity vs. energy transferred to the sample, $\hbar\omega$, is directly proportional to the dynamic structure factor, $S(\mathbf{Q},\omega)$. For crystalline samples, one must choose the orientation of the crystal properly, as IXS is sensitive to a phased sum of atomic motions projected onto the momentum transfer (see discussion in sections 2.8, 2.9 and 2.11) while for disordered materials (liquids, glasses) the spectra depend only on the magnitude of the momentum transfer.

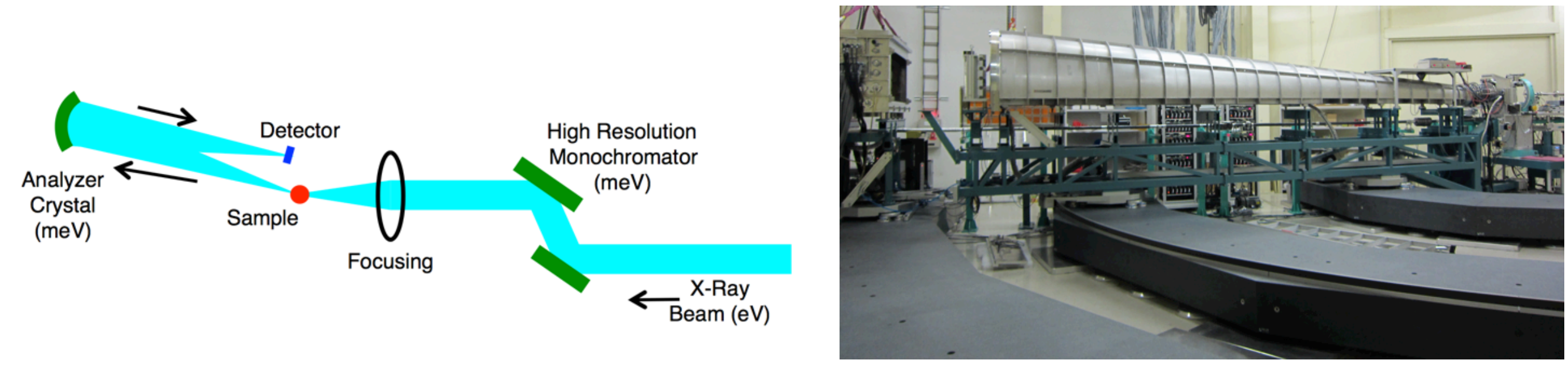

**Figure 2. Schematic of IXS spectrometer layout (left) and photograph of the 10m arm of the spectrometer of BL43LXU of SPring-8 (right).** The schematic indicates main components conceptually, and is not to scale. For the photo, the beam is incident from the right side with the sample also on the far right in the center of the green chi circle and the analyzers crystals sit inside the chamber at the far left.

Figure 3 shows examples from several classes of materials that have been investigated using IXS. Disordered materials such as liquids (figure 3a) and glasses (3b), are usually characterized by spectra with a strong peak near zero energy transfer, and, at least at smaller (say, <10 $nm^{-1}$) momentum transfers, a dispersing longitudinal acoustic (sound) mode. At larger momentum transfers (and sometimes also even at low momentum transfers) the acoustic mode can be broad and effectively merge with the quasi-elastic peak (though one can define the mode energy by the peak of the current-current correlation function, J~$\omega^2 S(\mathbf{Q},\omega)/Q^2$). The clearest difference in the spectra of liquids and glasses is often the width of the peak at zero energy transfer, which, for liquids is usually broadened by diffusion while, for glasses, where the characteristic relaxation is much slower ($\geq$100s, by definition), the peak is resolution limited on the meV scale. In contrast, spectra for crystalline materials (e.g. figure 3c) generally have well-defined peaks corresponding to specific phonon modes at finite energy transfers. No strong peak at zero energy transfer is generally expected, except at Bragg reflections, though often there is some diffuse scattering due to sample imperfections, or superstructure, and, at higher temperatures, multi-phonon contributions. In most cases, low-energy phonon modes have higher intensity than high-energy modes, due to both temperature-dependent occupation factors and the fact that high-frequency modes have smaller displacements. Acoustic modes usually have the largest intensity, followed by a band of mid-energy-range (say 10-40 meV) optic modes with less intensity dueteite to both their higher energy and more complex polarizations, followed by a high-energy band of optical modes (say, 40 to as much as 200 meV) that can be extremely weak due to their high frequency and the fact that their high frequency often is the result of the modes being mostly motion of light

---

[+] The magnitude of the momentum transfer, $|\mathbf{Q}| = (4\pi/\lambda)\sin(2\Theta/2)$ is determined by the scattering angle, $2\Theta$, and the incident x-ray wavelength, $\lambda$, and there is no coupling between energy transfer and momentum transfer so, different than a neutron triple axis spectrometer, scans at fixed Q are done without moving the sample or the two-theta arm.

atoms, which have reduced x-ray scattering. In many cases, the line-width of phonon modes will be resolution limited as intrinsic phonon line widths tend to be << meV (lifetimes >> ps) though in some cases, such as strong phonon-phonon scattering, or strong electron-phonon coupling, line-widths can be measurable. The final spectrum (figure 3d) shows a relaxor, where there is strong phonon-phonon coupling (anharmonicity) leading to broad phonon modes.

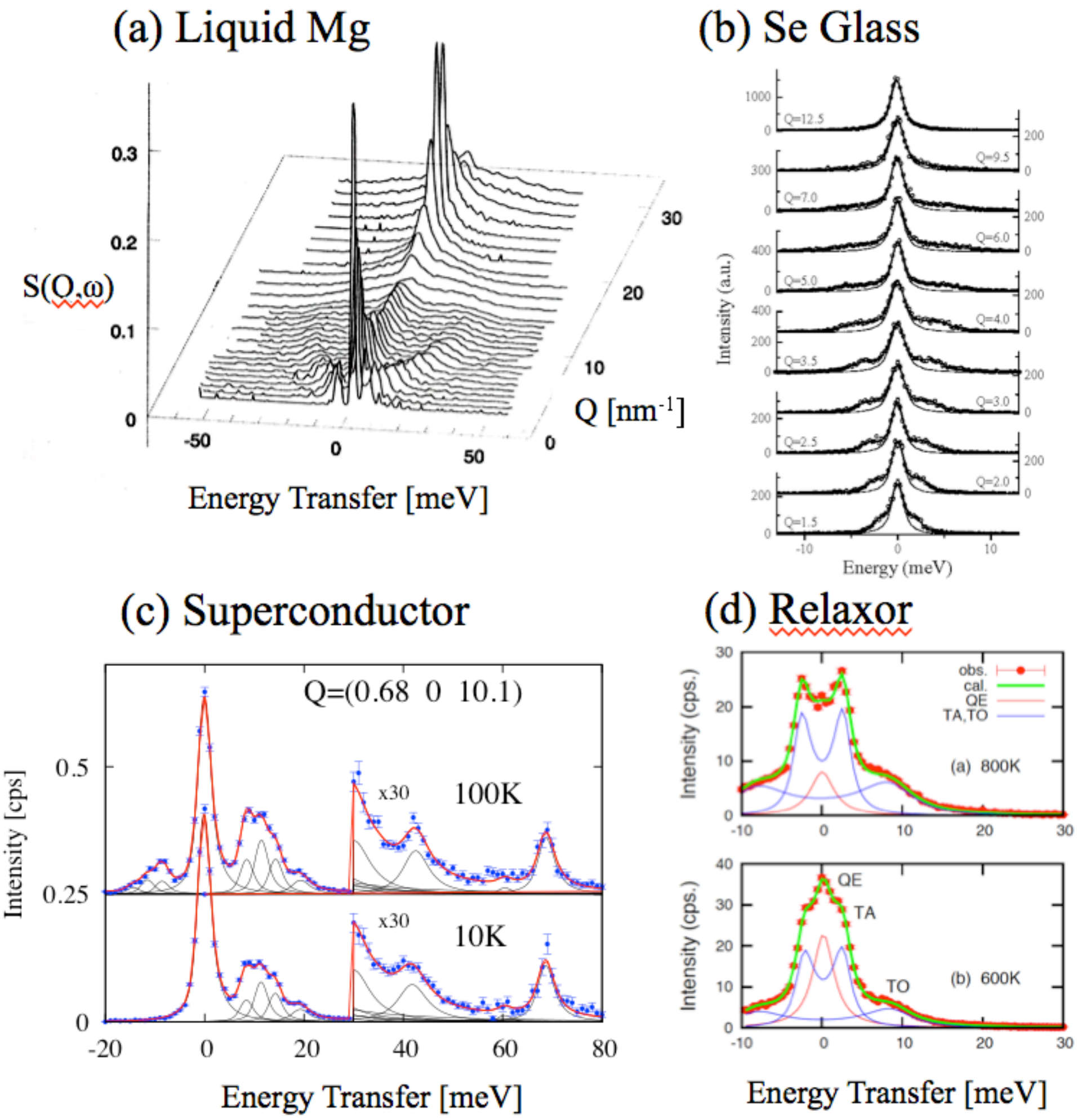

**Figure 3. Gallery of IXS Spectra.** Figures show typical IXS data sets including (a) liquid Mg (based on data in (Kawakita, et al., 2003) - note a resolution deconvolution has been done using a maximum-entropy method, and (b) glassy Se (after (Scopigno, et al., 2004), Copyright (2004) by the American Physical Society) as examples of disordered materials. Panels (c) and (d) show data from crystalline materials: a superconductor (after (Baron, et al., 2008)) and a relaxor (based on data in (Ohwada, et al., 2008)). For the disordered materials, the acoustic mode can be seen dispersing out from small momentum transfers, and eventually broadening and becoming difficult to distinguish from the quasi-elastic peak at zero energy transfer. For the crystalline materials, several distinct peaks corresponding to the phonon modes are easily observed, with, strong broadening being visible for the relaxor TO mode.

## 1.3 Overview of Facilities and Future Options

There are, as of early 2018, 6 relatively stable facilities world-wide doing meV IXS measurements: two beamlines, sector 3 and sector 30, at the Advanced Photon Source (APS) at Argonne National Laboratory near Chicago, Illinois, in the United States, one beamline, ID28, at the European Synchrotron Radiation Facility (ESRF) in Grenoble, France, two beamlines, BL35XU and BL43LXU, at SPring-8 in Harima Science Garden City of Hyogo Prefecture, Japan and one beamline, 10ID, at BNL's NSLS-II on Long Island in New York. While, to a first approximation, all of these facilities perform similar measurements, there are significant differences in the optics and capabilities of each instrument. Table 2 presents some of the parameters of these instruments, and the instruments will be discussed in more detail in section 4 below. One notes also that the ESRF source was upgraded with it reopening in 2020 with a smaller horizontal source size.

At this time, to the best of the author's knowledge, there has been no attempt to carry out conventional meV-resolution IXS at the newest types of x-ray sources, the x-ray free electron lasers (XFELs). XFELs are strong pulsed sources, with the time-averaged flux (as measured in photons/sec/meV) often much lower than that of the present, 3rd, generation of storage rings (excepting the European XFEL in Hamburg, and the upgrade of the LCLS). Given that, and the severe over-subscription of these facilities, and the few beamlines available, a meV spectrometer has not yet been set up. However, a spectrometer at such a source might be extremely interesting for time resolved measurements, to investigate changes in dynamics associated with intermediate states in reactions, or to investigate responses to coherent phonons, or non-equilibrium conditions. Also, at these facilities a time resolved method has been employed (see discussion in section 1.4.6 below) that while not really an equilibrium spectroscopy, can provide related information about phonons.

| | | Res. / Energy meV / keV (GHz) | Focus at Sample μm Method (Throughput) | Source # x Len. x Period | Analyzer Setup | Comments |
|---|---|---|---|---|---|---|
| 1 | **APS Sector 3** | 2.1 / 21.67 (7) | 120x120 Toroid<br>17x20 Toroid+KB (0.8) | 2x 2.4m x 27 mm | Si (18 6 0)<br>4x1 Array<br>4x10 cm$^2$, R=6m | In Line HRM |
| 2 | **APS Sector 30** | 1.5 / 23.72 (5) | 35 x 15 KB | 2x 2.4m x17.2 mm | Si (12 12 12)<br>9x1 Array<br>ϕ10 cm, R=9m | In Line HRM |
| 3 | **ESRF ID 28** | 3 / 17.79<br>1.5 / 21.75 (>3)<br>1.3 / 23.72 (3)<br>1.0 / 25.7 | 270x70 Toroid<br>13x60 ML+Toroid (0.7)<br>13x6 KB+Toroid (0.4) | 3x 1.6m x 32mm<br>or<br>3x 1.6m x 17.6mm | Si (nnn n=9,11,12,13)<br>9x1 Array<br>2x6 cm$^2$, R=7 m | Backscattering HRM |
| 4 | **SPring-8 BL35XU** | 3 / 17.79 (33)<br>1.3 / 21.75 (10)<br>> 0.9 / 25.7 | ϕ80 Bent Cyl.<br>ϕ<20 Bent Cyl.+KB (0.6) | 1x 4.5m x 20 mm | Si (nnn n=8,9,11,13)<br>4x3 Array<br>ϕ10 cm, R=9.8 m | Backscattering HRM |
| 5 | **SPring-8 BL43LXU** | 2.8 / 17.79 (98)<br>1.6 /21.75 (30)<br>1.3 / 21.75 (20)<br>>0.75 / 25.7 | ϕ50 Bent Cylinder<br>5 x 5, KB (0.6) | 3x 5m x 19 mm | Si (nnn n=8,9,11,13)<br>6x4 Array (11x4-2)<br>9x9cm$^2$ R=9.8 m | Backscattering HRM |
| | | 25 /15.82 (2000) | 35 x 25, Bent Cylinder | 3x 5m x 19 mm | Si (888)<br>3x 9x9cm$^2$ R=1.9 m | In Line MRM |
| 6 | **NSLS-II** | 1.3 to 2 / 9.13 (4)<br>(Sharp Tails) | 5 x 7 KB | 3.0m x 22mm | Single Analyzer<br>PSC optics<br>10x10 mrad$^2$ | In-Line / PSC |
| 7 | **UHRIX** (APS) | 0.62 / 9.13 (2)<br>(Sharp Tails) | 18 x 45 KB | 2x 2.4m x 30 mm | Single Analyzer<br>PSC optics<br>10 x 1 mrad$^2$ | In-Line / PSC<br>**Temporary** |

**Table 2:** Facilities for High-Resolution IXS. Fluxes onto the sample (column 2) are approximate, and are sometimes reduced by additional optics (e.g. the throughput of the focusing in col. 3). Light grey type indicates estimates, or otherwise non-measured conditions. The first 5 facilities are established spectrometers, the 6th is commissioning. The last two use a different (post-sample-collimation or PSC) analyzer setup, of which 7 was a temporary configuration and 6 is still improving. Both PSC setups have resolution functions with tails that are much sharper than the Lorentzian response typical of backscattering analyzers. Resolution and beam sizes are FWHM and sizes are generally given as a diameter, ϕ, or as HorizontalxVertical (likewise, analyzer array sizes and acceptances are HxV). The listed analyzer acceptance is the maximum allowed in normal operational conditions - additional slitting to reduce the size for better Q resolution is generally possible. Operating parameters are 7 GeV & 100 mA for APS, 6 GeV and 200 mA for ESRF, 8 GeV & 100 mA for SPring-8 and 3 GeV & 350 mA for NSLS-II. ESRF will upgrade its source in 2018/19. Information in this table is partly based on private communications from A. Said, A. Alatas, A. Bosak & Y. Cai. The large SPring-8 spectrometers can also operate at the Si (888) with 6 meV resolution, but this is not often requested.

The next generation storage ring sources, the hard-x-ray "ultimate" or "diffraction limited" storage rings (USRs, DLSRs), will provide improvement for all samples in IXS in as much as they provide increased flux (photons/s/meV) onto the sample, as flux is usually the limiting parameter in IXS. However, the largest improvement planned for the next generation storage rings is in source brilliance (photons/s/source-size/source-divergence/bandwidth) achieved mostly by reducing the source (electron beam) size in the horizontal. The optics for IXS are not, in most cases, primarily brilliance limited so while improved brilliance will allow a smaller focus at the sample position, it does not imply similar gains in flux on the sample for typical sample sizes. Thus, while it is sure that experiments on (sub-) micron size samples will benefit at DLSRs, more general improvement for samples of size >10 microns, if present at all, will probably not be as large as might be expected based the numbers describing improvements in brilliance. In fact, changing other parameters, can even abrogate those improvements, with for example, the planned reduction of operating energy and insertion device length of SPring-8 when it is "upgraded" requiring also a significant beam current increase to keep a similar flux (photons/s/meV) for IXS.

The greatest potential over-all improvement for IXS (up to radiation damage issues) may be a cavity oscillator based source such as an x-ray free-electron laser oscillator, X-FELO (Kim et al.2008) where the available flux (photons/s/meV) might increase as much as 4 orders of magnitude beyond present levels, though such a source remains to be demonstrated.

## 1.4 Comparison of IXS with Other Methods

This section compares IXS with other methods of measuring atomic dynamics, and, especially, to inelastic neutron scattering (INS) which has long been the method of choice for measuring atomic dynamics over atomic-scale correlation lengths. These comparisons are useful to highlight the advantages and limits of the IXS technique, and may be valuable for people who have experience with, or want to consider, alternative measurement techniques. More discussion about alternative techniques can also be found in texts (e.g. (Brüesch, 1982), (Srivastava, 1990)) and the references given below.

### 1.4.1 Inelastic Neutron Scattering (INS)

Inelastic neutron scattering (INS) has long been the method of choice for measuring atomic dynamics over atomic-scale correlation lengths. The differences between INS and IXS can be traced to the fundamental differences in the probes (massless x-rays vs. massive neutrons) and their interaction cross-section with mater, and to differences in the source properties. Table 3 lists some of the aspects of IXS in the context of INS, and the underlying source of the difference.

| Characteristic of IXS (Relative to INS) | Reason | Comment |
|---|---|---|
| Access to small (<micro-gram) samples | Source | X-Ray Focus ~0.005 mm |
| Decoupling of Energy and Momentum Transfer | Massless Probe | X-Rays > 9000 eV & Neutrons < 0.1 eV |
| Energy Resolution Independent of Energy Transfer | Massless Probe | |
| Small Intrinsic Background | Cross Section | No Incoherent, No Multiple |
| Simple & Good Momentum resolution | Massless Probe & Source | Effective Slitting |
| Limited Energy Resolution | (Massless Probe & Cross Section) | X-rays >~ meV |
| High Radiation Damage | Cross Section | Large photo-electric absorption |

**Table 3:** Comparative points for IXS and INS

The most important characteristic of IXS as compared to INS is probably the ability of IXS to probe small samples. This is directly the result of the high flux and high brilliance of modern x-ray sources, where the small source size and small divergence (~ 0.01 x 0.5 $mm^2$ and ~10x30 $\mu rad^2$, vertical x horizontal, for a 3rd generation storage ring) for the x-ray beam makes it easy to focus the x-rays to a transverse size of ~0.1 mm or even, with some losses, to ~0.005 mm. This allows comparable size samples to be investigated, including samples in extreme (e.g. near earth's core) conditions in diamond anvil cells (DACs), and small crystals of new materials. The latter is especially important as one can investigate phonons without having to invest huge effort to grow large samples, making it relatively easy to do IXS on a newly discovered material. Up to the complexity of interpreting phonon spectra (which, of course, will not stop serious scientists), IXS is an easy spectroscopy, available nearly immediately after a reasonable quality (~0.1 degree mosaic) single crystal of new material is fabricated (and sometimes even a powder can be sufficient, for density of states measurements).

The massless nature of the probe means that the energy of Å-wavelength x-rays (~12400 eV) is much larger than the ~0.08 eV of Å-wavelength neutrons (where the Å scale is set by the interest in investigating correlations of motions over atomic-scale distances). While this creates a severe resolution requirement to achieve meV resolution with x-rays, it also leads to a complete decoupling of momentum and energy transfer: changing the energy of a >10,000 eV photon by the, typically, <0.1 eV energy of a phonon, makes a negligible change in its momentum on the scale of the Brillouin zone. This has important consequences for measurements of disordered materials where one would often like access to large (say 20-50 meV) energy transfer at small (~nm$^{-1}$) momentum transfers, which can be difficult with neutrons. This lack of kinematic constraints in IXS is responsible for an explosion in the investigation of the atomic dynamics of disordered materials over the last 20 years or so during which IXS has been available. (For crystalline materials, where one usually works in higher Brillouin zones, kinematic constraints are less of an issue). Other consequences of the massless probe/high x-ray energy relate to the way Bragg-reflection based optics work. Generally, Bragg reflections have a fixed fractional bandwidth so that the large absolute x-ray energy means there is little (practically, no) change in the absolute bandwidth over phonon-scale energy transfers. Meanwhile, the limits on the resolution stem from inability to lower the energy to get better resolution, combined with the generally high absorption cross-section for x-rays making improved resolution at high energies difficult without large losses (see section 4).

The difference in the x-ray and neutron cross-sections leads to several notable features. The largest cross-section for 10-20 keV energy x-rays interacting with matter is that for photo-electric absorption - the Thomson scattering cross section (which generates the signal of interest) is typically an order of magnitude, or more, smaller (see also figure 16). This effectively prevents multiple scattering backgrounds from appearing in most x-ray measurements (it also means X-rays often see only the top 0.01 or 0.1 mm of a sample, even if the total sample thickness is larger). In contrast, for neutrons the scattering cross-section is usually the dominant one, so as samples are made thicker to improve INS count-rates, multiple scattering becomes an issue. Further, the incoherent cross section for x-rays, Compton scattering, leads to large energy transfers (>eV) due to the small electron mass, whereas the neutron incoherent scattering is usually on exactly the meV scale of interest for atomic dynamics (see eqn 9). IXS is then essentially background free, excepting the multi-phonon contribution (see sections 2.8 and 2.11.4). However, the large absorption cross section also means that, with x-rays, radiation damage can be a serious issue, even with the *relatively* (as compared to other synchrotron work) weak meV-bandwidth beams used for IXS: protein crystals and polymers show visible damage on the hour time scales of typical scans, and the author has observed changes in elastic intensities in IXS spectra on these time scales. (Note, the trick of freezing a sample as used in structural studies is of arguable efficacy for IXS, as, in principle, the dynamics can change as soon as particles are ionized). As the relationship of the isotopic species and neutron scattering cross-section is complex, as opposed to the ~$Z^2$ scaling for x-ray scattering, there also can be advantages for each (x-rays or neutrons) depending on the atoms/isotopes in the sample. However, the x-ray analogue of isotope replacement in INS, which is tuning to atomic resonances to change cross-sections to help identify features of the scattering related to particular atomic species, is not yet possible with meV resolution (see the section on RIXS, below).

Neutrons are advantageous when very high-energy resolution is needed, as backscattering spectrometers provide ~ 10 μeV level resolution, at least for smaller energy transfers, and spin echo spectrometers can reach neV levels, sometimes even ~10 μeV for larger energy transfers (Aynajian et al.2008). In addition, whereas the energy resolution for *most* x-ray spectrometers is approximately Lorentzian and so has long tails, the resolution with neutrons often has shorter tails. This can make neutrons advantageous for observing weak modes near to stronger ones, or for measuring modes in the presence of strong elastic backgrounds. However, a new spectrometer design (see sections 4.7, 4.8) may also improve the situation for x-rays. Neutrons are extremely interesting when large single crystals of heavier materials are available, whereas x-rays are limited by the short penetration length into the sample due to the high photoelectric absorption. Also, modern time-of-flight neutron spectrometers (Kajimoto et al.2011)(Abernathy et al.2012) allow collection of a huge swath of momentum space at one time, and assuming modest (~0.1 cc) size samples are available, may offer advantages for survey measurements of all phonons in a material.

### 1.4.2 Inelastic Ultra-Violet Spectroscopy (IUVS)

It is interesting to extend IXS work to small, but still finite, momentum transfers: to go below the ~0.5 nm$^{-1}$ that is a practical limit in many IXS spectrometers today, down to the level 0.1 nm$^{-1}$, or lower. This is can be particularly interesting for disordered materials, where the transition from a continuum hydrodynamic limit to atomistic behavior can, depending on system, be expected to occur gradually as one increases momentum transfer from inverse-microns to inverse-nm. Extending the IXS measurements toward low Q has been possible using a synchrotron based facility (see (Masciovecchio, et al., 2006) and references there-in) with either an insertion device or laser based source: measurements at momentum transfers slightly below 0.1 nm$^{-1}$ have been achieved.

### 1.4.3 meV-Resolution *Resonant* IXS

In principle, resonant x-ray scattering, where the incident beam energy is tuned near to an atomic transition or edge may be done with meV resolution to investigate atomic dynamics, both in the hard x-ray regime (RIXS) with crystal optics and the soft x-ray regime (SIXS) where gratings are employed. This then probes the response in the presence of, usually, a core hole, which in turn can affect the dynamics, providing potentially different information than the non-resonant IXS that is the main topic of this review. At present, the resolution in hard x-ray RIXS is on the verge of reaching levels that are generally interesting for atomic dynamics, and will probably get to the few meV level soon, while soft x-ray approaches, even with rather large, 15m, two-theta arms, appear to be limited to more like 10-30 meV resolution, as might be interesting for atomic dynamics for some specific cases. We discuss some of the practical issues for meV-scale hard x-ray RIXS in section 4.3. Generally, one would expect meV-RIXS to make several things possible, including, most directly, anomalous scattering (the analogue of isotope replacement in INS) which may be useful to help identify the atomic species contributing to a given spectral feature, and to take advantage of resonant enhancements to increase count-rates, and provide more precise information on phonon eigenvectors. However, some caution is also needed, as, for the non-resonant case one is probing the equilibrium properties and the theory is well evolved, while for the resonant case there may be significant differences. This is evident, for example, in the understanding of the effects of electron phonon coupling (epc) on the spectra: for non-resonant scattering, epc generally leads to phonon energy shifts and line-width increases that are, in principle, directly related to the interesting quantity, $\lambda$, for estimating, e.g., the onset of superconductivity. For resonant scattering one can isolate a quantity related to electron-phonon coupling (Ament & Brink, 2011), but the relation to conventional properties is not as straightforward as for the non-resonant case (see also (Beye & Föhlisch, 2011)). There is already some indication of this in the presence of Frank-Condon sidebands [see e.g. (Henderson & Imbusch, 1989)] in some resonant studies [(Gretarsson, et al., 2013; Lee, et al., 2013)]. It can be possible to relate the observed phonon intensity to the electron-phonon coupling (Devereaux, et al., 2016)(Braicovich, et al., 2020).

### 1.4.4 Nuclear Resonant Scattering

Nuclear resonant scattering (NRS) from low-lying narrow (typically neV to μeV bandwidth) nuclear resonances can be used to investigate atomic dynamics on the meV scale, [(Seto, et al., 1995)(Sturhahn, et al., 1995)(Chumakov, et al., 1996)], using specific isotopes of materials such as Fe, Eu, Sn, Sm, Dy, Kr, and Tm to name a few of the more accessible resonances (for a review of the larger nuclear resonant scattering field see papers in (Gerdau & de Waard, 2000)). The most robust of the NRS techniques, sometimes called Nuclear Inelastic Scattering (NIS), uses a ~meV resolved beam incident on a sample containing the resonant material (Seto, et al., 1995). Scanning the energy of this beam about the nuclear transition energy near 14.4 keV, and selecting only the nuclear scattered events, yields the distribution of excitations in the sample that can make up the difference between the incident beam energy and the nuclear resonance energy. This is essentially an absorption measurement (incoherent scattering with momentum transfer equal to the incident photon momentum), and yields the partial density of states of resonant nuclear motion. While the method is limited to probing the motions of modes including motion of the resonant atom, the data is often of high quality compared to the DOS provided by INS or IXS. The nuclear/element-selectivity can also be useful to focus on motions of particular atomic species, and the large nuclear cross-section can allow access to relatively small samples (e.g. thin, even single atomic, layers). However, NIS does not allow measurement of phonon dispersion and selection of specific phonon modes is not possible using NIS, unless they are isolated from all other modes in energy. An alternative to the NIS method, which is not limited to resonant atoms in the sample, is to use a nuclear analyzer (NA) (Chumakov, et al., 1996). This is similar in concept to the IXS technique discussed in the bulk of this paper, except that nuclear scattering is used to analyze the scattered radiation. This method allows simultaneous measurement of a large area of momentum space (large solid angle of acceptance) as determined primarily by the detector solid angle. However, the effective bandwidth of the nuclear analyzer is ~1 microvolt, or less. Thus, the NA method is severely flux limited, and probably not practical in cases where good momentum resolution is required (Baron, 2013).

### 1.4.5 Thermal Diffuse Scattering (TDS)

Thermal diffuse scattering (TDS) is another probe of atomic dynamics: momentum resolved but energy-integrated measurements generally show structure due to the momentum-dependence of phonon polarization vectors and energies. Information can be derived from TDS (see, e.g. (Holt, et al., 1999) (Xu & Chiang, 2005)(Bosak, et al., 2009),(Wehinger, et al., 2017) and references therein) by fitting results from simple samples with models, and/or through temperature dependent measurements, or careful analysis. However, due to the energy-integrated nature of these measurements, TDS is of limited value as a direct probe of phonons in more complex materials. TDS is probably most useful as a complementary technique to IXS - it can be used to quickly investigate large volumes of momentum space to, for example, pinpoint the momentum transfers where intensity changes are present (e.g. across phase transitions) warranting further investigation by IXS. One might also consider combining TDS with resonant scattering, taking advantage of the tuneability of most modern x-ray sources, to help differentiate between the motions of different atoms, but, to this author's knowledge, this has not yet been done.

### 1.4.6 Fourier-Transform IXS (Pump-Probe TDS)

Recently it has been demonstrated (Trigo, et al., 2013) that it is possible to investigate TDS on a ~0.1 ps time scale in a pump-probe experiment - a laser pump sets the zero time and an x-ray free-electron laser (XFEL) pulse acts as the probe. One then can measure, stroboscopically, the oscillations in the TDS intensity as a function of time after the laser pulse, and relate them to phonon frequencies, at least for lower frequency modes. Further, given the availability of area detectors at XFELs, signal from a wide range of momentum transfers, several Brillouin zones, can be collected simultaneously with good Q resolution and effectively sub-meV energy resolution (Zhu, et al., 2015) (but not along exact symmetry directions - see the discussion in section 4.6 about using an analyzer array at a fixed sample orientation). This allows collection of data that can be used to determine the dispersion of low-frequency phonons. However, the pump pulse disturbs the thermal equilibrium of the sample, so the samples' thermodynamic state is complicated, similar to that of a coherent phonon (Fahy, et al., 2016). Of particular note is the recent result where the time domain nature of the technique allows direct visualization of phonon anharmonic decay: as the x-ray probe is delayed relative to the pump, the intensity of one mode is seen to decay while that of another increases (Teitelbaum, et al., 2018). That type of time domain result has no analogue in the frequency domain spectroscopy discussed in this paper. However, one should emphasize that the pump in this Fourier-Transform method means it is intrinsically a probe of *non-equilibrium* conditions so that, while potentially providing specific and unique information, results should be interpreted with care.

### 1.4.7 Atom and Electron Scattering

Surface and near-surface vibrational modes can be probed using atom (typically, helium) and electron scattering which both can measure meV-scale energy transfers at atomic scale momentum transfers. Recent work includes (Kostov, et al., 2011)(Tamtögl, et al., 2013)(Steurer, et al., 2008) and see (Benedek, et al., 2010) and references therein. In principle, a grazing incidence geometry with x-rays can also probe surface dynamics of solids (Murphy, et al., 2005) (and even liquids, (Reichert, et al., 2007)) but those are hard experiments: for solids amenable to UVH environments, atom or electron scattering may be easier than IXS for probing surface vibrations. Also, recently, ultrafast time resolved electron diffuse scattering (UEDS) has become possible (René de Cotret, et al., 2019) that provides data that is very similar to the Fourier Transform IXS discussed in the immediately preceding section.

### 1.4.8 X-ray Photon Correlation Spectroscopy

X-ray photon correlation spectroscopy (XPCS, also sometimes called intensity fluctuation spectroscopy, XIFS) is an x-ray scattering method used to probe atomic dynamics directly in the time domain by measuring the time-evolution of the fine-structure, the speckles, in the x-ray scattering from a sample. The speckle pattern responds to the detailed atomic-scale charge-distribution of the sample, and changes in the atomic positions then lead to changes in the speckle pattern, assuming appropriate conditions are met for the setup. This method is available at storage ring sources to investigate dynamics down to ~ms time scale, and probably to the µs scale after the next round of machine upgrades. There is now effort to do similar measurements using pulsed x-ray free electron lasers with split-and-delay (S&D) setups. Instrumental resolution at the level of ~10 ps has been demonstrated (see (Roseker, et al., 2012) (Robert, et al., 2013) (Osaka, et al., 2014) and references therein) which should be straightforward, technically, to extend to the meV (<1 ps) IXS scale in principle ($\hbar$=0.658 meV-ps). While presently under heavy development, so the situation is subject to rapid change, we note two caveats. First, the meV scale of interest here is comparable to the usual sample temperatures so the Bose factor makes the scattering strongly non-classical (not symmetric about zero energy transfer). This is a somewhat new situation in XPCS where energies are usually much less than $k_BT$ and the sample response is considered classically, though it may covered by some treatments (Sutton, 2002). Further, and perhaps more seriously, the experiments with the S&D crystal optics remain flux limited and it may take a large number of shots to get sufficient quality data (see, e.g., the related discussion in appendix B of (Martin, 2017)). An alternative method using a broader incident bandwidth and varying the length of a single x-ray pulse (without S&D optics) may offer practical advantages for a limited (short, ~0.1 ps) time range in some cases (Perakis, et al., 2018).

### 1.4.9 Raman and IR Spectroscopy

Raman and IR spectroscopy are two well-known methods of probing atomic dynamics. Both, using near-visible light, are effectively at nearly zero momentum transfer, and probe only the gamma point modes ($Q_{max}=4\pi/\lambda$ = 0.0012 Å$^{-1}$ for λ=1 micron). In Raman scattering, like IXS, the phonons introduce an energy transfer to the incident light. The intensity is then measured as a function of that energy transfer, with filtering to remove the elastic component. This process relies on interactions with (tails of the) electronic resonances in the system, so that interpretation of relative intensities can be more complex than for IXS. In IR measurements, the photon energy matches the phonon energy, and one does what is effectively an absorption (or reflection) measurement. Both of these methods, at Q~0, obey strong selection rules relating to the symmetry of the modes/material, the incident beam direction, and the polarization of the incident and scattered radiation. In fact, phonon modes are usually characterized as Raman-active, IR-active, or silent, based on their symmetry. These strong selection rules can lead to large affects being visible in spectra when passing through a phase transformation that changes the symmetry of a crystalline sample, with modes appearing or disappearing. IXS, in contrast, has selection rules (see section 2.8) that are smoothly dependent on momentum transfer (and therefore depend also on momentum resolution and crystal quality) so that, while phase transformations certainly affect the phonons in IXS, the effects can be more subtle than those sometimes seen in Raman scattering. The clarity of results in Raman scattering is also aided by the relatively good energy resolution, ~0.1 meV in many cases. Another difference is that for Raman and IR spectroscopies, there is the potential for interference between the phonon scattering and that directly from electronic resonances, thus, for example, Raman scattering can show Fano-type asymmetries in the spectral line-shape. Such interference is generally too weak to be seen with x-rays (the direct electronic scattering cross section is too small).

### 1.4.10 Other Methods for Acoustic Modes

IXS is often used as a probe of acoustic mode dispersion, as extrapolation to low Q allows one to estimate the speed of sound in materials, and eventually, elastic constants. There is particular interest in this from a geological standpoint, where precise seismic information about the sound velocity in the interior of the earth exists, but the interpretation of these velocities requires making controlled laboratory measurements to, in essence, create a data-base matching sound-speed with material composition, structure, pressure and temperature. In a different context, the frequencies, and especially, damping, of long wavelength acoustic vibrations in crystalline materials can provide information that is almost calorimetric in nature. Alternative methods to measure sound speeds include Brillouin Light Scattering (BLS), Resonant Ultrasound (RUS - see (Migliori & Sarrao, 1997)) and shockwave measurements (Brown & McQueen, 1986)(Jeanloz, et al., 2007). BLS is similar in principle to Raman scattering but is done in transmission, so is limited to transparent samples. RUS involves shaking a small sample at ultrasonic frequencies, preferably one having a very well defined (e.g. parallelepiped) shape. In general, both of these methods, when possible, yield rather high-quality data on sound velocities and damping. However, especially as regards RUS work, one notes that these methods can be sensitive to things that IXS is completely blind to: for example RUS experiments on $SrTiO_3$ show a remarkable near-global loss of phonon signal at the structural phase transition (at ~105K) due to the formation of domain structure (see (Scott, et al., 2011)) and similarly for $EuTiO_3$ (Spalek, et al., 2014). Thus, it can be very interesting, with due care, to compare the results of these methods with IXS. Finally, we mention that, for high pressure and temperature work, especially as related to geology, macroscopic measurements of shockwave propagation may be used to determine the sound velocity at a specific set of P, T and density values. This requires macroscopic samples.

# 2. Scattering Theory, Phonons.

## 2.1 Introduction

It is valuable to consider some theoretical background: this allows us to define many of the concepts relevant to IXS, and also helps to indicate at what point in the development various approximations are made. Here we start rather generally, but then emphasize the case of harmonic phonons, as that model provides the basis for discussing dynamics in crystalline materials. Other issues that are explicitly addressed include multi-phonon effects, non-harmonicity*, detailed balance, and sum rules. We take some small additional space to try to explain and motivate the progression of ideas and to try to emphasize practical issues related to IXS measurements.

The conditions for inelastic x-ray scattering, with the distance the beam propagates being much larger than the sample size and the x-ray wavelength, are appropriate to apply scattering theory to relate the measured scattered intensity to the microscopic properties of the sample. In effect, scattering theory allows one to go from the solution of the condensed matter physics problem where one solves for the atomic motions in some material, to the results of a measurement. There are (at least) two different ways that this can be approached, depending on what is most relevant to an experiment: one can either focus on constructing the scattered wave from the detailed sample charge distribution via the scattering amplitude, or one can focus on calculating probability of scattering the x-ray from one (usually, plane-wave) state to another with an accompanying change in the state of the sample via Fermi's golden rule. The first is essentially modified diffraction theory [see, e.g., (Jackson, 1999)], while the second is more classical scattering theory largely developed to describe neutron investigations [see, e.g., (Squires, 1978)]. The scattering amplitude approach lends itself immediately to considering the shape and the detailed distribution of atomic positions and electronic charge within the sample, and is especially useful in the context of coherent x-ray scattering (correlation spectroscopy, coherent diffraction imaging (CDI), pytography) as has now become very much in vogue given the high brilliance of x-ray sources. The golden-rule approach lends itself more directly to thinking about excitations within the sample and a quantum-mechanical treatment there-of. In principle, the two might be combined at the level of the first line of eqn (23) below, however, it is probable that when such a case might be interesting, say, for small, nano-scale, samples, the phonon model of the sample motions should be revisited to account for the exact shape and size of the sample. We also note that most diffraction or scattering amplitude based approaches assume fixed or very slowly moving scatterers, implicitly assuming the scattered intensity is symmetric about zero energy transfer. This violates the detailed balance symmetry discussed below and should probably be revisited when, as is the case for most of this review, energy scales are comparable to $k_BT$ (see (Sutton, 2002) for a possible form for generalized expressions).

In this paper, we emphasize the Fermi golden rule approach leading to formal scattering theory. The scattering cross-section is then the quantity of interest, being proportional to the experimental intensities. We will introduce the cross-section, the dynamic structure factor, $S(\mathbf{Q},\omega)$, and some useful relationships, with an emphasis on the development relating to (harmonic) phonons in crystals. We also continue the phonon expansion to second order, as is important for IXS. References developing scattering theory in detail include (Squires, 1978) and (Lovesey, 1984) ((Lovesey, 1984) is broader in scope than (Squires, 1978) but the restriction of many formulas to monatomic Bravais lattices can limit its utility), while (Sinha, 2001) relates various forms of the cross section used in different viewpoints on x-ray scattering. The original papers, especially (van Hove, 1954) and, for sum rules, (Placzek, 1952) are also very much worth reading. The present discussion follows most closely the work of (Squires, 1978)*.

We emphasize here that the present discussion takes place entirely within the Born approximation, where one ignores multiple scattering (similar to the kinematic limit of diffraction theory) since a later section of this review will discuss dynamical diffraction theory. The Born approximation is, in general, an excellent approximation for x-ray scattering, especially given the high absorption cross section in most cases prevents multiple scattering, as discussed in section 1.4.1. The only exception discussed so far in terms of IXS is the possibility of exciting a Bragg reflection in a crystalline sample while measuring at a momentum transfer far from the Bragg beam. Then the measured scattering is contaminated by having, effectively, a new incident direction from the Bragg beam, and the measured spectra can then reflect two momentum transfers (Bosak & Krisch, 2007)(Alatas, et al., 2008). The more interesting case of using dynamical scattering to control the spatial dependence of a wave field and generate phase information on an atomic scale, reminiscent of standing wave measurements, might be used to learn about phonon eigenvectors (Kohl, 1985) but, to this authors' knowledge, has not yet been tried with meV-scale energy resolution (though it has been done using TDS, without energy resolution (Spalt, et al., 1988)).

---

* The present review will use the term "non-harmonic" in an effort to distinguish between a general violation of the simplest harmonic phonon model, and "anharmonicity" that is often used to refer specifically to violations of harmonicity arising from phonon-phonon scattering. "Non-harmonic" will refer to any behavior that is not harmonic of which anharmonicity is only one example. e.g. electron-phonon coupling leads to non-harmonicity but is not anharmonic.

* For the purposes of considering atomic dynamics, formulas for neutron scattering can generally be converted to x-ray scattering by replacing the coherent neutron scattering length, b, by $r_e f(Q)$, where $r_e$ is the classical radius of the electron, $r_e=e^2/mc^2 \sim 2.81$ fm, and f(Q) is the x-ray form factor for the relevant atom or ion at momentum transfer $Q = |\mathbf{Q}|$.

## 2.2 The Dynamic Structure Factor, $S(\mathbf{Q},\omega)$

The main quantity used to discuss IXS results is the dynamic structure factor, $S(\mathbf{Q},\omega)$. It is directly related to the cross section by factoring out a scale factor related to the probe-sample interaction (the Thomson scattering of an x-ray by an electron). This allows one to focus more directly on the dynamics, and, to some extent, to compare results with different probes. For scattering of an x-ray from an initial state given by photon momentum, $\mathbf{k}_1$, and polarization, $\varepsilon_1$, to a final state given by $\mathbf{k}_2$, $\varepsilon_2$ the cross-section for scattering the photon into some solid angle $d\Omega$ about a momentum transfer of $\mathbf{Q} = \mathbf{k}_2 - \mathbf{k}_1$ and into some energy bandwidth, dE, is written

$$\left(\frac{d^2\sigma}{d\Omega dE}\right)_{\mathbf{k}_1\varepsilon_1 -> \mathbf{k}_2\varepsilon_2} = \frac{k_2}{k_1} r_e^2 \left|\varepsilon_2^* \bullet \varepsilon_1\right|^2 S(\mathbf{Q},\omega) \quad \mathbf{(1)}$$

Assuming a non-resonant scattering process allows us to concentrate on the energy transfer $E \equiv \hbar\omega \equiv \hbar(\omega_1 - \omega_2)$ (note, as is common in IXS work, we take positive energy transfer to correspond to the higher intensity side of the spectra corresponding to giving energy to the sample or Stokes scattering) and the scattering geometry, approximated as plane-wave in to plane-wave out, promotes emphasis on the momentum transfer. For photons, one has the usual relations between energy, wavelength, and momentum, $E_i \equiv \hbar\omega_i \equiv \hbar c|\mathbf{k}_i| = hc/\lambda_i$. The over-all scale is the Thomson cross-section for a single electron, $r_e^2\left|\varepsilon_1^* \bullet \varepsilon_2\right|^2$, and $S(\mathbf{Q},\omega)$ has units electrons squared per unit energy transfer. For the *non-resonant* scattering discussed in this paper, the *photon* polarizations $\varepsilon_1, \varepsilon_2$ enter only in the scale factor - there is no interesting dependence on x-ray photon polarizations. Further, for high-resolution IXS, the wavelength change is small compared to the wavelength, $\Delta\lambda/\lambda < 10^{-4}$, so the magnitude of the momentum transfer depends only on the scattering angle $2\Theta$, $|\mathbf{Q}| = (4\pi/\lambda)\sin(2\Theta/2)$, and the phase space factor can be taken as unity, $k_2/k_1 \to 1$.

The connection to the golden rule can be seen from writing

$$S(\mathbf{Q},\omega) = \sum_{\substack{States \\ \lambda\lambda'}} \sum_{\substack{Electrons \\ a,b}} p_\lambda \langle\lambda|e^{-i\mathbf{Q}\bullet\mathbf{r}_a}|\lambda'\rangle\langle\lambda'|e^{i\mathbf{Q}\bullet\mathbf{r}_b}|\lambda\rangle \; \delta(E_\lambda - E_{\lambda'} - \hbar\omega) \quad \mathbf{(2)}$$

where the exponentials come from the plane wave expansion of the vector potential in the $A^2$ term of the interaction Hamiltonian. In the usual way, one sums over final states and averages over initial states whose probabilities are given by $p_\lambda$ with the delta-function insuring energy conservation. In thermal equilibrium one has $p_\lambda = e^{-\beta E_\lambda}/Z$ at temperature $T = [k_B\beta]^{-1}$ where Z is the partition function and $k_B$ Boltzmann's constant ($k_B$ =0.0862 meV/K). Sometimes scattering is also described in terms of linear response with the dynamic susceptibility given by (see (Sinha, 2001), (Lovesey, 1984))

$$S(\mathbf{Q},\omega) = \frac{1}{\pi}\frac{1}{1-e^{-\beta\omega}}\mathrm{Im}\left[\chi(\mathbf{Q},\omega)\right] \quad \mathbf{(3)}$$

This is a version of the fluctuation-dissipation theorem, $S(\mathbf{Q},\omega)$ playing the role of the dissipation or scattering and $\chi(\mathbf{Q},\omega)$ the fluctuation or linear response.

## 2.3 The Static Structure Factor, $S(\mathbf{Q})$

Structural measurements, where there is no energy analysis of the outgoing beam, effectively integrate over energy transfer. The relation of the dynamic structure factor, $S(\mathbf{Q},\omega)$, to the (dimensionless) structure factor, $S(\mathbf{Q})$, is

$$S(\mathbf{Q}) = \hbar \int_{-\infty}^{+\infty} S(\mathbf{Q},\omega)\, d\omega = 1 + \rho \int d^3r \, [g(\mathbf{r}) - 1] \, e^{-i\mathbf{Q}\bullet\mathbf{r}} \tag{4}$$

where the second equality introduces the pair distribution function, g(**r**), sometimes used in structural studies (e.g. "pdf" analysis of powder diffraction) with ρ the average particle number density. One should, perhaps, emphasize that while $S(\mathbf{Q})$ is often called the "static" structure factor to differentiate it from the dynamic structure factor, $S(\mathbf{Q},\omega)$, the integral over all energy transfer in eqn. (4) means that $S(\mathbf{Q})$ really responds to the average (temporal, thermal) positions of the atoms: the intensity of the elastic scattering, $S(\mathbf{Q},\omega=0)$ is not, generally, related to structure in a simple way. However, measurements of $S(\mathbf{Q},\omega=0)$ *combined* with those of $S(\mathbf{Q})$ (say, with and without an analyzer crystal) can be useful to determine if a correlation, a peak in **Q** space, is due to static or dynamic order (e.g. (Shirane, et al., 1987)). Properly normalized, the static structure factor goes to unity at large Q and is proportional to the isothermal compressibility at small Q.

## 2.4 The X-Ray Form Factor and Charge Density Operator

At this point, one can introduce an explicit model, e.g. phonons, to describe the dynamical states of the system $\lambda$ , $\lambda'$ and calculate the relevant cross-sections. However, it is useful, momentarily, to keep more generality and introduce the momentum-space charge-density operator given as

$$\rho(\mathbf{Q},t) = \sum_{\substack{i \\ Electrons}} e^{-i\mathbf{Q}\bullet\mathbf{r}_i(t)} = \sum_{\substack{j \\ Atoms}} f_j(Q)\, e^{-i\mathbf{Q}\bullet\mathbf{R}_j(t)} \tag{5}$$

where $\mathbf{r}_i(t)$ is the position of the i$^{th}$ electron and $\mathbf{R}_j(t)$ is the nuclear position of the j$^{th}$ atom. The second equality introduces the atomic form factor, which is just the Fourier transform of the atomic charge density*. Tabulations of the form factor for free atoms and ions (as a function of Q) are available in the literature (Waasmaier & Kirfel, 1994) and some databases (e.g. DABAX in XOP (Sánchez del Río & Dejus, 2011)).

The introduction of the form factor, as commonly done in x-ray scattering, has additional implication in the present case: the form of f(Q), where all time dependence and electron and nuclear coordinates have been dropped, effectively includes the assumption that the atomic electrons move instantaneously with the nucleus, and that the shape of the electron cloud does not change when the atom moves. This bears some relation to the adiabatic or Born-Oppenheimer approximation (see section 2.9), but, in principle, is much more restrictive, and, certainly, when thinking about the solution to the condensed matter problem of solving for the dynamical excitations of the scatterer, is just wrong - the valence electrons certainly do change their spatial distribution as atoms move in a solid. However, as most of the scattering at the momentum transfers of interest is from inner, core, electrons, and as these do mostly move with the nucleus and, to a good approximation, do not change shape, using the from factor in this way is reasonable *for the purpose of calculating the scattered intensity*. It is also the basis for suggesting that IXS probes essentially the same nuclear motion as is probed by INS.

## 2.5 Coherent and Incoherent Scattering

Using the charge density operator, we introduce the idea of coherent and incoherent scattering, as used in the context of inelastic scattering measurements. This is a division that has great utility in neutron scattering where chemically equivalent atoms can have different scattering amplitudes due to, e.g., isotope disorder, or the presence and random orientation of nuclear spins. These are parameters that can be difficult to control experimentally, and are averaged over. In x-ray scattering, especially from core electrons as we are discussing here, there is no incoherent scattering at the same fundamental level: the scattering amplitude is well defined as soon as one specifies the element doing the scattering. The reason that we discuss this distinction is then partly historical (to explain, more precisely, the reason for distinction between coherent and incoherent scattering) partly utility (in some cases, such as disordered or doped materials there can be, effectively, incoherent scattering, even for x-rays) and also as a point of nomenclature (to distinguish the use of the term coherent in the present context, inelastic scattering, from that often used in x-ray scattering (such as "coherent diffraction") which means something else. The concept of incoherent scattering remains

* We take f to be normalized to give the number of electrons in the atom at Q=0.

useful, even for x-rays, in that, if it can be identified, the incoherent part of the scattering does not depend directly on the microscopic atomic arrangement, while the coherent part generally does.

The charge density operator may be used to write the dynamic structure factor as

$$S(\mathbf{Q},\omega)=\frac{1}{2\pi\hbar}\int_{-\infty}^{+\infty} dt\, e^{-i\omega t}\left\langle \rho(\mathbf{Q},0)\,\rho^{+}(\mathbf{Q},t)\right\rangle \qquad (6)$$

where the brackets indicate configurational averaging. The transformation of eqn. (2) into eqn (6) requires some work, with the introduction of Heisenberg time-dependent operators allowing removal of the explicit sum over intermediate states (see (Sinha, 2001)(Squires, 1978) (Lovesey, 1984)). Using the definition of the charge density operator, eqn. (5), one can expand the product as

$$\begin{aligned}\left\langle \rho(\mathbf{Q},0)\rho^{+}(\mathbf{Q},t)\right\rangle &= \left\langle \sum_{jk} f_j f_k\, e^{-i\mathbf{Q}\bullet\mathbf{R}_j(0)} e^{i\mathbf{Q}\bullet\mathbf{R}_k(t)} \right\rangle = \left\langle \sum_{jk} f_j f_k\, \alpha_{jk} \right\rangle \\ &= \left\langle \bar{f}^2 \sum_{jk} \alpha_{jk} + \bar{f}\sum_{jk}\left(\Delta_j+\Delta_k\right)\alpha_{jk} + \sum_{jk}\Delta_j\Delta_k\,\alpha_{jk} \right\rangle\end{aligned} \qquad \textbf{(7a)}$$

where we have dropped the explicit **Q** dependence from the atomic form factors and the second equation in the first line just replaces the complex exponential with $\alpha_{jk}$ for notational convenience. The last equation then follows if one writes $f_j = \bar{f} + \Delta_j$ and we define an average form factor $\bar{f} = \frac{1}{N_\ell}\sum_j f_j$ with the corollary that $\sum_j \Delta_j = 0$. If we now assume that the variation in the form factor for each atom is not correlated with its position - something that is *not* correct in general, but is reasonable (for neutrons) for a monatomic Bravais lattices or monatomic disordered materials, then the configurational average will cause terms linear in $\Delta$ to drop out, and the term quadratic in delta to reduce to diagonal terms only giving

$$\begin{aligned}\left\langle \rho(\mathbf{Q},0)\rho^{+}(\mathbf{Q},t)\right\rangle &= \left\langle \bar{f}^2 \sum_{jk} \alpha_{jk} + \sum_{j}\Delta_j^{\,2}\,\alpha_{jj} \right\rangle \\ &= \bar{f}^2\left\langle \sum_{jk}\alpha_{jk}\right\rangle + \left(\overline{f^2}-\bar{f}^2\right)\left\langle \sum_{j}\alpha_{jj}\right\rangle\end{aligned} \qquad \textbf{(7b)}$$

Equation (7b) is then the usual division of the cross section used in inelastic neutron scattering. The first term is the "coherent" part of the scattering and includes interference terms responding to the positions of all of the atoms. The second term is called the "self-" or "incoherent-" part of the cross section and responds just to single-particle motion, averaged over the sample. For neutrons, the weighting of the two contributions in (7b) is material and isotope dependent, and one speaks of separate contributions from the incoherent and coherent scattering even for monatomic samples. For x-rays, generally specifying the atom type is enough to specify the scattering amplitude so, for monatomic samples, one has $\overline{f^2} - \bar{f}^2 = 0$ and there is no incoherent scattering. However, there can be similar contributions arising from, e.g., doping a material so that one atom is randomly replaced by another with a different form factor, leading to a random fluctuation in the scattering amplitude, and an effective incoherent scattering contribution. Furthermore, any process (such as most types of absorption, scattering at very larger momentum transfers, or scattering from randomly arranged atoms - perfect fluids, ideal gases - section 2.7) that tends to emphasize the diagonal (self) terms of correlation function is often called incoherent.

In the context of modern x-ray scattering measurements, the term "coherent" (or "incoherent") is often used in a different way. Here, for IXS, the usage focuses on the effects of disorder in the microscopic scattering properties of the material - the impact of the *sample* disorder on the averaging of experiments over many scattering events. However, often, in x-ray scattering one talks about "coherent" scattering as, effectively, related to the possibility to observe speckles in the scattered radiation, as is determined, primarily, by the source, by the preparation of the incoming x-ray beam (it must present a well defined phase front over the illuminated sample - i.e. it must be transversely coherent - and longitudinally coherent over relevant path-length difference) and the spatial resolution of the detector. The latter usage appears in the context of imaging or intensity fluctuation (photon correlation, XPCS) spectroscopy experiments. Conceptually there is some overlap in that both refer to the effects of averaging, but, really, they are distinct: in one case the relevant averaging is over sample properties and, in the other, over source (and detector) properties. Unless specifically stated otherwise, we will use coherence in the former sense, referring to averaging over sample properties.

## 2.6 Other Correlation Functions, Sum Rules

Texts on scattering theory describe several correlation functions related to $S(\mathbf{Q},\omega)$ by Fourier transform - these include the intermediate scattering function, sometimes called $I(\mathbf{Q},t)$ or $S(\mathbf{Q},t)$ but more often $F(\mathbf{Q},t)$, and the time-dependent pair correlation function, $G(\mathbf{r},t)$ (see, e.g., chapter 4 of (Squires, 1978), chapter 3 of (Lovesey, 1984)) G is also sometimes called the van Hove correlation function after the beautiful work in (van Hove, 1954) and is given by

$$G(\mathbf{r},t)=\frac{\hbar}{(2\pi)^3}\int d\omega\; e^{i\omega t}\int d\mathbf{r}\; e^{-i\mathbf{Q}\bullet\mathbf{r}}\,S(\mathbf{Q},\omega) \qquad \textbf{(8)}$$

These functions can be useful in developing the general theory further, looking at the symmetries of the scattering, and considering sum rules. They frequently are discussed in more detail in the context of scattering from disordered materials, where, as there is no primitive cell, it can be difficult to simplify the dynamic structure factor further. While we do not go into detail in the present paper, it is worth noting the sum rule that the first moment of $S(\mathbf{Q},\omega)$ for a monatomic sample is just the recoil energy:

$$\int_{-\infty}^{\infty}\hbar\omega\; S(\mathbf{Q},\omega)\; d(\hbar\omega)=\frac{\hbar^2|\mathbf{Q}|^2}{2M}\equiv E_r \qquad \textbf{(9)}$$

where M is the atomic mass. This is analogous to the f- or oscillator-strength- sum rule from optics, and sets the energy scale of the scattering as the recoil energy (e.g. $E_r$ = 2.1 meV / M[amu] at Q=1 $Å^{-1}$). This sum rule may be derived by taking commutators with the Hamiltonian (Placzek, 1952) or time derivatives of the intermediate scattering function, and will apply as long as there is no velocity dependence to the inter-atomic potential - generally a good approximation (one possible exception is bcc He (Gov, 2003)). This first moment sum rule applies to both the incoherent and the total scattering - i.e.: the first moment of the incoherent part of the dynamic structure factor is also the recoil energy. This sum rule (and other related ones) can be useful to derive average properties of materials from spectra measured at high Q (in the incoherent limit - see section 2.7). It (and other related rules) are also very useful in nuclear inelastic scattering (section 1.4.4). A modified form for *1-phonon* scattering of a Bravais lattice is given by (Ambegaokar, et al., 1965) with the recoil energy reduced by a factor of the Debye-Waller factor ( $e^{-2W}$ ). For materials where there is more than one type of atom doing the scattering, there is, to this author's knowledge, no rigorously justified generalization. Also, the above discussion assumes that the energy transfer integration range is limited to the range over which the sample motions are well described by motions of the atoms - if one goes to higher (eV-scale) energy transfers where one primarily observes scattering from valence electrons, the atomic mass should be replaced by the electron mass.

## 2.7 The Incoherent Limit: Ideal Gas, Density of States, Impulse Approximation

There are cases, even with x-rays, when it is reasonable to neglect the interatomic correlations of atoms in a material and the scattering becomes effectively incoherent (as discussed in the present section, this does not come from fluctuations in the scattering from different atoms, but the cancellation of the interatomic terms in the first term of eqn 7, leaving only the diagonal part, which then has the dependence of the incoherent term of 7b -> hence "incoherent"). Two immediate cases include the scattering from an ideal gas (or "perfect fluid") and scattering from any material at sufficiently large momentum transfer. Solution for the case of the ideal gas or perfect fluid (a collection of colliding but otherwise non-interacting, uncorrelated atoms at some temperature T) is just a Gaussian

$$S(\mathbf{Q},\omega)_{Ideal\,Gas}=\sqrt{\frac{\beta}{4\pi E_r}}\;\exp\left\{-\frac{\beta}{4E_r}(\hbar\omega-E_r)^2\right\} \qquad \textbf{(10)}$$

centered at the recoil energy defined above.

The high Q limit is the case where small displacements strongly influence the phase of interatomic contribution to eqn. 7b, and will lead to reduction of the cross terms between different atoms. This limit, for monatomic scatterers allows one to get some information, with sum rules analogous to eqn. (9) providing fairly direct access to various average properties such as kinetic energy and derivatives of the potential. In the case of crystalline materials, incoherent scattering can be related, at least approximately, to the phonon density of states, though, the different form factors of different atoms, and their momentum dependence, with the resulting formal lack of sum rules, makes the calculations of the density of states less rigorous than for nuclear inelastic scattering (see section 1.4.4). Going further, to extremely large Q, if the energy transfer is large compared to the energies of the excitations in a material (i.e. Q is large enough so that recoil energy is larger than phonon energies), one can reach the impulse approximation limit (Hohenberg & Platzman, 1966) of deep inelastic scattering. In this limit, the scattered intensity is directly proportional to the atomic momentum density (e.g. Compton scattering) - essentially one takes $R(t) \to \frac{p}{m}t$ where p is the atomic momentum and m the atomic mass. Carefully speaking, IXS is usually not in this limit, however, this limit has been demonstrated to be mostly acceptable for a case with relatively small atomic-mass scatterer, liquid neon (Monaco, et al., 2002), and, might, given sufficient interest, be extended to other light materials (see also discussion in (Glyde, 1994)).

## 2.8 The Phonon Expansion: Formulas for $S(\mathbf{Q},\omega)$

Considerable simplification of, or at least a change in perspective regarding, the dynamic structure factor is possible for crystalline solids. We take advantage of the assumed periodicity to write

$$\rho(\mathbf{Q},t) = \sum_{\substack{\ell \\ \text{Primitive Cells}}} \sum_{\substack{d \\ \text{Atoms/Cell}}} f_d(\mathbf{Q})\, e^{-i\mathbf{Q}\bullet\mathbf{r}_{\ell d}}\, e^{-i\mathbf{Q}\bullet\mathbf{u}_{\ell d}(t)} \qquad \begin{pmatrix}\text{Crystalline} \\ \text{Materials}\end{pmatrix} \quad \textbf{(11)}$$

where the $N_\ell$ repeating primitive[+] cells are indexed by $\ell$ and the r atoms in each primitive cell are indexed by d (in some treatments, the index $\kappa$ is used instead of d). The displacements of the atoms from their time-averaged positions, $\mathbf{r}_{\ell d}$, are given by $\mathbf{u}_{\ell \mathbf{d}}(t)$ so that we have taken $\mathbf{R}_j(t) \equiv \mathbf{R}_{\ell \mathbf{d}}(t) = \mathbf{r}_{\ell d} + \mathbf{u}_{\ell d}(t) = \mathbf{r}_\ell + \mathbf{r}_d + \mathbf{u}_{\ell d}(t)$. Using this to expand the dynamic structure factor gives

$$S(\mathbf{Q},\omega) = \frac{N_\ell}{2\pi\hbar} \sum_{\substack{\ell \\ \text{Prim. Cells}}} e^{i\mathbf{Q}\bullet\mathbf{r}_\ell} \sum_{\substack{dd' \\ \text{Atoms/Cell}}} f_d(\mathbf{Q}) f_{d'}(\mathbf{Q})\, e^{-i\mathbf{Q}\bullet(\mathbf{r}_{d'}-\mathbf{r}_d)} \int_{-\infty}^{+\infty} dt e^{-i\omega t} \left\langle e^{-i\mathbf{Q}\bullet\mathbf{u}_{0d'}(t=0)}\, e^{+i\mathbf{Q}\bullet\mathbf{u}_{\ell d}(t)} \right\rangle \quad \textbf{(12)}$$

where we have assumed that we can take the sum over $\ell,\ell'$ as depending only on the difference, so we take $\ell' = 0$ and one sum to a factor of $N_\ell$.

Proceeding further is possible by assuming some form for dynamical excitations in the sample - making assumptions about the from of $\mathbf{u}_{\ell \mathbf{d}}(t)$. This, in general, is a complex problem, but, for the purposes of a first understanding of phonon measurements, many of the essential ideas can be introduced by assuming a harmonic (or pseudo-harmonic[+]) model of the lattice dynamics. While, strictly speaking, this eliminates many interesting cases, in fact, for most of the interesting cases, one considers the non-harmonicity by introducing corrections to the harmonic results: the harmonic model is a useful, nearly necessary, starting point. (Developments of $S(\mathbf{Q},\omega)$ that do *not* assume harmonic motion at the present stage may be found in e.g. (Baym, 1961; Ambegaokar, et al., 1965)) We take (see chapter 3 and appendix G of (Squires, 1978))

$$\mathbf{u}_{\ell \mathbf{d}}(t) = \left(\frac{\hbar}{2M_d N_q}\right)^{1/2} \sum_{\substack{\mathbf{q} \\ \text{First Zone}}} \sum_{\substack{j=1 \\ \text{Modes}}}^{3r} \frac{1}{\sqrt{\omega_{\mathbf{q}j}}} \left[ \mathbf{e}_{\mathbf{q}jd}\, a_{\mathbf{q}j}\, e^{i(\mathbf{q}\bullet\mathbf{r}_\ell - \omega_{\mathbf{q}j}t)} + \mathbf{e}_{\mathbf{q}jd}^{*}\, a_{\mathbf{q}j}^{+}\, e^{-i(\mathbf{q}\bullet\mathbf{r}_\ell - \omega_{\mathbf{q}j}t)} \right] \quad \textbf{(13)}$$

---

[+] We emphasize the distinction between a *primitive* cell as (one of) the smallest repeatable unit(s) whose translation can generate the entire structure, as frequently different from a *unit* cell that often is a larger and is chosen for convenience. For example, the cubic unit cell used to describe diamond or silicon contains 8 atoms, while the proper primitive cell contains only 2 atoms. The phonon dispersion is then described by 6=3x2 phonon branches, not 3x8=24. Using the larger cell in a model would cause a lot of unnecessary computation and lead to many phonon branches with zero intensity.

[+] The term "pseudo-harmonic" is often used to indicate a harmonic model where the atomic interaction terms are scaled with temperature to account for effect of, say, thermal expansion. This leads to harmonic phonons at any fixed temperature, but, since, strictly speaking, a harmonic solid does not undergo thermal expansion, is termed pseudo-harmonic.

where we have expanded the motions in terms of the allowable normal modes (indexed by **q**,j, with frequencies $\omega_{\mathbf{q}j}$ and polarizations $\mathbf{e}_{\mathbf{q}jd}$ ) with the creation and annihilation operators $a_{\mathbf{q}j}^{+}\ a_{\mathbf{q}j}$ . The assumption of harmonic motion allows application of the Baker-Hausdorf theorem to give

$$\begin{aligned}\left\langle e^{+i\mathbf{Q}\bullet\mathbf{u}_{\ell'\mathbf{d}'}(0)}\ e^{-i\mathbf{Q}\bullet\mathbf{u}_{\ell\mathbf{d}}(t)}\right\rangle &= e^{-\frac{1}{2}\left\langle\left|\mathbf{Q}\bullet\mathbf{u}_{\ell'\mathbf{d}'}\right|^2\right\rangle} e^{-\frac{1}{2}\left\langle\left|\mathbf{Q}\bullet\mathbf{u}_{\ell\mathbf{d}}\right|^2\right\rangle} e^{\left\langle\left(\mathbf{Q}\bullet\mathbf{u}_{\ell'\mathbf{d}'}(0)\right)\left(\mathbf{Q}\bullet\mathbf{u}_{\ell\mathbf{d}}(t)\right)\right\rangle} \\ &= e^{-W_d}e^{-W_{d'}}\left(1+\langle UV\rangle+\tfrac{1}{2}\langle UV\rangle^2+\ldots\right)\end{aligned} \qquad \textbf{(14)}$$

where $W_d \equiv \frac{1}{2}\left\langle\left|\mathbf{Q}\bullet\mathbf{u}_{\ell\mathbf{d}}\right|^2\right\rangle$ has been assumed to be independent of primitive cell, and we have taken $U \equiv \left(\mathbf{Q}\bullet\mathbf{u}_{\ell'\mathbf{d}'}(0)\right)$ and $V \equiv \left(\mathbf{Q}\bullet\mathbf{u}_{\ell\mathbf{d}}(t)\right)$. Counting the number of creation and annihilation operators in each term one finds that the terms in parentheses correspond, respectively, to no-phonon, 1-phonon and 2-phonon scattering, so we write

$$S(\mathbf{Q},\omega) = S(\mathbf{Q},\omega)_{0p} + S(\mathbf{Q},\omega)_{1p} + S(\mathbf{Q},\omega)_{2p} + \ldots \quad \textbf{(15)}$$

We now discuss the form of each term as can be derived by applying the results of (Squires, 1978). We emphasize that the picture here is of a collection of phonon modes that is completely defined - polarization and frequency - at momentum transfers **q** within the first Brillouin zone (c.f. the expansion of eqn. (13)) but probed at some momentum transfer **Q** that can be (and usually is) outside the first zone. More discussion about practical interpretation of these terms, may also be found in section 2.11.

**The zero-phonon, elastic, term is**

$$\begin{aligned}S(\mathbf{Q},\omega)_{0p} &= N_\ell^2\left|\sum_d f_d(\mathbf{Q})\ e^{-W_d}\ e^{i\mathbf{Q}\bullet\mathbf{r}_d}\right|^2 \delta_{\mathbf{Q},\tau}\ \delta(\hbar\omega) \\ &= N_\ell^2\left|F_\tau\right|^2\ \delta_{\mathbf{Q},\tau}\ \delta(\hbar\omega)\end{aligned} \qquad \textbf{(16)}$$

which is just the usual expression for Bragg scattering. The second equality defines the structure factor $F_\tau$ (in units of electrons - recall factoring out $r_e$ in eqn. (1)) in the usual way for the Bragg reflection at reciprocal lattice vector $\tau$ .

**The first order, one-phonon, term is**

$$S(\mathbf{Q},\omega)_{1p} = N_\ell \sum_{\substack{\mathbf{q}\\ 1st\ Zone}}\ \sum_{\substack{j\\ Modes}} \left|\mathrm{F}_{1\mathrm{p}}(\tau,\mathbf{q},j)\right|^2 \delta_{\mathbf{Q}-\mathbf{q},\tau}\begin{Bmatrix}\left\langle n_{\mathbf{q}j}+1\right\rangle\delta\left(\omega-\omega_{\mathbf{q}j}\right)\\ +\left\langle n_{\mathbf{q}j}\right\rangle\delta\left(\omega+\omega_{\mathbf{q}j}\right)\end{Bmatrix} \quad \textbf{(17)}$$

where the two terms in the last bracket are due to phonon creation and phonon annihilation, respectively. The occupation factors are from the expectation values of the number operator $\left\langle n_{\mathbf{q}j}\right\rangle = \left\langle a_{\mathbf{q}j}^{+}a_{\mathbf{q}j}\right\rangle$ and its Hermitian conjugate $\left\langle a_{\mathbf{q}j}a_{\mathbf{q}j}^{+}\right\rangle = \left\langle n_{\mathbf{q}j}+1\right\rangle_{\text{(bosons)}}$. In thermal equilibrium one then has $\left\langle n_{\mathbf{q}j}\right\rangle = \left(e^{+\beta\hbar\omega_{\mathbf{q}j}}-1\right)^{-1}$ and detailed balance (see section 2.13) is given by $\left\langle n_{\mathbf{q}j}+1\right\rangle / \left\langle n_{\mathbf{q}j}\right\rangle = e^{+\beta\hbar\omega_{\mathbf{q}j}}$ . The one-phonon structure factor $F_{1p}$ (units of electrons) is given by

$$\left|F_{1\mathrm{p}}(\tau,\mathbf{q}j)\right|^2 = \frac{1}{\omega_{\mathbf{q}j}}\left|\sum_d \frac{f_d(\mathbf{Q})}{\sqrt{2M_d}}\ e^{-W_d}\ \mathbf{Q}\bullet\mathbf{e}_{\mathbf{q}jd}\ e^{i\mathbf{Q}\bullet\mathbf{r}_d}\right|^2 \qquad \textbf{(18)}$$

where we have decomposed the total momentum transfer, **Q,** as required by the delta function

$$\mathbf{Q} = \tau + \mathbf{q} \quad \begin{pmatrix}\tau \text{ a reciprocal lattice vector}\\ \mathbf{q}\text{ in the first brillouin zone}\end{pmatrix} \quad \textbf{(19)}$$

The generalization of eqn (17) to allow for finite phonon line-width may be found in eqns. (31)-(33).

**The second order, 2-phonon term is**

$$S(\mathbf{Q},\omega)_{2p} = \frac{1}{2}\sum_{\mathbf{qq}'}\sum_{jj'}\left|F_{2p}(\tau,\mathbf{q}j,\mathbf{q}'j')\right|^2 \delta_{\mathbf{Q}-\mathbf{q}-\mathbf{q}',\tau} \begin{Bmatrix} \langle n_{\mathbf{q}j}+1\rangle\langle n_{\mathbf{q}j}'+1\rangle \;\; \delta(\omega-\omega_{\mathbf{q}j}-\omega_{\mathbf{q}j}') \\ +2\langle n_{\mathbf{q}j}\rangle\langle n_{\mathbf{q}j}'+1\rangle\delta(\omega+\omega_{\mathbf{q}j}-\omega_{\mathbf{q}j}') \\ + \;\; \langle n_{\mathbf{q}j}\rangle\langle n_{\mathbf{q}j}'\rangle \;\; \delta(\omega+\omega_{\mathbf{q}j}+\omega_{\mathbf{q}j}') \end{Bmatrix} \quad \textbf{(20)}$$

where

$$\left|F_{2p}(\tau,\mathbf{q}j,\mathbf{q}'j')\right|^2 = \frac{\hbar}{\omega_{\mathbf{q}j}\;\omega_{\mathbf{q}'j'}}\left|\sum_d \frac{f_d(\mathbf{Q})}{2M_d} e^{-W_d}\;\mathbf{Q}\bullet\mathbf{e}_{\mathbf{q}jd}\;\mathbf{Q}\bullet\mathbf{e}_{\mathbf{q}'j'd}\; e^{i\mathbf{Q}\bullet\mathbf{r}_d}\right|^2 \quad \textbf{(21)}$$

and $\mathbf{Q} = \tau + \mathbf{q} + \mathbf{q}'$. The terms in the brackets correspond to two-phonon creation, phonon creation and annihilation, and two-phonon annihilation, respectively. This form has been simplified from (Baron, et al., 2007) by assuming that the eigenvectors and frequencies have been chosen so that $\omega_{\mathbf{q}j} = \omega_{-\mathbf{q}j}$ and $\mathbf{e}_{\mathbf{q}jd}^{\;*} = \mathbf{e}_{-\mathbf{q}jd}$.

The Debye-Waller factor is

$$\begin{aligned} W_d &= \tfrac{1}{2}\left\langle(\mathbf{Q}\bullet\mathbf{u}_{\ell d})^2\right\rangle = \frac{\hbar}{4M_d}\frac{1}{N_q}\sum_{\mathbf{q}}\sum_{j=1}^{3r}\frac{\left|\mathbf{Q}\bullet\mathbf{e}_{\mathbf{q}jd}\right|^2}{\omega_{\mathbf{q}j}}\coth\left(\hbar\omega_{\mathbf{q}j}\beta/2\right) \\ &\approx \frac{3\hbar^2|\mathbf{Q}|^2}{2M\beta(k_B\Theta_D)^2} \quad (\text{Debye Spectrum}) \end{aligned} \quad \textbf{(22)}$$

where the last expression is for a monatomic sample with a Debye spectrum (acoustic modes only, to a maximum energy of $\hbar\omega_D = k_B\Theta_D$, $\Theta_D$ the "Debye temperature"- see (Lovesey, 1984), chapter 4) as is sometimes used as a first approximation. The mode frequencies, $\omega_{\mathbf{q}j}$, in all these expressions are assumed to be real and positive definite, as is required for a stable lattice in the harmonic approximation. The presence of the Kroneker delta function (other treatments use a Dirac delta) in the various expressions comes from writing

$$\begin{aligned} \sum_{\substack{\ell \\ Prim\,Cells}} e^{i\mathbf{Q}\bullet\mathbf{r}_\ell} &= \prod_{i=1}^{3}\frac{\sin(N_i\varphi_i/2)}{\sin(\varphi_i/2)}e^{-i(N_i-1)\varphi_i} \quad (\text{Parallelepiped}) \\ &\to N_\ell\,\delta_{\mathbf{Q},\tau} \quad (\text{Large } N_\ell) \end{aligned} \quad \textbf{(23)}$$

The first equation assumes the sample is a parallelepiped with sides of $N_i$ primitive cells parallel to each primitive lattice vector $\mathbf{a}_i$ (with $\varphi_i = \mathbf{a}_i \bullet \mathbf{Q}$ and the number of cells $N_\ell = N_1N_2N_3$) and is easily derived by considering the geometric series implied by the sum. This form explicitly includes the effects of the diffraction from the finite size of the sample. The second equation neglects the geometric effects of the sample extent, with the Kroneker delta-function giving unity when **Q** is any vector of the reciprocal lattice, $\tau$. It is worth emphasizing, in the context of XPCS (x-ray photon correlation spectroscopy) or CDI (coherent diffractions imaging) that the approximation of the second part of eqn 23, while allowing the theory to proceed easily is, in effect, where one discards all information about the impact of the sample shape on the scattering process.

## 2.9 Harmonic Phonons: The Dynamical Matrix, Eigenpolarizations, Acoustic & Optical Modes, Participation Ratio

We now discuss harmonic phonons in more detail. The solution postulated above comes from applying periodic boundary conditions to the problem and expanding the potential of the interaction to lowest non-vanishing order in atomic displacements - see (Born & Huang, 1954) and (Maradudin, 1974). The chain of argument can be broken down into steps including (1) the adiabatic approximation (2) neglect of the effect of atomic displacements on the electronic states, and (3) assumption of simple lowest-order (spring-constant) interactions between pairs of atoms, neglecting the higher order terms coupling different normal modes. The steps are discussed in detail in (Maradudin, 1974) including estimates of the magnitude of the various terms. Assumptions (2) and (3) are rather limiting, with (2), for example, effectively removing electron-phonon or spin-phonon coupling from consideration and (3) removing phonon-phonon scattering (anharmonicity). However, the relatively simple, closed-form, of the harmonic solution is the starting point for perturbation theory to consider more complex and interesting interactions.

The importance and generality of the adiabatic or Born-Oppenheimer approximation, make it worthy of further mention. Formally, it amounts to assuming the problem is separable, with the full wave function, a function of all nuclear coordinates, $\{\mathbf{R}_i\}$ and all electronic coordinates $\{\mathbf{r}_j\}$,rewritten as a product of functions of the nuclear positions and one of the electronic positions at a particular set of nuclear coordinates $\Psi(\mathbf{R},\mathbf{r}) \to \chi(\mathbf{R})\psi_{\mathbf{R}}(\mathbf{r})$. The nuclear positions then are just taken as parameters for the second function. This amounts to assuming the electrons arrange themselves instantly to accommodate any nuclear configuration or that the nuclei move slowly compared to electronic time scales so that the electrons follow the nuclei adiabatically - and then leads to simplification of the equations of motion. This simplification is used in almost all calculations of vibrational behavior, especially those described in section 3, and is generally a good approximation. However, the approximation can fail, with specific examples including doped graphene (Lazzeri & Mauri, 2006), $MgB_2$ (Boeri, et al., 2005)(d'Astuto, et al., 2016) or boron doped diamond (Caruso, et al., 2017). In general, one might expect the adiabatic approximation can fail when changes in electron energy by phonon-scale energies correspond to significant changes in Fermi-surface shape, e.g. when electron bands have flat regions near the Fermi-surface - or at very high temperatures where thermal energies are comparable to electronic levels.

For the harmonic approximation, the Hamiltonian is

$$H = \sum_{\ell d} \frac{\mathbf{P}_{\ell d}{}^2}{2M_d} + \tfrac{1}{2} \sum_{\ell d \ell' d'} \mathbf{u}_{\ell d} \overset{\rightrightarrows}{\Phi}_{\ell d \ell' d'} \mathbf{u}_{\ell' d'} \quad \textbf{(24)}$$

where $\mathbf{P}_{\ell d}$ is the momentum operator for the $\ell d$'th atom and $\overset{\rightrightarrows}{\Phi}_{\ell d \ell' d'}$ are the 3x3 real-valued inter-atomic force constant matrices linking displacements of the atom $\ell' d'$ to the force on atom $\ell d$.* They are the second-derivative of the potential function, and generally would be considered a function of all atomic and electronic coordinates (and, likewise, the full Hamiltonian would include electronic kinetic energies), but (see, e.g., (Maradudin, 1974)), assumptions (1) and (2) above allow reducing the problem to just as a function of the nuclear coordinates while assumption (3) allows truncation at second order. The corresponding equations of motion are

$$M_d \ddot{\mathbf{u}}_{\ell d} = -\sum_{\ell' d'} \overset{\rightrightarrows}{\Phi}_{\ell d \ell' d'} \mathbf{u}_{\ell' d'} \quad \textbf{(25)}$$

Considering solutions to be of the form of a superposition of normal modes with well-defined periodicities, **q**, and frequencies shows that the sufficient condition for a solution is

$$\omega_{\mathbf{q}}{}^2 \vec{\mathbf{e}}_{\mathbf{q}} = \overset{\rightrightarrows}{\mathbf{D}}_{\mathbf{q}} \vec{\mathbf{e}}_{\mathbf{q}} \quad \textbf{(26)}$$

where $\vec{\mathbf{e}}_{\mathbf{q}}$ is the vector made up of the r 3-vectors $\mathbf{e}_{\mathbf{q}d}$ which give the direction, and relative phase, of the motion of each of the r atoms in the primitive cell. The dynamical matrix, $\overset{\rightrightarrows}{\mathbf{D}}_{\mathbf{q}}$ is the Fourier-transform of the force constant matrix and is made up of 3x3 matrices for each possible pair of atoms within the primitive cell, $\overset{\rightrightarrows}{\mathbf{D}}_{\mathbf{q}dd'}$ given by

$$\overleftrightarrow{\mathbf{D}}_{\mathbf{q}dd'} = \frac{1}{\sqrt{M_d\, M_{d'}}} \sum_{\ell'} \overleftrightarrow{\Phi}_{\ell d \ell' d'}\, e^{-i\mathbf{q}\bullet(\mathbf{r}_\ell - \mathbf{r}_{\ell'})} \quad \mathbf{(27)}$$

---

* We use boldface quantities to indicate 3-vectors (in either real space or reciprocal space) and dual arrow superscripts to indicate matrices such as $\overset{\rightrightarrows}{\Phi}_{\ell d \ell' d'}$. In some cases, such as the eigenvector matrix for all atoms, $\vec{\mathbf{e}}_{\mathbf{q}}$, a single vector over a boldface quantity indicates an extended 1-dimension matrix of many vectors

The $\ell$ dependence has been dropped on the left since, in the limit of a large crystal, the sum is independent of $\ell$. In the usual way, this reduces to an eigenvalue problem with 3r solutions at any fixed **q**, which we index by j=1...3r. (sometimes a subscript "s" is used to represent the combined index "$\mathbf{q}j$"). One notes that the reality of the force-constants insures D is Hermitian $\overleftrightarrow{\mathbf{D}}_{\mathbf{q}dd'} = \overleftrightarrow{\mathbf{D}}_{\mathbf{q}d'd}^{\;*}$ and therefore the eigenvalues, $\omega_{\mathbf{q}}^{\;2}$, are real, with the eigenfrequencies either real or purely imaginary. Imaginary eigenfrequencies correspond to blow up of the atomic motion with time indicating that the lattice is unstable: in the harmonic limit, a lattice is only stable if all eigenvalues of the dynamical matrix are real.

Eigenpolarizations corresponding to non-degenerate eigenvalues will be orthogonal, while those for degenerate eigenvalues *can be chosen* orthogonal, so one usually takes $\sum_d \mathbf{e}_{\mathbf{q}jd}^{\;*} \bullet \mathbf{e}_{\mathbf{q}j'd} = \delta_{j,j'}$. The eigenpolarizations also satisfy the closure relation $\sum_{j=1}^{3r} e_{\mathbf{q}jd}^{\alpha\;*} e_{\mathbf{q}jd'}^{\beta} = \delta_{d,d'}\;\delta_{\alpha,\beta}$ where the superscripts refer to components in an orthonormal coordinate system. Noting that the reality of the force-constant matrices gives that $\overrightarrow{\mathbf{D}}_{-\mathbf{q}} = \overrightarrow{\mathbf{D}}_{\mathbf{q}}^{\;*}$ so that the eigenvalues at q and at **-q** are the same, $\{\omega_{\mathbf{q}j}\} = \{\omega_{-\mathbf{q}j}\}$, and we can choose the eigenvectors so that $\mathbf{e}_{-\mathbf{q}jd} = \mathbf{e}_{\mathbf{q}jd}^{\;*}$.

Acoustic modes may be identified by investigation of the eigenmodes of the solution either using a perturbation series expansion for small wave vectors (section 26 of (Born & Huang, 1954)), or considering symmetry properties (Maradudin, 1974)): in general there are 3 modes for which the frequency approaches zero linearly at small wave vectors ($|\mathbf{q}| \to 0$). They obey

$$\begin{aligned} &\omega_{\mathbf{q}j} \propto |\mathbf{q}| \\ &\mathbf{e}_{\mathbf{q}jd} \propto \sqrt{M_d}\;\mathbf{u}_{\mathbf{q}j} \\ &\left|F_{1\mathrm{p}}(\tau,\mathbf{q}j)\right|^2 \propto \frac{1}{\omega_{\mathbf{q}j}} \left|\mathbf{Q}\bullet\mathbf{u}_{\mathbf{q}j}\right|^2 \left|F_\tau\right|^2 \end{aligned} \qquad \left(\begin{array}{c} \text{Acoustic Modes,} \\ |\mathbf{q}| \to 0 \\ F_\tau \equiv \sum_d f_d(\tau)\, e^{-W_d}\, e^{i\tau\bullet\mathbf{r}_d} \end{array}\right) \quad \mathbf{(28)}$$

where $\mathbf{u_j}$ is a vector giving the direction of motion, which for these modes at small **q,** is the same for all atoms (independent of d**).** The third equation just is from substituting the second into eqn. (18) for the 1-phonon structure factor. This shows that the intensity of these modes scales as the product of the structure factor of the nearby Bragg reflection, $F_\tau$, and a term related to the mode polarization. These modes are the acoustic modes, where all atoms oscillate in phase near q=0. The remaining (3r-3) modes are referred to as optical modes. Materials with only one atom per primitive cell, true Bravais lattices, without a basis, have no optical modes. (Practically, near the Brillouin zone boundary the polarizations for optical and acoustic modes can appear very similar). Depending of the version of this paper, supplemental materials may be available showing examples of phonon motions including a longitudinal acoustic mode (ESM1), a transverse acoustic (ESM2) mode, a longitudinal optic mode (ESM3) and a transverse optic mode (ESM4), all near zone center (q=0 or Γ). Finally, we note that for acoustic modes with $\hbar\omega_{\mathbf{q}j} \ll k_B T$, the cross section or measured mode intensity can be expected to scale roughly as $1/|\mathbf{q}|^2 \propto 1/(\hbar\omega_{\mathbf{q}j})^2$, where one factor of $1/\hbar\omega_{\mathbf{q}j}$ comes from the structure factor, above, and the other from the Bose occupation factor in thermal equilibrium.

In addition to the above relations, the phonon eigenvectors are strongly constrained by the symmetry of the lattice, as reflected, in the present discussion, in the symmetry of the force constant matrices, $\overleftrightarrow{\Phi}_{\ell d \ell' d'}$. This may be investigated using a group-theoretical approach and, practically, can allow large simplifications of the dynamical matrix, which, when recognized, can simplify algebra, reduce the time for computations, and give constraints on allowable atomic motions. Further discussion about symmetry and its impact on phonons can be found in, e.g., (Maradudin & Vosko, 1968), (Lax, 2012). The web page (Bilbao Crystallographic Server) is also a useful tool. We note that even a very simple nearest-neighbor (NN) or next-nearest-neighbor (NNN) force-constant model can be useful to get a rough idea of what the phonon eigenvectors may look like.

Finally in this section, we note that it can sometimes be useful to consider the contribution of different atoms to a given phonon mode. Looking directly at eqn (18), this is immediately seen to be related to the magnitude of the phonon eigenvector $\mathbf{e}_{\mathbf{q}jd}$, possibly scaled by the atomic mass. However, it can be useful to make the relation more formally, especially when the primitive cell may be large. One then sometimes discusses a phonon "participation ratio" or "inverse phonon participation ratio (IPR)" (see (Hafner & Krajci, 1993)(Finkemeier & von Niessen, 1998) and references therein) given as

$$P_{\mathbf{q}j} = \left( \sum_d \left| \mathbf{e}_{\mathbf{q}jd} \right|^2 \Big/ M_d \right)^2 \Bigg/ \left( r \sum_d \left| \mathbf{e}_{\mathbf{q}jd} \right|^4 \Big/ M_d^{\,2} \right)$$

$$= \left( r \sum_d \left| \mathbf{e}_{\mathbf{q}jd} \right|^4 \right)^{-1} \qquad \begin{pmatrix} \text{Monatomic} \\ \text{Sample} \end{pmatrix}$$

where the second relationship uses the orthonormality of the eigenvectors. If, for example, a mode consists of motion of a single atom only in the r-atom primitive cell, one has P=1/r, while if the motions are uniformly spread across all atoms, P will tend toward unity. While being somewhat ad-hoc, this can be a useful way to identify possible localization of modes in complex materials such as cage compounds (e.g. clathrates, skutterudites), quasi-crystals, or glasses (some other metrics are also given in the references of (Finkemeier & von Niessen, 1998) that, e.g., include the distances between moving atoms).

## 2.10 The Real-Space Force Constant Matrix & Acoustic Sum Rule

The real space force constant matrix of eqns. (24)-(27) (also called the Interatomic Force Constant (IFC) matrix) takes on added importance as a transferrable version of the results of phonon calculations. Most crystal-based *ab-initio* methods calculate phonons on a small grid of selected momentum transfers in the first Brillouin zone, as each calculation often requires significant CPU time. This is extended to arbitrary momentum transfers by Fourier interpolation through calculation of the real-space force constant matrices $\overleftrightarrow{\Phi}_{\ell d \ell' d'}$ of eqn. (27). Thus, the real-space force constant matrices (essentially an extended Born-von-Karman model) become the portable form of the phonon calculation result, allowing the frequencies and eigenvectors of modes to be calculated for arbitrary momentum transfers. A similar matrix can be defined for estimating the electron-phonon line-width at arbitrary momentum transfers.

The $\overleftrightarrow{\Phi}_{\ell d \ell' d'}$ matrices, as noted in the previous section, must obey a host of symmetry relations based on the lattice structure. Of these, the invariance of the problem to a uniform translation of the lattice as a whole, as applicable to all materials, gives an often-quoted condition on the force constant matrices, the acoustic sum rule. Namely

$$\sum_{l'd'} \overleftrightarrow{\Phi}_{\ell d \ell' d'} = \overleftrightarrow{0} \qquad \begin{pmatrix} \text{Acoustic Sum Rule} \\ \text{Translational Invariance} \end{pmatrix} \quad \textbf{(29)}$$

or the sum over force-constant matrices for one atom with all other atoms should be zero in all components. This effectively defines the self-term of the force constant matrix, $\overleftrightarrow{\Phi}_{\ell d \ell d}$, and also ensures that the acoustic mode frequencies approach 0 as **q** approaches zero. If one attempts to manually modify the force-constant matrices, as is discussed in some of the subsequent sections of this paper, one must *both* preserve the symmetry of the lattice, and insure that eqn (29) is always satisfied by appropriate modification of the self-terms.

## 2.11 Practical (Harmonic) Phonons

There are a variety of practical issues that we would now like to introduce as regards phonon measurements, including constraints on where to measure, some nomenclature issues, the idea of anti-crossings, and some comments on the relative intensities of the terms in the phonon expansion.

### 12.11.1 One-Phonon Cross Section, Transverse and Longitudinal Geometries, Q vs. q

The form of the one-phonon cross section, eqns. (17-19), is relatively simple in principle, but the dependence on the phased sum of the components of the phonon eigenvectors in the direction of the total momentum transfer adds complexity, effectively providing selection rules for observable modes. Generally, for considering single-phonon scattering, one decomposes the total momentum transfer, **Q**, into a sum (eqn. (19)), $\mathbf{Q} = \tau + \mathbf{q}$ where $\tau$ is a vector of the reciprocal lattice (sometimes also called **G** or **H**) and **q** is in the first Brillouin zone and is called the propagation direction of the phonon mode. The general requirement for non-zero intensity at **Q** is that the mode must have some net component of the atomic motions parallel to **Q** - but estimating the intensity without detailed calculations is generally not straightforward. Considering the discussion of section 2.9, a generally reliable statement is that, near zone center, **q**=0, the intensity of the acoustic modes will scale as the intensity of the nearby Bragg reflection. Similar considerations suggest that, for simple materials, the optic modes, where atoms move out of phase with each other, scale inversely with nearby Bragg peaks. (e.g. going to a forbidden Bragg reflection may be a good way to measure an optic mode, but, while it may be tempting to remove possible elastic backgrounds, it is a poor way of measuring acoustic modes unless one is specifically interested in violations of eqn. (28)) However, as materials become more complex, or as one moves away from zone center, phonon intensities can quickly become non-obvious and calculations are needed both to determine where a mode of interest may be strong, and where less-interesting modes may be weak and not obscure the mode of interest.

The terms "longitudinal" and "transverse" are often used to refer to phonon modes, and/or measurement geometries. For modes, the terms refer to the directions of atomic motion, $\mathbf{e}_{\mathbf{q}jd}$ (for some j and **q**) relative to the propagation direction, **q**: an acoustic shear mode is transverse, while the compression or pressure mode is longitudinal. (Movies of longitudinal modes may be available in ESM1 and ESM3 while transverse modes may be available in ESM2 and ESM4). For measurement geometries the terminology refers to direction of the reduced momentum transfer, **q**, relative to the total momentum transfer, **Q,** or, more precisely, relative to the nearest reciprocal lattice vector $\tau$ **:** a longitudinal geometry has **q** parallel to τ, and a transverse geometry has **q** perpendicular to τ. If one wants to observe a transverse phonon mode, a transverse geometry is generally favorable, and, likewise a longitudinal geometry for a longitudinal mode. Most phonon modes, however, can *not* be classified as simply longitudinal or transverse. Especially, as one moves away from the center of the Brillouin zone, or as materials become complex, or low dimensional, or if one moves off of high-symmetry directions, mode polarizations become complicated. In some cases it can be clearer to refer to modes by their eigenvectors (e.g. plane polarized, c-axis polarized), without reference to the direction of propagation. In other cases, the atomic motions may not be linear, but circular or elliptical (e.g. modes with complex polarization vectors) so are not simply related to a real (linear) momentum transfer (for example, this can be seen in the iron pnictides where the lack of reflection symmetry of the As locations about the plane of Fe atoms leads quickly to elliptical motions as one moves away from the gamma point. A movie of this can be seen in ESM11).

### 2.11.2 On the intensity of the terms in the phonon expansion.

The various terms in the phonon expansion scale as the product of a structure factor, the Debye-Waller factor and the thermal occupation factors. To a first approximation, the Debye-Waller factor can often be neglected, as, especially at low temperature, it is usually near to unity and independent of atom. This means that first estimates of phonon intensities are often done with $W_d$ =0, the more so as calculation of $W_d$ requires integration over the Brillouin zone. However, in the presence of soft modes that may preferentially lead to large (low frequency) motions of specific atoms, $W_d$ can become atom dependent and require some care*. Also, as Q or T is increased, $W_d$ can increase quickly. In those cases, the one-phonon intensity will be reduced, and the Debye-Waller factor can easily become atom dependent, so the intensity of specific modes or motions can be suppressed. Increasing $W_d$ is also generally an indication that multi-phonon contributions become stronger.

The structure factors $F_{1p}$, $F_{2p}$ , etc., generally scale as $Q^{2n}$ where n is the number of phonons involved in the scattering, and will increase with Q, until the exponential die off of the atomic form factors or Debye-Waller factor begins to dominate. Practically, one finds product of the magnitude of the momentum transfer and the atomic form factor, $Q\ f(Q)$, which determines the scale of $F_{1p}$, tends, for most atoms, to increase out to Q ~ 10Å$^{-1}$, or so, before first flattening out and then falling off. The $1/\omega_j$ in front of the structure factors can be seen to come from the amplitude of the phonon motion - large amplitude (or, equivalently, given phonon energies are quantized, low frequency) motions lead to larger scattering. The temperature dependence appears in the occupation factors: modes with $|\hbar\omega_j| < k_B T$ (26 meV at T=300K) will be enhanced, increasing as $1/|\omega_j|^2$ , while the signal at $\hbar\omega < -k_B T$ will be strongly reduced, and the occupation factor for $\hbar\omega > k_B T$ approaches unity.

---

* For example, in the case of CaAlSi the presence of a soft mode introduces large differences in the Debye-Waller factor for different atoms, with factors of two easily possible at room temperature (Kuroiwa, et al., 2008).

Finally, investigation of scaling of the terms with the number of primitive cells shows that the one-phonon intensity, and higher order terms as well, scale as the number of unit cells, $N_\ell$ (The $N_\ell$ scaling is explicit in the 1-phonon term, and for the 2-phonon term is implied by the double-sum over q, q' with only one delta-function). This is in contrast to the Bragg scattering which scales as $N_\ell^2$ (Though, as visible in eqn. 23, the Bragg peaks narrow as $N_\ell$ is increased so that the scattered intensity is really being concentrated into a narrow angular range). Given that all terms discussed are coherent (in the sense of section 2.5) it is clear the remarkable scaling of the Bragg intensity is due to having coherent and elastic scattering from a periodic medium: all three are needed to get the enhancement.

### 2.11.3 Anti-crossings

Often the frequency of two dispersing phonon modes can become close and even tend to cross at some momentum transfers. If the modes have the same symmetry, then this will create an "avoided crossing", or an "anti-crossing" in the dispersion, where the dispersion curves repel each other. It is characteristic of such an anti-crossing that the mode polarizations will become mixed near the point of close approach, and that, while the mode frequencies will not cross, the mode polarizations do cross. Figure 4 shows an example of this for a calculation of phonons in CaAlSi where a soft optical mode (corresponding to anti-phase motions of Al and Si along the c-axis) has an avoided crossing with the longitudinal acoustic (LA) mode (essentially in-phase motions of all atoms). The intensity calculated for the (0 0 L) zone (indicated by the height of the bars in the figure) tracks the polarization: the optical mode pattern, essentially equal and opposite motions Si and Al atoms along the c-axis has weak intensity and jumps across the gap: the low energy mode at zone boundary (0 0 4.5) shows the same anti-phase motion character as the optical mode at zone center. This anti-phase motion leads to correspondingly weak intensity for momentum transfers that are purely in the c-axis (00L) direction. Movies of the relevant mode motions are shown in ESM5 and ESM6 for the low and high energy modes, respectively near Γ=(004) and ESM7 and ESM8 for the low and high energy modes near to zone boundary (0 0 4.5). ESM9 and ESM10 show the two modes near the anti-crossing (0 0 4.22). The useful rule is then that frequencies can anti-cross, while polarizations (and intensities in IXS) cross, and the polarization mixes in the region of close approach.

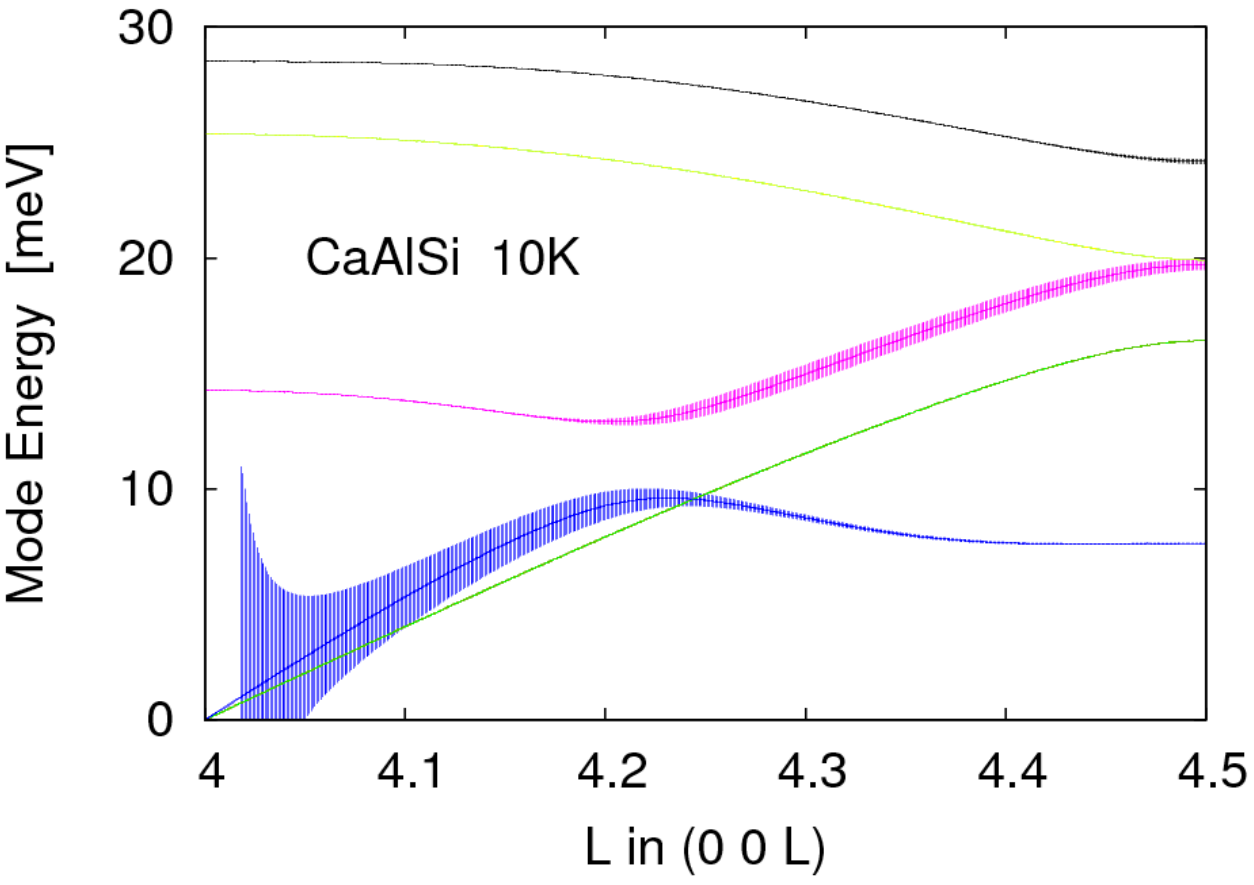

**Figure 4.** Example of an anti-crossing (after (Kuroiwa, et al., 2008)). Vertical bars are proportional to the calculated phonon intensity in at the indicated momentum transfer. See text for discussion. (Note intensities are not plotted for the acoustic mode at L<~4.02)

### 2.11.4 Two-Phonon Scattering

Two phonon scattering is the largest intrinsic source of background in most IXS measurements since incoherent (Compton) scattering occurs on large, > eV, scales, out of the experimental energy window, while multiple scattering is usually prevented by the large intrinsic x-ray absorption cross-section. Equations (20-21) give the magnitude of the effect within a harmonic model - thus the relative contribution of the two-phonon scattering, as compared to the one-phonon intensity, is directly calculable. It scales as the product of phonon contributions at two different reduced momentum transfers within the first Brillouin zone, but, with polarization dot products (and form factors) as calculated at the *total* momentum transfer, **Q**.

Figure 5a shows a measurement of the scattering from $MgB_2$ where the two-phonon contribution leads to a peak in the spectrum that can not be explained by, otherwise, extremely good calculations: the calculation of the two-phonon contribution gives, as shown, a peak with the correct energy and intensity. In this case, the contribution was important because the peak mimics the $E_{2g}$ mode that is so important for superconductivity. Figure 5b shows the temperature dependence of the two-phonon contribution, which, in detail, can have a complex structure. In general, the high-energy part from two-phonon creation is always present, even at low temperature, and weakly temperature dependent until high

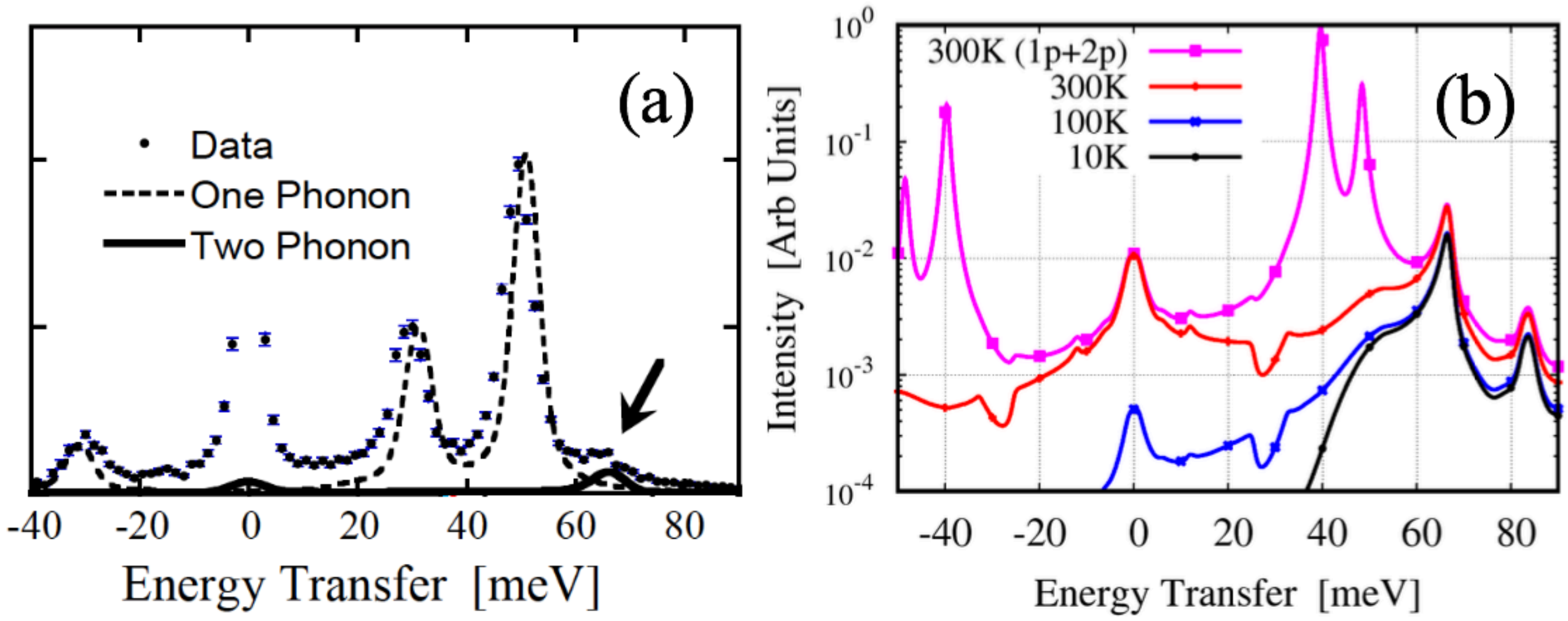

**Figure 5. Two-Phonon contributions in $MgB_2$** (after (Baron, et al., 2007)). The left panel, (a) shows data and calculations at room temperature near Q= (0 0 3.32). The two-phonon-contribution, with no free parameters, scales to exactly explain the additional peak at about 65 meV. (b) shows calculations on log scale for several temperatures for Q=(0 0 3.5) covlved with a 1.5 meV lorenzian. Note that the result at 300K is shown twice, once with the inclusion of the 1-phonon intensity. See the discussion in the text. (Panel (a) of the figure is Copyright (2007) by the American Physical Society)

temperatures are reached. Meanwhile the near-zero energy part is from simultaneous creation and annihilation and increases rapidly with temperature. It also usually has a finite (non-zero) line width. There can be additional structure and for the particular case of $MgB_2$, the main contribution to the 65 meV two-phonon peak was found to be simultaneous excitation of two TA modes, which each have energies of about 30 meV near zone boundary. Practically, it is worth noting that while often it is possible to reasonably estimate the Debye-Waller factor from a calculation including $\sim 10^4$ or $\sim 10^5$ points that are randomly distributed over the Brillouin zone, good estimation of the two-phonon part typically requires more, $>\sim 10^6$ points to avoid artifacts. Possible a more judicious choice of points in the zone (similar to that in (Gilat & Raubenheimer, 1966) or (Monkhorst & Pack, 1976)) could reduce the number of points needed and speed calculations.

## 2.12 Non-Harmonic Phonons

The assumption of harmonic lattice behavior is useful to introduce the normal modes and the phonon expansion. However, solids are not harmonic - with some degree of non-harmonic behavior needed, for example, to account for thermal expansion, finite heat conductivity, and any case where the phonons may interact with themselves (phonon-phonon scattering) or with other systems (electron-phonon coupling, spin-phonon coupling). Thus, in fact, most of the interesting properties of materials currently under investigation involve non-harmonicity. (As mentioned above, we will use the term "non-harmonicity" to refer to any process beyond the harmonic model, and "anharmonicity" to refer specifically to the effects of phonon-phonon scattering). In principle, non-harmonicity can lead to modification of the phonon energies, line shapes, and even polarizations[x]. It also can add interference terms to the phonon expansion, eqns. (14) (15), as the conditions for the re-arrangement of terms using the Baker-Hausdorf lemma are violated [(Ambegaokar, et al., 1965) while, for experimental work, see (Meyer, et al., 1976) and references therein]. However, the most well-known effect of non-harmonicity is a shift in phonon energies and an increase in phonon line width. This leads, as a first approximation, to the damped-harmonic oscillator (DHO) line shape as given in this context by [(Fak & Dorner, 1992, 1997)] and as we will discuss here. We consider this from the point of view of Green's functions, but note equivalent expressions can also be derived via rate-equations, where one explicitly considers the pumping of the phonons by an external system and vice versa and then applying the Kramers-Kronig relation (see the appendix of (Allen, et al., 1997)).

Considering the strongest, one-phonon, term, non-harmonicity may be introduced through the phonon self-energy: the phonon Green's function is modified from the non-interacting one, $D^0_{\mathbf{Q}j}(\omega)$, to the interacting one, $D_{\mathbf{Q}j}(\omega)$, by the inclusion of a self-energy term $\Pi_{\mathbf{Q}j}(\omega)$ (electrons are described by are $G, \Sigma$ and phonons by $D, \Pi$, similar to (Mahan, 2000) and, by writing **Q** instead of **q,** we allow for the possibility of coupling to some perturbation that does not obey the periodicity of the lattice). The interacting Green's function then becomes $D_{\mathbf{Q}j}(\omega) = 2\omega_{\mathbf{q}j} \Big/ \left(\omega^2 - \omega_{\mathbf{q}j}{}^2 - 2\omega_{\mathbf{q}j}\,\Pi_{\mathbf{Q}j}(\omega)\right)$. The self-energy is causal, obeying a Kramers-Kronig relation, with the real part even and the imaginary part being odd. So we then write

$$\Pi_{\mathbf{Q}j}(\omega) \equiv \Delta_{\mathbf{Q}j}(\omega) - i\,\gamma_{\mathbf{Q}j}(\omega)\,\frac{\omega}{\omega_{\mathbf{q}j}} \qquad \textbf{(30)}$$

where $\Delta(\omega)$ and $\gamma(\omega)$ are both real and even functions of frequency. The dynamic structure factor is proportional to the imaginary part of the Greens function (cf. eqn (3)) so that the dynamic structure factor becomes

$$S(\mathbf{Q},\omega)_{1p} = N_\ell \sum_{\mathbf{q}} \sum_{j} \left|\mathrm{F}_{1\mathrm{p}}(\tau,\mathbf{q},j)\right|^2 \delta_{\mathbf{Q}-\mathbf{q},\tau}\, L_{\mathbf{Q}j}(\omega,T) \qquad \textbf{(31)}$$

where the line shape function

$$L_{\mathbf{Q}j}(\omega,T) = \frac{1}{\pi}\frac{1}{1-e^{-\hbar\omega/kT}}\ \frac{4\,\omega\,\omega_{\mathbf{q}j}\,\gamma_{\mathbf{Q}j}(\omega)}{\left(\omega^2 - \Omega_{\mathbf{Q}j}(\omega)^2\right)^2 + 4\omega^2\gamma_{\mathbf{Q}j}(\omega)^2} \qquad \textbf{(32)}$$

has replaced the sum of delta functions in eqn (17) and the shifted frequency is given by

$$\begin{aligned}\Omega_{\mathbf{Q}j}(\omega)^2 &= \omega_{\mathbf{q}j}{}^2 + 2\omega_{\mathbf{q}j}\Delta_{\mathbf{Q}j}(\omega) \\ &\approx \left[\omega_{\mathbf{q}j} + \Delta_{\mathbf{Q}j}(\omega)\right]^2 \quad \left(\Delta_{\mathbf{Q}j} \ll \omega_{\mathbf{q}j}\right)\end{aligned} \qquad \textbf{(33)}$$

[x] Most work, including that here, emphasizes the effect of non-harmonicity on lineshapes and frequency, neglecting the effect on polarization. However, such effects must be present, if only in that a non-harmonic shift in the frequency may move a mode closer or further from an anti-crossing, leading to a change in polarization. One would also expect such effects to appear more generally, but the usual treatment separating the intensity and the line-shape does not facilitate its discussion. Lovesey, (Lovesey, 1984), also suggests that direct non-harmonic effects on polarization tend to vanish at high symmetry points of the Brillouin zone.

In the limit where $|\Pi(\omega)| \ll \omega_{\mathbf{q}j}$ these terms have the largest effect when $\omega \approx \omega_{\mathbf{q}j}$ with $\Delta(\omega_{\mathbf{q}j})$ the frequency shift of the mode and $\gamma(\omega_{\mathbf{q}j})$ the half-width at half maximum (HWHM) of the phonon line. Thus, these quantities, $\Delta(\omega_{\mathbf{q}j})$ and $\gamma(\omega_{\mathbf{q}j})$, are often interpreted as the effect of screening of the phonon mode by the relevant perturbation and the (inverse) lifetime of the phonon mode to decay into the perturbing system (e.g. other phonons, or electron-hole pairs). In this context, the general dependence of $\Delta$ and $\gamma$ on the probe frequency, $\omega$ , may not be immediately obvious (why should the lifetime of the phonon depend on the probe frequency?), however, one needs to recall that, *in this frequency domain picture*, one is really, steady-state, gently pumping the entire system at the probe frequency (the phonon is *not* being excited and then freely decaying) and, in that context, the pump frequency certainly affects the response of the system. However, often, as a first approximation, γ and Ω are taken as constants, which case, the lineshape L, is a bose-weighted damped harmonic oscillator (DHO) function (Fak & Dorner, 1992, 1997), Finally we note that the same DHO lineshape is often used in the analysis of liquid response with the assumption that $\Delta_{\mathbf{Q}j}(\omega) = \gamma_{\mathbf{Q}j}^2 \big/ 2\omega_{\mathbf{Q}j}$ or $\Omega_{\mathbf{Q}j}{}^2 = \omega_{\mathbf{Q}j}{}^2 + \gamma_{\mathbf{Q}j}{}^2$.

More generally, one can investigate $\Delta$ and $\gamma$ using the Kramers-Kronig relation. Some relevant aspects of this can be seen by taking a very simple model (see also (Gunnarsson & Rösch, 2008)) having the line width as constant below some cutoff frequency:

$$\gamma(\omega') = \begin{Bmatrix} \gamma & \omega' < \omega_c \\ 0 & \text{otherwise} \end{Bmatrix} \qquad \textbf{(34)}$$

which, via the Kramers-Kronig transform, gives the energy shift

$$\Delta(\omega) = \frac{2}{\omega_{\mathbf{q}j}\pi} P \int_0^{\omega_c} d\omega' \, \gamma(\omega') \frac{\omega'^2}{\omega^2 - \omega'^2}$$
$$= -\gamma \frac{\omega_c}{\omega_{\mathbf{q}j}} \frac{2}{\pi} \left( 1 - \frac{\omega}{\omega_c} \tanh^{-1}\left[ \frac{\omega}{\omega_c} \right] \right) \quad \textbf{(35)}$$
$$\approx -\gamma \frac{\omega_c}{\omega_{\mathbf{q}j}} \frac{2}{\pi} \quad [\omega \ll \omega_c]$$

In this model, the phonon is softened by the interaction, and, given that the cutoff energy can be large, for, e.g. electronic interactions, the softening can be much larger than line width: the coupling serves primarily to screen the interaction, with the ratio of the softening to the broadening being given by the bandwidth of the coupling over the phonon energy. Thus Kohn anomalies (Kohn, 1959) (Stedman, et al., 1967) which occur at nesting vectors of the Fermi-surface where phonons couple strongly to the electronic system, are most often observed as changes in dispersion, and not line width (though, also, energy shifts are easier to measure than line-widths).

It is interesting to look at the integrated intensity of a broadened phonon. In a harmonic model, after correction for the Bose temperature factor, the integrated intensity of a mode responds only to the phonon polarization, and, at higher temperatures, the Debye-Waller factor as given in eqn (18). However, in the event there is significant non-harmonicity, it is not obvious what may be conserved. Consideration of a series of models for non-harmonicity of an isolated phonon that each satisfy the Kramers-Kronig relation shows that, in general the integral of the lineshape (eqn. 32) remains constant to a the level of ~1% and better than 10% for even extreme non-harmonicity. Thus the integrated phonon intensity, in the limit where anharmonicity does not affect the phonon polarization, is reasonably conserved (at least at the level of typical measurements of phonon intensity). In the limit of strong phonon-phonon (anharmonic) coupling, however, this should be reexamined.

It is worth noting the there are cases where more complicated self-energy functions have been calculated. These include, for example, the effect of electron-phonon coupling on phonons with an energy near the superconducting gap (Allen, et al., 1997) which leads, e.g., to asymmetric tails or near splitting of a phonon line (see the discussion in section 5.2.3), and the more complicated self energies from anharmonicity as can become more important at higher temperature (e.g. (May, et al., 1998)). The DHO lineshape of eqn (32) is an excellent approximation for many simple cases, but real self energies require understanding the full phonon decay path, including coupling to other systems, and other phonons, so are complicated. Given this complexity, and the fact that in many cases a phonon still mostly appears as a well-defined peak in $S(\mathbf{Q},\omega)$ with the self-energy contributions being relatively small, the complex shape of the self-energy is often ignored. Possibly, like with multi-phonon terms, the relative cleanliness of IXS data, will allow self energies to be investigated more thoroughly and routinely.

## 2.13 Detailed Balance, Symmetrizing, and Scaling

Inspection of equation (17) above, and the conditions on the eigenvectors will show the positive and negative energy (Stokes and anti-Stokes) sides of the $S(\mathbf{Q},\omega)$ spectra are not independent, with the anti-Stokes side being a lower-intensity copy of the Stokes side. This is a general property of a sample in thermal equilibrium, where the population of dynamical states depends on the energy of the state and the temperature, and is formally expressed as the detailed balance condition for $S(\mathbf{Q},\omega)$ (see, e.g., chapter 4 of (Squires, 1978)) usually given as

$$S(\mathbf{Q},-\omega) = e^{-\hbar\omega\beta} S(-\mathbf{Q},\omega) \quad \textbf{(36)}$$

which relates the intensity of the energy loss side of the measured spectrum to the energy gain part of the spectrum *for the opposite momentum transfer*. IXS, however, most easily measures the energy gain and loss at fixed momentum transfer, so, for IXS, the more interesting symmetry relation is between positive and negative frequency at fixed **Q**. A moments consideration, as noted by (Squires, 1978), shows that eqn. (36) applies also for positive **Q** on the right-hand-side, if the material is either disordered or centrosymmetric. *

$$S(\mathbf{Q},-\omega) = e^{-\hbar\omega\beta} S(\mathbf{Q},\omega) \quad \begin{pmatrix} \text{Disordered Materials or} \\ \text{Centrosymmetric Crystals} \end{pmatrix} \quad \textbf{(37)}$$

* In a previous version of this review, a result of Lovesey (Lovesey, 1984), that negligible absorption is also a sufficient condition for eqn (37) to hold was mentioned. However, it is only correct for Bravais lattices, and is not true in general.

This relation is useful experimentally as an internal check on the temperature of a sample - though care must be taken as regards possible effects of finite energy-resolution (tails of strong peaks arising from finite energy resolution do not obey detailed balance - one must do a resolution de-convolution in some fashion *before* applying detailed balance). If the temperature for a sample in thermal equilibrium is well known, the author has found that violation of this symmetry can be taken as an indication of experimental error[†].

Given the simplicity of the detailed balance relation, one sometimes considers a "classical" or symmetrized version of the spectra,

$$S^{cl}(\mathbf{Q},\omega) = e^{-\hbar\omega\beta/2}\, S(\mathbf{Q},\omega) \qquad (38a)$$

or

$$S^{cl}(\mathbf{Q},\omega) = \frac{1-e^{-\hbar\omega\beta}}{\hbar\omega\beta} S(\mathbf{Q},\omega) \qquad (38b)$$

Generally, such symmetrized versions are most useful for comparing measured spectra of liquids or glasses to calculations (e.g. molecular dynamics results with classical trajectories) which are symmetric about zero energy transfer. The first equation, (38a), is reasonable to describe uncorrelated motions of ideal gasses or perfect fluids (e.g. the classical response of an perfect fluid, based on eqn. (10), is a Gaussian centered at zero energy transfer) or samples at high temperatures or high momentum transfers where multi-phonon scattering is dominant. However, the choice of symmetrizing function is not unique, and the second equation, (38b), corresponds to the case when the response is dominated by single phonons, as can be seen from comparing to eqn. (32), and is often more generally useful, even for liquids. The two do deviate significantly from each-other as $|\hbar\omega\beta|$ increases above 1. Other forms can also be considered (as is evident from considering an expansion of incoherent scattering in terms of the phonon density of states where each order of multi-phonon contribution has a different temperature scaling, (Sjölander, 1958)), however, the two listed are the more common ones. Additional discussion about the non-classical nature of $S(\mathbf{Q},\omega)$ can be found in chapter 3 of (Lovesey, 1984).

A related issue is the practical comparison of phonon spectra at different temperatures. Here we only point out that while one can attempt to apply scaling, e.g. from eqns. (17) or (37) or (38b) for single phonons, there is no general (model independent) way to do this if multi-phonon contributions are significant. Even in the case that single phonon response dominates, one needs to care for the effects of tails of the resolution function. Practically, in the one-phonon limit, comparison of spectra different temperatures can be done by fitting the spectra with a series of resolution functions and then scaling the intensity of each function to a common temperature (essentially: de-convolution, scale to common T, re-convolution, compare).

# 3. Calculations of Atomic Dynamics

Calculation of phonons, especially in crystalline materials with more than one atom per primitive cell, are needed both before an experiment to understand where to measure, and afterwards to interpret the results. The former bears emphasis, as, too often, people focus only on the periodicity of interest (i.e.: little **q**, in the first Brillouin zone), and do not appreciate the complexity of the eigenvectors which have a huge impact on the choice of the total momentum transfer, **Q,** where the measurement takes place. Given this complexity, calculations are required before an experiment, with the only possible exception being measurements of acoustic modes near to zone center. Calculations for disordered materials are less crucial before an experiment, since the measurements are fundamentally one-dimensional in Q, but they are still useful to interpret results. We describe first model calculations for crystalline materials, then *ab-initio* approaches for crystalline solids, and then move to molecular dynamics, as may be used for both disordered and crystalline materials. We also discuss where, practically, one seems to approach the limit of reliable calculations. This section is meant to briefly indicate calculations that are broadly available, and is not comprehensive.

Model calculations, with user-input parameters, are useful because they can be relatively simple and have parameters that can be optimized to fit dispersion relations, or even spectra. They are usually also relatively fast. The Born-von Karman (BvK) model, where one directly chooses the force constants (or matrices) over some limited range (nearest neighbors, or next-nearest neighbors, etc.) is conceptually the simplest, a "ball and spring" model. Assuming appropriate care is taken to be consistent with the crystal symmetry, and to get a stable result, this sort of model can be useful to get a first understanding of phonon behavior. A next level of sophistication is the addition of long-range

[†] In one case a sample with an interesting magnetic structure appeared to show such a violation, but the lack of symmetry in the spectra was eventually traced to a software bug in a counter card driver. The card was actually in use in in many places, but there were very few that used it at the low rates of the IXS experiment, and, precisely in the low-rate region, the software error became noticeable.

Coulomb interactions by placing charges on different atoms (ions), using a rigid ion model (where "rigid" is used in the sense of non-deformable, to be contrasted with the shell model mentioned presently). The 1/r Coulomb potential adds complexity in that the sums are slowly (and not uniformly) convergent. However, Ewald's method can be applied* (see e.g. (Brüesch, 1982)(Gonze & Lee, 1997)). The long-range interaction can cause splitting between longitudinal and transverse optical modes at zone center (LO-TO or Lydane-Sachs-Teller (LST) splitting), as is often observed in ionic insulators. A next level of sophistication may be found in a shell model. Here one considers each atom or ion to consist of a massive charged nucleus surrounded by a massless charged "shell"+ with the two coupled together by a spring (usually assumed to be isotropic). This introduces the possibility to dynamically polarize each lattice site. The shell model does a surprisingly good job of fitting the dispersion of many materials including ionic insulators, covalent materials, and also metals, with *relatively* few free parameters. The parameters - e.g. charges on atoms, atomic polarizabilities - are also, in principle, more physically motivated than choosing force constant matrices of the BvK model, but practically, the best fit to phonon dispersion can lead to values outside the range of ones normal intuition. Available shell model codes include OpenPhonon (Mirone), Unisoft (Eckhold, et al., 1987), and GULP (Gale, 1997). While the number of free parameters can initially appear daunting, it is often possible to build on models that are already published in the literature. There are also other models at a similar level - the deformation dipole model, the valence overlap shell model, valence force model, the bond-charge model (see e.g. (Karo & Hardy, 1969)(Kunc & Bilz, 1976)(Brüesch, 1982)(L. J. Sham, 1974), and references therein) - which focus on different aspects of the problem. It may be worth noting that, even in a harmonic limit, a model predicting a given dispersion is not unique: more than one model can, in principle, generate the same dispersion. For example, a unitary transformation acting on the dynamical matrix preserves eigenvalues or mode frequencies, but mixes polarizations (Leigh, et al., 1971).

Electronic structure codes allow an "*ab-initio*" approach to phonons by considering the forces on the atoms as being the result of changes in electronic structure and system energy when atoms are displaced. These calculations most often use pseudo-potentials, where only the outer electrons are explicitly considered, and also take advantage of the repeating structure of a crystal - Bloch's theorem. "Direct" methods rely on calculating changes in energy, and resulting forces on atoms, when atoms are explicitly displaced from the equilibrium positions. These can either be done considering displacements of one atom at a time (small displacement method) or choosing specific phonon displacement patterns as governed by symmetry conditions (frozen phonon method), and, in both cases, they can be built on repetitive calls to a conventional electronic structure code. The use of a primitive cell and Bloch's theorem, as is typical for these calculations, however, means that if one wants results away from q=0, displacements must be carried out using super-cells. Proper inclusion of long-range forces for insulators also requires care. Generally available codes that allow direct methods to be built on top of established electronic structure codes include Phon (Alfè, 2009) and Phonopy (Togo, et al., 2008). An alternative to the direct method is density functional perturbation theory (DFPT) as originally described using a plane wave-approach (Baroni, et al., 2001) but also be formulated for ultra-soft pseudo-potentials (e.g. PWscf) and in a mixed basis approach (Heid & Bohnen, 1999). The DFPT approach builds a perturbation calculation for the phonons into the code, and can be applied at arbitrary momentum transfers without using super-cells. Generally available (open source) codes include Quantum Espresso/PWscf (Giannozzi, et al., 2009) and ABINIT [(Gonze, et al., 2009)]. Others codes are available based on collaborative or financial agreements. One notes that both direct and DFPT methods, by default, include the effect of electron-phonon coupling on the phonon frequencies for metals, as the calculations are based on the changes of the electronic structure accompanying the atomic displacements. (In some cases the coupling may be may be significantly under-estimated, especially in correlated materials - see discussion in (Yin, et al., 2013)). Inclusion of anharmonicity due to phonon-phonon scattering is also possible.

Molecular Dynamics (MD) calculations, where one follows atomic trajectories over small time steps, offer another approach to calculations of atomic dynamics. These can either be done with classical potentials that are static (configuration independent), or via *ab-initio* methods employing pseudo-potentials where the total effective potential is a function of the atomic configuration. MD methods are mostly employed for calculations of disordered materials such as liquids, but, can also yield information about phonons in crystals. Due, in part, to the large computation time required, MD calculations are used much less for periodic materials than the above-mentioned crystal-based methods. However, as computation speeds increase, MD methods may become more appealing, especially to consider phonon-phonon interactions (anharmonicity) (see discussion in (Hellman, et al., 2013) and, e.g., (Lan, et al., 2014) (Li, et al., 2014) for examples of simpler materials), dynamics close to melting points or other phase transitions, or for time dependent phenomena, such as valence fluctuations, or cases when a there may be fluctuations between different structures (e.g. cage compounds with large cages). Examples of codes to convert from MD trajectory information to phonon dispersions via the velocity autocorrelation function, include (Ramirez-Cuesta, 2004; Hinsen, et al., 2012). A recent code that uses a mixture of *ab-initio* direct (crystal) and MD calculations with a focus on anharmonicity is ALAMODE (Tadano, et al., 2014).

---

* As an exercise, one can replace the Ewald sum by calculating many force constant matrices out, say, to some range R - one finds, as might be expected, the phonon behavior is generally well predicted until you approach small momentum transfers (~1/R).

+ This usage of the term "shell" in the shell model should not be confused with the shells (1st neighbors, second neighbors, etc.) sometimes referred to in BvK models.

From an experimentalist's viewpoint, for simple (uncorrelated, non-magnetic, low-T-stable) metals and insulators, the crystal based *ab-initio* calculations can do an amazing job of estimating both dispersion and, for metals, electron-phonon line-width. However, as soon as one adds correlation, they can fail. Another limit is that *ab-initio* calculations are generally done at what is effectively zero temperature - thus high-temperature phases of materials can be challenging, though, sometimes, it is possible to gain information by forcing the atomic configuration to that of the higher temperature structure (e.g. not relaxing the calculation) or other methods of achieving finite temperatures ((Souvatzis, et al., 2009)(Hellman, et al., 2011)(Chen, et al., 2014)). Finally, the methods discussed above all rely on the adiabatic approximation, and while this is generally good, it can fail as noted in section 2.9, and then more complex calculations (e.g. time dependent density functional theory) may be needed. Here we should point out that there is also a "multicomponent" version of density functional theory, where the nuclear coordinates are also treated as variables much like the electronic wave functions (Kreibich & Gross, 2001), which then allows DFT without the Born-Oppenheimer approximation. Returning to the general issue, one summarizes by saying that, for simple, non-correlated materials, without phase transitions, or magnetism, ab-initio pseudo-potential calculations often do an excellent job of estimating phonon dispersion and electron-phonon line-width. For other materials, special efforts still need to be made, and, in many cases, especially with correlated materials, the differences between calculation and measurement remain significant, with one particular area of present interest being the interaction of magnetism, and especially fluctuating (para-)magnetism with lattice dynamics (see, e.g., the sections on the iron pnictides 5.4.5) (Murai, et al., 2016) (Pradip, et al., 2016)(Tóth, et al., 2016). A recent review focusing on the status of calculations of paramagnetic materials is (Abrikosov, et al., 2016).

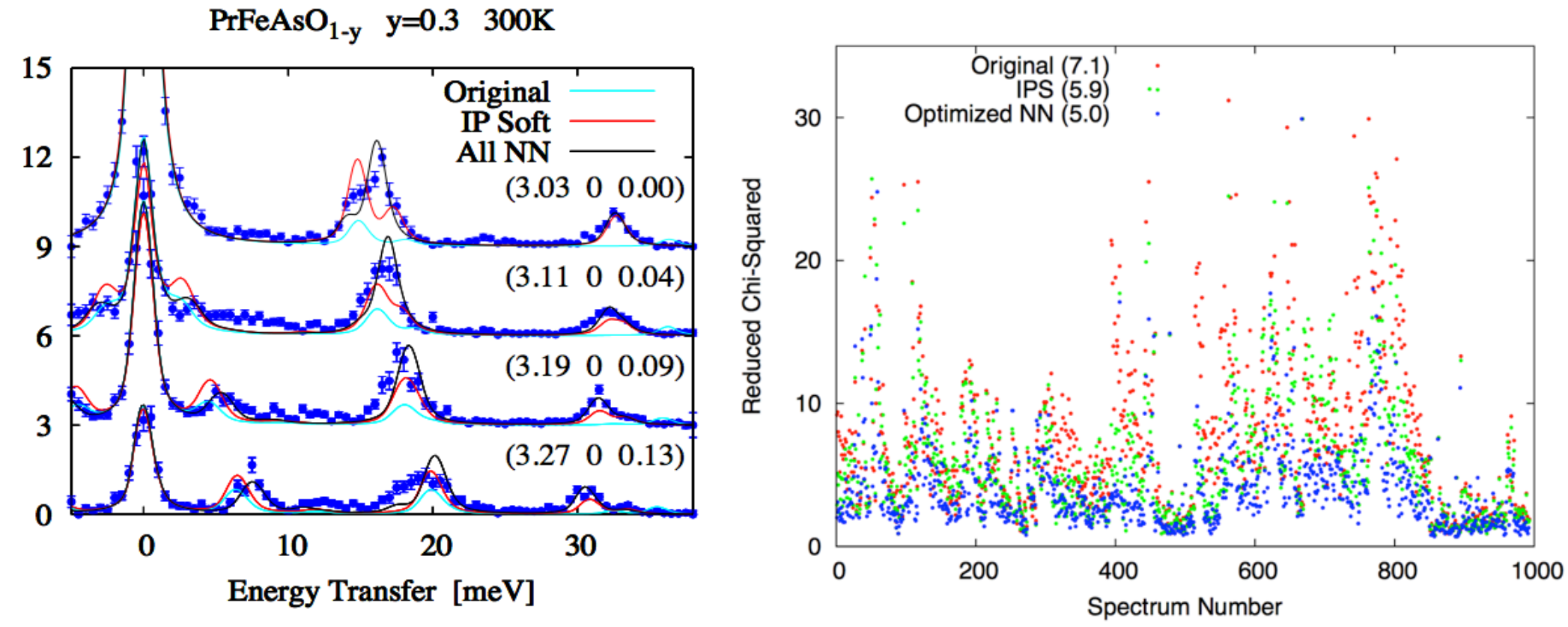

**Figure 6. Optimization of a model for phonons in $PrFeAsO_{1-y}$ by modifying nearest-neighbor force constant matrices**. Left shows the fits to the spectra, while the change in the reduced chi-squared for all spectra is shown at right for the original ab-initio matrices, an improvement were the in-plane component of the iron-arenic bond is softened and a model where all nearest-neighbor (NN) interactions are optimized. No two-phonon contribution is included. Based on data from (Fukuda, et al., 2011) and (Baron, *et al,* unpublished)

Finally, we note that it is possible to modify the force constant matrices from the *ab-initio* methods in an attempt to fit measured dispersion (similar to what has been done for model calculations). This has been done, e.g., to try to understand the softening of the modes in iron-arsenide superconductors (Fukuda, et al., 2008, 2011). However, the cleanliness of the IXS spectra, also offers, in principle, the opportunity to go further and fit the full spectra to develop an optimized microscopic model of the vibrations that, e.g., then might be used to precisely calculate other parameters. Figure 6, shows such an attempt to fit the $PrFeAsO_{1-y}$ by relaxing all nearest neighbor force-constant matrices (subject to appropriate symmetry conditions and, especially, the acoustic sum rule). Improvement with modeling is possible. However, it quickly becomes clear that, at least near room temperature, the 2-phonon contribution can not be neglected, which, because that contribution requires an integral over momentum space, then requires significant computational resources.

# 4. IXS Spectrometers

## 4.1 Spectrometer Layout and Design Issues

Understanding the design of IXS spectrometers is important on, at least, two levels: (1) the design concepts and implementation directly shape the practice and interpretation of measurements across the field and (2) the underlying optical principles determine the limits of the method. The first is relevant to anyone either reading about results from, or thinking to participate in, IXS experiments, while the second is important for an informed dialogue about expanding the boundaries of the method. The following section discusses spectrometer design with the goal of indicating the main conceptual issues that arise in such a design - we do not discuss the detailed design of specific components. The use of a high-resolution spectrometer was discussed in section 1.2: it is essentially a 4-cricle x-ray diffractometer with an energy-scan option, conceptually rather similar to a neutron triple-axis spectrometer (except simpler[+]). A schematic of the main components was shown in figure 2 and an overview of operation is presented there as well. Energy scans are usually done by fixing the analyzer energy and changing the incident beam energy (scanning the HRM), as this is easier than the reverse, because the analyzers tend to be larger and harder to scan, and, often, one has an array of analyzer crystals: scanning a single monochromator is easier than scanning an array of analyzers. The result of such scan is a spectrum that is directly proportional to $S(\mathbf{Q},\omega)$. (Measurement as a function of momentum transfer, with fixed energy transfer is possible, but not often done[*]).

Spectrometer design hinges on a variety of issues that we describe quickly here, and then re-visit in more detail after introduction of dynamical diffraction. Specific issues include the choice of high-resolution monochromator (HRM) (backscattering or in-line), the focusing, the clear aperture around the sample, and the type of analyzer optics (spherical or post-sample-collimation, PSC) and the number of analyzers. The choices made at different facilities reflect a combination of the constraints and expertise and interests at each facility. Issues and impact are listed in table 4, and are discussed both in the present section, and, somewhat, revisited in section 4.8 in comparing spherical backscattering analyzers (SAs) and post-sample collimation PSC (optics).

| Design Issue | Impact | Options |
|---|---|---|
| High Resolution Monochromator (HRM) | Beamline Layout<br>Flexibility<br>(Scan Range) | Single Backscattering<br>In-Line |
| Focusing | Minimum Sample Size<br>Q, E Resolution<br>Sample Clear Aperture | Compound<br>Single-Element |
| Sample Clear Aperture | Sample Environments<br>Energy Resolution | - |
| Analyzer Setup | Energy Resolution<br>Parallelization<br>X-Ray Energy | Spherical Backscattering<br>PSC optics |

**Table 4: Spectrometer Design Issues**

The beamlines at ESRF and SPring-8 use single backscattering monochromators and spherical backscattering analyzers, while those at APS use in-line monochromators followed by spherical backscattering analyzers. Meanwhile the UHRIX setup at APS, and the beamline at NSLS-II both use in-line monochromators and post-sample-collimation (PSC) based analyzers, where the beam is collimated and flat analyzers are used, instead of figured analyzers. Sketches of the various beamline setups are shown in figure 7 while more details can also be found in table 2. In addition to the items in the table, since most IXS experiments are flux limited, and, indeed, scan times in a given set of conditions can sometimes be days, the entire beamline design needs to emphasize both high through-put (efficiency) and stability.

---

[+] Momentum transfer and energy transfer are completely decoupled with x-rays so, for IXS, scanning energy transfer at constant momentum transfer requires no motion of the sample or analyzers.

[*] Scanning Q at a fixed energy transfer is possible, as are more sophisticated correlated scans, but can be dangerous, except in well-defined simple cases. This danger is because (1) IXS energy resolution is typically Lorentzian, with long tails, so that the intensity at a fixed energy can respond to changes in intensity of spectral features that are far away - thus it can be necessary to know the entire energy spectrum at a particular momentum transfer to properly interpret the intensity at one energy and (2) dynamical spectra are generally complex, with many modes for materials with larger unit cells, so even in the absence of long tails, it can be important to know the local shape of the spectrum to evaluate intensity changes at a particular energy.

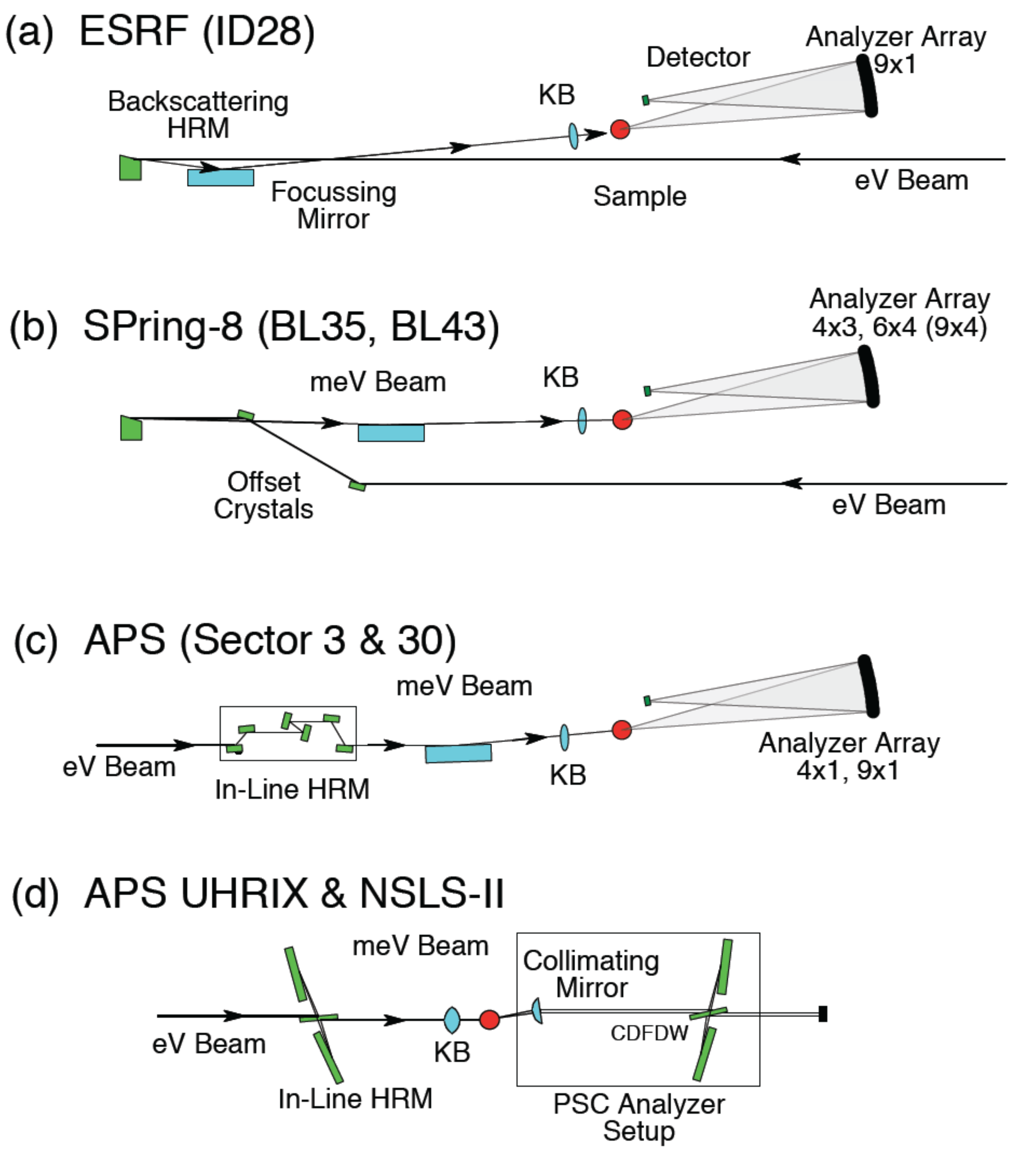

**Figure 7. Spectrometer conceptual layout.** (Dimensions are not to scale). When multiple beamlines are listed, the details can be slightly different than shown - e.g. in (c) sector 30 of APS has only a KB (no separate mirror) and in (d) the sketch of the HRM is appropriate for UHRIX, but at NSLS-II is more similar to the design for APS (c). Various components are as labeled, with "KB" being an abbreviation for the use of a Kirkpatrik-Baez focusing pair of mirrors, or a similar focusing element.

The choice of high-resolution-monochromator (HRM) design has a large impact on beamline layout: an "in-line" option (where the beam out of the HRM is in the same direction as the incident beam) preserves the order of hutches and is conceptually easiest to integrate in a beamline design. This is the design favored at APS, where there is considerable experience in making such in-line monochromators. A single-reflection backscattering design, as used at SPring-8 and ESRF, is much simpler as regards the operation of the HRM, but the outgoing beam is nearly on top of the incident beam in space. This requires careful beamline design, but, in this author's opinion, is easier to operate than in-line HRMs, and, also, given the layout of the experimental floor at most synchrotron radiation facilities, can allow a larger two-theta scan range for the spectrometer. The PSC optical setup requires an in-line monochromator as it operates at resolution that is better than simple symmetric backscattering can provide, in silicon, at the low x-ray energy. The comment about flexibility in the table for the HRM regards the possibility to operate at different resolutions: for backscattering HRMs, it is relatively easy to change resolution, e.g., from 2.8 meV at the Si (999) at 17.79 keV to 1.3 meV at the Si(11 11 11) at 21.75 keV as no change in monochromator geometry or analyzer crystals is needed, while in-line monochromators must be entirely replaced (see also table 2). The comment about scan range in table 4 refers to the fact that, in general, the scan range of the in-line monochromators can be ~10's of eV, while that of a backscattering monochromator is limited by the range of thermal expansion to ~eV in silicon - but given that phonons are mostly on energy scales <0.1 eV this is generally not an issue (though see also section 4.9).

Focusing is accomplished by any number of optics - cylindrical or toroidal mirrors, dynamic or static KB setups, or even lenses, but there are two separate and competing constraints on the focal spot size. On one hand, the beam size should be small enough as dictated either by the optical acceptance of the spectrometer (~0.3 mm for spherical analyzers, depending somewhat on resolution, ~0.01 mm for a PSC setup) or the sample size: if the beam size at the sample is too large, it may partly miss the sample, or be partly outside the spectrometer acceptance, reducing rates and/or degrading energy resolution. On the other hand, the incident beam divergence must not be too large in order to keep good momentum resolution: the best achievable momentum resolution for a spectrometer, in the limit of small beam sizes and small analyzers, is given by the incident angular divergence: $\Delta Q_{\min} = \Delta\theta_{in}\ 2\pi / \lambda$ (at small Q). Considering these two competing issues, spectrometers often use compound focusing, allowing one to choose between small beam size and larger divergence for very small samples and a larger beam size with smaller divergence for better momentum resolution. For example, at BL35 of SPring-8, the incident divergence in the scattering plane with just the standard bent cylindrical mirror is ~0.35 mrad. FWHM (beam size of ϕ<80 μm) giving $\Delta Q_{\min}$ ~ 0.038 nm$^{-1}$ at 21.7 keV, which increase to ~2 mrad (~0.22 nm$^{-1}$) when the KB setup is installed to reduce the beam size to ϕ<20 μm. The main mirror alone is then used for liquid experiments, and most crystalline experiments, but, when the samples are very small (e.g. in a diamond anvil cell at high, >50 GPa, pressure) the KB setup is added. Another advantage of a compound focusing setup is that the size of the KB mirrors can be reduced. A disadvantage, however, with compound focusing is that the source size demagnification is also usually reduced, so at BL43 a single-focusing KB setup was chosen to push to few-micron scale beam size, even though it does require significant setup time to re-align the beamline. For the PSC setups, the requirements for the beam size are more severe for spectrometer operation than with spherical analyzers, so it can be difficult to relax the beam size much, and slitting may be needed upstream of the focusing element for the best momentum resolution. We will not discuss the details of focusing further in the present paper, but refer the interested reader to references (Alatas, et al., 2011) (Ishikawa, et al., 2013)(Baron, et al., 2018).

The free aperture around the sample has a large impact on possible sample environments. Generally, the default configuration of a beamline (without micro-focusing / KB optics) has relatively large space upstream of the sample. However, the space downstream of the sample tends to be limited: when spherical analyzers are used, it is desirable to place the detectors for each analyzer close to sample in order to reduce geometric contributions to the energy resolution, while, in the PSC setup, it is advantageous to place the collimating mirror close to the sample to accept a maximum solid angle of the scattered radiation but still keep the beam size small. Thus, in both cases, one practically only has a free aperture downstream of the sample of <~100mm, except in special cases (e.g. the spectrometer of BL43LXU is built to allow up 150 mm clear aperture downstream of the sample for a 7T magnet). In contrast, upstream of the sample, one can, in many cases, have 200 or 300 mm (or more) free aperture. The space above the sample is usually not limited (except by ceiling height, which can be an issue for a magnet), while the space below the sample is set by the beamline layout - including default stage configuration. (In principle, for backscattering monochromator beamlines, the beam flight path location is another limit. But, at, SPring-8, the free sample space below the sample is close to 400 mm and has not limited any experiments.) Installing focusing, such as a KB setup, where placing focusing elements close to the sample desirable to improve the demagnification, can limit the free aperture upstream of the sample, and, can be an issue. especially when combined with laser heating for high-pressure setups.

The analyzers are the heart of the IXS beamline, being the most complex and difficult to fabricate optical component. This is because they should accept a large solid angle from the sample to preserve flux so their angular acceptance, approaching ~10x10 mrad$^2$, needs to be much larger than the typical angular acceptance of Bragg reflections. They will be discussed in detail below, as will the optical concerns that lead to the rather large (6 to 12 m) two-theta arms of most spectrometers. In order to improve efficiency, beamlines with spherical analyzers use an array of between 4 and 24 analyzers (eventually 42 at BL43LXU) which allows parallelization of data collection. For disordered materials, where only the magnitude of the momentum transfer is relevant, the gain is linear in the number of analyzers. For solids, where one often focuses on measurements along high symmetry directions, the value of multiple analyzers is more complex, as only some of the analyzers may be close enough to the symmetry direction (this is discussed in detail in section 4.6). The choice of analyzer can also impact flexibility, as spherical analyzers can be operated at any allowed harmonic of the fundamental backscattering reflection (e.g. the same analyzer may be used for 6 meV resolution at the Si(888) as for 1.5 meV resolution at the Si(11 11 11) or sub-meV resolution at the Si(13 13 13)). For the PSC setup, the present instruments are still at the level of a single analyzer per instrument, and it is not yet clear what level of parallelization will be possible in the longer term, as the cost per channel remains high, and the space needed for the collimating mirrors, and their mechanics, close to the sample is a limiting constraint.

## 4.2 Dynamical Bragg Diffraction

IXS spectrometers rely heavily on diffraction of x-rays by perfect crystals for good energy resolution. We now take a brief interlude to introduce some of the main relations important for these optics, including estimates of the angular and energy acceptance of perfect crystals based on dynamical theory. Dynamical diffraction differs from the scattering theory discussed in section 2 by concentrating on only the elastic term (the first term in eqn. (15)), and extending the treatment to explicitly include multiple scattering and boundary conditions based on Maxwell's equations. Some of the main results may be obtained using a largely heuristic approach to multiple scattering related to the work of Darwin and Prins (see, e.g.,(James, 1962)) while a wave based treatment (Batterman & Cole, 1964) (Authier, 2003) can be more general. Other useful references include (Pinsker, 1978) and, especially as regards high-resolution monochromatization, (Shvyd'ko, 2004). In the context of diffraction theory, the single-scattering or Born-approximation treatment, equivalent to that in section 2, is called "kinematic" diffraction: in kinematic theory, the multiple scattering - the interference between the incident and scattered waves within the crystal - is neglected. However, for efficient optics, one requires that crystal reflectivity be near unity, and, in that limit, dynamical theory is required. Here we largely follow (Batterman & Cole, 1964)[+]. We treat only the case of a single excited Bragg reflection (the "two-beam" case) neglecting multi-beam cases (where the Bragg condition is simultaneously satisfied, or nearly satisfied, for more than one set of lattice planes). Multiple beams (Colella, 1974) can be important near backscattering (Sutter, et al., 2001) but we assume they have been deliberately avoided.

The most useful relationship, beyond Bragg's law, $\lambda = 2d\sin(\theta_B)$, in diffraction, is the relation between a small angle change, $\Delta\theta$, and a small energy shift, $\Delta E$, that follows immediately from differentiation of Bragg's law, namely,

$$\Delta\theta = \tan(\theta_B)\,\frac{\Delta E}{E} \qquad \textbf{(39)}$$

This also shows that, for a given energy bandwidth, the largest angular acceptance can be obtained by going close to backscattering, $\theta_B \approx \pi/2$, hence the reliance of most spectrometers on a backscattering geometry. In that case, one can write,

$$\varepsilon \approx \delta\,\Delta\theta \qquad \left(\varepsilon = \frac{\Delta E}{E} \quad \theta_B \equiv \frac{\pi}{2} - \delta,\ \delta \ll 1\right) \qquad \textbf{(40)}$$

i.e. the contribution of an angular divergence $\Delta\theta$ to the fractional energy resolution $\varepsilon$ is just the product of that divergence with the deviation from exact backscattering, $\delta$. Dynamical diffraction theory allows one to calculate the absolute reflectivity and bandwidth of Bragg reflections. The ratio of reflected to incident intensity (the absolute squares of the amplitudes of the plane waves[*]) for a wave diffracted from a thick crystal near to the Bragg condition (incident and reflected beams on the same sides of the crystal) can be written (Batterman & Cole, 1964)

$$R = \left|\eta \pm \sqrt{\eta^2 - 1}\right|^2 \left|\frac{F_\tau}{F_{-\tau}}\right| \qquad \textbf{(41)}$$

where $\eta$ is a parameter that indicates the deviation from the Bragg condition (and is complex due to absorption in the material). The sign in front of the root is chosen to preserve $|R| \le 1$. Specifically, and referring to figure 8,

$$\begin{aligned}\eta &= \frac{b(\theta - \theta_B)\sin(2\theta_B) + \frac{1}{2}\Gamma F_0(1-b)}{\Gamma\sqrt{|b|\,F_\tau F_{-\tau}}} \\ &= \frac{\Gamma F_0 - 2\varepsilon_B}{\Gamma\sqrt{F_\tau F_{-\tau}}} \qquad \left(\text{Exact Backscattering}^\dagger,\ \text{b=-1}\right)\end{aligned} \qquad \textbf{(42)}$$

---

[+] Note that this reference takes $k = 1/\lambda$, dropping the factor of $2\pi$ that is commonly used today.

[*] Sometimes the reflectivity is defined in terms of intensity *per unit area* and this will lead to an additional factor of |b| in eqn. (41).

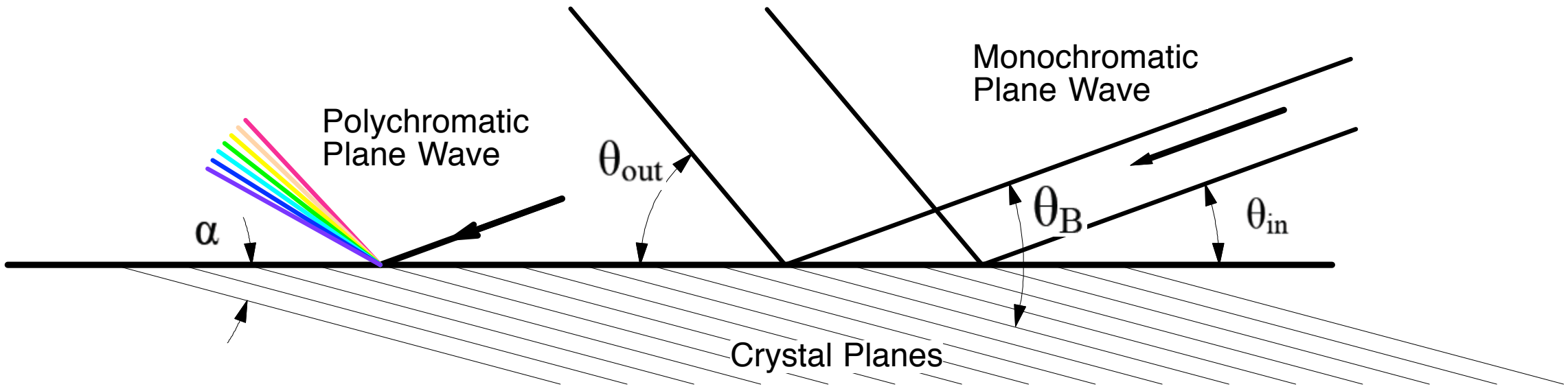

**Figure 8. Bragg diffraction.** An asymmetric case is shown, with the asymmetry angle α giving the inclination of the crystal planes as compared to the crystal surface. The right side shows the reflection of a monochromatic plane wave, where the beam size is increased by the asymmetry factor, -1/b (-1<b<1 for the case shown). The left side shows the reflection of polychromatic wave, which, for an asymmetric reflection leads to dispersion.

where $b = -\sin(\theta_{in})/\sin(\theta_{out}) = -\sin(\theta_B - \alpha)/\sin(\theta_B + \alpha)$ is the asymmetry parameter for a crystal when the surface makes an angle of $\alpha$ with respect to the Bragg planes, $F_0, F_\tau, F_{-\tau}$ are the structure factors in the forward, Bragg and reverse-Bragg directions and $\Gamma = r_e\lambda^2/\pi V$ (V the volume of the unit cell). Here we will focus exclusively on Bragg reflections, where the incident and reflected beams are on the same side of the crystal (b<0) as, for the present considerations, the Bragg case is more efficient than the Laue case. The structure factors are essentially those given in eqn. (16), or the phased sum of the charge density over the unit cell. However, now, with the multiple scattering and interest in absolute reflectivity, one can not neglect the small imaginary parts due to photoelectric absorption - the imaginary parts of $F_0, F_\tau$ must be included, usually by including imaginary contributions to the form factors (e.g. $f = f_1 + i f_2$ or $f \to f + f' + i f''$ depending on notation). Note, often $F_\tau \approx F_{-\tau}$ is a good approximation, and is rigorously true for centrosymmetric crystal, but can fail in more complex materials such as quartz. The equations are for sigma-polarized radiation - electric field perpendicular to the scattering plane - for pi polarization one takes $F_\tau \to \cos(2\theta_B)F_\tau$ (cf. the polarization dot-product in the Thomson cross section, eqn. 1). Finally, the first eqn in (42) is only valid if the Bragg angle is less than 90 degrees, and at incident, and outgoing, angles greater than the critical angle for external reflection. A version that is good at exact backscattering‡ can be derived using eqn (39) to change the free variable from angle to the fractional energy shift, $\varepsilon_B = (E - E_B)/E_B$, $E_B = hc/2d$ giving the second equation of (42). We also note for comparison to other work, that sometimes the scattering is expressed in terms of a dielectric susceptibility or polarizability given by $\chi = -\Gamma F$ and the relation with the index of refraction for x-rays is given by $n = 1 - \delta - i\beta = \sqrt{\varepsilon} = \sqrt{1 + \chi_0} \approx 1 + \chi_0/2$ or $\delta = -\chi_0/2 = \Gamma F_0/2$ and the (1/e) absorption length is related to the imaginary part of the scattering by $\lambda_{abs} = 1/\mu = \lambda/4\pi\beta = \lambda/2\pi\left|\mathrm{Im}\{\chi_0\}\right|$ and the critical angle is given by $\Theta_c = \sqrt{2\delta}$ .

The region of high reflectivity, $|\eta| < 1$, is called the **Darwin width** ( $\Delta\Theta_D$ or $\Delta E_D$ in angle or energy) of the reflection - the full width at half maximum of the reflectivity curve tends to be slightly, about 6%, larger than the Darwin width, though the precise ratio depends on the magnitudes of the real and imaginary parts of $F_0, F_\tau$. The Darwin width is given by

$$\Delta\Theta_D = \frac{2\Gamma\sqrt{F_\tau F_{-\tau}}}{\sqrt{|b|}\sin(2\theta_B)} \quad or \quad \frac{\Delta E_D}{E} = \frac{\Gamma\sqrt{F_\tau F_{-\tau}}}{\sqrt{|b|}\sin^2(\theta_B)} = \frac{4r_e d^2\sqrt{F_\tau F_{-\tau}}}{\pi V\sqrt{|b|}} \qquad \textbf{(43)}$$

‡ Carefully speaking, exact backscattering is usually a multi-beam situation, with more than one Bragg reflection excited. The second equation in (42) should then be interpreted as an easy way to estimate the rocking curve shape, bandwidth and peak reflectivity *near* to backscattering when multi-beam excitations have been avoided.

where the fractional energy width is seen to be mostly a constant for any given reflection. The second equation of (43) can be obtained from the first using eqn (39). At large deviation from the Bragg condition one has

$$R \sim 1/4\eta^2 \qquad \left(|\eta| \gg 1\right) \qquad \textbf{(44)}$$

Thus, the tails of the reflectivity are approximately Lorentzian (but the Darwin width, Δη=2, is double what one would expect from the decay of the tails). In general, this means that the resolution of any single-reflection monochromator or analyzer will have Lorentzian tails, though shaper tails can be created by passing through multiple reflections.

"Asymmetric diffraction" refers to the case when the crystal planes are not parallel to the surface of the crystal (|α|>0 in figure 8) and has $|b| \neq 1$. In this case, the Bragg reflection can both change the divergence of the beam and introduce an energy-angle correlation, dispersion, into the output beam[§]. If one considers perfectly monochromatic radiation, the asymmetry leads to change in the size of the beam normal to its propagation direction by a factor of -1/b after the reflection. The asymmetry also changes the acceptance of the Bragg reflection, the Darwin width (in energy or angle) by a factor of $1/\sqrt{|b|}$ for the input beam and a factor of $\sqrt{|b|}$ on the output side. This means that, for monochromatic radiation whose divergence is smaller than the Darin width, an asymmetric Bragg reflection with |b|<1 can be used to collimate in angle and expand the size, or, with |b|>1, to reduce size and increase the divergence of an x-ray beam. There is a conservation law in that the product of the change in divergence and the beam size is unity (for monochromatic beams within the acceptance of the crystal). One notes that with increased angular acceptance (|b|<1) comes also increased energy acceptance (the b-dependence is the same in both parts of eqn. (43)).

One can also consider the effect of asymmetry on a parallel beam of *polychromatic* radiation. Here one finds that the divergence of the output beam will always be increased as different energies will be reflected with different scattering angle. Consideration of the equation $\mathbf{k}_{out}=\mathbf{k}_{in}+\tau$, where τ is the Bragg momentum transfer (|τ|=2π/d) directed normal to the crystal planes, gives $\Delta\theta = (1+b)\Delta\lambda/\lambda$, however, as pointed out by Shvyd'ko (see section 2.2.4 of (Shvyd'ko, 2004)) the shift of the vacuum wave vector is not $\tau$ but includes a small added component normal to the crystal surface. The correct relation then becomes (eqn. 2.140 of (Shvyd'ko, 2004))

$$\Delta\theta = \frac{\Delta\lambda}{\lambda}(1+b)\tan(\theta_{in}) = -\frac{\Delta E}{E}(1+b)\tan(\theta_{in}) \qquad \textbf{(45)}$$

For example, a parallel beam having broad ($\Delta E > \Delta E_D$) bandwidth incident on an asymmetric crystal, on output, will have an angle-energy correlation over a bandwidth $\Delta E_D$ and a divergence $|\Delta\theta| = \tan(\theta_{in})|1+b|\Delta E_D/E$. The angle-energy correlation, or induced dispersion, of the beam can then be used to aid monochromatization of a beam using a narrow angular aperture (another crystal) after the asymmetric crystal (see discussion in section 4.7).

Wave fields in the crystal are exponentially damped as one penetrates inward, with a characteristic length of damping (normal to the crystal surface) for the intensity given by the **extinction length** which, for $-1 \le \eta \le 1$, is

$$\lambda_{ext}(\eta) = \frac{\lambda \sin(\theta_B)}{2\pi\ \Gamma F_\tau}\left(1-\eta^2\right)^{-1/2} \qquad \textbf{(46)}$$

Then noting that the average value of $\left(1-\eta^2\right)^{-1/2}$ over the range $-1 \le \eta \le 1$ is just $\pi/2$ one has $\bar{\lambda}_{ext} = \lambda \sin(\theta_B)/4\ \Gamma F_\tau$. If one then defines the number of planes contributing to the reflection as $N_p = \bar{\lambda}_{ext}/d$, one can write Darwin width in energy (eqn 43) as [*]

$$\frac{\Delta E_D}{E} = \frac{1}{2\,N_p} \quad \left(\text{Symmetric Bragg Reflection}\right) \qquad \textbf{(47)}$$

---

§ For a symmetric reflection, b=-1, for any plane wave, the angle of incidence will be the same as the angle of reflection, with the Bragg reflection only reducing the intensity.

* Note that sometimes the extinction length is just taken as the value at η=0 which will lead to replacing the factor of 2 by π in eqn (44),

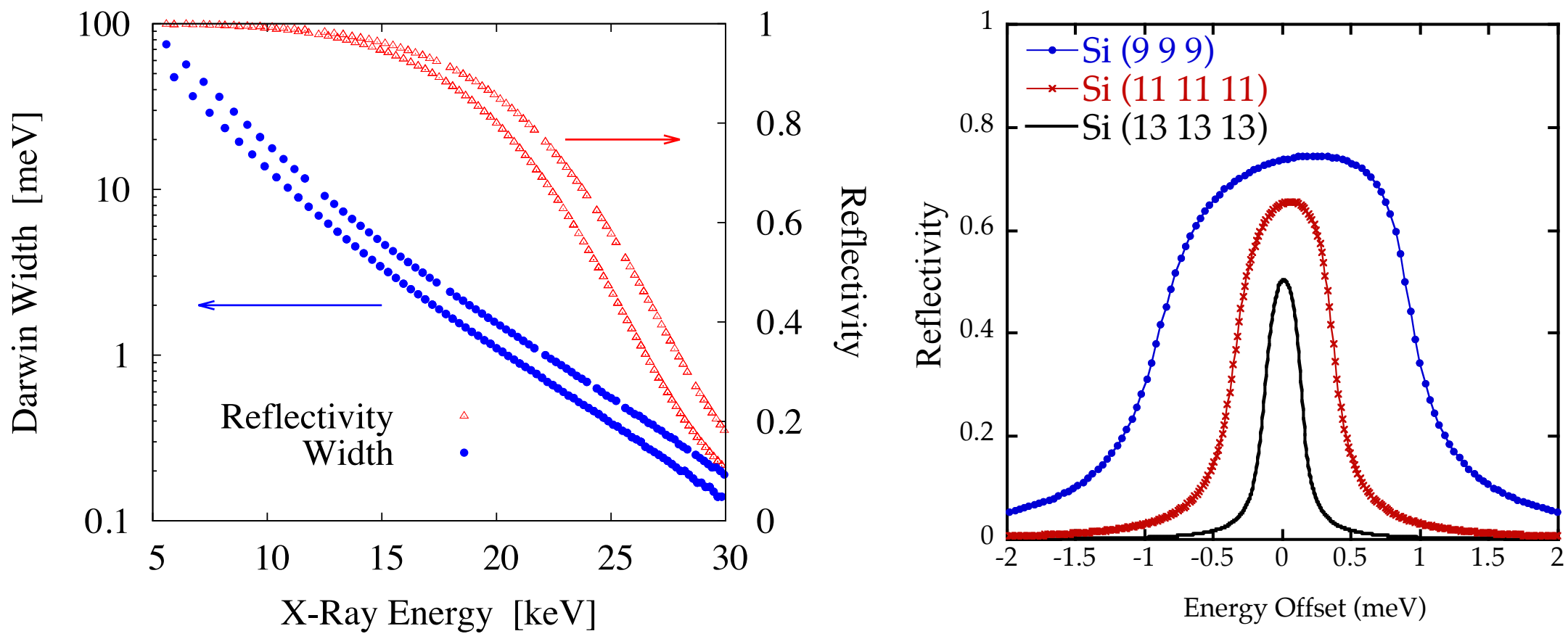

**Figure 9: Calculated results for symmetric backscattering reflections in silicon (at 300K)**: the peak ( $\eta = 0$ ) reflectivity and energy width (left panel) and reflectivity as a function of energy (right panel) for several commonly used reflections. The two traces on the left are from the two classes of reflection in silicon - even reflections where all atoms in a unit cell contribute in phase and odd where the phasing is not perfect, reducing the reflectivity and the bandwidth. Even reflections generally have larger bandwidth and peak reflectivity.

i.e.: within a factor of two, the fractional bandwidth of a symmetric Bragg reflection is just $1/N_p$ where $N_p$ is the average number of planes contributing to the reflection. Thus, to achieve good energy resolution, one would like a reflection with a large number of planes contributing, or weak scattering for each plane (see also the next section).

## 4.3 Silicon & Non-Silicon Optics

The following sections discuss optics for IXS, and, in every case they will be silicon. The reason is that excellent quality crystals are available for modest prices: highly perfect Si crystals (lattice spacing variability $\Delta d/d<10^{-8}$) over macroscopic - 100 mm scales) can be purchased for ~$1 to $5/g. Furthermore, silicon is a light, low-Z, material, so the x-ray absorption is not severe in the 9-26 keV range, and the thermal conductivity (about 1/3 of copper) is also good. There is also a huge amount of experience in processing silicon: crystal growth, cutting, etching, strain-free polishing of flat surfaces, etc. Of course, especially for high-resolution optics, there is considerable know-how and experience required to preserve the high-resolution, but the essential crystal quality is sufficient over large lengths scales, at modest prices. Thus, practically, essentially all optics, and especially those for ~ 1 meV resolution, are silicon.

Figure 9 and table 5 show some of the properties of backscattering reflections in silicon as these are the most relevant to our discussion of high-resolution spectrometers. (Here, as is typically done, the reflections are indexed in units of the 8-atom cubic unit cell, a=5.43Å, as opposed to the 2-atom primitive cell). Generally, as the d-spacing becomes small (corresponding to a larger backscattering energy), the scattering per plane becomes less and the bandwidth gets narrower. At the same time, the extinction length becomes larger, becoming comparable to the absorption length, and the peak reflectivity falls. Typical reflectivity curves are shown in figure 9b. One should note that at the high momentum transfers of high-energy backscattering, the effect of the Debye-Waller factor is not negligible. In the calculations that are shown, the factor has been calculated using a Debye model (second line of eqn. (22), $\Theta_D$=534K) however careful measurements (Deutsch, et al., 1989) have suggested this slightly under-estimates W at higher momentum transfers, and inclusion of that level of correction can make small (~10%) differences in expected energy widths in figure 9b and a slight reduction in peak reflectivity.

| Reflection | E (keV) | $\Delta E_D$ (meV) | Spectrometer Resolution (meV) | $\overline{\lambda}_{ext}$ (mm) | $\lambda_{abs}$ (mm) |
|---|---|---|---|---|---|
| (8 8 8) | 15.816 | 4.1 | 6 | 0.075 | 0.51 |
| (9 9 9) | 17.793 | 1.8 | >2.7 | 0.17 | 0.73 |
| (18 6 0) | 21.657 | 1.1 | 2.1 | 0.27 | 1.31 |
| (11 11 11) | 21.747 | 0.77 | >1.25 | 0.40 | 1.32 |
| (12 12 12) | 23.724 | 0.71 | 1.3 | 0.44 | 1.71 |
| (13 13 13) | 25.701 | 0.32 | >0.75 | 0.96 | 2.15 |
| (15 15 15) | 29.656 | 0.13 | - | 2.41 | 3.22 |
| (8 0 0) | 9.132 | (25.9) | >0.62 | (0.012) | (0.10) |

**Table 5:** Backscattering reflections in silicon (300K, a=5.43108Å, $\Theta_{Debye}$=534K) used in some high-resolution spectrometers and properties in a symmetric Bragg reflection geometry. The last line is for a reflection used in a PSC geometry (section 3.6) with extreme asymmetries so that the nominal Darwin width, etc., are not relevant and are given in parenthesis.

Non-silicon optics are interesting, primarily as analyzers, for at least two reasons. On the one hand, materials with different Debye temperatures and structure factors are interesting to change the scattering per plane and thus the energy resolution at a given energy. On the other, materials with different lattice constants, or lower symmetry, than silicon, allow access to backscattering at different energies. The first is interesting to help tailor a desired energy resolution to the available energy (e.g. at a low energy storage ring, fluxes can be higher at low energy) while the latter is particularly interesting in the context of potential high-resolution resonant IXS (RIXS) where one would like to choose a backscattering energy very close to a particular atomic resonance or edge. Crystals that have generated significant interest are sapphire (Shvyd'ko & Gerdau, 1999) and quartz (Sutter, et al., 2005), which are both trigonal materials, with relatively large c-axis lattice constants, and offer many reflections to choose from. Quartz has a lower Debye temperature, so is probably more interesting at lower energies (say <~15 keV) where the bandwidth can be chosen to be at the ~meV level and the reflectivity is not too small, while sapphire, being harder with a higher Debye temperature, is more interesting at higher energies (of course, cooling or heating the crystal can also change the reflection properties). Of critical importance, few-cm-scale highly perfect quartz crystals are commercially available with better than 4 meV resolution over ~10 cm$^2$ areas, as demonstrated more than a decade ago (Sutter, et al., 2006) whereas sapphire crystals tend to be non-uniform (see, e.g. (Sergueev, et al., 2011) where after careful examination ~2meV resolution was possible over areas of a few square mm). Spherical analyzer crystals have been fabricated with sapphire (Yavas, et al., 2007) with 31 meV resolution, and quartz with 25 (Ketenoglu, et al., 2014) and 10 meV (Said, et al., 2018) resolution, while a PSC design using quartz has achieved ~10 meV resolution (Kim, et al., 2018). Additional discussion of materials can be found in (Yavaş, et al., 2017)). It is this authors belief that, while not yet implemented, a PSC based design with multiple backscattering reflections may be an effective instrument for a few, 3-10, meV resolution spectrometer.

## 4.4 High Resolution Monochromator (HRM) Design

IXS requires a meV-scale bandwidth beam incident on the sample. However, the bandwidth of radiation from an undulator insertion device is usually several hundred eV*. This means that at least two steps, and sometimes more, are used to reduce the bandwidth to the meV level needed for IXS. For the first optical element, where the power is ~0.1 to 1 kW in a transverse beam size of ~1 mm$^2$, the power load problem is especially severe (1 kW/mm$^2$ is sufficient, at normal incidence, to quickly drill a hole in copper), but, this problem is common across nearly all synchrotron beamlines and nearly-standard, if facility specific, "high-heat-load" monochromator solutions have been developed. We do not discuss them in detail here, but mention the solution has usually either been to cryogenically cool silicon to approach the zero in its thermal expansion at ~120K or to use thin diamonds which have low absorption and excellent thermal conductivity (so most of the power is transmitted through the diamond, and the power that is absorbed does not severely distort the crystal). After the high-heat-load monochromator one typically has an x-ray beam that is ~eV in bandwidth with a power of ~ 0.1 to 1 W. The power in this "monochromatic" or "eV" beam is much more manageable than the kW levels of the pink beam, but still an issue.

Two approaches are applied to reducing the bandwidth of the eV beam to the meV level needed for IXS: an in-line monochromator or a single backscattering monochromator. They are different approaches to dealing with the fact the divergence of x-ray beams, from even very high brilliance synchrotron radiation sources, is significantly larger than the typical angular acceptance of a meV-resolution Bragg reflection. The required level of energy resolution is $\varepsilon \sim 10^{-7}$ and, usually, smaller, so, considering eqn. (40), the angular acceptances of interesting Bragg reflections are at the (sub-) micro-radian level for δ~100 mrad, while synchrotron x-ray beam divergences are ~10 μrad. The acceptance increases as δ becomes small (as the Bragg angle approaches 90 degrees or exact backscattering) and the different monochromator solutions correspond to choosing either (1) to push δ as small as possible, δ~0.3 mrad, which then allows one to accept the full x-ray divergence from the source, or (2) choosing to keep δ ~ 100 mrad and use additional optics to further collimate the x-ray beam before it impinges on the high-resolution crystal. It is also possible to use an extreme angular dispersion scheme (as discussed in section 4.8 below). We discuss each option in more detail below.

Backscattering, as is used at ESRF and SPring-8, is the conceptually simpler option (see, e.g. (Graeff & Materlik, 1982) as an early demonstration of backscattering). The choice to work at δ~3x10$^{-4}$ gives angular acceptances of >~15 μrad, even for the most severe case (~0.3 meV monochromator resolution at 25.7 keV) and is generally comfortably larger. However, energy scans are then only possible by changing the Bragg plane spacing, d, using thermal expansion. (Scanning the Bragg angle, changing δ, changes the bandwidth, and indeed the entire beamline geometry).

* Beam from an undulator usually has a complex frequency spectrum with several peaks and is called "pink" to distinguish it from the broader "white" beam (without such narrow peaks) that is provided by wigglers or bending magnets

Fortunately, the thermal expansion of silicon is conveniently sized, α=2.6 ppm/K near room temperature, corresponding, for example, to ~18 mK/meV (56 meV/K) at 21.75 keV. Thus, with ~mK stability and few degree scan ranges, one can effectively scan the interesting region of energy transfer with a backscattering monochromator. The other immediate issue regards the geometry, as when δ is small the reflected beam lies nearly on top of the incident beam. This means additional optics must be used to separate the two beams to allow enough space at the sample position to install samples, sample environments, and motion stages. At ESRF (figure 7a) this is done by the using focusing mirror (operating at few mrad grazing angle, just downstream of the backscattering mono) and a long travel distance. At SPring-8 (figure 7b), a pair of "offset" crystals are used to allow as much as ~400 mm between the two beams. Using offset crystals introduces an additional alignment component, but then makes it possible to install large sample setups, and to allow stronger focusing in the default setup. Also, the use of offset crystals also allows one to reduce the power load on the backscattering monochromator, as is discussed below.

One should mention a few important practical details about the backscattering setup: a backscattering monochromator must be thermally isolated (e.g. mounted in vacuum) on top of stages that allow one to carefully control the angle and temperature. Some care in design is needed both to allow reliable temperature control and to minimize the angular drift of the crystal when its temperature is changed: the sample position is typically several 10's of meters from the backscattering mono so that ~microradian stability is required. Similarly, this level control is needed over the crystal angles for initial alignment. Finally, the power load of the monochromatic beam incident on this monochromator can adversely affect performance in two ways: it introduces a thermal gradient across the crystal so that the temperature measured by the sensor (usually at least 20 mm away from the beam position) is different than the temperature where the beam hits, and, if the power is large enough, there may even be a significant temperature gradient *within* the size of the x-ray beam spot. The first means that the incident power on the crystal must be considered when determining the temperature at the beam spot, while the second can lead to worsening of the resolution as the silicon lattice spacing may change over the x-ray-illuminated volume. In practice these problems are avoided by reducing the power density on the backscattering crystal by first going to a grazing incidence setup (as seen in figure 3) as employed both at SPring-8 (Baron, et al., 2002) and then at ESRF (Verbeni, et al., 2003) and/or using additional optics. At SPring-8, the bandwidth is also reduced by using an asymmetrically cut higher order reflection in the offset crystal (e.g. Si(220) or Si(400)), and/or other optics.

In-line monochromators build on designs and design concepts that were largely developed for nuclear resonant scattering (Ishikawa, et al., 1991)(Toellner, et al., 1992)(Chumakov, et al., 1996)(Yabashi, et al., 2001)(Baron, et al., 2001)(Toellner, et al., 2002). For those experiments one must build optics to match the nuclear resonance energies, so operating within <~ 1 milliradian of exact backscattering, at least in silicon, is generally not possible. The monochromator then usually uses crystals that are asymmetrically cut to manipulate the divergence and bandwidth of the x-ray beam (c.f. the dependence of the energy and angular bandwidths on the symmetry parameter, b, in equation 43). We do not track those developments in detail here, but focus instead on how this has been done for the APS IXS instruments - as this will introduce most of the main issues. We also note that a highly dispersive HRM design is possible (e.g. UHRIX at APS (Shvyd'ko, et al., 2014)) and, as this design is very similar to the PSC analyzer design, we refer the reader to that section (essentially, remove the collimating mirror and optimize for reduced incident divergence). The APS monochromator design (Toellner, et al., 2011) uses 6 bounces that are arranged in pairs, with two bounces collimating the beam to the 0.3 μrad level, two bounces to reduce the bandwidth and two bounces then undoing the collimation (and returning the beam size to near the incident size). The layout can be seen in figure 10. The first and last pair of crystals are mounted in a channel-cut geometry, but are in fact flat crystals connected to a flexure based stage, forming an "artificial" channel cut (Shu, et al., 2000). This is important to allow compensation for un-even heat load on the crystals, as the beam intensity (bandwidth) changes from one reflection to the next. It is also convenient because a fixed channel cut crystal is difficult to polish as it must be largely be done by hand, while flat crystals can be polished by machine. However, the flexure stage introduces a possible point of instability, and the APS designs are actively stabilized. The high-resolution crystal is a conventional channel-cut, held at low temperature (again, near the zero in thermal expansion) to avoid thermal effects, and to improve its peak reflectivity (by making the Debye-Waller factor, $e^{-2W}$ closer to unity*). Each pair of crystals is mounted on very high (<50 nrad) resolution angular stages, with, again, stabilization.

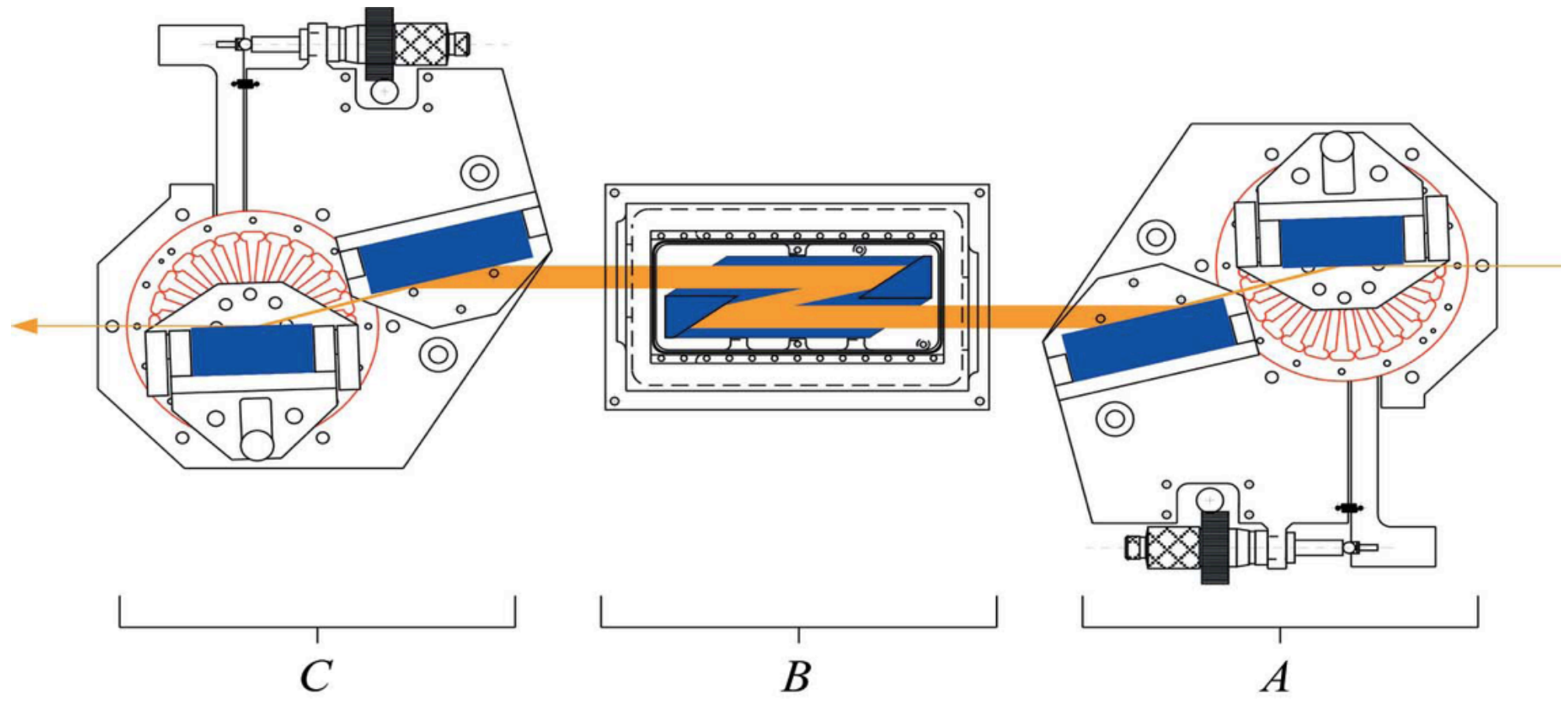

**Figure 10. In-line monochromator design** (after (Toellner, et al., 2011)). Beam travels from right to left through (A) an artificial channel-cut collimating crystal at room temperature, (B) a high resolution channel-cut crystal at cryogenic temperature and (C) an artificial channel cut crystal, at room temperature, to recover the incident beam size. (reproduced with the permission of the International Union of Crystallography, http://journals.iucr.org/)

## 4.5 Spherical Analyzers

The analyzers are the heart of the IXS beamline, and the most difficult optic to fabricate. The goal of the analyzer is essentially the same as the high-resolution monochromator - to select a very narrow bandwidth of radiation - *but,* instead of accepting a highly collimated beam from an insertion device, the analyzer sees a highly divergent beam from the sample. The desired angular acceptance is set by the experiment, but, practically is usually between about 0.5 and 10 mrad in each direction, with smaller acceptances being interesting for experiments where momentum resolution is especially important, but, often, larger acceptances are desirable to increase count-rates. However, even the small end of this angular range is much larger than the acceptance of a Bragg reflection, even at Bragg angles near to backscattering. Then, one of two options can be considered: either a figured analyzer where the shape is chosen so that the Bragg angle does not vary significantly over the desired angular acceptance, or one can collimate the divergent beam from the sample. We will discuss the first of these here, which accounts for most of the presently operating user spectrometers, and will discuss the second, "post-sample collimation" (PSC) setup later.

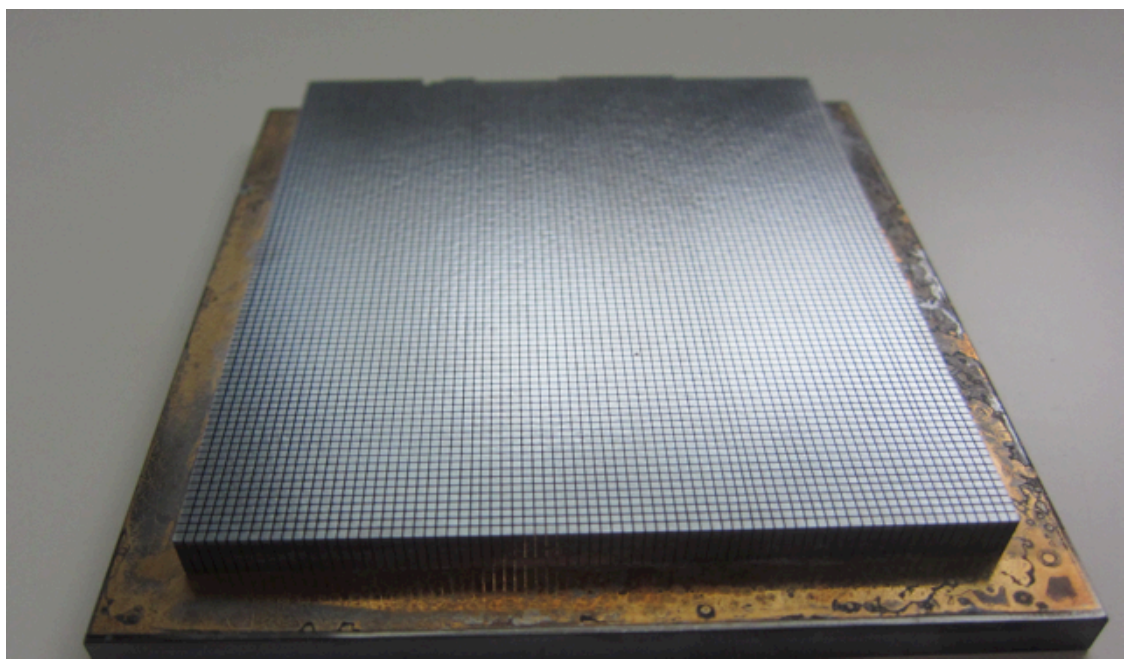

**Figure 11.** A pixelated analyzer crystal (9x9cm2). Often circular substrates are used, however, for the temperature gradient, discussed below, a rectangular substrate is easier.

The figured analyzer dates back at least to the discussion of IXS in the 80's (Dorner, et al., 1980)(Fujii, et al., 1982), where, considering the necessity of operating nearly at backscattering, the appropriate shape is immediately spherical. Bending a crystal introduces strain so that this is not an option, and the universal solution has been to bond many small (mm-scale) crystallites to a curved substrate (figure 11) - though the details of his process vary significantly from instrument to instrument. Published papers describing fabrication include (Masciovecchio, et al., 1996)(Masciovecchio, et al., 1996)(Baron, et al., 2001)(Sinn, 2001)(Verbeni, et al., 2005)(Said, et al., 2011) while additional discussion can be found in (Burkel, 1991). The design of spherical analyzers requires an intimate understanding of both the silicon crystal processing and all the optical parameters of the setup, as it is the detailed interplay of these that determine the

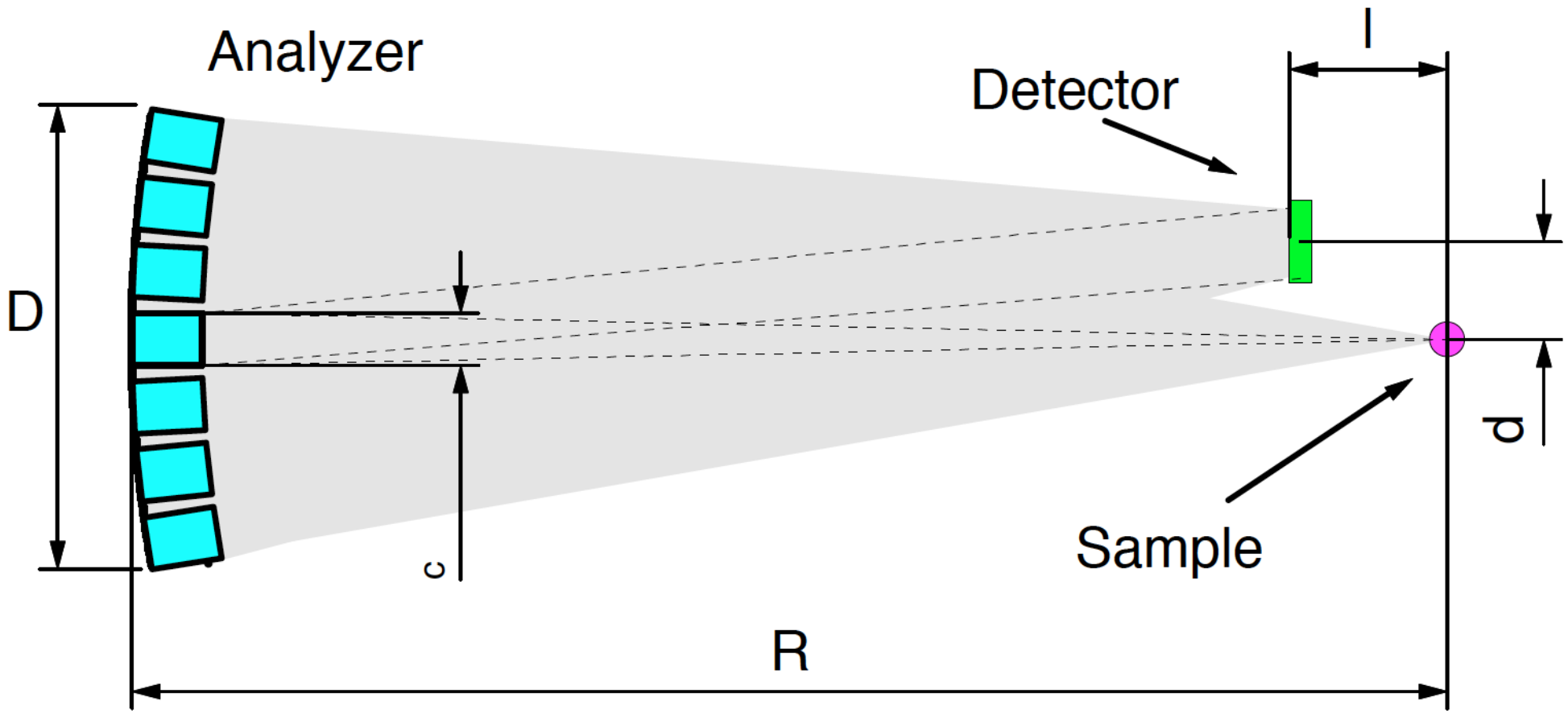

**Figure 12. Spherical analyzer geometry** - not to scale: typically R~10m, d~5mm and l~200 mm, c~1mm). See text.

* Reducing temperature increases both reflectivity, and the bandwidth. However, the can be a net gain in reflectivity/bandwidth.

performance, especially as one begins to focus on the limits of achievable resolution. While a comprehensive discussion is beyond the scope of the present paper (and, indeed, design is now often accompanied by sophisticated ray-tracing codes) we will try to indicate the main issues.

Consideration of Bragg's law (e.g. eqn. (39)) and figure 12 immediately leads one to a geometric contribution to the analyzer resolution that scales as

$$\left(\frac{\Delta E}{E}\right)_1 \approx \frac{cd}{2R^2} \qquad \textbf{(48)}$$

where $d/2R$ is the deviation from exact backscattering, δ, and $c/R$ is the divergence accepted by a single crystallite. Thus, from the point of view of minimizing this contribution to the energy resolution it is clearly desirable to have R large, and the transverse crystallite size, c, and the detector offset, d, both small. However, the analyzers are usually 3 to 7 mm thick[+] so that limits of silicon processing[*] (e.g. saw blade thickness typically > 50 μm, etching to remove strain after sawing > 2x30 μm) typically set groove widths >~0.1 mm, so that the crystallite size, c, is typically not much smaller than 1 mm to prevent loss of a very large fraction of the solid angle to the grooves (0.15 mm groves on a 1mm pitch lead to surface area loss of $1\text{-}0.85^2$ = 28%). Meanwhile, a desire to have the space around the sample, and requiring some entrance to the vacuum flight path (not shown) and slitting typically puts l~200 mm, which forces, e.g., d~5 mm for a 10 mrad acceptance analyzer (to prevent blocking the outgoing beam). Then taking R~10m one finds this contribution to the resolution is $\Delta E/E \approx 2.5x10^{-8}$ or 0.5 meV at 20 keV. This sets the scale then for the spectrometer. The $R^2$ dependence of this term strongly favors large arm radii, however the practical issues of space on the experimental floor, as well as expense, tend to push for smaller radii, with the operating spectrometers having arm radii varying from R=6 to 10 m (the spectrometer at ESRF ID28 once had an option for R=12m), pixel sizes of c~1mm and detector offsets, d~few mm. (Carefully speaking, one might add a source size contribution to eqn 48, however, in most spectrometers the beam focal spot sizes tend to be a small fraction of a mm, which, except for the effect of a projection through a thick sample, tends to be negligible compared to the ~mm scale crystallite sizes).

Another contribution to the energy resolution comes from the choice of l>0, as is needed to keep some space available for sample environments. This stems from the small variation in Bragg angle over the analyzer surface that is induced by violation of the exact Roland-circle (l=0) focusing condition. It has been called the "demagnification" contribution (Burkel, 1991) and scales as (Burkel, 1991) (Ishikawa & Baron, 2010)

$$\left(\frac{\Delta E}{E}\right)_2 \approx \Omega \frac{dl}{4R^2} \qquad \textbf{(49)}$$

where $\Omega = D/R$ is the solid angle of the analyzer in the scattering plane. Taking $\Omega$=10 mrad for a 10 cm analyzer at R=10m radius, and, again, l, d, and c = 200, 5 and 1mm, respectively, one finds this also gives a contribution of

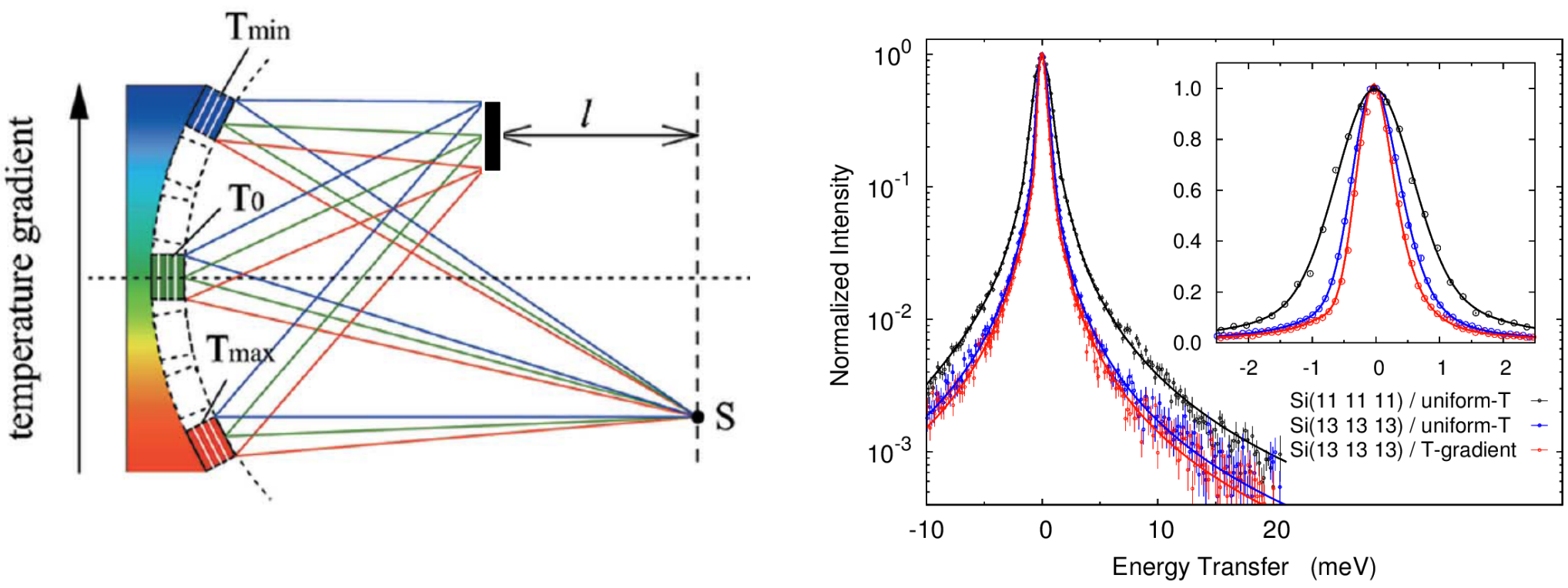

**Figure 13**: Spherical analyzer with temperature gradient and resolution functions (based on results in (Ishikawa & Baron, 2010) (Ishikawa, et al., 2015)) . The three resolution correspond to 1.5 meV at the (11 11 11) without the gradient (this can be improved to 1.25 meV with a gradient - not shown) and the 0.9 meV improved to 0.75 meV resolution at the (13 13 13))

[+] This thickness is needed to preserve good resolution. See (Said, et al., 2011).

[*] For thinner (<~0.5 mm thick) analyzers, one can do the "dicing" using reactive ion etching, but this is not possible for thicker where severe aspect ratios (~50:1) are needed to keep the grove width small. (Finkelstein, 2005)

$\Delta E/E \approx 2.5x10^{-8}$ or 0.5 meV at 20 keV. However, this can largely be removed by applying a temperature gradient over the surface of the analyzer to change the lattice spacing to compensate for the variation in Bragg angle, allowing a record spectrometer resolution of 0.75 (2) meV to be obtained using the Si(13 13 13) back reflection at 25.7 keV for a total measured resolution, including monochromator, analyzer and geometric contributions of $\Delta E/E_{total} < 3x10^{-8}$ (Ishikawa, et al., 2015).

It is possible to consider using a position sensitive detector instead of a single element detector, as has been applied to setups operating at lower (~10 to 100 meV) resolution, in a so called "dispersion compensation" setup (Huotari, et al., 2005). This allows one to replace the angular scale given by the crystallite size in eqn (48) with that of the detector pixel size, c/R -> p/2R, where p is the detector pixel size (convolved with the source size) in the scattering plane. Taking a relatively conservative value of p~0.2 mm would allow a factor of 10 increase in d from 5 to 50 mm for the same contribution, $\Delta E/E \approx 2.5x10^{-8}$, to the resolution as is discussed above. However, this method also requires one strictly obey the Roland circle condition, l=0, and it is not clear that d=50 mm is then enough for interesting sample environments. Of course, if one looks to the future, where one might hope for p~0.02 mm, and a corresponding d=500 mm, then this becomes much more attractive. However, it does require that the spot size on the sample, as projected onto the direction of the analyzed beam, be of similar (really, smaller) scale, and may also require careful reconsideration of the required perfection of the analyzer crystallite alignment, and, indeed the entire optical setup, with, for example, deviation of the alignment of the analyzer crystallites out of the nominal scattering plane possibly becoming a serious limitation.

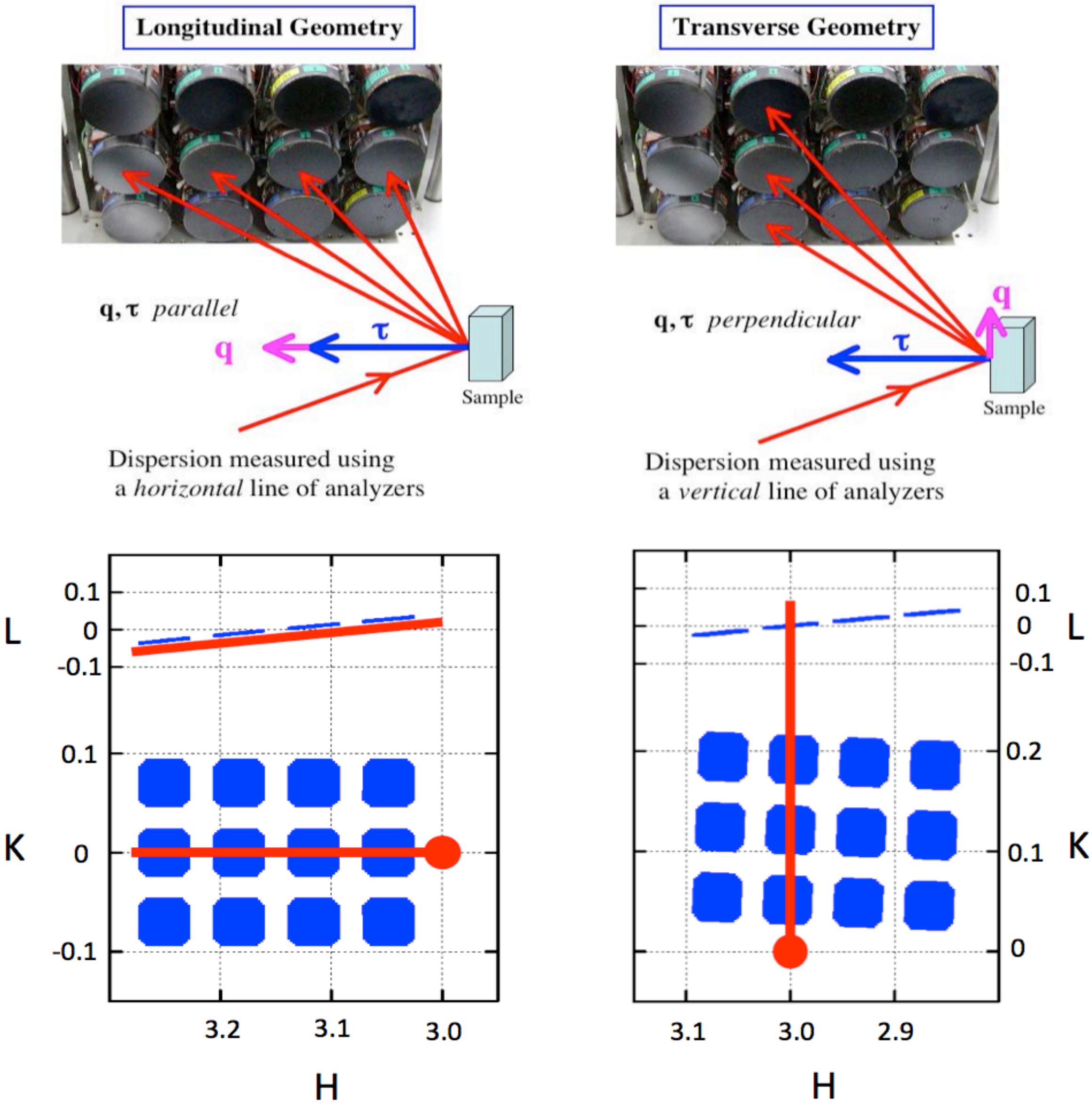

**Figure 14: Using an analyzer array** (after (Baron, et al., 2008)). A 2-Dimensional array can be used to parallelize measurements of longitudinal (left) and transverse phonons (right). In each case, the top panel shows a schematic while the lower panel shows and exact calculation for the BL35 array for a material with a 4Å lattice constant at 17.8 keV. See text for discussion.

## 4.6 Efficient use of an Analyzer Array

Given the flux-limited nature of IXS experiments, every facility with spherical analyzers uses an array to speed, parallelize, data collection. Linear arrays in the horizontal scattering plane are used at ESRF and APS and 2-dimensional arrays at SPring-8. For *disordered materials*, each analyzer adds either a new momentum transfer (for analyzers in the scattering plane) or additional data at a nearly equivalent momentum transfers (for those out of the plane). In-plane analyzers then extend the range of measurements while out-of-plane analyzers and add redundancy (for low rate data) and checks (e.g. for texture in powder measurements). For *crystalline materials*, using an array is more complicated, as, for a fully 3-dimensional sample with a fixed orientation relative to the incident beam, it is generally only possible to put one analyzer *precisely* at a high symmetry point* and most phonon measurements continue to emphasize high symmetry directions. However, it is possible to put several analyzers *near* the symmetry direction, as shown in figure 14, with a line of analyzers in the scattering plane lying nearly along the symmetry direction for a longitudinal mode dispersing from Γ, and a vertical line being on the symmetry direction for a transverse mode. The analyzers will increasingly move off the symmetry direction as one moves from the central analyzer, but, it can be possible to put several (~4 or 5) near to a symmetry line depending somewhat on the tolerance of the experiment. Given the long scan times, such parallelization is often critical. Further, it is generally useful to see the results for nearby momentum transfers, to, for example, help identify the nature of the modes in a spectrum (e.g. contamination by a transverse mode in a longitudinal spectrum due to imperfect momentum resolution can be easy to spot) or determination the localization of interesting effects. In addition, when one considers fitting a microscopic model to the data (as mentioned in section 3), all analyzer crystals are then useful. One should note that the parallelization for transverse modes requires careful choice of the azimuthal angle about the nominal momentum transfer, so should be considered in advance. For samples with reduced dimensionality, layered or 1-dimensional, it is often possible to arrange the system so the deviation from the high symmetry line is along a direction where there is weak (nearly no) dispersion, effectively keeping all analyzers on the symmetry direction.

## 4.7 Post-Sample-Collimation (PSC) Optics

Given the tricks that are possible using asymmetric dynamical diffraction with highly collimated beams, it is immediately interesting, in the context of building a high-resolution spectrometer, to consider the possibility of collimating the radiation scattered from the sample and then making use of those tricks to analyze the scattered radiation - in essence replacing the spherical analyzer discussed above with a collimator and a suitably optimized high resolution monochromator. This, of course, requires a sufficiently high-quality and efficient collimating device, but also quickly leads to two other issues: (1) the spot size on the sample must be small, as the best collimation that can be obtained from a reflective or refractive collimator without losses is the spot size on the sample over the distance to the collimator, and (2) the beam size out of the collimator must be small, to prevent the sizes of the subsequent optical elements from becoming too large. In the past decade, there has been significant improvement in optics, and, while a PSC scheme seemed beyond reach 10 or 15 years ago, it is now becoming a viable alternative, at least for some classes of experiments.

The push for the PSC setup for high-resolution IXS was in large part due to the work of Shvyd'ko and co-workers (Shvyd'ko, 2004)(Shvyd'ko, et al., 2006)(Shvyd'ko, et al., 2014) and, beyond the collimating element, they pushed the limits in two directions with respect to the high-resolution monochromators mentioned above: (1) the potential to achieve ~meV and even substantially sub-meV resolution at relatively low (9.1 keV) x-ray energy (0.62 meV spectrometer resolution has been demonstrated (Shvyd'ko, et al., 2014) and 0.1 meV suggested (NSLS-II Conceptual Design Report, 2006)), and (2) the possibility to provide energy resolution with extremely sharp tails. These are, in-principle, both step-wise improvements over previous HRM designs, as most of the earlier designs already have relatively small tails/sharp response (compared to spherical analyzers) and, if one pushes, can have good resolution (see also discussion in (Sturhahn & Toellner, 2011)). However, the use of extreme dispersion (pushing eqn. (45) to the limit) in backscattering made the improvements both more practical, up to needing rather long perfect crystals (approaching 1m in some cases) and more dramatic. Operation at lower x-ray energy is also particularly interesting at low-energy storage rings (e.g. Diamond, NSLS-II) due to the higher flux available at ~10 keV, compared to 20 keV , but should be considered with care as the lower energy also strongly reduces the penetration into samples and sample environments, decreasing signal rates and potentially increasing radiation damage (see also section 4.8). However, the good resolution (FWHM) coupled with dramatically improved tails compared to spherical analyzers, makes the PSC setup of interest, as do some more exotic designs discussed at the end of this section. This approach is now being pursued most earnestly at the NSLS-II (Cai, et al., 2013), in a new user facility.

---

* The same restriction exists also for INS, but, often, exemplary plots of the Q-E plane are shown for low-dimensional samples - so the limit is not always obvious. Meanwhile, the new chopper spectrometers can also rotate the sample and do post-selection to choose only those events in a high-symmetry plane.

There are essentially 3 main conceptual steps in the PSC analyzer setup as embodied in the first spectrometer (Shvyd'ko, et al., 2014): (1) the use of a collimating mirror just after the sample to reduce the divergence of the scattered radiation, (2) the use of a crystal in a reflection-transmission geometry to allow operation very close to backscattering (Baron, et al., 2001), and then highly asymmetric backscattering to create large dispersion, and (3) the combination of the dispersion of (2) with an angular filter (based on Bragg reflection and anomalous transmission) to reduce the tails of the energy response function. The collimating mirror acts essentially as the inverse of the usual focusing setup, with the location being the result of two competing constraints: the desire to have a small collimated beam size pushes one to place the optic close to the sample while the desire to have small impact of the (projected) beam spot on the sample pushes for larger distances. However, in addition, there is also the desire, long-term, to allow multiple analyzers, which leads to tight constraints on the extent of the collimator transverse to the beam direction. The choice, so far, has been a figured Montel multilayer optical mirror (Mundboth, et al., 2014) which is essentially a pair of mirrors at right angles, conceptually similar to combining a KB pair into one monolithic unit. This setup accepts ~10 mrad vertical divergence at ~20 cm from the sample, collimating it to ~100 μrad divergence. The first reflection from the central crystal then collimates the beam by an additional factor of ~20, reducing the beam divergence to the ~5 μrad level, and blowing the beam size up ~20 times in the diffracting plane. This beam is incident at 1.9 degree grazing angle onto to a dispersive backscattering crystal (Shvyd'ko, et al., 2011). At this point, the beam size (after being increased by the collimating crystal) becomes an issue, with (Shvyd'ko, et al., 2014) being limited to <0.8 mrad acceptance of the beam from the sample in the analyzer scattering plane due to a limited, 90 mm, length of the dispersive backscattering crystal. The design for NSLS-II envisions a set of 5 independent dispersing backscattering crystals to obtain larger angular acceptance (Cai, et al., 2013). The beam then passes through the collimating crystal, allowing operation of the dispersing crystal very close to back-reflection and reflects off of a second dispersive backscattering crystal, before being reflected out of the system by the "back" side of the thin collimating crystal. One should also note that the use of the dispersing crystal in near backscattering also requires relatively good collimation of the beam incident on that crystal *perpendicular to the scattering plane* - the collimator described above must thus collimate in two dimensions, albeit with a (slightly) relaxed tolerance out of the scattering plane. The resolution from one PSC system is shown in figure 15, with clear improvement in the tail of the response compared to the Lorentzian response expected from a spherical analyzer.

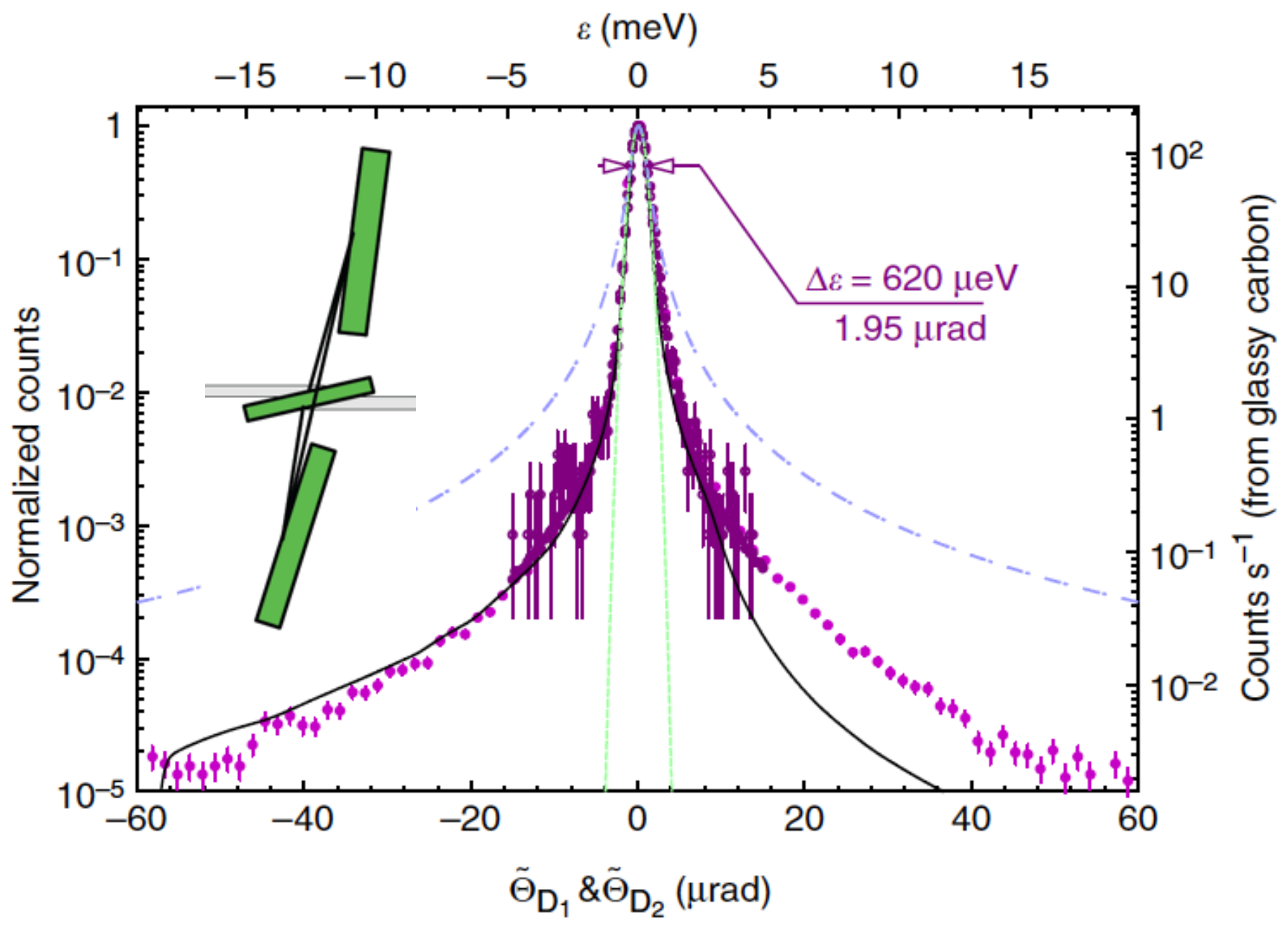

**Figure 15. Resolution of the PSC analyzer setup** after (Shvyd'ko, et al., 2014). Dark solid points measured from glassy carbon at Q=1nm$^{-1}$ and light circles in forward scattering. The solid line is a the calculated response and the dashed line Lorentzian matched at the FWHM, which is approximately what one would expect from a spherical analyzer with the same FWHM - the improvement is clear. (This figure has been modified from the original by including the schematic of the monochromator at left, as is acceptable under license. http://creativecommons.org/licenses/by/4.0/legalcode)

Two extensions of the PSC approach that have been discussed in the context of high-resolution spectrometers. One of them, a "spectrograph" (Shvyd'ko, et al., 2013)(Shvyd'ko, 2015), essentially adds dispersion analysis on the beam from the sample. It creates a position-energy-transfer correlation on the final detector that effectively allows a larger bandwidth (say 50+ meV) to be collected from the sample while retaining good (e.g. meV or potentially sub-meV) resolution for the energy transfer, though, possibly it may effectively sacrifice the extremely sharp tails of the PSC setup described above. The optics layout has been demonstrated (though *not* with a sample) in (Shvyd'ko, et al., 2013). To a first approximation, this allows a gain of a factor of ~N = total output bandwidth over energy-transfer resolution, which can be a ~5 to ~100, for the same flux hitting the sample (so *without* increasing radiation damage). Another option (a so called "echo" spectrometer) sets up a position-energy correlation on the sample and a position energy-transfer correlation at the detector, allowing one to open up the incident bandwidth onto the sample while keeping good resolution in the energy transfer (Shvyd'ko, 2016). Assuming an appropriate geometry and a sufficiently uniform sample this then allows a ~MN enhancement in rates, in principle, with a factor of M coming from the increased rate (bandwidth) on the sample and the other factor of N is gained in the spectrograph. The concept originated from an idea now used in the soft x-ray regime (Lai, et al., 2014) but has yet to be demonstrated for hard x-rays. One notes that both of these methods are sophisticated optical designs and may be difficult to realize in a user instrument, but might offer a way of achieving ~0.1 meV resolution with acceptable count rates.

## 4.8 Comparison of 9.1 keV PSC and Spherical Analyzers

We compare post-sample-collimation (PSC) optics operating at 9.1 keV, and spherical analyzers (SAs) operating between 15.8 and 25.7 keV. The previous of this review (Baron) was optimistic regarding the potential of the PSC setup for high resolution, but cautious about count-rates, and this seems so far, to be borne out by experience. Table 6 lists several points of comparison.

The simplest point of comparison, a-priori, is the operating energy. Using spherical analyzers at energies near 20 keV strongly reduces the photo-electric absorption of x-rays as compared to the 9.1 keV operating energy of the PSC setups. Thus with SAs it is easier to penetrate further into samples and into sample environments, often increasing rates. The former can be particularly important for heavier sample materials and can be quantified by comparing the ratio of the Thomson scattering cross section that is responsible for the signal in IXS to the photo-electric absorption cross section which leads to absorption and radiation damage. The elemental breakdown of this is shown in figure 13. We consider

| | Characteristic | Comment | Advantage |
|---|---|---|---|
| 1 | Operating Energy | The high, ~20 keV, energy of most SA setups allows easier penetration into sample environments and samples, relative to the 9.1 keV operating energy of the PSC, facilitating higher rates in many cases. | SA |
| 2 | Energy Resolution (FWHM) | Presently SAs seem to provide, practically, better energy resolution, with present operating setups having 1.3 and 0.8 meV resolutions. | SA |
| 3 | Energy Resolution (Tails) | The strong dispersion of the PSC setup with the angular selection allows a sharp resolution function with small tails. | PSC |
| 4 | Parallelization (Multiple Momentum Transfers) | Spherical analyzers are relatively inexpensive and easy to install in arrays. | SA |
| 5 | Energy Resolution Flexibility | Changing the resolution from 2.8 to 1.3 to 0.8 meV at different orders of backscattering (e.g. (999) or (11 11 11) or (13 13 13)) can be relatively straightforward with spherical analyzers. | SA |
| 6 | Sample Volume Selectivity | The collimating optic of the PSC has limited acceptance, collecting scattered radiation from a region of ~10's of microns near the sample. This can reduce backgrounds from scatterers, e.g. windows, near the sample. | PSC |
| 7 | Thicker Samples | The acceptance of SA's tends to be ~mm. Thus if a sample can be made thick but remain transmitting (e.g. low Z materials) then the SA can see a thick portion of the sample, giving higher rates. | SA |
| 8 | Potential 0.1 meV Resolution | It might be easier to reduce the resolution to <0.5 meV or even to ~0.1 meV using PSC than it is with a SA, assuming a sufficiently strong source. | PSC |

**Table 6:** Comparison of operation with ~meV resolution spherical analyzers (SA) and post-sample collimation (PSC) optics with respect to several issues. The table mentions main points, but there are many caveats so please see the text for details.

YBCO as an example: the penetration into $YBa_2Cu_3O_7$ is about 66 microns at 21.7 keV, and only 8.4 microns at 9.1 keV, so, with equal analyzer solid angle, efficiency, and flux (photons/s/meV) onto the sample and similar geometry, the scattered intensity would be about 8 times larger for 21.7 keV x-rays if the sample is ~0.1 mm thick or more. Of course, using a lower, 9.1 keV, energy means that the momentum transfer resolution is improved for a given solid angle, so that, for a fixed momentum transfer resolution one might gain back a factor of 2.4 (one dimension) or $2.4^2$ (two dimensions) in rate increase at 9.1 keV (compared to 21.7 keV), *assuming* the solid angle of the PSC analyzer is large enough. However, the PSC setups are now mostly aiming at about 10 mrad angular acceptance, similar to spherical analyzers, so in most flux limited experiments, where the analyzer apertures are large, the PSC option will yield improved momentum resolution at reduced rates. One also notes that the number of scattered photons per absorbed photon is much smaller for lower energy x-rays (see figure 13), so that radiation damage, as can be an issue even for IXS, may be more severe with the PSC setup. For the issue of penetration into/out-of sample environments, we consider a diamond anvil cell as an example: transmission through 3 mm of diamond is about 69% at 17.8 keV but only 13% at 9.1 keV or, through a 6mm Be gasket, 70 to 76% transmission at 17.8 keV and 26 to 43% at 9.1 keV (where the exact values are sensitive to the impurity content of the Be).

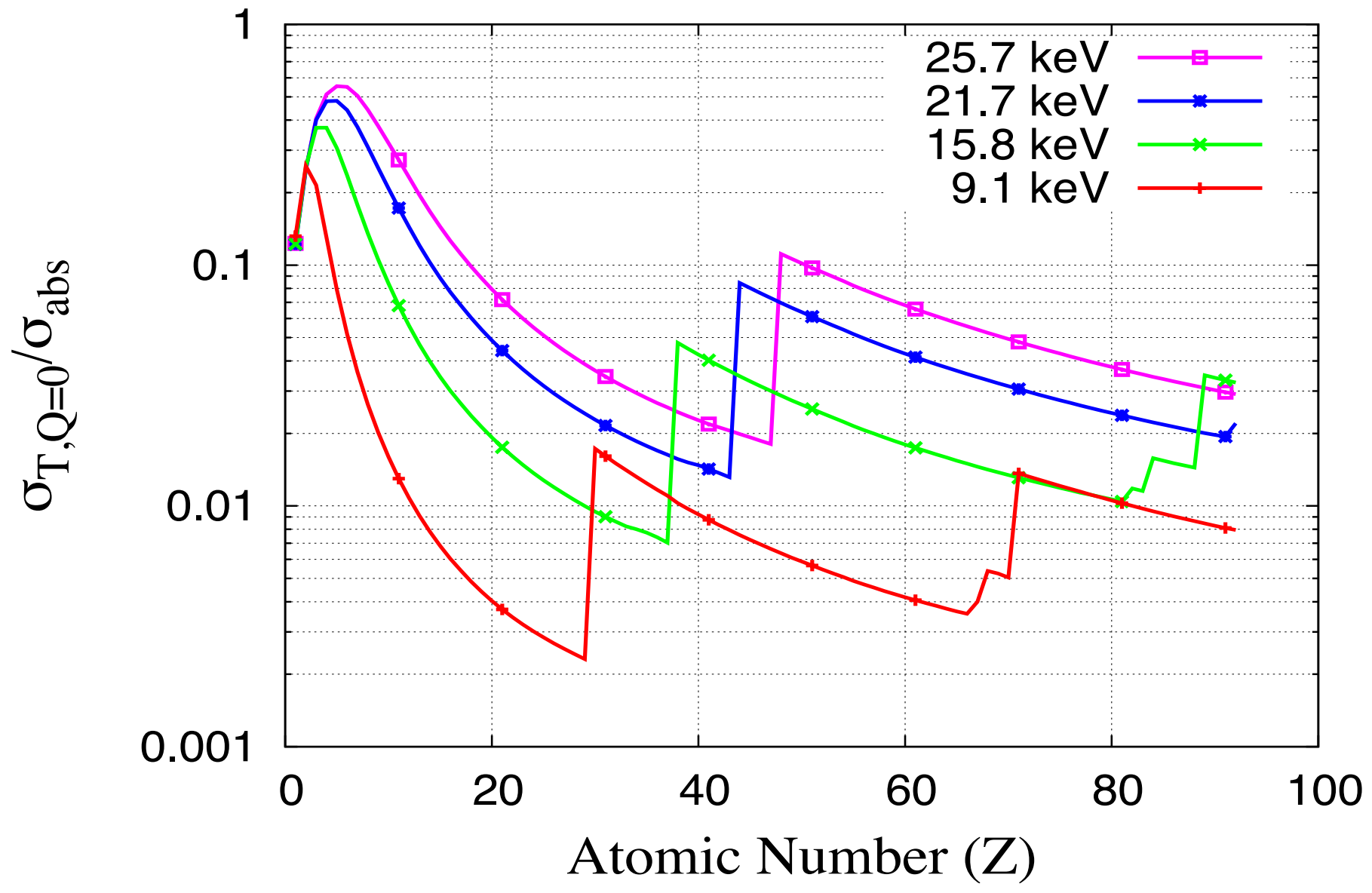

**Figure 16.** Ratio of the cross section for Thomson scattering (at Q=0) to the absorption cross section for several x-ray energies. This is effectively signal per incident photon in the thick sample limit. See text for discussion.

The SA geometry lends itself to relatively easy parallelization (see the discussion of the arrays in section 4.6) as, once the recipe for fabricating analyzers is known, it is not too hard to make more, and the mechanics for installation (usually two tilt axes) is not too extensive. In contrast, each channel in a PSC setup requires a new multilayer mirror collimating optic, precise positioning of that mirror near the sample, in addition to several crystals and the associated mechanics. Also, with SAs one can use the same analyzers (even high resolution monochromator in some cases) for several different resolutions: when advantageous, one can trade larger flux at 17.79 keV for slightly poorer, 2.8 meV, resolution instead of 1.3 meV resolution at 21.747 keV with lower flux.

Items 6 and 7 in table 6, sample volume selectivity and thickness, are different sides of the same issue. The SA setups are not extremely sensitive to the source position, whereas the PSC optics are very sensitive to the source position. This means the SA's can see scattering from large samples, but, at the same time, will also see scattering from, e.g., windows near a sample. Meanwhile the PSC setup is comparatively sensitive to the source position so may not accept scattering from all of a thick sample, but will also not be as sensitive to nearby windows. The length of sample observed scale roughly given by $a/\sin(\Theta_{scatt})$ with a~1mm for SAs and for the PSC setup at NSLS-II (Cai, 2018) a~0.04 mm.

Finally we return to the issue of energy resolution. The original demonstration of PSC optics gave a 0.62 meV resolution, FWHM, as used to measure a glycerol sample (Shvyd'ko, et al., 2014), but more recent work at NSLS-II has been at closer to 2 meV resolution (Bolmatov, et al., 2017) which, while also early in the operation, highlights the practical difficulty of obtaining good resolution in the PSC setup*. Meanwhile SA optics have been demonstrated to have a resolution as good at 0.75 meV for a standard plastic sample (Ishikawa, et al., 2015) while 0.8 meV has been used in practical experiments, and 1.3 meV is the workhorse resolution of the SPring-8 instruments. Thus, as indicated in the table, with the exception of the tails, which can be better with PSC optics, there may be an advantage for the SA's in many cases. However, there is, in principle, an upgrade path for the PSC setup that one expects can probably push to 0.5 meV or maybe even 0.4 meV or less, given a sufficiently strong source, and possibly as good as 0.1 meV (NSLS-II Conceptual Design Report, 2006) Improving the resolution using spherical analyzers to better than 0.75 meV would probably require a new instrument with a longer arm. However, given recent tendencies for rather specialized instrumentation, such as soft x-ray spectrometers with ~15m arms scannable over nearly 180 degrees in two-theta, this may be considered, especially as the resolution scales as the inverse square of the arm length (see eqn 48): e.g a 15m arm probably could give better than 0.5 meV resolution using special analyzers.

The picture that arises from these considerations is one of spherical analyzer main-line instruments with the PSC instruments having a niche assuming the count-rates are sufficiently high. Longer term development of PSC optics is also interesting in the context of future, higher flux, sources, or possibly the now more exotic "spectrograph" or "echo" configurations.

* The NSLS-II PSC setup has achieved 1.3 meV resolution, FWHM, but in fact is now aiming at 2 meV resolution to help improve rates (Private communications from Y.Q. Cai).

## 4.9 Electronic Excitation with a High Resolution Spectrometer

It is interesting to consider the possibility to extend high-resolution method to higher energy transfers to measure, for example, electronic excitations, or high-energy vibrational modes. In a very high flux limit (e.g. the XFEL-O - see section 1.3) one could dream about measuring phonons, localized excitations and band structure, all on the same instrument with ~meV resolution (practically, probably, radiation damage issues may impose limits so force one to tailor the resolution to the particular science of interest). Along these lines, there has been some work to extend the usual ~0.1 eV scale scans in backscattering to the few eV level. This is straightforward for a typical in-line HRM design, but requires some effort for a backscattering monochromator. However, one can replace standard coolants with a specialized liquid (e.g. silicone oils) viable down to about -80C and then cool the backscattering crystal: this gives one a scan range of $\Delta E/E = \Delta d/d$ ~ 20 ppm, between +30 and -60 degrees C in silicon - or about 3 eV at 15.816 keV. Figure 17 shows an example of measuring a d-d excitation in nickel oxide NiO and phonons with the same over-all ~7 meV resolution. The reduced intensity of the d-d excitation, scattering from a fraction of an electron per unit cell, as compared to the LA phonon, where full atoms of electrons move together, is very clear: the d-d excitation is down by almost 4 orders of magnitude in peak intensity compared to the phonon, and about 3 orders in integrated intensity.

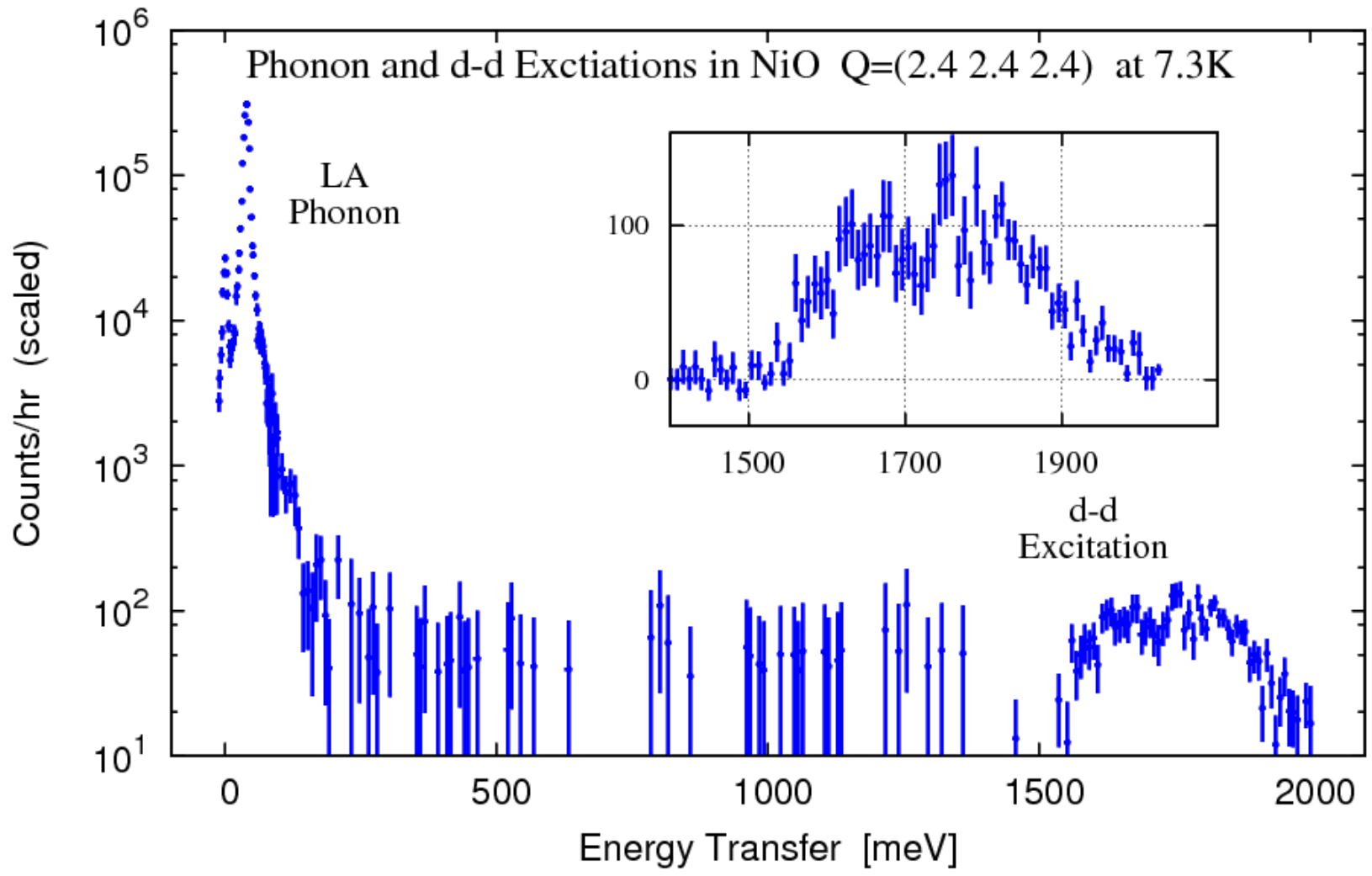

**Figure 17. Phonon and d-d excitation in NiO with 7 meV resolution.** (Baron, et al, unpublished)

None-the-less, the experiments are possible, with more complex structure visible in the d-d excitation. However, attempts to move beyond this demonstration experiment using a high-resolution spectrometer have not yet been successful due to the limited flux: NiO appears to be a particularly high count-rate case. Thus this work will be continued using a smaller (2 m) radius analyzer crystals to gain solid angle and improve count-rates with ~25 meV resolution (Ishikawa, et al., 2017). Considering high-energy transfer more generally, 3 meV-resolution IXS has also been used to measure the hydrogen vibron/diatomic stretching mode (at about 500 meV) in both liquid and solid hydrogen in a DAC (Liu, et al.), and somewhat similarly, the stretching mode in liquid nitrogen with ~10 meV resolution at slightly lower (~300 meV) energy transfer (Monaco, et al., 2001). It is perhaps worth emphasizing that in many cases there are intrinsic sources of broadening at larger energy transfers (e.g. phonons for electronic excitations (Ishikawa, et al., 2017), or recoil for liquids, see eqn (10)) that may allow larger bandwidths to be used without strongly negatively impacting the measurement result.

# 5. Sample Science

## 5.1 Introduction and Overview

IXS has been used to investigate the atomic dynamics of many different types of samples, in different conditions, and also to look some more basic physical quantities. While the field was initially dominated by studies of liquids and glasses, the possibility to investigate small samples coupled with generally clean data (see section 1.4) has led to work on many types of crystalline materials. Probably the largest areas of work have been investigations of ***superconductors***, ***charge density wave (CDW***) materials and ***ferrorelectric*** (& related) materials - these are fields in which phonons are, classically, known to play an important role in the material properties. We will treat superconductors in some depth below, providing motivation for the measurements and examples of the data obtained. As regards the other materials, tables 6a and 6b give a list of materials and references. The listing is broken up into larger categories (**superconductors, CDW materials, ferroelectric, multiferroic and relaxor materials, cage compounds, quasicrystals, actinides**) and a list of samples that do not fall so easily into one of the above categories. This is followed, at the bottom of table 6b, by a list of, mostly, **high-pressure measurements** carried out in diamond anvil cells (DACs). The majority of the DAC work is on geologically relevant materials, with the goal of determining acoustic wave velocities for comparison with seismic measurements of earth's interior, but there are also some where pressure is more just taken as another thermodynamic variable to drive the system of interest into different phases. For the tables 6a and 6b the listed references are examples, and are not exhaustive: readers are strongly encouraged to supplement this list with their own literature searches.

Finally we mention a few papers with a different, more methodological, aspect. These include studies aiming to detect the phonons in the first ~nm of a surface using an extreme grazing incidence geometry (Murphy, et al., 2005) (Reichert, et al., 2007) (Wehinger, et al., 2011), investigation of an OH stretching mode at very high , ~300 meV, energy transfer (Winkler, et al., 2008), investigation of x-ray form factors at low Q (Alatas, et al., 2008), investigation of two-phonon contributions (Baron, et al., 2007) and a way of determining elastic constants by fitting to Christoffel's equation (Fukui, et al., 2008). Issues relating to measurements of densities of states and powders, as opposed to single crystals are also discussed in (Bosak & Krisch, 2005)(Fischer, et al., 2009) and application of sum rules in a nearly atomic-Compton limit in (Monaco, et al., 2002).

| | | |
|---|---|---|
| Conventional Superconductors | $MgB_2$ & $Mg(B_{1-x}C_x)_2$, $Mg_{1-x}AlB_2$ | (Shukla, et al., 2002) (Baron, et al., 2004)(d'Astuto, et al., 2007)(Baron, et al., 2007)(Baron, et al., 2007)(d'Astuto, et al., 2016) |
| | $NaCoO_2$ | (Rueff, et al., 2006) |
| | Boron Doped Diamond | (Hoesch, et al., 2007) (Caruso, et al., 2017) |
| | $CaC_6$ | (Upton, et al., 2007) (d'Astuto, et al., 2010) |
| | CaAlSi | (Kuroiwa, et al., 2008) |
| | $MgCNi_3$ | (Hong et al. 2010) |
| | $Ba_{1-x}K_xBiO_3$ | (Khosroabadi, et al., 2011) |
| | $SrPt_3P$ | (Zocco, et al., 2015) |
| | $V_sSi$ | (Sauer, et al., 2019) |
| Cuprates & Related | $Nd_{1.86}Ce_{0.14}CuO_{4+\delta}$ | (d'Astuto, et al., 2002) |
| | $Ca_{2-x}CuO_2Cl_2$ | (D'Astuto, et al., 2013)(Lebert, et al., 2020) |
| | $La_{2-x}Sr_xCuO_4$ | (Fukuda, et al., 2005)(Ikeuchi, et al., 2006) (Graf, et al., 2007) (Park, et al., 2014) |
| | $La_{2-x}Ba_xCuO_{4+\delta}$ | (d'Astuto, et al., 2008) |
| | $La_{1.48}Nd_{0.4}Sr_{0.12}CuO_4$ | (Reznik, et al., 2008) |
| | $Bi_2Sr_{1.6}La_{0.4}Cu_2O_{6+\delta}$ | (Graf, et al., 2008) |
| | $HgBa_2CuO_{4+\delta}$ | (d'Astuto, et al., 2003) (Uchiyama, et al., 2004) |
| | $YBa_2Cu_3O_{7-\delta}$ | (Baron, et al., 2008) (Blackburn, et al., 2013) (Le Tacon, et al., 2013) (Souliou, et al., 2018) |
| Iron Pnictides & Nematicity & Related | "1111" Materials $LaFeAsO_{1-y}$, etc. | (Fukuda, et al., 2008) (Le Tacon, et al., 2008)(Le Tacon, et al., 2009)(Fukuda, et al., 2011)(Hahn, et al., 2013) |
| | "122" Materials $BaFe_2As_2$, etc. | (Reznik, et al., 2009) (Mittal, et al., 2010) (Lee, et al., 2010) (Niedziela, et al., 2011) |
| | "11" Materials, FeSe, etc. | (Murai, et al., 2020) (Merritt, et al., 2020) |

**Table 6a: List of some samples investigated by IXS and references**. 6a. Focusses on superconductors, however, other papers on superconductors may also appear below in other categories (e.g. CDW materials in table 6b) Readers are encouraged to supplement the table with their own literature search.

| | | |
|---|---|---|
| Charge Density Wave (CDW) & Related | $NbSe_3$ | (Requardt, et al., 2002) |
| | $Rb_{0.3}MoO_3$, "blue bronze" | (Ravy et al. 2004) |
| | Te-III | (Loa, et al., 2009) |
| | $NbSe_2$, $NbS_2$ | (Murphy, et al., 2005)(Weber, et al., 2011), (Weber, et al., 2013) (Leroux, et al., 2012)(Leroux, et al., 2015)(Leroux, et al., 2018) |
| | $ZrTe_3$, $TbTe_3$, $DyTe_3$ | (Hoesch, et al., 2009) (Maschek, et al., 2015)(Maschek, et al., 2018) |
| | $TiSe_2$ | (Maschek, et al., 2016) |
| | Cuprates | (Blackburn, et al., 2013)(Le Tacon, et al., 2013)(Souliou, et al., 2018)(Miao, et al., 2018) |
| Ferroelectric, Multiferroic, Relaxors & Related | $Pb(In_{0.5}Nb_{0.5})O_3$ | (Ohwada, et al., 2008) |
| | $SrTiO_3$ | (Hong, et al., 2008)(Shih-Chang Weng, et al., 2014) |
| | $TbMnO_3$, | (Kajimoto, et al., 2009) |
| | $Sr_{1-x}B_xMnO_3$ | (Sakai, et al., 2011)(Sakai, et al., 2012) |
| | $EuTiO_3$ | (Ellis, et al., 2012) (Ellis, et al., 2014) |
| | $BiFeO_3$ | (Borissenko, et al., 2013) |
| | $PbZr_{1-x}Ti_xO_3$ | (Hlinka, et al., 2011)(Tagantsev, et al., 2013)(Burkovsky, et al., 2014) |
| | $PbHfO_3$ | (Burkovsky, et al., 2015) |
| | $TiO_2$ | (Wehinger, et al., 2016) |
| | SnTe | (O'Neill, et al., 2017) |
| | $CuCrO_2$ | (Bansal, et al., 2017) |
| | $Bi_2SiO_5$ | (Girard, et al., 2018) |
| | $(Ca,Sr)_3Ti_2O_7$ | (Bansal, et al., 2019) |
| | $Cu_3Bi(SeO_3)_2O_2Cl$ | (Milesi-Brault, et al., 2020) |
| Magneto-Elastic Coupling | $SrFe_2As_2$ | (Murai, et al., 2016) |
| | EuO | (Pradip, et al., 2016) |
| | $LiCrO_2$ | (Tóth, et al., 2016) |
| Thermal Conductivity & related, inc. Cage Compounds | Methane Hydrate | (Baumert, et al., 2003)(Baumert, et al., 2005) |
| | Skutterudites | (Rotter, et al., 2008)(Tsutsui, et al., 2008)(Koza, et al., 2011)(Tsutsui, et al., 2012) |
| | $LaRu_2Zn_{20}$ | (Wakiya, et al., 2016) |
| | Sodium Cobaltate | (Voneshen, et al., 2013) |
| | $PbTe_{1-x}Se_x$ | (Tian, et al., 2015)(Kuna, et al., 2016) |
| | $(CH_3NH_3)_3Bi_2I_9$ | (Ma, et al., 2019) |
| | $Si_{1-x}Ge_x$ | (Usuda, et al., 2020) |
| Quasicrystals & Related | i-AlPdMn, MgZnY, ZnMgSc, MgZn | (Brand, et al., 2001)(Krisch, et al., 2002)(Brand, et al., 2004)(De Boissieu, et al., 2007)(Euchner, et al., 2011) |
| Actinides | Pu-u, U, $PuCoGa_5$, NpFeAsO, $NpO_2$, $URu_2Si_2$, UO, $UO_2$ | (Wong, et al., 2003) (Manley, et al., 2003) (Wong, et al., 2005) (Manley, et al., 2006) (Raymond, et al., 2006) (Manley, et al., 2009) (Walters, et al., 2011)(Klimczuk, 2015)(Maldonado, et al., 2016) (Gardner, et al., 2016)(Rennie, et al., 2018)(Brubaker, et al., 2019) |
| Graphite & Related | Graphite | (Mohr, et al., 2007)(Bosak, et al., 2007)(Grüneis, et al., 2009) |
| | Graphite Intercalation | (d'Astuto, et al., 2010)(Upton, et al., 2010)(Walters, et al., 2011) |

**Table 6b: List of samples investigated by IXS and references, continued**. Readers are encouraged to supplement the table with their own literature searches.

| | | |
|---|---|---|
| Other Materials | $CeRu_2Si_2$ | (Raymond, et al., 2001) |
| | Benzoic Acid | (Plazanet, et al., 2005) |
| | SiC | (Strauch, et al., 2006) |
| | DNA | (Liu, et al., 2005) (Krisch, et al., 2006) |
| | Lipid Membranes | (Zhernenkov, et al., 2016)(Bolmatov, et al., 2017) |
| | Boron Nitride, BN | (Bosak, et al., 2006)(Serrano, et al., 2007) |
| | Boron Arsenide, BAs | (Ma, et al., 2016) |
| | Vanadium | (Bosak, et al., 2008) |
| | $NaBH_4$ | (Chernyshov, et al., 2008) |
| | BeO | (Bosak, et al., 2008) |
| | Asbestos | (Mamontov, et al., 2009) |
| | NiO | (Uchiyama, et al., 2010) |
| | $ReB_6$ Rare-Earth Hexaborides | (Iwasa, et al., 2011) (Iwasa, et al., 2012) (Iwasa, et al., 2014) |
| | Ce | (Krisch, et al., 2011) (Loa, et al., 2012) |
| | Nd | (Waller, et al., 2016) |
| | $Fe_3O_4$, Magnetite | (Hoesch, et al., 2013) (Bosak, et al., 2014) |
| | $LaCoO_3$ | (Doi, et al., 2014) |
| | $Ba_3CuSb_2O_9$ | (Wakabayashi, et al., 2016) |
| | $RTi_2O_7$ R=Tb, Dy, Ho | (Ruminy, et al., 2016) |
| | $La_{0.7}Sr_{0.3}MnO_3$ | (Maschek, et al., 2016) |
| | Quantum Well InGaAs/GaAsP | (Xia, et al., 2017) |
| | ScN | (Uchiyama, et al., 2018) |
| | $ScF_3$ , $HgI_2$ | (Occhialini, et al., 2017) |
| | $M_2Mo_6Se_6$ (M=K,Rb,Cs,In,Tl) | (Gannon, et al., 2018) |
| | FeSi | (Miao, et al., 2018) |
| | $MoB_2$ | (Zhang, et al., 2019) |
| | $Ta_2NiSe_5$ | (Nakano, et al., 2018) |
| | $(Pr_{.85}Y_{.15})_{0.7}Ca_{0.3}CoO_{3-d}$ | (Phelan, et al., 2019) |
| | $MoS_2$ | (Tornatzky, et al., 2019) |
| High Pressure, Elasticity, Geoscience, & Related | Pure Fe & Related | (Fiquet, et al., 2001) (Antonangeli, et al., 2004)(Mao, et al., 2008)(Hosokawa, et al., 2008) (Ohtani, et al., 2013) (Mao, et al., 2012) (Liu, et al., 2014)(Nakajima, et al., 2015) (Jin, et al., 2016) |
| | FeO, FeSi, FeS, $FeS_2$ | (Badro, et al., 2007) |
| | FeNi | (Kantor, et al., 2007) |
| | $Mg_{1-x}Fe_xO$ | (Fukui, et al., 2017)(Finkelstein, et al., 2018) |
| | FeH | (Shibazaki, et al., 2012) |
| | $Fe_3S$ | (Kamada, et al., 2014) |
| | l-FeNiS | (Kawaguchi, et al., 2017)(Umemoto, et al., 2014) |
| | $FeCO_3$, $CaCO_3$, $MgCO_3$ | (Stekiel, et al., 2017) (Stekiel, et al., 2019) |
| | Bridgmanite, $MgSiO_3$ | (Fukui, et al., 2016) (Björn, et al., 2016) |
| | $CaIrO_3$ | (Yoneda, et al., 2014) |
| | MgO | (Ghose, et al., 2006) (Fukui, et al., 2008) (Antonangeli, et al., 2011)(Giura, et al., 2019) |
| | Ar | (Occelli, et al., 2001) |
| | Co | (Antonangeli, et al., 2004) (Antonangeli, et al., 2005) (Antonangeli, et al., 2008) |
| | Mo | (Farber, et al., 2006) |
| | Ta | (Antonangeli, et al., 2010)(Zhang, et al., 2019) |
| | Au | (Yoneda, et al., 2017) |
| | Sb | (Minelli, et al., 2019) |

**Table 6c: List of samples investigated by IXS and references, continued**. Readers are encouraged to supplement the table with their own literature searches.

## 5.2 Superconductors

### 5.2.1 Background

Superconductors and related materials are an active area of IXS investigation, in part due to the fact that they are a broad and active field in materials science, but also specifically because of the phonon-mediated nature of conventional (BCS (Bardeen-Cooper-Schrieffer) or Migdal-Eliashberg) * superconductors. However, there is also great interest in the phonon behavior of non-conventional superconductors, including cuprates and iron-related materials (pnictides). While the wider field of superconductivity is certainly dominated by a focus on electronic properties, phonon investigations do provide insight, with the normal state phonon properties being an indicator of the potential for phonon-mediated superconductivity, and the general behavior of the phonons providing support (or not) for the same calculations that sometimes indicate the nature of the superconductivity. Also, in some, rare, cases, the onset of superconductivity can be observed to affect phonon spectral properties.

The relationship between phonon spectral properties and classical, phonon-mediated, superconductivity was explored in a series of papers by Allen and collaborators, largely in the 1970s, ((Allen, 1972; Allen & Cohen, 1972; Allen & Silberglitt, 1974; Allen & Dynes, 1975) (Allen & Dynes, 1975), see the review (Allen, 1980) and also the book (Grimvall, 1981)) with more recent work of note also (Allen, et al., 1997). One of the main results was to establish a relationship between the electron-phonon coupling, epc, parameter, λ (also called the mass-enhancement parameter), and the normal-state phonon line-width (half-width at half maximum - HWHM), $\gamma_{\mathbf{q}j}$,

$$\begin{aligned} \lambda &= \frac{2}{\pi N_0} \frac{1}{N_{\mathbf{q}}} \sum_{\mathbf{q}j} \frac{\gamma_{\mathbf{q}j}}{\omega_{\mathbf{q}j}{}^2} \\ &= \frac{1}{N_{\mathbf{q}}} \sum_{\mathbf{q}j} \lambda_{\mathbf{q}j} \qquad \textbf{(49)} \\ &= 2\int_0^{\infty} \alpha^2 F(\omega)/\omega \end{aligned}$$

where $N_0$ is the density of electronic states at the Fermi surface. The second equation defines the mode and momentum-transfer-specific electron-phonon coupling parameter and the third equations relates λ to the $\alpha^2 F$ as it appears in Eliashberg theory (see, e.g. (Allen & Mitrovic, 1982)). Carefully speaking, the line-width mentioned here is only that part due to electron-phonon coupling - other contributions, such as that from phonon-phonon scattering must be subtracted out before applying these formulas.

The electron-phonon coupling parameter, $\lambda_{\mathbf{q}j}$, can be considered an average over $\lambda_{\mathbf{q}j\mathbf{k}n\to n'\mathbf{k}'=\mathbf{k}+\mathbf{q}}$ linking interaction of a phonon mode, j, with momentum **q** to a transition of an electronic state with momentum **k** in band n to a momentum $\mathbf{k}' = \mathbf{k} + \mathbf{q}$ in band n'. Phonon measurements by IXS then allow selection of a particular phonon mode and momentum transfer, **q,** but integrate over the electronic system, providing partially resolved information. Thus, assuming the line width due to epc can be clearly determined, IXS can complement angle-resolved photoemission spectroscopy (ARPES): ARPES, with some experimental caveats, allows investigation of particular electron bands and energies, but integrates over the phonon structure, while IXS investigates phonons but integrates over the electronic structure. In the context of band structure and ARPES work, the phonon momentum should be considered as a candidate for a nesting vector, and not confused with a real electronic momentum. Meanwhile tunneling spectra provide a more direct measurement of $\alpha^2 F$.

The connection between the electron-phonon coupling parameter and the critical temperature for the onset of superconductivity can be made using Eliashberg theory. A numerical approximation to solutions of the Eliashberg equations is possible with the results modeled using one of a number of "McMillan" formulas (McMillan, 1968) (Allen & Dynes, 1975) such as

$$T_c \approx \frac{\langle \omega \rangle}{1.2} \exp\left\{ -\frac{1.04(1+\lambda)}{\lambda - \mu^* - 0.62\lambda\mu^*} \right\} \quad \textbf{(50)}$$

where $\langle \omega \rangle$ is an average phonon frequency (with pre-factors - see discussion in (Allen & Dynes, 1975)). $\mu^*$ describes the effective Coulomb repulsion which is not easy to calculate and is usually taken as $\mu^* \approx 0.1-0.15$ (see discussion in (Bauer, et al., 2013) and an attempt in (Floris, et al., 2005)), though a method of calculation allowing solution of the Eliashberg equations directly in a density functional approach (Lüders, et al., 2005) circumvents this, as has been applied (Akashi, et al., 2015), recently, to estimate the observed (Drozdov, et al., 2015) record (~200K) superconducting transition temperature of sulfur hydride under pressure.

---

* Here we will use "BCS" to refer to the original discussion (Bardeen, et al., 1957) where phonons were introduced in a weak-coupling limit without an explicit relation between the coupling and the detailed phonon spectra, while Eliashberg theory (Eliashberg, 1960) allowed a direct relation based on application of Migdal's approximation (Migdal, 1958) ignoring vertex corrections - effectively assuming interesting electronic states have energies are much larger than phonon energies.

The relation between phonon line width and the onset of superconductivity is part of what makes IXS measurements of superconductors interesting. In principle, based on the above discussion, if one can measure the line width of all the phonons of a sample in the normal state, and isolate the part due to epc, one can determine λ, or, more precisely, $\lambda N_0$ and eventually $T_c$. However, this is generally hard, as line widths are generally difficult to measure, and the electron-phonon line widths are often not large (sub-meV), and there can be line width contributions from other processes such as phonon-phonon scattering. Thus, practical phonon investigations of superconductors tend to fall into one of several categories: (1) measurements of phonon dispersion and line widths over targeted regions of momentum space to spot-check (or invalidate) calculations; (2) general investigations of phonons to look for anomalous dispersion and/or broad lines or (3) measurements of doping dependence to observe possible softening and/or line broadening that may suggest changes in electron-phonon coupling with doping. It is also possible to measure the temperature dependence of phonon modes as one passes through $T_c$, however, while an intriguing experiment conceptually, often there are no noticeable effects. We will discuss several examples below, focusing first on conventional superconductors and then on high $T_c$ materials (cuprates and pnictides).

## 5.2.2 Conventional Superconductors

Inelastic x-ray scattering measurements on conventional superconductors began, to a large extent, after Akimitsu's group (Nagamatsu, et al., 2001) demonstrated the surprisingly high superconducting transition temperature, $T_c = 39K$, of $MgB_2$ and then calculations (Bohnen, et al., 2001; Kong, et al., 2001; Yildirim, et al., 2001) suggested $MgB_2$ was a phonon-mediated superconductor, with extremely large electron-phonon coupling between the "sigma" Fermi-surface sheets (two warped cylinders about the Γ-A axis - from (000) to (00 1/2)) and the in-plane boron-boron stretching phonon modes with $E_{2g}$ symmetry at Γ. Thus it was expected that huge effects should be visible in phonon spectra. However, $MgB_2$ is difficult to grow, requiring high-pressure synthesis to obtain single crystals (see, e.g., (Lee, et al., 2001)), so samples are small, a fraction of a cubic mm. First phonon measurements at zone center, gamma, were carried out by Raman scattering and the Raman-active $E_{2g}$ mode was found to have line-widths that were large - possibly even larger than theory - and a somewhat unusual shape (Bohnen, et al., 2001; Goncharov, et al., 2001; Hlinka, et al., 2001; Quilty, et al., 2002). While apparently confirming the main broadening, the interpretation of these spectra is not simple, as, on the one hand, part of the electron phonon coupling might not be allowed exactly at gamma (Calandra & Mauri, 2005) and, on the other, interactions with the entire electron-phonon spectrum might significantly affect the line shape (Cappelluti, 2006).

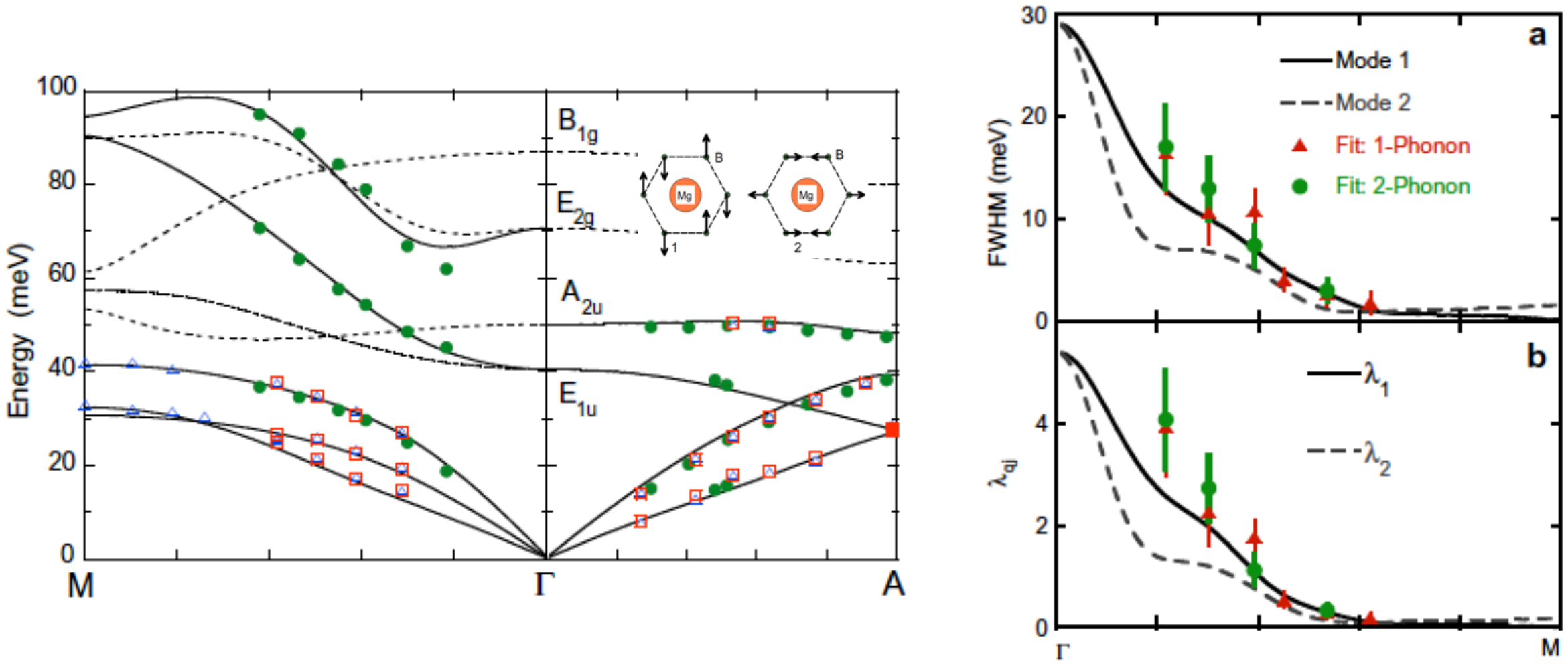

**Figure 18. Dispersion and linewidth of $MgB_2$.** The left panel shows the dispersion and the softening along the MΓ direction through the large Kohn anomaly while the right panel shows the change in linewidth, and the re-scaled to $\lambda_{qj}$. In each case, the lines are results LDA-based *ab-initio* calculations, with dashed lines indicating modes that are symmetry forbidden in the chosen geometry. The inset to the left panel shows mode polarizations for the two $E_{2g}$ modes at Γ (mode 1 is the one that was investigated by IXS) After (Baron, et al., 2004) (Baron, et al., 2007)

Given the above issues for the Raman scattering, and the limited sample size, IXS is a very interesting probe for phonons in $MgB_2$. Initial work confirmed the line width of the $E_{2g}$ mode was broad along Γ-A (Shukla, et al., 2002), followed by demonstration that the line width along Γ-M (essentially through the huge Kohn anomaly in the (100) direction), went from resolution limited for wave vectors too large to span the sigma sheets to nearly 20 meV as one moved inside the sigma Fermi-surfaces and the mode softened (Baron, et al., 2004). This can be seen in the fit results of figure 18, or even directly in the spectra in figure 19. The highest energy mode (that is not symmetry forbidden) shows both extreme softening and simultaneously a very large increase in width, as one moves inward from the M point to the gamma point. The importance of this softening and broadening for superconductivity is corroborated by the results on carbon-doped $Mg(B_{1-x}C_x)_2$. Here the superconducting transition temperature can be reduced from 39K to 2.5 K by doping in C in place of B, which tends to fill the sigma surface. The very nice correlation of the phonon dispersion with the change in $T_c$ as seen in figure 19 then helps to cement the picture of electron-phonon coupling graphically, though the detailed behavior is not yet in accordance with theoretical estimates (see discussion in (Baron, et al., 2007)).

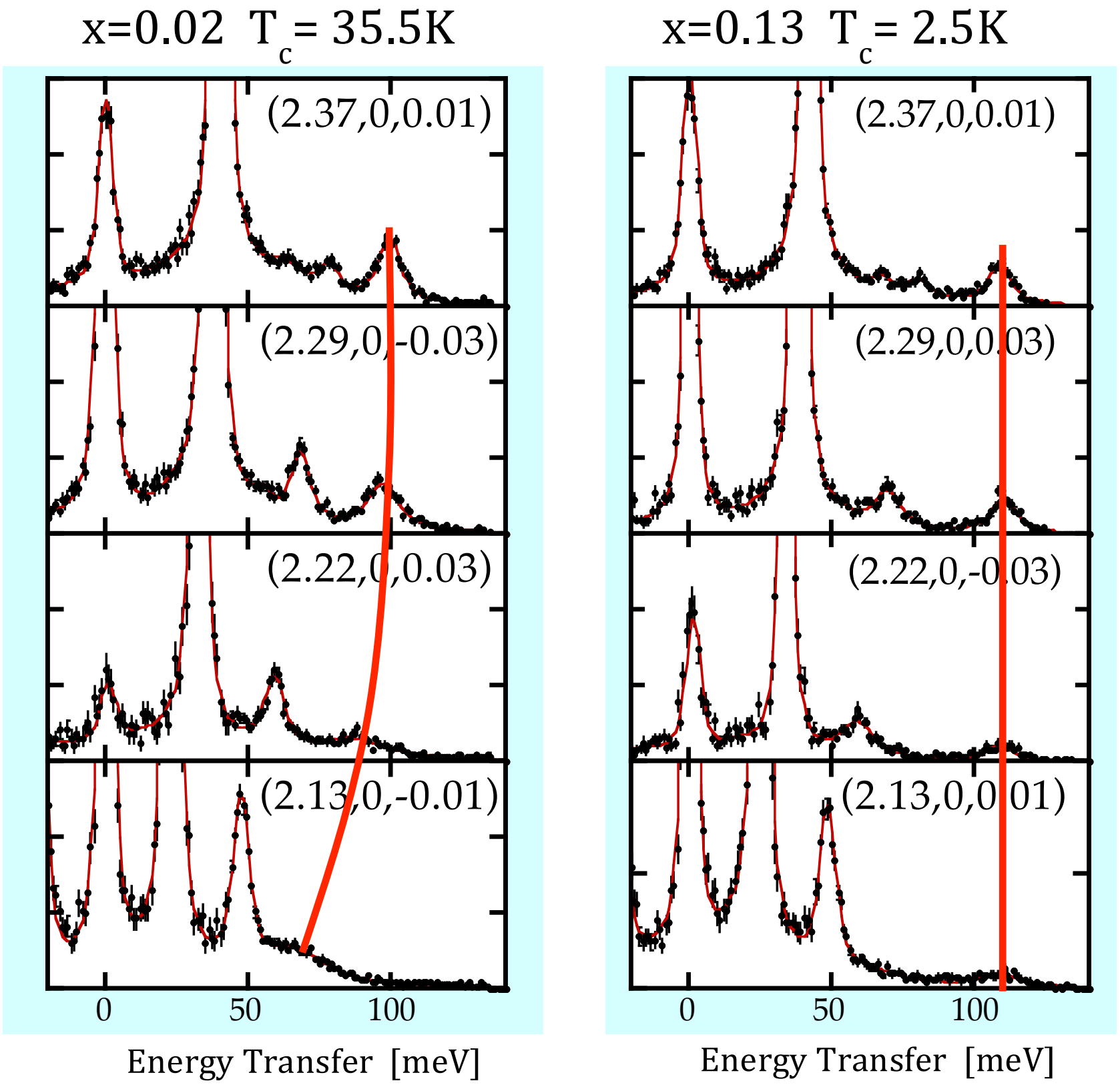

**Figure 19. Phonon spectra of carbon doped $MgB_2$ ($Mg(B_{1-x}C_x)_2$) at room temperature.** The red line highlights the softening of the $E_{2g}$ mode as the gamma (200) point is approached. There is no softening in the carbon-doped material, qualitatively consistent with the severely reduced critical temperature. After (Baron, et al., 2007).

Boron doped diamond is another superconductor that garnered significant interest after its discovery in 2004 (Ekimov, et al., 2004). Here one can dream about marrying some of the excellent material properties of diamond (hardness, exceptionally high thermal conductivity) with superconductivity, and further, since the nominal phonon energies are very high (>~150 meV for diamond optical modes) it might be possible to have extremely strong electron-phonon coupling (and the resulting softening) without introducing structural instability. The conceptual picture, based on LDA pseudo-potential calculations, was that the boron doping creates Fermi-surface pockets about the Γ point with, somewhat similarly to $MgB_2$, strong epc for small-momentum phonon wave vectors (Boeri, et al., 2004). However, the suggestion was also that the 3-dimentional character of the epc would probably limit the achievable maximum $T_c$ For this sample, like $MgB_2$, there was again, an issue with making large sample, with the largest samples only available as thin films, ~0.1 mm thick. However, by etching away the substrate, it was possible to isolate these films and measure the phonon response using IXS, and indeed the expected softening of the phonon mode was observed (see figure 20) confirming the nature of the superconductivity (Hoesch, et al., 2007).

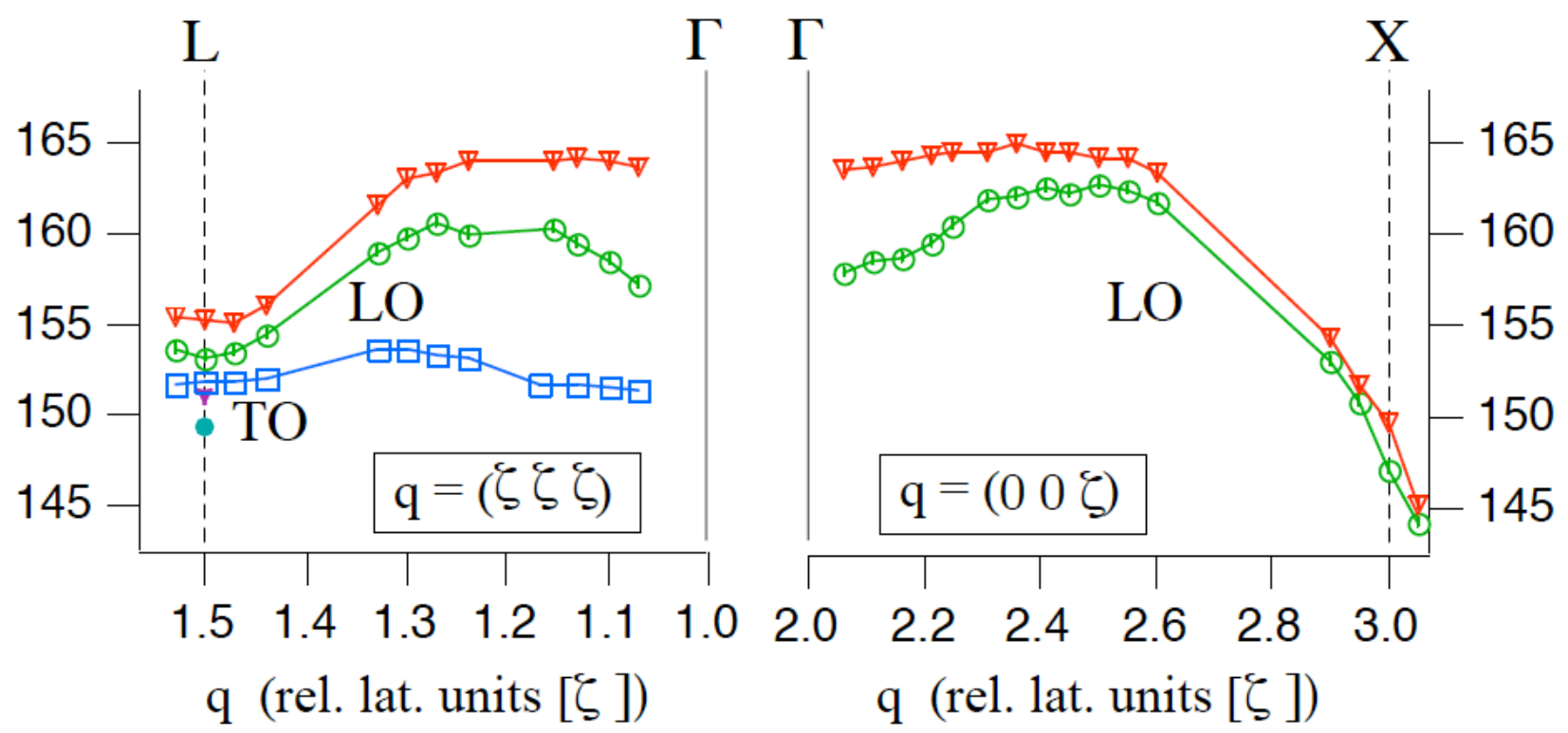

**Figure 20. Phonon softening in boron doped diamond** (after (Hoesch, et al., 2007)**).** Measurements from thin epitaxially grown ~0.1 mm films. The vertical scale in energy transfer in meV. The red triangles are a nominally undoped material while the green circles (and blue squares) are boron doped samples. The softening of the LO mode (red to green) and its increase in the neighborhood of the gamma point are in good qualitative agreement with estimates from (Boeri, et al., 2004).

## 5.2.3 Conventional Superconductors - Temperature Dependence

The immediately obvious experiment of investigating the temperature dependence of phonon spectra as one cools a superconductor below the superconducting transition temperature, has not yet been done (with a measured effect) using IXS, though several experiments have been carried out using neutron scattering. Here we discuss some of the INS results in part to round out the discussion of phonons in conventional superconductors, in part to set the scene for discussions to follow, and in part because the results are pretty and the, mostly, good agreement with theory supports the microscopic picture of electron-phonon coupling in conventional superconductors. In general, the measured phonon changes tend to be small (meV scale or smaller) and the largest effects appear for low energy phonons (generally those with energy <~ twice the gap energy, E<~2Δ) so that it is understandable that neutron scattering, with the potential for better energy resolution, has been the investigative method of choice for these experiments. Clear effects were observed using INS in the 70s, first with $Nb_3Sn$ with its relatively high $T_c$=18K (Axe & Shirane, 1973) and later even the sub 100-μeV effects in Sn ($T_c$=9.2K) became observable (Shapiro, et al., 1975) allowing careful mapping to be done. The observed changes can be understood based on the idea that opening of the superconducting gap in the electronic density of states prevents a phonon with energy less than 2Δ from decaying into an electron hole pair. Thus, phonon modes with energy smaller 2Δ lose a decay path and have a resulting increase in lifetime - the line width of low-energy phonon modes tends to narrow when the temperature is reduced. At the same time, there will be a build-up in the electronic density of states at energies just above the gap, leading to enhanced coupling, and increased broadening, for modes with energy just above 2Δ.

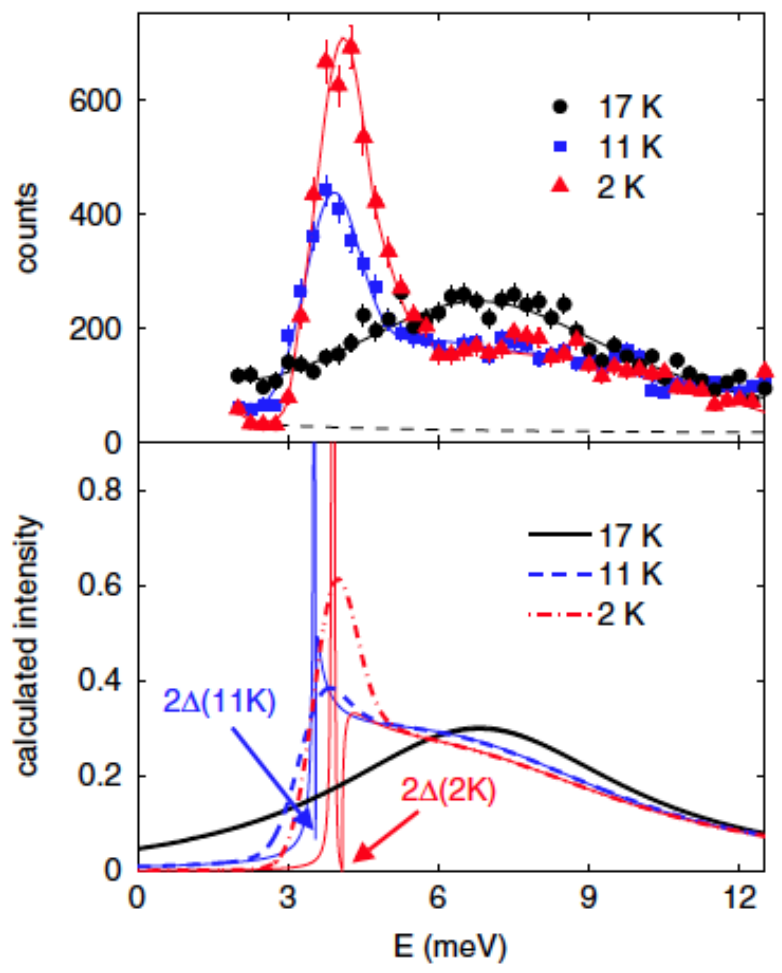

Figure 21. Superconductivity induced phonon lineshape changes in $YNi_2B_2C$ after (Weber, et al., 2008). The top panel shows the spectra measured by inelastic neutron scattering while the lower panel the calculations based on (Allen, et al., 1997) both before and after convolving in the resolution. Figure Copyright (2008) by The American Physical Society.

The interesting case of what happens when the phonon energy is near 2Δ was investigated in the nickel borocarbides, with essentially similar effects observed first for $YNi_2B_2C$ ($T_c$~14.2) (Kawano, et al., 1996) and then in $LuNi_2B_2C$ ($T_c$~16.5K) (Stassis, et al., 1997). Theory (Allen, et al., 1997), at the level of eqn 30, then very beautifully explained the rather complex line shapes with essentially no free parameters These experiments were then confirmed more recently (Weber, et al., 2008), with the data and calculations shown in figure 21.

### 5.2.4 Cuprate Superconductors

The relevance of phonons to superconductivity in the cuprates remains debatable: most workers focus on other mechanisms with, depending on the paper (and sometimes the phonon mode), the phonons being superfluous, or pair enhancing, or pair breaking. In some cases, even the suggestion that there might be a relation between superconductivity and phonons can lead to immediate disregard of subsequent discussion. Thus, here we try to take the conservative viewpoint that phonons are one component of a complex interacting system, that they have an energy scale that is approximately that of the superconducting gap, and that, very generally, one can hope to learn something about a system from its phonon spectra. As before, one looks for anomalies in the form of unusual dispersion, broad line widths or, perhaps, unusual temperature dependence. A theoretically oriented review of phonons in cuprates can be found in (Gunnarsson & Rösch, 2008) while a recent experimental review including some results from both IXS and INS is (Reznik, 2012).

There have been, mostly, two types of phonons that have generated interest in the cuprates: "buckling" modes corresponding to a buckling of the Cu-O planes (having d-type symmetry at Γ) (see the movie in ESM13) and "stretching" (or half-breathing) modes which change the Cu-O bond distance (movie in ESM12). The buckling mode, typically at 35 to 43 meV (depending on material and momentum transfer) is accessible by Raman scattering and shows temperature dependence, with changes in line width and energy on ~meV scales appearing, most strongly, near and below $T_c$ (Macfarlane & Rosen, 1987; Cooper, et al., 1988; Thomsen, et al., 1988). Meanwhile the longitudinal bond stretching mode (typically 60 to 80 meV in energy) is accessible only by INS and IXS with the interest being primarily being rapid softening and increased in line-width as one moves in the (x00) or (0x0) directions in reciprocal space (Pintschovius & Reichardt, 1994). The interesting region is generally x~0.3[+]. More recently, the appearance of a charge density wave in under-doped YBCO (Ghiringhelli, et al., 2012), which also occurs also along (x00) with x~0.3 has revitalized interest in phonons.

Naively, cuprates are immediately interesting candidates for IXS. This is because it is difficult to produce the large size samples needed for neutron scattering measurements, with, arguably, the only good sample available in "bulk" single crystal form being $La_{2-x}Sr_xCu_2O_4$, and some closely related materials, which have a relatively low $T_C$<40K. The YBCO family of materials are available, with effort, for some doping levels, but, for example, optimally doped samples are generally twinned, while the Hg family is difficult to grow and dope, the Bi family tends to have strong long-range structural modulations that blurs out the phonon response and the Tl family remains relatively unexplored. Thus, with access to small samples, one might expect that IXS could make a large impact for cuprate investigations. However, IXS experiments on cuprates are severely hampered by low scattering rates coupled with a large unit cell size: the interesting phonons are copper oxygen modes in a material with many heavier (Ba, Hg, Bi, Y) atoms that limit the penetration into the sample (see discussion in section 4.8) so intensities are generally weak; while primitive cell sizes start at 8 atoms per cell for $HgBa_2CuO_{4+d}$ but are usually in fact rather larger (e.g. 13 atoms for $YBa_2Cu_3O_{7-d}$) so separating out individual modes can be challenging. In fact, studies of cuprates has been part of the motivation driving the construction of the long insertion device beamline in RIKEN at SPring-8 (Baron, 2010).

Considering the bond-stretching mode first, there have been several studies including $Nd_{1.86}Ce_{0.14}CuO_4$ (d'Astuto, et al., 2002), $HgBa_2CuO_{4+d}$, (Uchiyama, et al., 2004), $Bi_2Sr_{1.6}La_{0.4}Cu_2O_{6+d}$ (Graf, et al., 2008) $La_{2-x}Sr_xCuO_4$ (Fukuda, et al., 2005) (Ikeuchi, et al., 2006)(Graf, et al., 2007)(Park, et al., 2014), $La_{2-x}Ba_xCuO_{4+d}$ (d'Astuto, et al., 2008) and $La_{1.48}Nd_{0.4}Sr_{0.12}CuO_4$ (Reznik, et al., 2008). These have typically been done with either 3 or 6 (as opposed to 1.5) meV resolution in order to maximize count-rates. The interesting points about this mode were that, at least in LSCO, it tended to soften more, as the material is doped into the superconducting state and usually there is added broadening as one moves out into the Brillouin zone in the (x00) direction with x>~0.25 (as previously seen by INS (Pintschovius & Reichardt, 1994)). The line width of the mode, typically ~5 meV or more, tends to be large on the scale of typical electron-phonon line widths (excepting extreme cases, like $MgB_2$) and may be interesting. Initial interest in the softening of the mode with increasing x, however, has been reduced somewhat based on the results of LDA calculation for both $YBa_2Cu_3O_{7-d}$ (Bohnen, et al., 2003), and $La_{2-x}Sr_xCuO_4$ (Giustino, et al., 2008) which show that the LDA predicts the mode softens approximately in agreement with measurement - except that the measurements sometimes show a sharp kink in dispersion near to 0.3. The calculations, however, do not predict the observed large line width of the mode. There has also been discussion of this mode (see e.g. (Graf, et al., 2008) in favor and (Park, et al., 2014) against) as being related to kinks in photo-emission data (Lanzara, et al., 2001), as the energy of the kink is about that of the bond-stretching phonon and the momentum transfer of the fast dispersion and onset of large line width agrees, qualitatively, with a possible nesting vector of the of Fermi surface. While this idea remains to be well verified

[+] We use tetragonal notation where the [100] direction in real space is along the Cu-O bond and the [001] direction is normal to the Cu-O plane. For the moment, and as is often done, we neglect differences between (100) and (010) though, for most materials, these are not completely equivalent.

quantitatively, it is interesting to consider - the more so as the momentum transfer corresponds to an apparent CDW instability in some materials (Ghiringhelli, et al., 2012) as has also been verified by investigating acoustic phonon modes using IXS (Blackburn, et al., 2013) (Le Tacon, et al., 2013). We note that, in this context, the LDA calculations are internally consistent in that the same calculations that fail to predict the large phonon line width also fail to predict the kinks (Heid, et al., 2008)(Giustino, et al., 2008) - with the calculations under-estimating both the bond-stretching phonon line width and the kink magnitude by about an order of magnitude. We note that the mode assignment in $Nd_{1.86}Ce_{0.14}CuO_4$ (d'Astuto, et al., 2002) has been called into question in (Braden, et al., 2005) while other work has suggested that there might have been sample contamination for the over-doped $La_{2-x}Sr_xCuO_4$ (x=0.3) in (Fukuda, et al., 2005). Finally, we note the relatively good momentum resolution of the IXS work served to clarify (Reznik, et al., 2008) an interpretation based on INS (Reznik, et al., 2006).

IXS work on the buckling mode is limited compared to that on the stretching mode. The mode is at lower energy than the stretching mode (~42 meV in YBCO) and, while visible with Raman scattering, is somewhat difficult to isolate by either IXS or INS. So far, IXS and INS explorations have mainly been limited to bi-layer materials where, by a nice application of the phasing of phonon eigenvectors, one can, partly, isolate the mode by going to rather specific momentum transfers (Fong, et al., 1995). The only published IXS study to date (Baron, et al., 2008), figure 3c, is in rough qualitative agreement with previous INS studies (Reznik, et al., 1995). However, the IXS studies would probably benefit strongly from improved resolution, say, 1.5 meV instead of the 3 meV used previously, as should be relatively straightforward at present-day spectrometers.

### 5.2.5 Iron Pnictide Superconductors

After the discovery of the high superconducting transition temperature in $LaFeAsO_{1-y}$ by Hosono's group (Kamihara, et al., 2008), a huge amount of work, including IXS, has been carried out to investigate these materials. Work has been done on the "1111" family including $LaFeAsO_{1-y}$, $PrFeAsO_{1-y}$ and $NdFeAsO_{1-y}$ where the La material has a relatively low $T_c$~26K and the others, if optimally doped, are in the neighborhood of 50K (Ren, et al., 2008) and on the "122" materials such as (doped varieties of) $BaFe_2As_2$ and $SrFe_2As_2$ which tend to have slightly lower $T_c$s (e.g. $T_c$=38K for $Ba_{1-x}K_xFe_2As_2$ (Rotter, et al., 2008)) and increasingly the "11" materials including FeSe. Sample growth is easiest for 122 and 11 materials where relatively large single crystals are readily available. The issues related to phonons and IXS include (1) a tendency of the detailed superconducting properties to be rather sensitive to structure with, e.g. pressure making large changes in $T_c$, suggesting strong lattice-superconductivity interaction and (2) the interaction of the phonons with magnetic order and, on some levels, the failure of density functional pseudo-potential calculations (3) the possible impact of nematic (1-D) order. Notably, phonon measurements of optical modes in iron pnictide and related materials are much easier than in the cuprates, since (1) the interesting modes tend to involve higher-Z elements (iron and arsenic, as compared to oxygen in the cuprates) so have higher intensities, and (2) the structures are often simpler so with fewer modes at each momentum transfer. Thus good IXS spectra can often be readily be obtained in few-hour scans.

Work in phonons in the iron based superconductors has tended to either be broader studies or tended to focus specifically on the acoustic modes in the neighborhood of structural or super-conducting transitions (Niedziela, et al., 2011) (Parshall, et al., 2015): the TA modes tend to show anomalous (non-linear) dispersion at small momentum transfers that has been associated with domain structure. Meanwhile the broader studies of optical modes show no obvious anomalies but have frustratingly poor agreement with calculations. There are no modes with large line width or strange dispersion (say, comparable to the bond-stretching mode of the curates) but the DFT calculations using non-magnetic models generally have mode frequencies that are often higher than experiment, even when the experiments are on materials without static magnetic order (e.g. the parent materials above the onset of magnetic order, or the superconducting materials that show no static magnetic order). Inclusion of static magnetic order in the calculations then reduces the energy of some modes to be closer to measurement, but also predicts strong splitting of phonon modes that does not agree with experiment. When comparing to calculations, the experimental results look both generally soft, and relatively isotropic, with only small differences (compared to the calculated ones) being induced by the onset of magnetic order, and essentially no difference with the onset of superconductivity.

These issues are visible figure 22 where several different calculations are compared against an extensive data set on $PrFeAsO_{1-y}$. The "original" calculation (tetragonal, no magnetism) in (a) clearly over-estimates the energy of the highest optical mode shown. Meanwhile, an effort to simulate the doping by adding an ordered oxygen vacancy and a larger (x4) primitive cell in (b) introduces many modes which are not observed. Adding static magnetic ordering to the calculations in (c) with an orthorhombic structure leads improvement for one mode, but also strong splitting, e.g., of the mode at ~30-35 meV, that is not observed, while removal of the *structural* orthorhombicity, but preservation of the magnetic order/orthorhombicity (d) does not make significant changes (i.e. the calculated splitting responds primarily to the magnetic order). Trying a series of ad-hoc modifications led to some improvement, but was not entirely satisfactory (Fukuda, et al., 2011),. Recently some progress has been made by measuring a de-twinned sample (so the effect of the onset of magnetic order on the phonons could be clearly identified) and considering fast magnetic fluctuations to effectively collapse phonon lines, as is possible if the fluctuations occur more quickly than the time scale set by the inverse (static) splitting of the lines (Murai, et al., 2016). This allows a calculated phonon that, if not in perfect agreement with the data, is much better, and begins to approach the level of agreement of calculation for other materials.

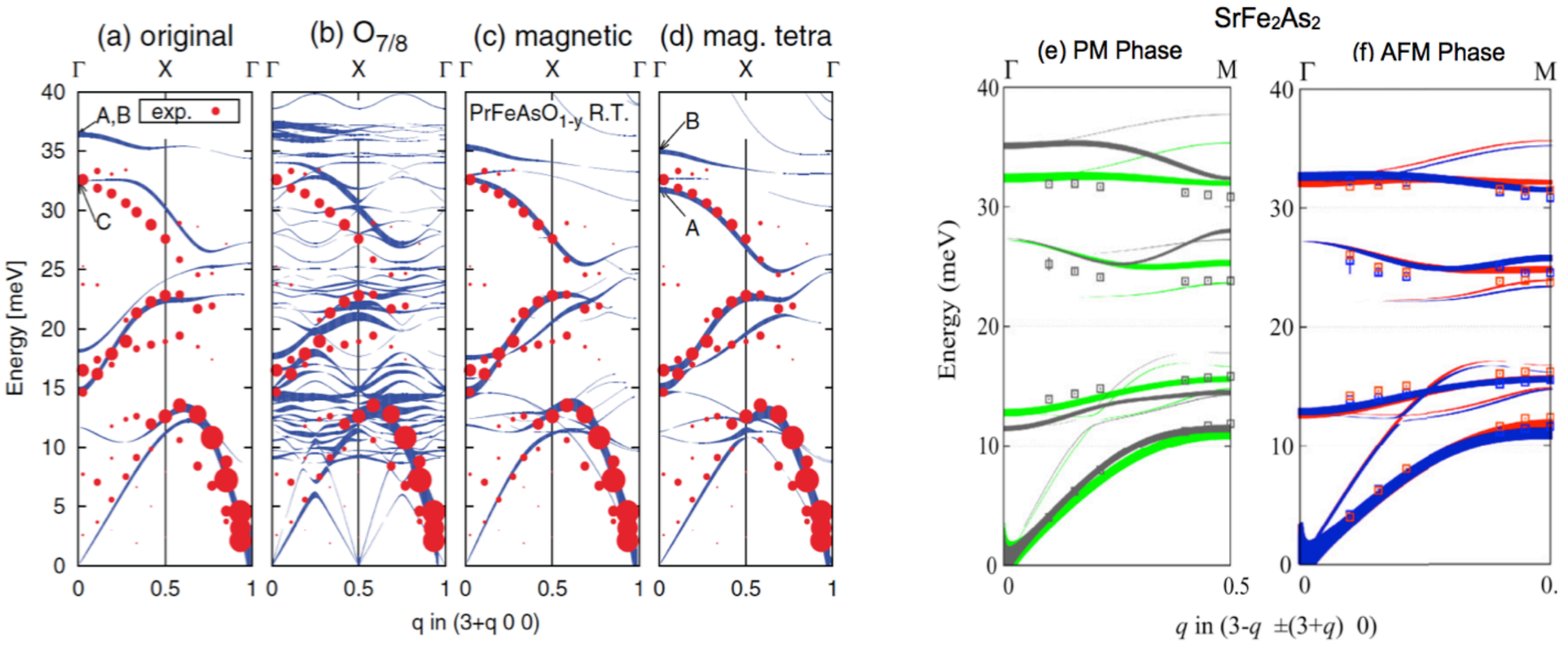

**Figure 22. Comparison of measured dispersion and calculations for $PrFeAsO_{1-y}$ (left) and $SrFe_2As_2$ (right).** The lines are calculated from one of several *ab-initio* models, with thickness of the lines indicating the expected phonon intensity. The points are measured mode energies, with, for the $PrFeAsO_{1-y}$, the point size indicating the mode intensity. In general the agreement between measurements and calculation is not very good until one allows averaging over magnetic calculations, as was done for the $SrFe_2As_2$, with the improvement from the grey to light green lines (panel e) to model the paramagnetic (PM) response being especially clear, and the improvement by reducing the splitting from (f to g) also notable in the anti-ferromagnetic (AFM) phase (for panels f&g the two colors, red and blue, represent the two different orthorhombic dispersion diretions). ($PrFeAsO_{1-y}$ after (Fukuda, et al., 2011), $SrFe_2As_2$ after (Murai, et al, 2016). Copyright by the American Physical Society)

# 6. Concluding Comments

This paper has summarized many of the issues relevant to using high-resolution inelastic x-ray scattering, IXS, to measure phonons in crystalline materials. The field is operating an advanced level, with several established spectrometers presently available. For users, the main advantage of these experiments is the relatively easy availability of clean spectra from small samples. This makes it possible to measure phonons of newly discovered materials, without heroic efforts being needed to grow large single crystals. From the viewpoint of instrumentation, these flux-limited experiments do push the limits of crystal optics, and remain challenging. At this time there is also a large opportunity for improvement in data treatment, with the cleanliness of the data offering the possibility to fit the full spectra to microscopic models.

# 7. Acknowledgements

I am grateful to several scientists who kindly read and offered comments on preliminary versions of this paper including Sunil Sinha, Rolf Heid, Aleksandr Chumakov, Hiroshi Fukui, Kazuyoshi Yamada and Yuri Shvyd'ko. I thank the scientists who kindly shared some of the details of the beamlines with me, including Ahmet Alatas, Ayman Said, Alexey Bosak, Claudio Masciovecchio and Yong Cai. I also thank people for useful relevant comments including F. Weber and M. Sutton. The paper is based on work carried out at SPring-8. I would like to express my deep appreciation to the many people in all parts of SPring-8 that I have had the pleasure of working with over the last two decades, as well as collaborators outside SPring-8. This work is based on experience gained during many proposals including 2001B 0203 0481 0482 0508 0575 3607, 2002A 0182 0279 0280 0520 0537 0559 0560 0561 0562 0627, 2002B 0151 0178 0179 0180 0243 0248 0249 0287 0382 0383 0529 0539 0565 0593 0594 0632 0668 0709, 2003A 0022 0081 0153 0175 0235 0284 0357 0555 0637 0638 0683 0716, 2003B 0019 0132 0206 0248 0359 0397 0574 0693 0743 0744 0745 0755 0766, 2004A 0322 0439 0510 0519 0577 0582 0590 0634, 2004B 0003 0070 0204 0343 0491 0597 0632 0635 0722 0730 0736 0752 , 2005A 0039 0061 0146 0147 0148 0157 0330 0369 0428 0475 0567 0596 0616 0712 0751, 2005B 0082 0093 0124 0253 0266 0295 0346 0441 0443 0484 0603 0623 0650 0731 0736, 2006A 1023 1039 1057 1081 1181 1226 1242 1272 1273 1291 1345 1376 1379 1417 1430 1453 1467 1502, 2006B 1053 1082 1089 1146 1186 1204 1235 1259 1299 1311 1337 1352 1356 1405 1417, 2007A 1109 1118 1125 1125 1222 1234 1279 1281 1301 1374 1436 1441 1473 1505 1507 1523 1539 1561 1612 1647 1671, 2007B 1053 1062 1099 1114 1118 1197 1198 1215 1322 1328 1336 1343 1375 1444 1538 1614 1640 1662, 2008A 1058 1064 1125 1140 1204 1205 1394 1456 1491 1522 1568 1582 1584 1587 1588 1626, 2008B 1381 1403 1473 1178 1108 1326 1584 1240 1144 1169 1491 1634, 2009A 1054 1093 1146 1189 1203 1224 1274 1290 1299 1358 1379 1436 1451 1492 1506 1548, 2009B 1074 1114 1126 1150 1165 1286 1323 1423 1439 1548 1555 1584 1609 1619, 2010B 1108 1112 1177 1185 1206 1353 1354 1392 1410 1453 1497 15271538 1575 1579 1593, 2011A 1051 1075 1104 1117 1136 1154 1180 1256 1271 1300 1304 1366 1373 1452 1502, 2011B 1122 1213 1215 1314 1332 1336 1353 1388 1406 1408 1423 1425 1536 1590, 2012A 1102 1115 1122 1156 1219 1237 1243 1250 1255 1354 1362 1390 1406 1417 1452 1506 1583, 2012B 1080 1125 1159 1196 1226 1236 1277 1283 1343 1356 1358 1364 1439 1577 1596 1658 , 2014A 1026 1059 1076 1086 1089 1100 1106 1122 1131 1154 1207 1231 1235 1240 1346 1368 1378 1385 1434 1678 1687, 2014B 1052 1066 1068 1130 1143 1159 1182 1222 1269 1271 1290 1365 1381 1465 1381 1465 1536 1545 1739 1760 1761 1175 1192.

## 8. Electronic Supplementary Materials

Depending on the version, this paper has electronic supplementary materials that may be available with electronic supplementary materials (ESM). These are a selection of movies of various phonons, designed to complement the text. If you would do not have access to them and would like them, please contact the author.

Specific figures include

**ESM1: Movie of a longitudinal acoustic (LA) mode in MgB$_2$ near to Γ.** The view is looking down the c-axis, small grey spheres are B atoms arranged in a hexagonal lattice while yellow spheres are represent Mg atoms.
*File: baron_esm1_mgb2_la.mov*

**ESM2: Movie of a transverse acoustic (TA) mode in MgB$_2$ near to Γ.** (see notes on ESM1). Note the motion is transverse to the direction of correlation.
*File: baron_esm2_mgb2_ta.mov*

**ESM3: Movie of a longitudinal optic (LO) mode in MgB$_2$ near to Γ.** (see notes on ESM1). Note the out-of-phase (anti-phase) motion of the atoms.
*File: baron_esm3_mgb2_lo.mov*

**ESM4: Movie of a Transverse optic (TO) mode in MgB$_2$ near to Γ.** (see notes on ESM1). Note the out-of-phase (anti-phase) motion of the atoms.
*File: baron_esm4_mgb2_to.mov*

**ESM5: Movie of the (low energy) LA mode near to Γ in CaAlSi.** The view is perpendicular to the c-axis, with the plane of Ca atoms (white) alternating with the Al-Si planes (blue and red, respectively). Note all motions are in phase.
*File: baron_esm5_cas_low_e_gamma.mov*

**ESM6: Movie of the (high energy) LO mode near to Γ in CaAlSi.** Same view and colors as for ESM5. Note the anti-phase motion of the adjacent Al and Si.
*File: baron_esm6_cas_high_e_gamma.mov*

**ESM7: Movie of the low energy mode near to zone boundary (0 0 0.5)** in CaAlSi, which, by following the dispersion from Γ, is the acoustic mode. Same view and colors as for ESM5. Note the anti-phase motion of the adjacent Al and Si.
*File: baron_esm7_cas_low_e_zb.mov*

**ESM8: Movie of the high energy mode near to zone boundary (0 0 0.5)** in CaAlSi, which, by following the dispersion from Γ, is the optic mode. Same view and colors as for ESM5. Note the in-phase motion of the adjacent Al and Si.
*File: baron_esm8_cas_high_e_zb.mov*

**ESM9: Movie of the low energy mode near to the anti-crossing at (0 0 0.21)** in CaAlSi, which, by following the dispersion from Γ, is the acoustic mode. Same view and colors as for ESM5. Note the polarization mixing.
File: *File: baron_esm9_cas_low_e_ac.mov*

**ESM10: Movie of the high energy mode near to the anti-crossing at (0 0 0.21)** in CaAlSi, which, by following the dispersion from Γ, is the optic mode. Same view and colors as for ESM5. Note the polarization mixing.
*File: baron_esm10_cas_high_e_ac.mov*

**ESM11: Movie of a Fe-As optical mode in PrFeAsO.** The lack of reflection symmetry about the Fe plane leads to elliptical atomic motions.
*File: baron_esm11_pr1111_0_p25_0_m12.mov*

**ESM12: Movie of one of the bond-stretching modes in $YBa_2Cu_3O_7$ near to Γ.** Cu atoms are red, O white, Y yellow, Ba blue. The c-axis is vertical.
*File: baron_esm12_ybco_bs_3p02_0_0_m39.mov*

**ESM13: Movie of one of the buckling modes in $YBa_2Cu_3O_7$ near to Γ.** (see notes on ESM12)
*File: baron_esm13_ybco_buck_0_0_0_m27.mov*

# 9. References

Abernathy, D. L., Stone, M. B., Loguillo, M. J., Lucas, M. S., Delaire, O., Tang, X., Lin, J. Y. Y., & Fultz, B. (2012). "Design and operation of the wide angular-range chopper spectrometer ARCS at the Spallation Neutron Source" *Rev. Sci. Instrum.* **83**, 15114.

Abrikosov, I. A., Ponomareva, A. V, Steneteg, P., Barannikova, S. A., & Alling, B. (2016). "Recent progress in simulations of the paramagnetic state of magnetic materials" *Curr. Opin. Solid State Mater. Sci.* **20**, 85–106. DOI: http://dx.doi.org/10.1016/j.cossms.2015.07.003.

Akashi, R., Kawamura, M., Tsuneyuki, S., Nomura, Y., & Arita, R. (2015). "First-principles study of the pressure and crystal-structure dependences of the superconducting transition temperature in compressed sulfur hydrides" *Phys. Rev. B.* **91**, 224513. DOI: 10.1103/PhysRevB.91.224513.

Alatas, A., Leu, B. M., Zhao, J., Yavaş, H., Toellner, T. S., & Alp, E. E. (2011). "Improved focusing capability for inelastic X-ray spectrometer at 3-ID of the APS: A combination of toroidal and Kirkpatrick-Baez (KB) mirrors" *Nucl. Inst. Meth. Phys. Res. A.* **649**, 166–168. DOI: http://dx.doi.org/10.1016/j.nima.2010.11.068.

Alatas, A., Said, A., Sinn, H., Bortel, G., Hu, M., Zhao, J., Burns, C., Burkel, E., & Alp, E. (2008). "Atomic form-factor measurements in the low-momentum transfer region for Li, Be, and Al by inelastic x-ray scattering" *Phys. Rev. B.* **77**, 64301. DOI: 10.1103/PhysRevB.77.064301.

Alfè, D. (2009). "PHON: A program to calculate phonons using the small displacement method" *Comput. Phys. Commun.* **180**, 2622–2633. DOI: http://dx.doi.org/10.1016/j.cpc.2009.03.010.

Allen, P. B. (1972). "Neutron Spectroscopy of Superconducors" *Phys. Rev. B.* **6**, 2577.

Allen, P. B. (1980). "Phonons and the Superconducting Transition Temperature" *Dynamical Properties of Solids*, , edited by G.K. Horton, & A.A. Maradudin, pp. 96–196. Amsterdam: North Holland.

Allen, P. B. & Cohen, M. L. (1972). "Supercondcutivity and Phonon Softening" *Phys. Rev. Lett.* **29**, 1593.

Allen, P. B. & Dynes, R. C. (1975). "Superconductivity and phonon softening: II. Lead Alloys" *Phys. Rev. B.* **11**, 1895.

Allen, P. B. & Dynes, R. C. (1975). "Transition Temperature of strong-coupled superconductors reanalyzed" *Phys. Rev. B.* **12**, 905.

Allen, P. B., Kostur, V. N., Takesue, N., & Shirane, G. (1997). "Neutron-scattering profile of $Q\ensuremath{\ne}0$ phonons in BCS superconductors" *Phys. Rev. B.* **56**, 5552–5558. http://link.aps.org/doi/10.1103/PhysRevB.56.5552.

Allen, P. B., Kostur, V. N., Takesue, N., & Shirane, G. (1997). "Neutron-scattering profile of Q>0 phonons in BCS superconductors" *Phys. Rev. B.* **56**, 5552–5558. http://link.aps.org/doi/10.1103/PhysRevB.56.5552.

Allen, P. B. & Mitrovic, B. (1982). "Theory of Superconducting Tc" *Solid State Phys.* **37**, 1–92.

Allen, P. B. & Silberglitt, R. (1974). "Some effects of phonon dynamics on electron lifetime, mass renormalization, and sueprconducting transition temperature" *Phys. Rev. B.* **9**, 4733.

Ambegaokar, V., Conway, J. M., & Baym, G. (1965). "Inelastic Scattering of Neutrons by Anharmonic Crystals" *Lattice Dynamics*, , edited by R.F. Wallis, p. 261. New York: Pergamon.

Ament, L. J. P. & Brink, J. van den (2011). "Determining the electron-phonon coupling strength from Resonant Inelastic X-ray Scattering at transition metal L-edges" *Europhys. Lett.* **95**, 27008. http://stacks.iop.org/0295-5075/95/i=2/a=27008.

Ament, L. J. P., van Veenendaal, M., Devereaux, T. P., Hill, J. P., & van den Brink, J. (2011). "Resonant inelastic x-ray scattering studies of elementary excitations" *Rev. Mod. Phys.* **83**, 705–767. http://link.aps.org/doi/10.1103/RevModPhys.83.705.

Antonangeli, D., Farber, D. L., Said, A. H., Benedetti, L. R., Aracne, C. M., Landa, A., Söderlind, P., & Klepeis, J. E. (2010). "Shear softening in tantalum at megabar pressures" *Phys. Rev. B.* **82**, 132101. http://link.aps.org/doi/10.1103/PhysRevB.82.132101.

Antonangeli, D., Krisch, M., Farber, D. L., Ruddle, D. G., & Fiquet, G. (2008). "Elasticity of Hexagonal-Closed-Packed Cobalt at High Pressure and Temperature: A Quasiharmonic Case" *Phys. Rev. Lett.* **100**, 85501. http://link.aps.org/doi/10.1103/PhysRevLett.100.085501.

Antonangeli, D., Krisch, M., Fiquet, G., Badro, J., Farber, D. L., Bossak, A., & Merkel, S. (2005). "Aggregate and single-crystalline elasticity of hcp cobalt at high pressure" *Phys. Rev. B.* **72**, 134303–134307. DOI: 10.1103/PhysRevB.72.134303.

Antonangeli, D., Krisch, M., Fiquet, G., Farber, D. L., Aracne, C. M., Badro, J., Occelli, F., & Requardt, H. (2004). "Elasticity of cobalt at high pressure studied by inelastic X-ray scattering" *Phys. Rev. Lett.* **93**, 215504–215505. DOI: 10.1103/PhysRevLett.93.215505.

Antonangeli, D., Occelli, F., Requardt, H., Badro, J., Fiquet, G., & Krisch, M. (2004). "Elastic anisotropy in textured hcp-iron to 112 GPa from sound wave propagation measurements" *Earth Planet. Sci. Lett.* **225**, 243–251. DOI: 10.1016/j.epsl.2004.06.004.

Antonangeli, D., Siebert, J., Aracne, C. M., Farber, D. L., Bosak, A., Hoesch, M., Krisch, M., Ryerson, F. J., Fiquet, G., & Badro, J. (2011). "Spin

Crossover in Ferropericlase at High Pressure: A Seismologically Transparent Transition?" *Sci.* . 331, 64–67. http://www.sciencemag.org/content/331/6013/64.abstract.

Authier, A. (2003). Dynamical Theory of X-Ray Diffraction Oxford: Oxford University Press. http://www.oxfordscholarship.com/view/10.1093/acprof:oso/9780198528920.001.0001/acprof-9780198528920.

Axe, J. D. & Shirane, G. (1973). "Inelastic Neutron Scatterin study of Acoustic Phonons in Nb3Sn" *Phys. Rev. B.* 8, 1965.

Aynajian, P., Keller, T., Boeri, L., Shapiro, S. M., Habicht, K., & Keimer, B. (2008). "Energy Gaps and Kohn Anomalies in Elemental Superconductors" *Science (80-. ).* 319, 1509–1512. http://www.sciencemag.org/content/319/5869/1509.

Badro, J., Fiquet, G., Guyot, F., Gregoryanz, E., Occelli, F., Antonangeli, D., & D'Astuto, M. (2007). "Effect of light elements on the sound velocities in solid iron: Implications for the composition of Earth's core" *Earth Planet. Sci. Lett.* 254, 233–238. DOI: 10.1016/j.epsl.2006.11.025.

Bansal, D., Niedziela, J. L., He, X., Lanigan-Atkins, T., Said, A., Alatas, A., Abernathy, D. L., Ren, Y., Gao, B., Cheong, S.-W., et al. (2019). "Lattice dynamics of the hybrid improper ferroelectrics ${(\mathrm{Ca},\mathrm{Sr})}_{3}{\mathrm{Ti}}_{2}{\mathrm{O}}_{7}$" *Phys. Rev. B.* 100, 214304. DOI: 10.1103/PhysRevB.100.214304.

Bansal, D., Niedziela, J. L., May, A. F., Said, A., Ehlers, G., Abernathy, D. L., Huq, A., Kirkham, M., Zhou, H., & Delaire, O. (2017). "Lattice dynamics and thermal transport in multiferroic CuCrO_{2}" *Phys. Rev. B.* 95, 54306. DOI: 10.1103/PhysRevB.95.054306.

Bardeen, J., Cooper, L. N., & Schrieffer, J. R. (1957). "Theory of Superconductivity" *Phys. Rev.* 108, 1175. http://link.aps.org/abstract/PR/v108/p1175.

Baron, A. Q. R. "High-Resolution Inelastic X-Ray Scattering I & II" *Synchrotron Light Sources and Free-Electron Lasers: Accelerator Physics, Instrumentation and Science*, , edited by E. Jaeschke, S. Khan, J.R. Schneider, & J.B. Hastings, pp. 1643–1757. Cham: Springer International Publishing. DOI: 10.1007/978-3-319-14394-1_52 see also https://arxiv.org/abs/1504.01098.

Baron, A. Q. R. (2009). "X線非弾性散乱による結晶中フォノンの研究 (Phonons in Crystals Using Inelastic X-Ray Scattering)" *分光研究 (Journal Spectrosc. Soc. Japan - Japanese, ArXiv 0910.5764 English).* 54, 205–214.

Baron, A. Q. R. (2010). "The RIKEN Quantum NanoDynamics Beamline (BL43LXU): The Next Generation for Inelastic X-Ray Scattering" *SPring-8 Inf. Newsl.* 15, 14–19. http://user.spring8.or.jp/sp8info/?p=3138.

Baron, A. Q. R. (2013). "Toward Sub-meV, Momentum-Resolved, Inelastic X-Ray Scattering using a Nuclear Analyzer" *J. Phys. Soc. Japan.* 82, SA029. http://journals.jps.jp/doi/abs/10.7566/JPSJS.82SA.SA029.

Baron, A. Q. R., Ishikawa, D., Fukui, H., & Nakajima, Y. (2018). "Auxiliary Optics For meV-IXS at SPring-8: KB, Analyzer Masks, Soller Slit & Screen, BPM" *Accept. Publ. Conf. Proc. SRI2018.* https://arxiv.org/abs/1807.03620.

Baron, A. Q. R., Sutter, J. P., Tsutsui, S., Uchiyama, H., Masui, T., Tajima, S., Heid, R., & Bohnen, K.-P. (2008). "First Study of the B1g Buckling Phonon Mode in Optimally Doped, De-Twinned, YBa2Cu3O7-δ by Inelastic X-Ray Scattering" *J. Phys. Chem. Solids.* 69, 3100. DOI: 10.1016/j.jpcs.2008.06.119.

Baron, A. Q. R., Tanaka, Y., Ishikawa, D., Miwa, D., Yabashi, M., & Ishikawa, T. (2001). "A compact optical design for Bragg reflections near backscattering" *J. Synch. Rad.* 8, 1127–1130. DOI: 10.1107/S0909049501010901.

Baron, A. Q. R., Tanaka, Y., Miwa, D., Ishikawa, D., Mochizuki, T., Takeshita, K., Goto, S., Matsushita, T., & Ishikawa, T. (2001). "Early Commissioning of the SPring-8 Beamline for High Resolution Inelastic X-Ray Scattering" *Nucl. Inst. Meth. Phys. Res. A.* 467–8, 627–630. DOI: 10.1016/S0168-9002(01)00431-4.

Baron, A. Q. R., Tanaka, Y., & Tsutsui, S. (2002). "Performance of BL35XU for High Resolution Inelastic X-ray Scattering" *SPring-8 Front.* 2002, 104. http://www.spring8.or.jp/en/news_publications/publications/research_frontiers/html/RF01B-02A.html.

Baron, A. Q. R., Uchiyama, H., Heid, R., Bohnen, K. P., Tanaka, Y., Tsutsui, S., Ishikawa, D., Lee, S., & Tajima, S. (2007). "Two-phonon contributions to the inelastic x-ray scattering spectra of MgB_2" *Phys. Rev. B.* 75, 20504–20505. DOI: 10.1103/PhysRevB.75.020505.

Baron, A. Q. R., Uchiyama, H., Tanaka, Y., Tsutsui, S., Ishikawa, D., Lee, S., Heid, R., Bohnen, K. P., Tajima, S., & Ishikawa, T. (2004). "Kohn Anomaly in MgB2 by Inelastic X-Ray Scattering" *Phys. Rev. Lett.* 92, 197004. http://link.aps.org/abstract/PRL/v92/e197004.

Baron, A. Q. R., Uchiyama, H., Tsutsui, S., Tanaka, Y., Ishikawa, D., Sutter, J. P., Lee, S., Tajima, S., Heid, R., & Bohnen, K.-P. (2007). "Phonon spectra in pure and carbon doped MgB2 by inelastic X-ray scattering" *Phys. C Supercond. Its Appl.* 456, 83–91. DOI: 10.1016/j.physc.2007.01.028.

Baroni, S., Gironcoli, S. de, Corso, A. D., & Giannozzi, P. (2001). "Phonons and related crystal properties from density-functional perturbation theory." *Rev. Mod. Phys.* 73, 515–562.

Batterman, B. W. & Cole, H. (1964). "Dynamical Diffraction of X-Rays by Perfect Crystals" *Rev. Mod. Phys.* 36, 681–717.

Bauer, J., Han, J., & Gunnarsson, O. (2013). "Retardation effects and the Coulomb pseudopotential in the theory of superconductivity" *Phys. Rev. B.* 87, 54507. DOI: 10.1103/PhysRevB.87.054507.

Baumert, J., Gutt, C., Krisch, M., Requardt, H., M¸ller, M., Tse, J. S., Klug, D. D., & Press, W. (2005). "Elastic properties of methane hydrate at high pressures" *Phys. Reveiw B.* 72, 54302–54305. DOI: 10.1103/PhysRevB.72.054302.

Baumert, J., Gutt, C., Shpakov, V. P., Tse, J. S., Krisch, M., Mueller, M., Requardt, H., Klug, D. D., Janssen, S., & Press, W. (2003). "Lattice dynamics of methane and xenon hydrate: Observation of symmetry-avoided crossing by experiment and theory" *Phys. Reveiw B.* 68, 174301–174307. DOI: 10.1103/PhysRevB.68.174301.

Baym, G. (1961). "Thermodynamic Green's Function Methods in Neutron Scattering by Crystals" *Phys. Rev.* 121, 741–747. DOI: 10.1103/PhysRev.121.741.

Benedek, G., Bernasconi, M., Chis, V., Chulkov, E., Echenique, P. M., Hellsing, B., & Toennies, J. P. (2010). "Theory of surface phonons at metal surfaces: recent advances" *J. Phys. Cond. Mat.* 22, 84020. http://stacks.iop.org/0953-8984/22/i=8/a=084020.

Beye, M. & Föhlisch, A. (2011). "A soft X-ray approach to electron–phonon interactions beyond the Born–Oppenheimer approximation" *J. Electron Spectros. Relat. Phenomena.* 184, 313–317. DOI: http://dx.doi.org/10.1016/j.elspec.2010.12.032.

Björn, W., Alexeï, B., Sabrina, N., Daniele, A., Alessandro, M., Lal, C. S., Ranjan, M., Eiji, O., Anton, S., Surendra, S., et al. (2016). "Dynamical and elastic properties of MgSiO3 perovskite (bridgmanite)" *Geophys. Res. Lett.* 43, 2568–2575. DOI: 10.1002/2016GL067970.

Blackburn, E., Chang, J., Said, A. H., Leu, B. M., Liang, R., Bonn, D. A., Hardy, W. N., Forgan, E. M., & Hayden, S. M. (2013). "Inelastic x-ray study of phonon broadening and charge-density wave formation in ortho-II-ordered YBa2Cu3O6.54" *Phys. Rev. B.* 88, 54506. http://link.aps.org/doi/10.1103/PhysRevB.88.054506.

Boeri, L., Cappelluti, E., & Pietronero, L. (2005). "Three-Dimensional MgB2 Type Superconductivity in Hole-Doped Diamond" *Phys. Rev. B.* 71, 12501. DOI: 10.1103/PhysRevB.71.012501.

Boeri, L., Kortus, J., & Andersen, O. (2004). "Three-Dimensional MgB2 Type Superconductivity in Hole-Doped Diamond" *Phys. Rev. Lett.* 93, 237002. DOI: 10.1103/PhysRevLett.93.237002.

Bohnen, K. P., Heid, R., & Krauss, M. (2003). "Phonon dispersion and electron-phonon interaction for YBa2Cu3O7 from first-principles

calculations" *Europhys. Lett.* . 64, 104–110. http://dx.doi.org/10.1209/epl/i2003-00143-x.

Bohnen, K. P., Heid, R., & Renker, B. (2001). "Phonon Dispersion and Electron-Phonon Coupling in MgB2 and AlB2" *Phys. Rev. Lett.* 86, 5771.

De Boissieu, M., Francoual, S., Mihalkovič, M., Shibata, K., Baron, A. Q. R., Sidis, Y., Ishimasa, T., Wu, D., Lograsso, T., Regnault, L.-P., et al. (2007). "Lattice dynamics of the Zn-Mg-Sc icosahedral quasicrystal and its Zn-Sc periodic 1/1 approximant" *Nat. Mater.* 6, 977. DOI: 10.1038/nmat2044.

Bolmatov, D., Zhernenkov, M., Sharpnack, L., Agra-Kooijman, D. M., Kumar, S., Suvorov, A., Pindak, R., Cai, Y. Q., & Cunsolo, A. (2017). "Emergent Optical Phononic Modes upon Nanoscale Mesogenic Phase Transitions" *Nano Lett.* 17, 3870–3876. DOI: 10.1021/acs.nanolett.7b01324.

Borissenko, E., Goffinet, M., Bosak, A., Rovillain, P., Cazayous, M., Colson, D., Ghosez, P., & Krisch, M. (2013). "Lattice dynamics of multiferroic BiFeO 3 studied by inelastic x-ray scattering" *J. Phys. Cond. Mat.* 25, 102201. http://stacks.iop.org/0953-8984/25/i=10/a=102201.

Born, M. & Huang, K. (1954). Dynamical Theory of Crystal Lattices Oxford: Clarendon press.

Bosak, A., Chernyshov, D., Hoesch, M., Piekarz, P., Le Tacon, M., Krisch, M., Kozłowski, A., Oleś, A. M., & Parlinski, K. (2014). "Short-Range Correlations in Magnetite above the Verwey Temperature" *Phys. Rev. X.* 4, 11040. http://link.aps.org/doi/10.1103/PhysRevX.4.011040.

Bosak, A., Hoesch, M., Antonangeli, D., Farber, D. L., Fischer, I., & Krisch, M. (2008). "Lattice dynamics of vanadium: Inelastic x-ray scattering measurements" *Phys. Rev. B.* 78, 20301. http://link.aps.org/doi/10.1103/PhysRevB.78.020301.

Bosak, A., Hoesch, M., Krisch, M., Chernyshov, D., Pattison, P., Schulze-Briese, C., Winkler, B., Milman, V., Refson, K., Antonangeli, D., et al. (2009). "3D Imaging of the Fermi Surface by Thermal Diffuse Scattering" *Phys. Rev. Lett.* 103, 76403. DOI: 10.1103/PhysRevLett.103.076403.

Bosak, A. & Krisch, M. (2005). "Phonon density of states probed by inelastic X-ray scattering" *Phys. Rev. B.* 72, 224305–224309. DOI: 10.1103/PhysRevB.72.224305.

Bosak, A. & Krisch, M. (2007). "Inelastic x-ray scattering from phonons under multibeam conditions" *Phys. Rev. B.* 75, 92302. DOI: 10.1103/PhysRevB.75.092302.

Bosak, A., Krisch, M., Mohr, M., Maultzsch, J., & Thomsen, C. (2007). "Elasticity of single-crystalline graphite: Inelastic x-ray scattering study" *Phys. Rev. B.* 75, 153404–153408. DOI: 10.1103/PhysRevB.75.153408.

Bosak, A., Schmalzl, K., Krisch, M., van Beek, W., & Kolobanov, V. (2008). "Lattice dynamics of beryllium oxide: Inelastic x-ray scattering and ab initio calculations" *Phys. Rev. B.* 77, 224303. http://link.aps.org/doi/10.1103/PhysRevB.77.224303.

Bosak, A., Serrano, J., Krisch, M., Watanabe, K., Taniguchi, T., & Kanda, H. (2006). "Elasticity of hexagonal boron nitride: Inelastic x-ray scattering measurements" *Phys. Rev. B.* 73, 41402–41404. DOI: 10.1103/PhysRevB.73.041402.

Braden, M., Pintschovius, L., Uefuji, T., & Yamada, K. (2005). "Dispersion of the high-energy phonon modes in Nd1.85Ce0.15CuO4" *Phys. Rev. B.* 72, 184517.

Braicovich, L., Rossi, M., Fumagalli, R., Peng, Y., Wang, Y., Arpaia, R., Betto, D., De Luca, G. M., Di Castro, D., Kummer, K., et al. (2020). "Determining the electron-phonon coupling in superconducting cuprates by resonant inelastic x-ray scattering: Methods and results on ${\mathrm{Nd}}_{1+x}{\mathrm{Ba}}_{2\ensuremath{-}x}{\mathrm{Cu}}_{3}{\mathrm{O}}_{7\ensuremath{-}\ensuremath{\delta}}$" *Phys. Rev. Res.* 2, 23231. DOI: 10.1103/PhysRevResearch.2.023231.

Brand, R. A., Krisch, M., Chernikov, M. A., & Ott, H. R. (2001). "Phonons in the icosahedral quasicrystal i-AlPdMn studied by coherent inelastic scattering of synchrotron radiation" *Ferroelectrics.* 250, 233–236.

Brand, R. A., Voss, J., Hippert, F., Krisch, M., Sterzel, R., Assmus, W., & Fisher, I. R. (2004). "Phonon dispersion curve of icosahedral MgZnY quasicrystals" *J. Non. Cryst. Solids.* 334-335 %0, 207–209. DOI: 10.1016/j.jnoncrysol.2003.11.040.

Brown, J. M. & McQueen, R. G. (1986). "Phase transitions, grüneisen parameter, and elasticity for shocked iron between 77 GPa and 400 GPa" *J. Geophys. Res.* 91, 7485–7494.

Brubaker, Z. E., Ran, S., Said, A. H., Manley, M. E., Söderlind, P., Rosas, D., Idell, Y., Zieve, R. J., Butch, N. P., & Jeffries, J. R. (2019). "Phonon dispersion of Mo-stabilized $\ensuremath{\gamma}$-U measured using inelastic x-ray scattering" *Phys. Rev. B.* 100, 94311. DOI: 10.1103/PhysRevB.100.094311.

Brüesch, P. (1982). Phonons:  Theory and Experiments 1 Berlin: Springer-Verlag.

Burkel, E. (1991). Inelastic Scattering of X-rays with Very High Energy Resolution Berlin: Springer.

Burkel, E. (2000). "Phonon Spectroscopy by Inelastic X-ray Scattering" *Reports Prog. Phys.* 63, 171–232.

Burkel, E., Peisl, J., & Dorner, B. (1987). "Observation of Inelastic X-Ray Scattering from Phonons" *Europhys. Lett.* 3, 957–961.

Burkovsky, R. G., Andronikova, D., Bronwald, Y., Krisch, M., Roleder, K., Majchrowski, A., Filimonov, A. V, Rudskoy, A. I., & Vakhrushev, S. B. (2015). "Lattice dynamics in the paraelectric phase of PbHfO 3 studied by inelastic x-ray scattering" *J. Phys. Cond. Mat.* 27, 335901. http://stacks.iop.org/0953-8984/27/i=33/a=335901.

Burkovsky, R. G., Tagantsev, A. K., Vaideeswaran, K., Setter, N., Vakhrushev, S. B., Filimonov, A. V., Shaganov, A., Andronikova, D., Rudskoy, A. I., Baron, A. Q. R., et al. (2014). "Lattice dynamics and antiferroelectricity in PbZrO3 tested by x-ray and Brillouin light scattering" *Phys. Rev. B.* 90, 144301. DOI: 10.1103/PhysRevB.90.144301.

Cai, Y. Q. (2018). "Presented at SRI2018" p.

Cai, Y. Q., Coburn, D. S., Cunsolo, A., Keister, J. W., Honnick, M. G., Huang, X. R., Kodituwakku, C. N., Stetsko, Y., Suvorov, A., Hiraoka, N., et al. (2013). "The Ultrahigh Resolution IXS Beamline of NSLS-II: Recent Advances and Scientific Opportunities" *J. Phys. Conf.* 425, 201001.

Calandra, M. & Mauri, F. (2005). "Electron-phonon coupling and phonon self-energy in MgB2: Interpretation of MgB2 Raman Sepctra" *Phys. Rev. B.* 71, 64501.

Cappelluti, E. (2006). "Electron-phonon effects on the Raman spectrum in MgB2" *Phys. Rev. B.* 75, 140505(R).

Caruso, F., Hoesch, M., Achatz, P., Serrano, J., Krisch, M., Bustarret, E., & Giustino, F. (2017). "Nonadiabatic Kohn Anomaly in Heavily Boron-Doped Diamond" *Phys. Rev. Lett.* 119, 17001. https://link.aps.org/doi/10.1103/PhysRevLett.119.017001.

Chen, Y., Ai, X., & Marianetti, C. A. (2014). "First-Principles Approach to Nonlinear Lattice Dynamics: Anomalous Spectra in PbTe" *Phys. Rev. Lett.* 113, 105501. DOI: 10.1103/PhysRevLett.113.105501.

Chernyshov, D., Bosak, A., Dmitriev, V., Filinchuk, Y., & Hagemann, H. (2008). "Low-lying phonons in NaBH4 studied by inelastic scattering of synchrotron radiation" *Phys. Rev. B.* 78, 172104. DOI: 10.1103/PhysRevB.78.172104.

Chumakov, A. I., Baron, A. Q. R., Ruffer, R., Grunsteudel, H., Grunsteudel, H. F., & Meyer, A. (1996). "Nuclear resonance energy analysis of inelastic x-ray scattering" *Phys. Rev. Lett.* 76, 4258–4261. http://link.aps.org/abstract/PRL/v76/p4258.

Chumakov, A. I. I., Metge, J., Baron, A. Q. R., Grünsteudel, H. F. F., Grünsteudel, H. F. F., Rüffer, R., & Ishikawa, T. (1996). "An X-ray Monochromator with 1.65 meV Energy Resolution" *Nucl. Inst. Meth.* 383, 642–644. DOI: 10.1016/S0168-9002(96)00924-2.

Colella, R. (1974). "Multiple Diffraction of X-rays and the Phase Problem. Computational Procedures and Comparison with Experiment" *Acta Crystallogr.* **A30**, 413–423.

Cooper, S. L., Klein, M. V, Pazol., B. G., Rice, J. P., & Ginzberg, D. M. (1988). "Raman scattering from superconducting gap except in single-crystal YBa2Cu3O7-d" *Phys. Rev. B.* **37**, 5920.

d'Astuto, M., Calandra, M., Bendiab, N., Loupias, G., Mauri, F., Zhou, S., Graf, J., Lanzara, A., Emery, N., Hérold, C., et al. (2010). "Phonon dispersion and low-energy anomaly in CaC6 from inelastic neutron and x-ray scattering experiments" *Phys. Rev. B.* **81**, 104519. http://link.aps.org/doi/10.1103/PhysRevB.81.104519.

d'Astuto, M., Calandra, M., Reich, S., Shukla, A., Lazzeri, M., Mauri, F., Karpinski, J., Zhigadlo, N. D., Bossak, A., & Krisch, M. (2007). "Weak anharmonic effects in MgB2: A comparative inelastic X-ray scattering and Raman study" *Phys. Rev. B.* **75**, 174508–174510. DOI: 10.1103/PhysRevB.75.174508.

d'Astuto, M., Dhalenne, G., Graf, J., Hoesch, M., Giura, P., Krisch, M., Berthet, P., Lanzara, A., & Shukla, A. (2008). "Sharp optical-phonon softening near optimal doping in La2-xBaxCuO4+delta observed via inelastic x-ray scattering" *Phys. Rev. B.* **78**, 140511. http://link.aps.org/doi/10.1103/PhysRevB.78.140511.

d'Astuto, M., Heid, R., Renker, B., Weber, F., Schober, H., De la Peña-Seaman, O., Karpinski, J., Zhigadlo, N. D., Bossak, A., & Krisch, M. (2016). "Nonadiabatic effects in the phonon dispersion of Mg_{1-x}Al_{x}B_{2}" *Phys. Rev. B.* **93**, 180508. DOI: 10.1103/PhysRevB.93.180508.

d'Astuto, M., Mang, P. K., Giura, P., SHukla, A., Ghigna, P., Mirone, A., Braden, M., Greven, M., Krisch, M., & Sette, F. (2002). "Anomalous Longitudinal Dispersion in Nd1.86Ce0.14)4+d Determined by Inelastic X-Ray Scattering" *Phys. Rev. Lett.* **88**, 167002.

d'Astuto, M., Mirone, A., Giura, P., Colson, D., Forget, A., & Krisch, M. (2003). "Phonon dispersion in the one-layer cuprate HgBa2CuO4+delta" *J. Phys. Condens. Matter.* **15**, 8827–8836. DOI: 10.1088/0953-8984/15/50/014.

D'Astuto, M., Yamada, I., Giura, P., Paulatto, L., Gauzzi, A., Hoesch, M., Krisch, M., Azuma, M., & Takano, M. (2013). "Phonon anomalies and lattice dynamics in the superconducting oxychlorides Ca2-xCuO2Cl2" *Phys. Rev. B.* **88**, 14522. http://link.aps.org/doi/10.1103/PhysRevB.88.014522.

Deutsch, M., Hart, M., & Sommer-Larsen, P. (1989). "Thermal motion of atoms in crystalline silicon: Beyond the Debye theory" *Phys. Rev. B.* **40**, 11666–11669. http://link.aps.org/doi/10.1103/PhysRevB.40.11666.

Devereaux, T. P., Shvaika, A. M., Wu, K., Wohlfeld, K., Jia, C. J., Wang, Y., Moritz, B., Chaix, L., Lee, W.-S., Shen, Z.-X., et al. (2016). "Directly characterizing the relative strength and momentum dependence of electron-phonon coupling using resonant inelastic x-ray scattering" *Phys. Rev. X.* **6**, 041019. DOI: 10.1103/PhysRevX.6.041019.

Doi, A., Fujioka, J., Fukuda, T., Tsutsui, S., Okuyama, D., Taguchi, Y., Arima, T., Baron, A. Q. R., & Tokura, Y. (2014). "Multi-spin-state dynamics during insulator-metal crossover in LaCoO3" *Phys. Rev. B.* **90**, 81109. DOI: 10.1103/PhysRevB.90.081109.

Dorner, B., Burkel, E., Illini, T., & Peisl, J. (1987). "First measurement of a phonon dispersion curve by inelastic X-ray scattering" *Zeitschrift Für Phys. B Condens. Matter.* **69**, 179–183. DOI: 10.1007/BF01307274.

Dorner, B., Burkel, E., & Peisl, J. (1986). "An X-ray backscattering instrument with very high energy resolution" *Nucl. Inst. Meth.* **A426**, 450.

Dorner, B., Fujii, Y., Hastings, J., Moncton, D., Siddons, D., & and others (1980). "Notes from 1980 Vienna Summer School provided by D. Moncton".

Drozdov, A. P., Eremets, M. I., Troyan, I. A., Ksenofontov, V., & Shylin, S. I. (2015). "Conventional superconductivity at 203 kelvin at high pressures in the sulfur hydride system" *Nature.* **525**, 73–76. http://dx.doi.org/10.1038/nature14964.

Eckhold, G., Stein-Arsic, M., & Weber, H. J. (1987). "UNISOFT - a program package for lattice dynamical calculations" *J. Appl. Crystallogr.* **20**, 134–139.

Ekimov, E. A., Sidorov, V. A., Bauer, E. D., Mel'nik, N. N., Curro, N. J., Thompson, J. D., & Stishov, S. M. (2004). "Superconductivity in diamond" *Nature.* **428**, 542–545. http://dx.doi.org/10.1038/nature02449.

Eliashberg, G. M. (1960). "Interactions between electrons and lattice vibrations in a superconductor" *Sov. Phys. JETP.* **11**, 696–702.

Ellis, D. S., Uchiyama, H., Tsutsui, S., Sugimoto, K., Kato, K., & Baron, A. Q. R. (2014). "X-ray study of the structural distortion in EuTiO3" *Phys. B Condens. Matter.* **442**, 34–38. https://doi.org/10.1016/j.physb.2014.02.041.

Ellis, D. S., Uchiyama, H., Tsutsui, S., Sugimoto, K., Kato, K., Ishikawa, D., & Baron, A. Q. R. (2012). "Phonon softening and dispersion in EuTiO3" *Phys. Rev. B.* **86**, 220301. http://link.aps.org/doi/10.1103/PhysRevB.86.220301.

Euchner, H., Mihalkovic, M., Gähler, F., Johnson, M., Schober, H., Rols, S., Suard, E., Bosak, A., Ohhashi, S., Tsai, A.-P., et al. (2011). "Anomalous vibrational dynamics in the Mg2Zn11 phase" *Phys. Rev. B.* **83**, 144202. DOI: 10.1103/PhysRevB.83.144202.

Fahy, S., Murray, É. D., & Reis, D. A. (2016). "Resonant squeezing and the anharmonic decay of coherent phonons" *Phys. Rev. B.* **93**, 134308. DOI: 10.1103/PhysRevB.93.134308.

Fak, B. & Dorner, B. (1992). "On the interpretation of phonon line shapes and excitation energies in neutron scattering experiments" *ILL Rep.* 92FA008S.

Fak, B. & Dorner, B. (1997). "Phonon line shapes and excitation energies" *Phys. B* . **234–236**, 1107–1108.

Farber, D. L., Krisch, M., Antonangeli, D., Beraud, A., Badro, J., Occelli, F., & Orlikowski, D. (2006). "Lattice dynamics of molybdenum at high pressure" *Phys. Rev. Lett.* **96**, 115502–115504. DOI: 10.1103/PhysRevLett.96.115502.

Finkelstein, G. J., Jackson, J. M., Said, A., Alatas, A., Leu, B. M., Sturhahn, W., & Toellner, T. S. (2018). "Strongly Anisotropic Magnesiowüstite in Earth's Lower Mantle" *J. Geophys. Res. Solid Earth.* **123**, 4740–4750. DOI: 10.1029/2017JB015349.

Finkelstein, K. (2005). "Private Communication about the Limits of RIE".

Finkemeier, F. & von Niessen, W. (1998). "Phonons and phonon localization in a-Si: Computational approaches and results for continuous-random-network-derived structures" *Phys. Rev. B.* **58**, 4473–4484. DOI: 10.1103/PhysRevB.58.4473.

Fiquet, G., Badro, J., Guyot, F., Requardt, H., & Krisch, M. (2001). "Sound velocities in iron to 110 gigapascals" *Science (80-. ).* **291**, 468–471. DOI: 10.1126/science.291.5503.468.

Fischer, I., Bosak, A., & Krisch, M. (2009). "Single-crystal lattice dynamics derived from polycrystalline inelastic x-ray scattering spectra" *Phys. Rev. B.* **79**, 134302. http://link.aps.org/doi/10.1103/PhysRevB.79.134302.

Floris, A., Profeta, G., Lathiotakis, N. N., L¸ders, M., Marques, M. A. L., Franchini, C., Gross, E. K. U., Continenza, A., & Massidda, S. (2005). "Superconducting Properties of MgB2 from First Principles" *Phys. Rev. Lett.* **94**, 37004. http://link.aps.org/abstract/PRL/v94/e037004.

Fong, H. F., Keimer, B., Anderson, P. W., Reznik, D., Dogan, F., & Aksay, J. A. (1995). "Phonon and magnetic neutron scattering at 41 meV in YBa2Cu3O7" *Phys. Rev. Lett.* **75**, 316–320.

Fujii, Y., Hastings, J., Ulc, S., & Moncton, D. (1982). "SSRPP Technical report VII-95".

Fukuda, T., Baron, A. Q. R., Nakamura, H., Shamoto, S., Ishikado, M., Machida, M., Uchiyama, H., Iyo, A., Kito, H., Mizuki, J., et al. (2011). "Soft and isotropic phonons in PrFeAsO1-y" *Phys. Rev. B.* **84**, 64504. http://link.aps.org/doi/10.1103/PhysRevB.84.064504.

Fukuda, T., Baron, A. Q. R. Q. R., Shamoto, S.-I., Ishikado, M., Nakamura, H., Machida, M., Uchiyama, H., Tsutsui, S., Iyo, A., Kito, H., et al. (2008). "Lattice Dynamics of LaFeAsO1-xFx and PrFeAsO1-y via Inelastic X-Ray Scattering and First-Principles Calculation" *J. Phys. Soc. Japan.* **77**, 103715. DOI: 10.1143/JPSJ.77.103715.

Fukuda, T., Mizuki, J., Ikeuchi, K., Yamada, K., Baron, A. Q. R., & Tsutsui, S. (2005). "Doping dependence of softening in the bond-stretching phonon mode of La-{2-x}Sr_xCuO_4 (0 <= x <= 0.29)" *Phys. Rev. B .* **71**, 60501–60504. http://link.aps.org/abstract/PRB/v71/e060501.

Fukui, H., Baron, A. Q. R. Q. R., Ishikawa, D., Uchiyama, H., Ohishi, Y., Tsuchiya, T., Kobayashi, H., Matsuzaki, T., Yoshino, T., & Katsura, T. (2017). "Pressure dependence of transverse acoustic phonon energy in ferropericlase across the spin transition" *J. Phys. Condens. Matter.* **29**, DOI: 10.1088/1361-648X/aa7026.

Fukui, H., Katsura, T., Kuribayashi, T., Matsuzaki, T., Yoneda, A., Ito, E., Kudoh, Y., Tsutsui, S., & Baron, A. Q. R. (2008). "Precise determination of elastic constants by high-resolution inelastic X-ray scattering" *J. Synch. Rad.* **15**, 618–623. DOI: 10.1107/S0909049508023248.

Fukui, H., Yoneda, A., Nakatsuka, A., Tsujino, N., Kamada, S., Ohtani, E., Shatskiy, A., Hirao, N., Tsutsui, S., Uchiyama, H., et al. (2016). "Effect of cation substitution on bridgmanite elasticity: A key to interpret seismic anomalies in the lower mantle" *Sci. Rep.* **6**, 33337. DOI: 10.1038/srep33337.

Gale, J. D. (1997). "GULP: A computer program for the symmetry-adapted simulation of solids" *J. Chem. Soc. Faraday Trans. .* **93**, 629–637.

Gannon, L., Boeri, L., Howard, C. A., Gougeon, P., Gall, P., Potel, M., Salloum, D., Petrović, A. P., & Hoesch, M. (2018). "Lattice dynamics of the cluster chain compounds ${M}_{2}{\mathrm{Mo}}_{6}{\mathrm{Se}}_{6}\phantom{\rule{4pt}{0ex}}(M=\mathrm{K},\text{Rb},\text{Cs},\text{In},\text{and Tl})$" *Phys. Rev. B.* **98**, 14104. DOI: 10.1103/PhysRevB.98.014104.

Gardner, D. R., Bonnoit, C. J., Chisnell, R., Said, A. H., Leu, B. M., Williams, T. J., Luke, G. M., & Lee, Y. S. (2016). "Inelastic x-ray scattering measurements of phonon dynamics in URu_{2}Si_{2}" *Phys. Rev. B.* **93**, 75123. DOI: 10.1103/PhysRevB.93.075123.

Gerdau, E. & de Waard, H. (2000). "Hyperfine Interactions" **123–125**,.

Ghiringhelli, G., Le Tacon, M., Minola, M., Blanco-Canosa, S., Mazzoli, C., Brookes, N. B., De Luca, G. M., Frano, A., Hawthorn, D. G., He, F., et al. (2012). "Long-Range Incommensurate Charge Fluctuations in (Y,Nd)Ba2Cu3O6+x" *Science (80-. ).* **337**, 821–825. DOI: 10.1126/science.1223532.

Ghose, S., Krisch, M., Oganov, A. R., Beraud, A., Bosak, A., Gulve, R., Seelaboyina, R., Yang, H., & Saxena, S. K. (2006). "Lattice dynamics of MgO at high pressure: Theory and experiment" *Phys. Rev. Lett.* **96**, 35504–35507. DOI: 10.1103/PhysRevLett.96.035507.

Giannozzi, P., Baroni, S., Bonini, N., Calandra, M., Car, R., Cavazzoni, C., Ceresoli, D., Chiarotti, G. L., Cococcioni, M., Dal Corso, A., et al. (2009). "QUANTUM ESPRESSO: a modular and open-source software project for quantum simulations of materials" *J. Phys. Cond. Mat. .* **21**, http://www.quantum-espresso.org.

Gilat, G. & Raubenheimer, L. J. (1966). "Accurate Numerical Method for Calculating Frequency-Distribution Functions in Solids" *Phys. Rev.* **144**, 390–395. DOI: 10.1103/PhysRev.144.390.

Girard, A., Taniguchi, H., Souliou, S. M., Stekiel, M., Morgenroth, W., Minelli, A., Kuwabara, A., Bosak, A., & Winkler, B. (2018). "Competing structural instabilities in ${\mathrm{Bi}}_{2}{\mathrm{SiO}}_{5}$" *Phys. Rev. B.* **98**, 134102. DOI: 10.1103/PhysRevB.98.134102.

Giura, P., Paulatto, L., He, F., Lobo, R. P. S. M., Bosak, A., Calandrini, E., Paolasini, L., & Antonangeli, D. (2019). "Multiphonon anharmonicity of MgO" *Phys. Rev. B.* **99**, 220304. DOI: 10.1103/PhysRevB.99.220304.

Giustino, F., Cohen, M. L., & Louie, S. G. (2008). "Small phonon contribution to the photoemission kink in the copper oxide superconductors" *Nature.* **452**, 975–978. http://dx.doi.org/10.1038/nature06874.

Glyde, H. R. (1994). "Momentum distributions and final-state effects in neutron scattering" *Phys. Rev. B.* **50**, 6726.

Goncharov, A. F., Struzhkin, V. V, Gregoryanz, E., Hu, J., Hemley, R. J., Mao, H., Lapertot, G., Bud'ko, S. L., & Canfield, P. C. (2001). "Raman spectrum and lattice parameters of MgB_{2} as a function of pressure" *Phys. Rev. B.* **64**, 100509. http://link.aps.org/abstract/PRB/v64/e100509.

Gonze, X., Amadon, B., Anglade, P.-M., Beuken, J.-M., Bottin, F., Boulanger, P., Bruneval, F., Caliste, D., Caracas, R., Cote, M., et al. (2009). "ABINIT : First-principles approach of materials and nanosystem properties." *Comput. Phys. Commun.* **180**, 2582–2615. http://www.abinit.org/.

Gonze, X. & Lee, C. (1997). "Dynamical matrices, Born effective charges, dielectric permittivity tensors, and interatomic force constants from density-functional perturbation theory" *Phys. Rev. B.* **55**, 10355–10368. http://link.aps.org/doi/10.1103/PhysRevB.55.10355.

Gov, N. (2003). "Velocity-dependent interactions and sum rule in bcc He" *Phys. Rev. B.* **67**, 52301. http://link.aps.org/doi/10.1103/PhysRevB.67.052301.

Graeff, W. & Materlik, G. (1982). "Millielectron Volt Energy Resolution in Bragg Backscattering" *Nucl. Inst. Meth.* **195**, 97.

Graf, J., D'Astuto, M., Giura, P., Shukla, A., Saini, N. L., Bossak, A., Krisch, M., Cheong, S. W., Sasagawa, T., & Lanzara, A. (2007). "In-plane copper-oxygen bond-stretching mode anomaly in underdoped La2-xSrxCuO4+delta measured with high-resolution inelastic x-ray scattering" *Phys. Rev. B.* **76**, 172504–172507. DOI: 10.1103/PhysRevB.76.172507.

Graf, J., d'Astuto, M., Jozwiak, C., Garcia, D. R., Saini, N. L., Krisch, M., Ikeuchi, K., Baron, A. Q. R., Eisaki, H., & Lanzara, A. (2008). "Bond Stretching Phonon Softening and Kinks in the Angle-Resolved Photoemission Spectra of Optimally Doped Bi_2Sr_1.6La_0.4Cu2O_6+delta Superconductors" *Phys. Rev. Lett.* **100**, 227002–227004. http://link.aps.org/abstract/PRL/v100/e227002.

Gretarsson, H., Clancy, J. P., Singh, Y., Gegenwart, P., Hill, J. P., Kim, J., Upton, M. H., Said, A. H., Casa, D., Gog, T., et al. (2013). "Magnetic excitation spectrum of Na2IrO" *Phys. Rev. B.* **87**, 220407. http://link.aps.org/doi/10.1103/PhysRevB.87.220407.

Grimvall, G. (1981). The Electron-Phonon Interaction in Metals Amsterdam: North Holland Publishing Co.

Grüneis, A., Serrano, J., Bosak, A., Lazzeri, M., Molodtsov, S. L., Wirtz, L., Attaccalite, C., Krisch, M., Rubio, A., Mauri, F., et al. (2009). "Phonon surface mapping of graphite: Disentangling quasi-degenerate phonon dispersions" *Phys. Rev. B.* **80**, 85423. http://link.aps.org/doi/10.1103/PhysRevB.80.085423.

Gunnarsson, O. & Rösch, O. (2008). "Interplay between electron-phonon and Coulomb interactions in cuprates" *J. Phys. Cond. Mat.* **20**, 43201. http://stacks.iop.org/0953-8984/20/043201.

Hafner, J. & Krajci, M. (1993). "Propagating and confined vibrational excitations in quasicrystals" *J. Phys. Cond. Mat.* **5**, 2489. http://stacks.iop.org/0953-8984/5/i=16/a=008.

Hahn, S., Tucker, G., Yan, J.-Q., Said, A., Leu, B., McCallum, R., Alp, E., Lograsso, T., McQueeney, R., & Harmon, B. (2013). "Magnetism-dependent phonon anomaly in LaFeAsO observed via inelastic x-ray scattering" *Phys. Rev. B.* **87**, 104518. DOI: 10.1103/PhysRevB.87.104518.

Heid, R. & Bohnen, K.-P. (1999). "Linear response in a density-functional mixed-basis approach." *Phys. Rev. B.* **60**, R3709.

Heid, R., Bohnen, K.-P., Zeyher, R., & Manske, D. (2008). "Momentum Dependence of the Electron-Phonon Coupling and Self-Energy Effects in Superconducting YBa2Cu3O7 within the Local Density Approximation" *Phys. Rev. Lett.* **100**, 137001–137004. http://link.aps.org/abstract/PRL/v100/e137001.

Hellman, O., Abrikosov, I. A., & Simak, S. I. (2011). "Lattice dynamics of anharmonic solids from first principles" *Phys. Rev. B.* **84**, 180301. http://link.aps.org/doi/10.1103/PhysRevB.84.180301.

Hellman, O., Steneteg, P., Abrikosov, I. A., & Simak, S. I. (2013). "Temperature dependent effective potential method for accurate free energy calculations of solids" *Phys. Rev. B.* **87**, 104111. http://link.aps.org/doi/10.1103/PhysRevB.87.104111.

Henderson, B. & Imbusch, G. F. (1989). Optical spectroscopy of inorganic solids Oxford, New York: Oxford University press.

Hinsen, K., Pellegrini, E., Stachura, S., & Kneller, G. R. (2012). "nMoldyn 3: Using Task Farming for a Parallel Spectroscopy-Oriented Analysis of Molecular Dynamics Simulations" *J. Comput. Chem.* **33**, 2043–2048. http://dirac.cnrs-orleans.fr/nmoldyn/home/.

Hlinka, J., Gregova, I., Pokrny, J., Plecenik, A., Kus, P., Satrapinsky, L., & Benacka, S. (2001). "Phonons in MgB2 by polarized Raman scattering on single crystals" *Phys. Rev. B.* **64**, 140503R.

Hlinka, J., Ondrejkovic, P., Kempa, M., Borissenko, E., Krisch, M., Long, X., & Ye, Z.-G. (2011). "Soft antiferroelectric fluctuations in morphotropic PbZr1-xTixO3 single crystals as evidenced by inelastic x-ray scattering" *Phys. Rev. B.* **83**, 140101. http://link.aps.org/doi/10.1103/PhysRevB.83.140101.

Hoesch, M., Bosak, A., Chernyshov, D., Berger, H., & Krisch, M. (2009). "Giant Kohn Anomaly and the Phase Transition in Charge Density Wave ZrTe3" *Phys. Rev. Lett.* **102**, 86402. DOI: 10.1103/PhysRevLett.102.086402.

Hoesch, M., Fukuda, T., Mizuki, J., Takenouchi, T., Kawarada, H., Sutter, J. P. P., Tsutsui, S., Baron, A. Q. R., Nagao, M., Takano, Y., et al. (2007). "Phonon Softening in Superconducting Diamond" *Phys. Rev. B.* **75**, 140508(R). DOI: 10.1103/PhysRevB.75.140508.

Hoesch, M., Piekarz, P., Bosak, A., Le Tacon, M., Krisch, M., Kozłowski, A., Oleś, A. M., & Parlinski, K. (2013). "Anharmonicity due to Electron-Phonon Coupling in Magnetite" *Phys. Rev. Lett.* **110**, 207204. http://link.aps.org/doi/10.1103/PhysRevLett.110.207204.

Hohenberg, P. C. & Platzman, P. M. (1966). "High-Energy Neutron Scattering from Liquid He4" *Phys. Rev.* **152**, 198–200. http://link.aps.org/doi/10.1103/PhysRev.152.198.

Holt, M., Wu, Z., Hong, H., Zschack, P., Jemian, P., Tischler, J., Chen, H., & Chiang, T. C. (1999). "Determination of Phonon Dispersions from X-Ray Transmission Scattering: The Example of Silicon" *Phys. Rev. Lett.* **83**, 3317. http://link.aps.org/abstract/PRL/v83/p3317.

Hong, H., Xu, R., Alatas, A., Holt, M., & Chiang, T.-C. (2008). "Central peak and narrow component in x-ray scattering measurementsnear the displacive phase transition in SrTiO3" *Phys. Rev. B.* **78**, 104121.

Hosokawa, S., Inui, M., Matsuda, K., Ishikawa, D., & Baron, A. Q. R. R. (2008). "Damping of the collective modes in liquid Fe" *Phys. Rev. B.* **77**, 174203. DOI: 10.1103/PhysRevB.77.174203.

Huotari, S., Vanko, G., Albergamo, F., Ponchut, C., Graafsma, H., Henriquet, C., Verbeni, R., & Monaco, G. (2005). "Improving the performance of high-resolution X-ray spectrometers with position-sensitive pixel detectors" *J. Synch. Rad.* **12**, 467–472. DOI: 10.1107/s0909049505010630.

Ikeuchi, K., Isawa, K., Yamada, K., Fukuda, T., Mizuki, J., Tsutsui, S., & Baron, A. Q. R. (2006). "Growth, Characterization, and Application of Single-Crystal La2-xSrCuO4 Having a Gradient in Sr Concentration" *Jpn. J. Appl. Phys.* **45**, 1594–1601.

Ishikawa, D. & Baron, A. Q. R. (2010). "Temperature gradient analyzers for compact high-resolution X-ray spectrometers" *J. Synch. Rad.* **17**, 12–24. DOI: 10.1107/S0909049509043167.

Ishikawa, D., Ellis, D. S., Uchiyama, H., & Baron, A. Q. R. (2015). "Inelastic X-ray scattering with 0.75meV resolution at 25.7keV using a temperature-gradient analyzer" *J. Synch. Rad.* **22**, 3–8. DOI: 10.1107/S1600577514021006.

Ishikawa, D., Haverkort, M. W., & Baron, A. Q. R. (2017). "Lattice and Magnetic Effects on a d–d Excitation in NiO Using a 25 meV Resolution X-ray Spectrometer" *J. Phys. Soc. Japan,.* **86**, 93706. DOI: 10.7566/JPSJ.86.093706.

Ishikawa, D., Uchiyama, H., Tsutsui, S., Fukui, H., & Baron, A. Q. R. (2013). "Compound focusing for hard-x-ray inelastic scattering" *Proc. SPIE.* **8848**, 88480. DOI: 10.1117/12.2023795.

Ishikawa, T., Hirano, K., & Kikuta, S. (1991). "Applications of Perfect Crystal X-Ray Optics" *Nucl. Inst. Meth. Phys. Res.* **A 308**, 356–362.

Iwasa, K., Iga, F., Yonemoto, A., Otomo, Y., Tsutsui, S., & Baron, A. Q. R. (2014). "Universality of anharmonic motion of heavy rare-earth atoms in hexaborides" *J. Phys. Soc. Japan.* **83**, DOI: 10.7566/JPSJ.83.094604.

Iwasa, K., Igarashi, R., Saito, K., Laulh√©, C., Orihara, T., Kunii, S., Kuwahara, K., Nakao, H., Murakami, Y., Iga, F., et al. (2011). "Motion of the guest ion as precursor to the first-order phase transition in the cage system GdB_{6}" *Phys. Rev. B.* **84**, 214308. DOI: 10.1103/PhysRevB.84.214308.

Iwasa, K., Kuwahara, K., Utsumi, Y., Saito, K., Kobayashi, H., Sato, T., Amano, M., Hasegawa, T., Ogita, N., Udagawa, M., et al. (2012). "Renormalized Motion of Dysprosium Atoms Filling Boron Cages of DyB6" *J. Phys. Soc. Japan.* **81**, 113601. DOI: 10.1143/JPSJ.81.113601.

Jackson, J. D. (1999). Classical Electrodynamics, Third Edition New York: John Wiley and Sons.

James, R. W. (1962). The Optical Principles of the Diffraction of X-rays London: G. Bell & Sons, Ltd.

Jeanloz, R., Celliers, P. M., Collins, G. W., Eggert, J. H., Lee, K. K. M., McWilliams, R. S., Brygoo, S., & Loubeyre, P. (2007). "Achieving high-density states through shock-wave loading of precompressed samples" *Proc. Natl. Acad. Sci. U. S. A.* **104**, 9172–9177. DOI: 10.1073/pnas.0608170104.

Jin, L., Jung-Fu, L., Ahmet, A., Y., H. M., Jiyong, Z., & Leonid, D. (2016). "Seismic parameters of hcp-Fe alloyed with Ni and Si in the Earth's inner core" *J. Geophys. Res. Solid Earth.* **121**, 610–623. DOI: 10.1002/2015JB012625.

Kajimoto, R., Nakamura, M., Inamura, Y., Mizuno, F., Nakajima, K., Ohira-Kawamura, S., Yokoo, T., Nakatani, T., Maruyama, R., Soyama, K., et al. (2011). "The Fermi Chopper Spectrometer 4SEASONS at J-PARC" *J. Phys. Soc. Japan.* **80**, SB025. DOI: 10.1143/JPSJS.80SB.SB025.

Kajimoto, R., Sagayama, H., Sasai, K., Fukuda, T., Tsutsui, S., Arima, T., Hirota, K., Mitsui, Y., Yoshizawa, H., Baron, A. Q. R., et al. (2009). "Unconventional Ferroelectric Transition in the Multiferroic Compound TbMnO[sub 3] Revealed by the Absence of an Anomaly in c-Polarized Phonon Dispersion" *Phys. Rev. Lett.* **102**, 247602–247604. DOI: 10.1103/PhysRevLett.102.247602.

Kamada, S., Ohtani, E., Fukui, H., Sakai, T., Terasaki, H., Takahashi, S., Shibazaki, Y., Tsutsui, S., Baron, A. Q. R., Hirao, N., et al. (2014). "The sound velocity measurements of Fe3S" *Am. Mineral.* **99**, 98–101. DOI: 10.2138/am.2014.4463.

Kamihara, Y., Watanabe, T., Hirano, M., & Hosono, H. (2008). "Iron-Based Layered Superconductor La[O1-xFx]FeAs (x = 0.05-0.12) with Tc = 26 K" *J. Am. Chem. Soc.* . **130**, 3296–3297. http://pubs3.acs.org/acs/journals/doilookup?in_doi=10.1021/ja800073m.

Kantor, A. P., Kantor, I. Y., Kurnosov, A. V, Kuznetsov, A. Y., Dubrovinskaia, N. A., Krisch, M., Bossak, A. A., Dmitriev, V. P., Urusov, V. S., & Dubrovinsky, L. S. (2007). "Sound wave velocities of fcc FeNi alloy at high pressure and temperature by mean of inelastic X-ray scattering" *Phys. Earth Planet. Inter.* **164**, 83–89. DOI: 10.1016/j.pepi.2007.06.006.

Karo, A. M. & Hardy, J. R. (1969). "Lattice Dynamics of NaF" *Phys. Rev.* **181**, 1272–1277. http://link.aps.org/doi/10.1103/PhysRev.181.1272.

Kawaguchi, S. I., Nakajima, Y., Hirose, K., Komabayashi, T., Ozawa, H., Tateno, S., Kuwayama, Y., Tsutsui, S., & Baron, A. Q. R. (2017). "Sound velocity of liquid Fe-Ni-S at high pressure" *J. Geophys. Res. Solid Earth.* **122**, DOI: 10.1002/2016JB013609.

Kawakita, Y., Hosokawa, S., Enosaki, T., Ohshima, K., Takeda, S., Pilgrim, W.-C., Tsutsui, S., Tanaka, Y., Baron, A. Q. R., Oshima, K., et al. (2003). "Coherent Dynamic Scattering Law of Divalent Liquid Mg" *J. Phys. Soc. Japan.* **L72**, 1603. DOI: 10.1143/JPSJ.72.1603.

Kawano, H., Yoshizawa, H., Takeya, H., & Kadowaki, K. (1996). "Anomalous Phonon Scattering Below Tc in YNi2B2C" *Phys. Rev. Lett.* **77**, 4628–4631. http://link.aps.org/doi/10.1103/PhysRevLett.77.4628.

Ketenoglu, D., Harder, M., Klementiev, K., Upton, M., Taherkhani, M., Spiwek, M., Dill, F.-U., Wille, H.-C., & Yavaş, H. (2014). "Resonant inelastic x-ray scattering spectrometer with 25 meV resolution at Cu K-edge" *Submitted for Publication.*

Khosroabadi, H., Miyasaka, S., Kobayashi, J., Tanaka, K., Uchiyama, H., Baron, A. Q. R., & Tajima, S. (2011). "Softening of bond-stretching phonon mode in Ba_{1-x}K_{x}BiO_{3} at the metal-insulator transition" *Phys. Rev. B.* **83**, 224525. http://link.aps.org/doi/10.1103/PhysRevB.83.224525.

Kim, J., Casa, D., Said, A., Krakora, R., Kim, B. J., Kasman, E., Huang, X., & Gog, T. (2018). "Quartz-based flat-crystal resonant inelastic x-ray scattering spectrometer with sub-10 meV energy resolution" *Sci. Rep.* **8**, 1958. DOI: 10.1038/s41598-018-20396-z.

Kim, K.-J., Shvyd'ko, Y., & Reiche, S. (2008). "A Proposal for an X-Ray Free-Electron Laser Oscillator with an Energy-Recovery Linac" *Phys. Rev. Lett.* **100**, 244802. http://link.aps.org/doi/10.1103/PhysRevLett.100.244802.

Klimczuk, A. C. W. and H. C. W. and R. S. and M. K. and A. B. and A. H. H. and C. E. Z.-W. and E. C. and J.-C. G. and D. B. and R. E. and R. C. and T. (2015). "Absence of superconductivity in fluorine-doped neptunium pnictide NpFeAsO" *J. Phys. Cond. Mat.* **27**, 325702. http://stacks.iop.org/0953-8984/27/i=32/a=325702.

Kohl, H. (1985). "Unambiguous Determination of Interatomic Force Consta.nts of Crystals by Coherent Inelastic Neutron Scattering" *Phys. Status Solidi.* **130**, 151.

Kohn, W. (1959). "Image of the Fermi Surface in the vibration Spectrum of a metal" *Phys. Rev. Lett.* **2**, 393.

Kong, Y., Dolgov, O. V, Jepsen, O., & Andersen, O. K. (2001). "Electron-phonon interacrion in the normal and superconducting states of MgB2" *Phys. Rev. B.* **64**, 20501.

Kostov, K. L., Polzin, S., & Widdra, W. (2011). "High-resolution phonon study of the Ag(100) surface" *J. Phys. Cond. Mat.* **23**, 484006. http://stacks.iop.org/0953-8984/23/i=48/a=484006.

Koza, M. M., Leithe-Jasper, A., Rosner, H., Schnelle, W., Mutka, H., Johnson, M. R., Krisch, M., Capogna, L., & Grin, Y. (2011). "Vibrational dynamics of the filled skutterudites M1-xFe4Sb12 (M=Ca, Sr, Ba, and Yb): Temperature response, dispersion relation, and material properties" *Phys. Rev. B.* **84**, 14306. http://link.aps.org/doi/10.1103/PhysRevB.84.014306.

Kreibich, T. & Gross, E. K. U. (2001). "Multicomponent Density-Functional Theory for Electrons and Nuclei" *Phys. Rev. Lett.* **86**, 2984–2987. http://link.aps.org/doi/10.1103/PhysRevLett.86.2984.

Krisch, M., Brand, R. A., Chernikov, M., & Ott, H. R. (2002). "Phonons in the icosahedral quasicrystal i-AlPdMn studied by inelastic X-ray scattering" *Phys. Rev. B.* **65**, 134201–134207. DOI: 10.1103/PhysRevB.65.134201.

Krisch, M., Farber, D. L., Xu, R., Antonangeli, D., Aracne, C. M., Beraud, A., Chiang, T.-C., Zarestky, J., Kim, D. Y., Isaev, E. I., et al. (2011). "Phonons of the anomalous element cerium" *Proc. Natl. Acad. Sci.* . **108**, 9342–9345. DOI: 10.1073/pnas.1015945108.

Krisch, M., Mermet, A., Grimm, H., Forsyth, V. T., & Rupprecht, A. (2006). "Phonon dispersion of oriented DNA by inelastic X-ray scattering" *Phys. Rev. E.* **73**, 61909–61910. DOI: 10.1103/PhysRevE.73.061909.

Krisch, M. & Sette, F. (2007). "Inelastic X-Ray Scattering from Phonons" *Light Scatt. Solids IX.* 317–369.

Kuna, R., Minikayeva, R., Trzyna, M., Gas, K., Bosak, A., Szczerbakow, A., Petit, S., Łazewskif, J., & Szuszkiewiczb, W. (2016). "Inelastic X-Ray Scattering Studies of Phonon Dispersion in PbTe and (Pb,Cd)Te Solid Solution" *ACTA Phys. Pol. A.* **130**, 1251. DOI: 10.12693/APhysPolA.130.1251.

Kunc, K. & Bilz, H. (1976). "Local approach to polarizabilities and trends in the Raman spectra of semiconductors" *Solid State Commun.* **19**, 1027–1030. DOI: http://dx.doi.org/10.1016/0038-1098(76)90091-0.

Kuroiwa, S., Baron, A. Q. R., Muranaka, T., Heid, R., Bohnen, K.-P., & Akimitsu, J. (2008). "Soft-phonon-driven superconductivity in CaAlSi as seen by inelastic x-ray scattering" *Phys. Rev. B.* **77**, 140503. DOI: 10.1103/PhysRevB.77.140503.

L. J. Sham (1974). "Theory of Lattice Dynamics of Covalent Crystlas" *Dynamical Properties of Solids*, , edited by G.K. Horton, & A.A. Maradudin, pp. 301–342. Amsterdam: North-Holland.

Lai, C. H., Fung, H. S., Wu, W. B., Huang, H. Y., Fu, H. W., Lin, S. W., Huang, S. W., Chiu, C. C., Wang, D. J., Huang, L. J., et al. (2014). "Highly efficient beamline and spectrometer for inelastic soft X-ray scattering at high resolution" *J. Synch. Rad.* **21**, 325–332. DOI: 10.1107/S1600577513030877.

Lan, T., Li, C. W., Niedziela, J. L., Smith, H., Abernathy, D. L., Rossman, G. R., & Fultz, B. (2014). "Anharmonic lattice dynamics of Ag_{2}O studied by inelastic neutron scattering and first-principles molecular dynamics simulations" *Phys. Rev. B.* **89**, 54306. DOI: 10.1103/PhysRevB.89.054306.

Lanzara, A., Bogdanov, P. V, Zhou, X. J., Keller, S. A., Feng, D. L., Le, E. D., Yoshida, T., Eisaki, H., Fujimori, A., Kishio, K., et al. (2001). "Evidence for ubiquitous strong electron-phonon coupling in high-temperature superconductors" *Nature.* **412**, 510.

Lax, M. (2012). Symmetry principles in solid state and molecular physics Dover Publications.

Lazzeri, M. & Mauri, F. (2006). "Nonadiabatic Kohn Anomaly in a Doped Graphene Monolayer" *Phys. Rev. Lett.* **97**, 266407. http://link.aps.org/doi/10.1103/PhysRevLett.97.266407.

Lebert, B. W., Yamamoto, H., Azuma, M., Heid, R., Tsutsui, S., Uchiyama, H., Baron, A. Q. R., Baptiste, B., & d'Astuto, M. (2020). "Doping-induced in-plane anisotropy of bond-stretching phonon softening in oxychloride ${\mathrm{Ca}}_{2\ensuremath{-}x}{\mathrm{CuO}}_{2}{\mathrm{Cl}}_{2}$ compounds" *Phys. Rev. B.* **101**, 20506. DOI: 10.1103/PhysRevB.101.020506.

Lee, C.-H., Kihou, K., Horigane, K., Tsutsui, S., Fukuda, T., Eisaki, H., Iyo, A., Yamaguchi, H., Baron, A. Q. R., Braden, M., et al. (2010). "Effect of K Doping on Phonons in $Ba_{1-x}K_xFe_2As_2$" *J. Phys. Soc. Japan* . **79**, 14714.

Lee, S., Mori, H., Masui, T., Eltsev, Y., Yamamoto, A., & Tajima, S. (2001). "Growth, Structure Analysis and Anisotropic Superconducting Properties of MgB2 Single Crystals" *J. Phys. Soc. Japan.* **70**, 2255–2258. DOI: 10.1143/JPSJ.70.2255.

Lee, W. S., Johnston, S., Moritz, B., Lee, J., Yi, M., Zhou, K. J., Schmitt, T., Patthey, L., Strocov, V., Kudo, K., et al. (2013). "Role of Lattice Coupling in Establishing Electronic and Magnetic Properties in Quasi-One-Dimensional Cuprates" *Phys. Rev. Lett.* **110**, 265502. http://link.aps.org/doi/10.1103/PhysRevLett.110.265502.

Leigh, R. S., Szigeti, B., & Tewary, V. K. (1971). "Force Constants and Lattice Frequencies" *Proc. R. Soc. London A Math. Phys. Eng. Sci.* **320**, 505–526. http://rspa.royalsocietypublishing.org/content/320/1543/505.abstract.

Leroux, M., Cario, L., Bosak, A., & Rodière, P. (2018). "Traces of charge density waves in NbS_{2}" *Phys. Rev. B.* **97**, 195140. DOI: 10.1103/PhysRevB.97.195140.

Leroux, M., Errea, I., Le Tacon, M., Souliou, S.-M., Garbarino, G., Cario, L., Bosak, A., Mauri, F., Calandra, M., & Rodière, P. (2015). "Strong anharmonicity induces quantum melting of charge density wave in 2H-NbSe_{2} under pressure" *Phys. Rev. B.* **92**, 140303. DOI: 10.1103/PhysRevB.92.140303.

Leroux, M., Le Tacon, M., Calandra, M., Cario, L., Méasson, M.-A., Diener, P., Borrissenko, E., Bosak, A., & Rodière, P. (2012). "Anharmonic suppression of charge density waves in 2H-NbS{}_{2}" *Phys. Rev. B.* **86**, 155125. DOI: 10.1103/PhysRevB.86.155125.

Li, C. W., Hellman, O., Ma, J., May, A. F., Cao, H. B., Chen, X., Christianson, A. D., Ehlers, G., Singh, D. J., Sales, B. C., et al. (2014). "Phonon Self-Energy and Origin of Anomalous Neutron Scattering Spectra in SnTe and PbTe Thermoelectrics" *Phys. Rev. Lett.* **112**, 175501. DOI: 10.1103/PhysRevLett.112.175501.

Liu, G., Mao, H. K., & Baron, A. Q. R. "Work in progress".

Liu, J., Lin, J.-F., Alatas, A., & Bi, W. (2014). "Sound velocities of bcc-Fe and Fe0.85Si0.15 alloy at high pressure and temperature" *Phys. Earth Planet. Inter.* **233**, 24–32. DOI: http://dx.doi.org/10.1016/j.pepi.2014.05.008.

Liu, Y., Berti, D., Baglioni, P., Chen, S.-H., Alatas, A., Sinn, H., Said, A., & Alp, E. (2005). "Inelastic X-ray scattering studies of phonons propagating along the axial direction of a DNA molecule having different counter-ion atmosphere" *J. Phys. Chem. Solids.* **66**, 2235–2245. DOI: http://dx.doi.org/10.1016/j.jpcs.2005.09.017.

Loa, I., Isaev, E. I., McMahon, M. I., Kim, D. Y., Johansson, B., Bosak, A., & Krisch, M. (2012). "Lattice Dynamics and Superconductivity in Cerium at High Pressure" *Phys. Rev. Lett.* **108**, 45502. http://link.aps.org/doi/10.1103/PhysRevLett.108.045502.

Loa, I., McMahon, M. I., & Bosak, A. (2009). "Origin of the Incommensurate Modulation in Te-III and Fermi-Surface Nesting in a Simple Metal" *Phys. Rev. Lett.* **102**, 35501. http://link.aps.org/doi/10.1103/PhysRevLett.102.035501.

Lovesey, S. W. (1984). Theory of Neutron Scattering from Condensed Matter Oxford: Clarendon Press.

Lüders, M., Marques, M. A. L., Lathiotakis, N. N., Floris, A., Profeta, G., Fast, L., Continenza, A., Massidda, S., & Gross, E. K. U. (2005). "\textit{Ab initio}  theory of superconductivity. I. Density functional formalism and approximate functionals" *Phys. Rev. B.* **72**, 24545. http://link.aps.org/doi/10.1103/PhysRevB.72.024545.

Ma, H., Li, C., Ma, Y., Wang, H., Rouse, Z. W., Zhang, Z., Slebodnick, C., Alatas, A., Baker, S. P., Urban, J. J., et al. (2019). "Supercompliant and Soft ${({\mathrm{CH}}_{3}{\mathrm{NH}}_{3})}_{3}{\mathrm{Bi}}_{2}{\mathrm{I}}_{9}$ Crystal with Ultralow Thermal Conductivity" *Phys. Rev. Lett.* **123**, 155901. DOI: 10.1103/PhysRevLett.123.155901.

Ma, H., Li, C., Tang, S., Yan, J., Alatas, A., Lindsay, L., Sales, B. C., & Tian, Z. (2016). "Boron arsenide phonon dispersion from inelastic x-ray scattering: Potential for ultrahigh thermal conductivity" *Phys. Rev. B.* **94**, 220303. DOI: 10.1103/PhysRevB.94.220303.

Macfarlane, R. M. & Rosen, H. (1987). "Temperature dependence of the Raman spectrum of the high Tc superconductor YBa2Cu3O7" *Solid State Commun.* **63**, 831.

Mahan, G. D. (2000). Many Particle Physics (3rd Edition) New York: Kluwer Academic Publishers-Plenum Publishers.

Maldonado, P., Paolasini, L., Oppeneer, P. M., Forrest, T. R., Prodi, A., Magnani, N., Bosak, A., Lander, G. H., & Caciuffo, R. (2016). "Crystal dynamics and thermal properties of neptunium dioxide" *Phys. Rev. B.* **93**, 144301. DOI: 10.1103/PhysRevB.93.144301.

Mamontov, E., Vakhrushev, S. B., Kumzerov, Y. A., Alatas, A., & Sinn, H. (2009). "Acoustic phonons in chrysotile asbestos probed by high-resolution inelastic x-ray scattering" *Solid State Commun.* **149**, 589–592. DOI: http://dx.doi.org/10.1016/j.ssc.2009.01.033.

Manley, M. E., Lander, G. H., Sinn, H., Alatas, A., Hults, W. L., McQueeney, R. J., Smith, J. L., & Willit, J. (2003). "Phonon dispersion in uranium measured using inelastic x-ray scattering" *Phys. Rev. B.* **67**, 52302. http://link.aps.org/doi/10.1103/PhysRevB.67.052302.

Manley, M. E., Said, A. H., Fluss, M. J., Wall, M., Lashley, J. C., Alatas, A., Moore, K. T., & Shvyd'ko, Y. (2009). "Phonon density of states of alpha- and delta -plutonium by inelastic x-ray scattering" *Phys. Rev. B.* **79**, 52301. http://link.aps.org/doi/10.1103/PhysRevB.79.052301.

Manley, M. E., Yethiraj, M., Sinn, H., Volz, H. M., Alatas, A., Lashley, J. C., Hults, W. L., Lander, G. H., & Smith, J. L. (2006). "Formation of a New Dynamical Mode in alpha-Uranium Observed by Inelastic X-Ray and Neutron Scattering" *Phys. Rev. Lett.* **96**, 125501. http://link.aps.org/doi/10.1103/PhysRevLett.96.125501.

Mao, W. L., Struzhkin, V. V, Baron, A. Q. R., Tsutsui, S., Tommaseo, C. E., Wenk, H.-R., Hu, M. Y., Chow, P., Sturhahn, W., Shu, J., et al. (2008). "Experimental determination of the elasticity of iron at high pressure" *J. Geophys. Res. Solid Earth.* **113**, DOI: 10.1029/2007JB005229.

Mao, Z., Lin, J.-F., Liu, J., Alatas, A., Gao, L., Zhao, J., & Mao, H.-K. (2012). "Sound velocities of Fe and Fe-Si alloy in the Earth's core" *Proc. Natl. Acad. Sci.* . **109**, 10239–10244. DOI: 10.1073/pnas.1207086109.

Maradudin, A. A. (1974). "Crystalline Solids, Fundamentals" *Dynamical Properties of Solids*, , edited by G.K. Horton, & A.A. Maradudin, p. Amsterdam: North-Holland.

Maradudin, A. A. & Vosko, S. H. (1968). "Symmetry Properties of the Normal Vibrations of a Crystal" *Rev. Mod. Phys.* **40**, 1–37. http://link.aps.org/doi/10.1103/RevModPhys.40.1.

Martin, A. V (2017). "Orientational order of liquids and glasses {\it via} fluctuation diffraction" *IUCrJ.* **4**, 24–36. DOI: 10.1107/S2052252516016730.

Maschek, M., Lamago, D., Castellan, J.-P., Bosak, A., Reznik, D., & Weber, F. (2016). "Polaronic metal phases in La_{0.7}Sr_{0.3}MnO_{3} uncovered by inelastic neutron and x-ray scattering" *Phys. Rev. B.* **93**, 45112. DOI: 10.1103/PhysRevB.93.045112.

Maschek, M., Rosenkranz, S., Heid, R., Said, A. H., Giraldo-Gallo, P., Fisher, I. R., & Weber, F. (2015). "Wave-vector-dependent electron-phonon coupling and the charge-density-wave transition in TbT_{3}" *Phys. Rev. B.* **91**, 235146. https://link.aps.org/doi/10.1103/PhysRevB.91.235146.

Maschek, M., Rosenkranz, S., Hott, R., Heid, R., Merz, M., Zocco, D. A., Said, A. H., Alatas, A., Karapetrov, G., Zhu, S., et al. (2016). "Superconductivity and hybrid soft modes in TiSe2" *Phys. Rev. B.* **94**, 214507. DOI: 10.1103/PhysRevB.94.214507.

Maschek, M., Zocco, D. A., Rosenkranz, S., Heid, R., Said, A. H., Alatas, A., Walmsley, P., Fisher, I. R., & Weber, F. (2018). "Competing soft phonon modes at the charge-density-wave transitions in $\mathrm{DyT}{\mathrm{e}}_{3}$" *Phys. Rev. B.* **98**, 94304. DOI: 10.1103/PhysRevB.98.094304.

Masciovecchio, C., Bergmann, U., Krisch, M., Ruocco, G., Sette, F., & Verbeni, R. (1996). "A perfect crystal X-ray analyser with 1.5 meV energy resolution" *Nucl. Inst. Meth. Phys. Res. B.* **117**, 339–340. https://doi.org/10.1016/0168-583X(96)00334-5.

Masciovecchio, C., Bergmann, U., Krisch, M., Ruocco, G., Sette, F., & Verbeni, R. (1996). "A perfect crystal X-ray analyser with meV energy resolution" *Nucl. Inst. Meth. Phys. Res. B.* **111**, 181–186. https://doi.org/10.1016/0168-583X(95)01288-5.

Masciovecchio, C., Gessini, A., & Santucci, S. C. (2006). "Ultraviolet Brillouin scattering as a new tool to investigate disordered systems" *J. Non-Crystalline Solids* . **352**, 5126–5129. http://www.sciencedirect.com/science/article/pii/S0022309306008933.

May, T., Müller, W., & Strauch, D. (1998). "Anharmonic lattice dynamics and neutron-scattering spectra in bcc transition metals" *Phys. Rev. B.* **57**, 5758–5763. DOI: 10.1103/PhysRevB.57.5758.

McMillan, W. L. (1968). "Transition Temperature of Strong-Coupled Sueprconductors" *Phys. Rev.* **167**, 331.

Merritt, A. M., Weber, F., Castellan, J.-P., Wolf, T., Ishikawa, D., Said, A. H., Alatas, A., Fernandes, R. M., Baron, A. Q. R., & Reznik, D. (2020).

"Nematic Correlation Length in Iron-Based Superconductors Probed by Inelastic X-Ray Scattering" *Phys. Rev. Lett.* 124, 157001. DOI: 10.1103/PhysRevLett.124.157001.

Meyer, J., Dolling, G., Scherm, R., & Glyde, H. R. (1976). "Anharmonic interference effects in potassium" *J. Phys. F Met. Phys.* 6, 943–956.

Miao, H., Ishikawa, D., Heid, R., Le Tacon, M., Fabbris, G., Meyers, D., Gu, G. D., Baron, A. Q. R., & Dean, M. P. M. (2018). "Incommensurate Phonon Anomaly and the Nature of Charge Density Waves in Cuprates" *Phys. Rev. X.* 8, 11008. DOI: 10.1103/PhysRevX.8.011008.

Miao, H., Zhang, T. T., Wang, L., Meyers, D., Said, A. H., Wang, Y. L., Shi, Y. G., Weng, H. M., Fang, Z., & Dean, M. P. M. (2018). "Observation of Double Weyl Phonons in Parity-Breaking FeSi" *Phys. Rev. Lett.* 121, 35302. DOI: 10.1103/PhysRevLett.121.035302.

Migdal, A. B. (1958). "No Title" *Zh. Eksp. Teor. Fiz.* 34, 1438.

Migliori, A. & Sarrao, J. L. (1997). Resonant Ultrasound Spectroscopy: Applications to Physics, Materials Measurements, and Nondestructive Evaluation New York: Wiley-VCH.

Milesi-Brault, C., Toulouse, C., Constable, E., Aramberri, H., Simonet, V., de Brion, S., Berger, H., Paolasini, L., Bosak, A., Íñiguez, J., et al. (2020). "Archetypal Soft-Mode-Driven Antipolar Transition in Francisite ${\mathrm{Cu}}_{3}\mathrm{Bi}({\mathrm{SeO}}_{3}{)}_{2}{\mathrm{O}}_{2}\mathrm{Cl}$" *Phys. Rev. Lett.* 124, 97603. DOI: 10.1103/PhysRevLett.124.097603.

Minelli, A., Souliou, S. M., Nguyen-Thanh, T., Romero, A. H., Serrano, J., Hernandez, W. I., Verstraete, M. J., Dmitriev, V., & Bosak, A. (2019). "Lattice dynamics and phase stability of rhombohedral antimony under high pressure" *Phys. Rev. B.* 100, 104305. DOI: 10.1103/PhysRevB.100.104305.

Mirone, A. "OpenPhonon: an open source computer code for lattice-dynamical calculations" http://sourceforge.net/projects/openphonon/.

Mittal, R., Heid, R., Bosak, A., Forrest, T. R., Chaplot, S. L., Lamago, D., Reznik, D., Bohnen, K.-P., Su, Y., Kumar, N., et al. (2010). "Pressure dependence of phonon modes across the tetragonal to collapsed-tetragonal phase transition in CaFe2As2" *Phys. Rev. B.* 81, 144502. http://link.aps.org/doi/10.1103/PhysRevB.81.144502.

Mohr, M., Maultzsch, J., Dobardzic, E., Reich, S., Milosevic, I., Damnjanovic, M., Bosak, A., Krisch, M., & Thomsen, C. (2007). "Phonon dispersion of graphite by inelastic x-ray scattering" *Phys. Rev. B.* 76, 35437–35439. DOI: 10.1103/PhysRevB.76.035439.

Monaco, G., Cunsolo, A., Pratesi, G., Sette, F., & Verbeni, R. (2002). "Deep inelastic atomic scattering of X-rays in liquid neon" *Phys. Rev. Lett.* 88, 227401–227404. DOI: 10.1103/PhysRevLett.88.227401.

Monaco, G., Nardone, M., Sette, F., & Verbeni, R. (2001). "Molecular vibrational spectroscopy by inelastic X-ray scattering: Experimental determination of the absolute vibrational cross section in liquid nitrogen" *Phys. Rev. B.* 64 %0 Gene, 212102–212104. DOI: 10.1103/PhysRevB.64.212102.

Monkhorst, H. J. & Pack, J. D. (1976). "Special points for Brillouin-zone integrations" *Phys. Rev. B.* 13, 5188–5192. https://link.aps.org/doi/10.1103/PhysRevB.13.5188.

Mundboth, K., Sutter, J., Laundy, D., Collins, S., Stoupin, S., & Shvyd'ko, Y. (2014). "Tests and characterization of a laterally graded multilayer Montel mirror" *J. Synch. Rad.* 21, 16–23. DOI: 10.1107/S1600577513024077.

Murai, N., Fukuda, T., Kobayashi, T., Nakajima, M., Uchiyama, H., Ishikawa, D., Tsutsui, S., Nakamura, H., Machida, M., Miyasaka, S., et al. (2016). "Effect of magnetism on lattice dynamics in SrFe2As2 using high-resolution inelastic x-ray scattering" *Phys. Rev. B.* 93, 20301. DOI: 10.1103/PhysRevB.93.020301.

Murai, N., Fukuda, T., Nakajima, M., Kawamura, M., Ishikawa, D., Tajima, S., & Baron, A. Q. R. (2020). "Lattice dynamics in FeSe via inelastic x-ray scattering and first-principles calculations" *Phys. Rev. B.* 101, 35126. DOI: 10.1103/PhysRevB.101.035126.

Murphy, B. M., Requardt, H., Stettner, J., Serrano, J., Krisch, M., Möller, M., & Press, W. (2005). "Phonon modes at the 2H-NbSe2 surface observed by grazing incidence inelastic X-ray scattering" *Phys. Rev. Lett.* 95, 256104–2561044. DOI: 10.1103/PhysRevLett.95.256104.

Nagamatsu, J., Nakagawa, N., Muranaka, T., Zenitani, Y., & Akimitsu, J. (2001). "Superconductivity at 39K in magnesium diboride" *Nature.* 410, 63.

Nakajima, Y., Imada, S., Hirose, K., Komabayashi, T., Ozawa, H., Tateno, S., Tsutsui, S., Kuwayama, Y., & Baron, A. Q. R. (2015). "Carbon-depleted outer core revealed by sound velocity measurements of liquid iron-carbon alloy" *Nat. Commun.* 6, 8942. DOI: 10.1038/ncomms9942.

Nakano, A., Hasegawa, T., Tamura, S., Katayama, N., Tsutsui, S., & Sawa, H. (2018). "Antiferroelectric distortion with anomalous phonon softening in the excitonic insulator ${\mathrm{Ta}}_{2}{\mathrm{NiSe}}_{5}$" *Phys. Rev. B.* 98, 45139. DOI: 10.1103/PhysRevB.98.045139.

Niedziela, J. L., Parshall, D., Lokshin, K. A., Sefat, A. S., Alatas, A., & Egami, T. (2011). "Phonon softening near the structural transition in BaFe2As2 observed by inelastic x-ray scattering" *Phys. Rev. B.* 84, 224305. http://link.aps.org/doi/10.1103/PhysRevB.84.224305.

O'Neill, C. D., Sokolov, D. A., Hermann, A., Bossak, A., Stock, C., & Huxley, A. D. (2017). "Inelastic x-ray investigation of the ferroelectric transition in SnTe" *Phys. Rev. B.* 95, 144101. DOI: 10.1103/PhysRevB.95.144101.

Occelli, F., Krisch, M., Loubeyre, P., Sette, F., Le Toullec, R., Masciovecchio, C., & Rueff, J. P. (2001). "Phonon dispersion curves in an argon single crystal at high pressure by inelastic X-ray scattering" *Phys. Rev. B.* 63, 224306–224308. DOI: 10.1103/PhysRevB.63.224306.

Occhialini, C. A., Handunkanda, S. U., Said, A., Trivedi, S., Guzmán-Verri, G. G., & Hancock, J. N. (2017). "Negative thermal expansion near two structural quantum phase transitions" *Phys. Rev. Mater.* 1, 70603. DOI: 10.1103/PhysRevMaterials.1.070603.

Ohtani, E., Shibazaki, Y., Sakai, T., Mibe, K., Fukui, H., Kamada, S., Tsutsui, S., & Baron, A. Q. R. (2013). "Sound velocity of hexagonal close-packed iron up to megabar pressures" *Geophys. Res. Lett.* 40, 50.

Ohwada, K., Hirota, K., Terauchi, H., Fukuda, T., Tsutsui, S., Baron, A. Q. R., Mizuki, J., Ohwa, H., & Yasuda, N. (2008). "Intrinsic ferroelectric instability in Pb(In[sub 1/2]Nb[sub 1/2])O[sub 3] revealed by changing B-site randomness: Inelastic x-ray scattering study" *Phys. Rev. B.* 77, 94136–94138. http://link.aps.org/abstract/PRB/v77/e094136.

Osaka, T., Hiran, T., Yabashi, M., Tono, Y. S., Yamauchi, K., Inubushi, Y., Sato, T., Ogawa, K., Matsuyama, S., Ishikawa, T., et al. (2014). "Development of split-delay x-ray optics using Si(220) crystals at SACLA" *Proc. SPIE.* 9210, 921009.

Park, S. R., Fukuda, T., Hamann, A., Lamago, D., Pintschovius, L., Fujita, M., Yamada, K., & Reznik, D. (2014). "Evidence for a charge collective mode associated with superconductivity in copper oxides from neutron and x-ray scattering measurements of La2,àíx" *Phys. Rev. B.* 89, 20506. http://link.aps.org/doi/10.1103/PhysRevB.89.020506.

Parshall, D., Pintschovius, L., Niedziela, J. L., Castellan, J.-P., Lamago, D., Mittal, R., Wolf, T., & Reznik, D. (2015). "Close correlation between magnetic properties and the soft phonon mode of the structural transition BaFe2As2 and SrFe2As2" *Phys. Rev. B.* 91, 134426. http://link.aps.org/doi/10.1103/PhysRevB.91.134426.

Perakis, F., Camisasca, G., Lane, T. J., Späh, A., Wikfeldt, K. T., Sellberg, J. A., Lehmkühler, F., Pathak, H., Kim, K. H., Amann-Winkel, K., et al. (2018). "Coherent X-rays reveal the influence of cage effects on ultrafast water dynamics" *Nat. Commun.* 9, 1917. DOI: 10.1038/s41467-018-04330-5.

Phelan, D., Krogstad, M. J., Schreiber, N. J., Osborn, R., Said, A. H., Zheng, H., & Rosenkranz, S. (2019). "Acoustic phonon dispersion and diffuse scattering across the valence transition of ${(\mathrm{P}{\mathrm{r}}_{0.85}{\mathrm{Y}}_{0.15})}_{0.7}\mathrm{C}{\mathrm{a}}_{0.3}\mathrm{Co}{\mathrm{O}}_{3\ensuremath{-}\ensuremath{\delta}}$" *Phys. Rev. B.* **100**, 54101. DOI: 10.1103/PhysRevB.100.054101.

Pinsker, Z. G. (1978). Dynamical scattering of x-rays in crystals New York: Springer-Verlag.

Pintschovius, L. & Reichardt, W. (1994). "Inelastic scattering studies of the lattice vibrations of high Tc compounds" *Physical Properties of High Temperature Superconductors IV*, , edited by D. Ginsberg, p. 295. Singapore: World scientific.

Placzek, G. (1952). "The scattering of Neutrons by Systems of Heavy Nuclei" *Phys. Rev.* **86**, 377.

Plazanet, M., Beraud, A., Johnson, M., Krisch, M., & Trommsdorff, H. P. (2005). "Probing vibrational excitations in molecular crystals by inelastic scattering: From neutrons to X-rays" *Chem. Phys.* **317**, 153–158. DOI: 10.1016/j.chemphys.2005.04.040.

Pradip, R., Piekarz, P., Bosak, A., Merkel, D. G., Waller, O., Seiler, A., Chumakov, A. I., Rüffer, R., Oleś, A. M., Parlinski, K., et al. (2016). "Lattice Dynamics of EuO: Evidence for Giant Spin-Phonon Coupling" *Phys. Rev. Lett.* **116**, 185501. http://link.aps.org/doi/10.1103/PhysRevLett.116.185501.

Quilty, J. W., Lee, S., Yamamoto, A., & Tajima, S. (2002). "Superconducting Gap in MgB2: Electronic Raman Scattering Measurements of Single Crystals" *Phys. Rev. Lett.* **88**, 87001.

Ramirez-Cuesta, A. J. (2004). "aCLIMAX 4.0.1, The new version of the software for analyzing and interpreting INS spectra" *Comput. Phys. Commun.* **157**, 226–238. DOI: http://dx.doi.org/10.1016/S0010-4655(03)00520-4.

Raymond, S., Piekarz, P., Sanchez, J. P., Serrano, J., Krisch, M., Janousov·, B., Rebizant, J., Metoki, N., Kaneko, K., Jochym, P. T., et al. (2006). "Probing the Coulomb interaction of the unconventional superconductor PuCoGa5 by phonon spectroscopy" *Phys. Rev. Lett.* **96**, 237003–237004. DOI: 10.1103/PhysRevLett.96.237003.

Raymond, S., Rueff, J. P., Shapiro, S. M., Wochner, P., Sette, F., & Lejay, P. (2001). "Anomalous lattice properties of the heavy fermion compound CeRu2Si2: An X-ray scattering investigation" *Solid State Commun.* **118 %0 Gen**, 473–477. DOI: 10.1016/s0038-1098(01)00162-4.

Reichert, H., Bencivenga, F., Wehinger, B., Krisch, M., Sette, F., & Dosch, H. (2007). "High-frequency subsurface and bulk dynamics of liquid indium" *Phys. Rev. Lett.* **98**, 96104. DOI: 10.1103/PhysRevLett.98.096104.

Ren, Z.-A., Lu, W., Yang, J., Yi, W., Shen, X.-L., Cai, Z., Che, G.-C., Dong, X.-L., Sun, L.-L., Zhou, F., et al. (2008). "Superconductivity at 55 K in Iron-Based F-Doped Layered Quaternary Compound Sm[O1-xFx] FeAs" *Chinese Phys. Lett.* **25**, 2215–2216.

René de Cotret, L. P., Pöhls, J.-H., Stern, M. J., Otto, M. R., Sutton, M., & Siwick, B. J. (2019). "Time- and momentum-resolved phonon population dynamics with ultrafast electron diffuse scattering" *Phys. Rev. B.* **100**, 214115. DOI: 10.1103/PhysRevB.100.214115.

Rennie, S., Lawrence Bright, E., Darnbrough, J. E., Paolasini, L., Bosak, A., Smith, A. D., Mason, N., Lander, G. H., & Springell, R. (2018). "Study of phonons in irradiated epitaxial thin films of UO_{2}" *Phys. Rev. B.* **97**, 224303. DOI: 10.1103/PhysRevB.97.224303.

Requardt, H., Lorenzo, J. E., Monceau, P., Currat, R., & Krisch, M. (2002). "Dynamics in the charge-density-wave system NbSe3 using inelastic X-ray scattering with meV energy resolution" *Phys. Rev. B.* **66**, 214303–214304. DOI: 10.1103/PhysRevB.66.214303.

Reznik, D. (2012). "Phonon anomalies and dynamic stripes" *Phys. C Supercond.* **481**, 75–92. http://www.sciencedirect.com/science/article/pii/S0921453412000469.

Reznik, D., Fukuda, T., Lamago, D., Baron, A. Q. R., Tsutsui, S., Fujita, M., & Yamada, K. (2008). "q-Dependence of the giant bond-stretching phonon anomaly in the stripe compound La1.48Nd0.4Sr0.12CuO4 measured by IXS" *J. Phys. Chem. Solids.* **69**, 3103.

Reznik, D., Keimer, B., Dogan, F., & Aksay, I. A. (1995). "q dependence and self energy effects of the plane oxygen vibrations of YBa2Cu3O7" *Phys. Rev. Lett.* **75**, 2396.

Reznik, D., Lokshin, K., Mitchell, D. C., Parshall, D., Dmowski, W., Lamago, D., Heid, R., Bohnen, K.-P., Sefat, A. S., McGuire, M. A., et al. (2009). "Phonons in doped and undoped BaFe2As2 investigated by inelastic x-ray scattering" *Phys. Rev. B.* **80**, 214534. http://link.aps.org/doi/10.1103/PhysRevB.80.214534.

Reznik, D., Pintschovius, L., Ito, M., Iikubo, S., Sato, M., Goka, H., Fujita, M., Yamada, K., Gu, G. D., & Tranquada, J. M. (2006). "Electron-phonon coupling reflecting dynamic charge inhomogeneity in copper oxide superconductors" *Nature.* **440**, 1170–1173. http://dx.doi.org/10.1038/nature04704.

Robert, A., Curtis, R., Flath, D., Gray, A., Sikorski, M., Song, S., Srinivasan, V., & Stefanescu, D. (2013). "The X-ray Correlation Spectroscopy instrument at the Linac Coherent Light Source" *J. Phys. Conf. Ser.* **425**, 212009.

Roseker, W., Lee, S., Walther, M., Schulte-Schrepping, H., Franz, H., Gray, A., Sikorski, M., Fuoss, P. H., Stephenson, G. B., Robert, A., et al. (2012). "Hard x-ray delay line for x-ray photon correlation spectroscopy and jitter-free pump-probe experiments at LCLS" *Proc. SPIE.* **8504**,.

Rotter, M., Rogl, P., Grytsiv, A., Wolf, W., Krisch, M., & Mirone, A. (2008). "Lattice dynamics of skutterudites: Inelastic x-ray scattering on CoSb3" *Phys. Rev. B.* **77**, 144301. http://link.aps.org/doi/10.1103/PhysRevB.77.144301.

Rotter, M., Tegel, M., & Johrendt, D. (2008). "Superconductivity at 38~K in the Iron Arsenide B_{1-x}K_{x})Fe_{2}As_{2}" *Phys. Rev. Lett.* **101**, 107006. http://link.aps.org/doi/10.1103/PhysRevLett.101.107006.

Rueff, J.-P. P., Calandra, M., D'Astuto, M., Leininger, P., Shukla, A., Bosak, A., Krisch, M., Ishii, H., Cai, Y., Badica, P., et al. (2006). "Phonon softening in NaxCoO2∑yH2O: Implications for the Fermi surface topology and the superconducting state" *Phys. Rev. B.* **74**, 20504. DOI: 10.1103/PhysRevB.74.020504.

Ruminy, M., Valdez, M. N., Wehinger, B., Bosak, A., Adroja, D. T., Stuhr, U., Iida, K., Kamazawa, K., Pomjakushina, E., Prabakharan, D., et al. (2016). "First-principles calculation and experimental investigation of lattice dynamics in the rare-earth pyrochlores R2Ti2O7 (R=Tb,Dy,Ho)" *Phys. Rev. B.* **93**, 214308. DOI: 10.1103/PhysRevB.93.214308.

Said, A. H., Gog, T., Wieczorek, M., Huang, X., Casa, D., Kasman, E., Divan, R., & Kim, J. H. (2018). "High-energy-resolution diced spherical quartz analyzers for resonant inelastic X-ray scattering" *J. Synch. Rad.* **25**, 373–377. DOI: 10.1107/S1600577517018185.

Said, A., Sinn, H., & Divan, R. (2011). "New developments in fabrication of high-energy-resolution analyzers for inelastic X-ray spectroscopy" *J. Synch. Rad.* **18**, 492.

Sakai, H., Fujioka, J., Fukuda, T., Bahramy, M. S., Okuyama, D., Arita, R., Arima, T., Baron, A. Q. R., Taguchi, Y., & Tokura, Y. (2012). "Soft phonon mode coupled with antiferromagnetic order in incipient-ferroelectric Mott insulators Sr1-xBaxMnO3" *Phys. Rev. B.* **86**, 104407. http://link.aps.org/doi/10.1103/PhysRevB.86.104407.

Sakai, H., Fujioka, J., Fukuda, T., Okuyama, D., Hashizume, D., Kagawa, F., Nakao, H., Murakami, Y., Arima, T., Baron, A. Q. R., et al. (2011). "Displacement-Type Ferroelectricity with Off-Center Magnetic Ions in Perovskite Sr1-xBaxMnO3" *Phys. Rev. Lett.* **107**, 137601. http://link.aps.org/doi/10.1103/PhysRevLett.107.137601.

Sánchez del Río, M. & Dejus, R. J. (2011). "XOP v2.4: recent developments of the x-ray optics software toolkit." *Proc. SPIE.* **8141**, 814115.

Sauer, A., Zocco, D. A., Said, A. H., Heid, R., Böhmer, A., & Weber, F. (2019). "Electron-phonon coupling and superconductivity-induced distortion of the phonon lineshape in ${\mathrm{V}}_{3}\mathrm{Si}$" *Phys. Rev. B.* **99**, 134511. DOI: 10.1103/PhysRevB.99.134511.

Schülke, W. (2007). Electron Dynamics by Inelastic X-Ray Scattering New York: Oxford University Press.

Scopigno, T., Di Leonardo, R., Ruocco, G., Baron, A. Q. R., Tsutsui, S., Bossard, F., & Yannopoulos, S. N. (2004). "High Frequency Dynamics in a Monatomic Glass" *Phys. Rev. Lett.* **92**, 25503–25504. http://link.aps.org/abstract/PRL/v92/e025503.

Scott, J. F., Bryson, J., Carpenter, M. A., Herrero-Albillos, J., & Itoh, M. (2011). "Elastic and Anelastic Properties of Ferroelectric SrTi18O3" *Phys. Rev. Lett.* **106**, 105502. http://link.aps.org/doi/10.1103/PhysRevLett.106.105502.

Sergueev, I., Wille, H.-C., Hermann, R. P., Bessas, D., Shvyd'ko, Y. V, Zajac, M., & Rüffer, R. (2011). "Milli-electronvolt monochromatization of hard X-rays with a sapphire backscattering monochromator" *J. Synch. Rad.* **18**, 802–810. DOI: 10.1107/S090904951102485X.

Serrano, J., Bosak, A., Arenal, R., Krisch, M., Watanabe, K., Taniguchi, T., Kanda, H., Rubio, A., & Wirtz, L. (2007). "Vibrational properties of hexagonal boron nitride: Inelastic X-ray scattering and ab initio calculations" *Phys. Rev. Lett.* **98**, 95503–95504. DOI: 10.1103/PhysRevLett.98.095503.

Seto, M., Yoda, Y., Kikuta, S., Zhang, X. W., & Ando, M. (1995). "Observation of Nuclear Resonant Scattering Accompanied by Phonon Excitation using Synchrotron Radiation" *Phys. Rev. Lett.* **74**, 3828–3831.

Shapiro, S. M., Shirane, G., & Axe, J. D. (1975). "Measurements of the electron-phonon interaction in Nb by inelastic neutron scattering" *Phys. Rev. B.* **12**, 4899–4908. http://link.aps.org/doi/10.1103/PhysRevB.12.4899.

Shibazaki, Y., Ohtani, E., Fukui, H., Sakai, T., Kamada, S., Ishikawa, D., Tsutsui, S., Baron, A. Q. R., Nishitani, N., Hirao, N., et al. (2012). "Sound velocity measurements in dhcp-FeH up to 70 GPa with inelastic X-ray scattering: Implications for the composition of the Earth's core" *Earth Planet. Sci. Lett.* **313–314**, 79–85. DOI: 10.1016/j.epsl.2011.11.002.

Shih-Chang Weng, Xu, R., Said, A. H., Leu, B. M., Ding, Y., Hong, H., Fang, X., Chou, M. Y., Bosak, A., Abbamonte, P., et al. (2014). "Pressure-induced antiferrodistortive phase transition in SrTiO 3 : Common scaling of soft-mode with pressure and temperature" *EPL (Europhysics Lett.* **107**, 36006. http://stacks.iop.org/0295-5075/107/i=3/a=36006.

Shirane, G., Endoh, Y., Birgeneau, R. J., Kastner, M. A., Hidaka, Y., Oda, M., Suzuki, M., & Murakami, T. (1987). "Two-dimensional antiferromagnetic quantum spin-fluid state in La2CuO4" *Phys. Rev. Lett.* **59**, 1613–1616. http://link.aps.org/doi/10.1103/PhysRevLett.59.1613.

Shu, D., Toellner, T. S., & Alp, E. E. (2000). "Design of a high-resolution high-stability positioning mechanism for crystal optics" *AIP Conf. Proc.* **521**,.

Shukla, A., Calandra, M., d'Astuto, M., Lazzeri, M., Mauri, F., Bellin, C., Krisch, M., Karpinski, J., Kazakov, S. M., Jun, J., et al. (2002). "Phonon Dispersion and Lifetimes in MgB2" *Cond-Mat/0209064.*

Shvyd'ko, Y. (2004). X-Ray Optics: High-Energy-Resolution Application New York: Springer-Verlag.

Shvyd'ko, Y. (2015). "Theory of angular-dispersive, imaging hard-x-ray spectrographs" *Phys. Rev. A.* **91**, 53817. http://link.aps.org/doi/10.1103/PhysRevA.91.053817.

Shvyd'ko, Y. (2016). "X-ray Echo Spectroscopy" *Phys. Rev. Lett.* **116**, 80801. http://link.aps.org/doi/10.1103/PhysRevLett.116.080801.

Shvyd'ko, Y., Lerche, M., Kuetgens, U., Rüter, H., Alatas, A., & Zhao, J. (2006). "X-Ray Bragg Diffraction in Asymmetric Backscattering Geometry" *Phys. Rev. Lett.* **97**, 235502. http://link.aps.org/doi/10.1103/PhysRevLett.97.235502.

Shvyd'ko, Y., Stoupin, S., Mundboth, K., & Kim, J. (2013). "Hard-x-ray spectrographs with resolution beyond 100 ueV" *Phys. Rev. A.* **87**, 43835. http://link.aps.org/doi/10.1103/PhysRevA.87.043835.

Shvyd'ko, Y., Stoupin, S., Shu, D., Collins, S. P., Mundboth, K., Sutter, J., & Tolkiehn, M. (2014). "High-contrast sub-millivolt inelastic X-ray scattering for nano- and mesoscale science" *Nat Commun.* **5**, http://dx.doi.org/10.1038/ncomms5219.

Shvyd'ko, Y., Stoupin, S., Shu, D., & Khachatryan, R. (2011). "Using angular dispersion and anomalous transmission to shape ultramonochromatic x rays" *Phys. Rev. A.* **84**, 53823. http://link.aps.org/doi/10.1103/PhysRevA.84.053823.

Shvyd'ko, Y. V & Gerdau, E. (1999). "Backscattering Mirrors for X-Rays and Mössbauer radiation" *Hyperfine Interact.* **123/124**, 741.

Sinha, S. K. (2001). "Theory of inelastic x-ray scattering from condensed matter" *J. Phys Condens. Matter.* **13**, 7511.

Sinn, H. (2001). "Spectroscopy with meV energy resolution" *J. Phys. Cond. Mat.* . **13**, 7525.

Sjölander, A. (1958). "Multi-Phonon processes in slow neutrons catetring by crystals" *Ark. Für Fys.* **14**, 315.

Souliou, S. M., Gretarsson, H., Garbarino, G., Bosak, A., Porras, J., Loew, T., Keimer, B., & Le Tacon, M. (2018). "Rapid suppression of the charge density wave in YBa_{2}Cu_{3}O_{6.6} under hydrostatic pressure" *Phys. Rev. B.* **97**, 20503. DOI: 10.1103/PhysRevB.97.020503.

Souvatzis, P., Eriksson, O., Katsnelson, M. I., & Rudin, S. P. (2009). "The self-consistent ab initio lattice dynamical method" *Comput. Mater. Sci.* **44**, 888–894. DOI: http://dx.doi.org/10.1016/j.commatsci.2008.06.016.

Spalek, L. J., Saxena, S. S., Panagopoulos, C., Katsufuji, T., Schiemer, J. A., & Carpenter, M. A. (2014). "Elastic and anelastic relaxations associated with phase transitions in EuTiO3" *Phys. Rev. B.* **90**, 54119. DOI: 10.1103/PhysRevB.90.054119.

Spalt, H., Zounek, A., Dev, B., & Materlik, G. (1988). "Coherent X-Ray Scattering by Phonons: Determination of Phonon Eigenvectors" *Phys. Rev. Lett.* **60**, 1868–1871. http://link.aps.org/doi/10.1103/PhysRevLett.60.1868.

Squires, G. L. (1978). Introduction to the Theory of Thermal Neutron Scattering New York: Dover Publications, Inc.

Srivastava, G. P. (1990). The Physics of Phonons Bristol: Adam Hilger.

Stassis, C., Bullock, M., Zarestky, J., Canfield, P., Goldman, A. I., Shirane, G., & Shapiro, S. M. (1997). "Phonon mode coupling in superconducting LuNi2B2C" *Phys. Rev. B.* **55**, R8678–R8681. http://link.aps.org/doi/10.1103/PhysRevB.55.R8678.

Stedman, R., Almqvist, L., Nilsson, G., & Raunio, G. (1967). "Fermi Surace of Lead from Kohn Anomalies" *Phys. Rev.* **163**, 567.

Stekiel, M., Girard, A., Nguyen-Thanh, T., Bosak, A., Milman, V., & Winkler, B. (2019). "Phonon-driven phase transitions in calcite, dolomite, and magnesite" *Phys. Rev. B.* **99**, 54101. DOI: 10.1103/PhysRevB.99.054101.

Stekiel, M., Nguyen-Thanh, T., Chariton, S., McCammon, C., Bosak, A., Morgenroth, W., Milman, V., Refson, K., & Winkler, B. (2017). "High pressure elasticity of FeCO3-MgCO3 carbonates" *Phys. Earth Planet. Inter.* **271**, 57–63. DOI: https://doi.org/10.1016/j.pepi.2017.08.004.

Steurer, W., Apfolter, A., Koch, M., Ernst, W. E., Søndergård, E., Manson, J. R., & Holst, B. (2008). "Vibrational excitations of glass observed using helium atom scattering" *J. Phys. Cond. Mat.* **20**, 224003. http://stacks.iop.org/0953-8984/20/i=22/a=224003.

Strauch, D., Dorner, B., Ivanov, A. A., Krisch, M., Serrano, J., Bosak, A., Choyke, W., Stojetz, B., & Malorny, M. (2006). "Phonons in SiC from INS, IXS, and Ab-Initio Calculations" *Mater. Sci. Forum.* **527–529**, 689–694.

Sturhahn, W. & Toellner, T. S. (2011). "Feasibility of in-line instruments for high-resolution inelastic X-ray scattering" *J. Synch. Rad.* **18**, 229–237. DOI: 10.1107/S0909049510053513.

Sturhahn, W., Toellner, T. S., Alp, E. E., Zhang, X., Ando, M., Yoda, Y., Kikuta, S., Seto, M., Kimball, C. W., & Dabrowski, B. (1995). "Phonon Density of States Measured by Inelastic Nuclear Resonant Scattering" *Phys. Rev. Lett.* **74**, 3832–3835.

Sutter, J. P., Alp, E. E., Hu, M. Y., Lee, P. L., Sinn, H., Sturhahn, W., Toellner, T. S., Bortel, G., & Colella, R. (2001). "Multiple-beam x-ray diffraction near exact backscatering" *Phys. Rev. B.* 63, 94111.

Sutter, J. P., Baron, A. Q. R., Miwa, D., Nishino, Y., Tamasaku, K., & Ishikawa, T. (2006). "Nearly perfect large-area quartz: 4 meV resolution for 10 keV photons over 10 cm2" *J. Synch. Rad.* 13, 278–280. DOI: 10.1107/S0909049506003888.

Sutter, J. P., Baron, A. Q. R., Yamazaki, H., Ishikawa, T., & Yamazaki, H. (2005). "Examination of Bragg backscattering from crystalline quartz." *J. Phys. Chem. Solids.* 66, 2306–2309. DOI: 10.1016/j.jpcs.2005.09.044.

Sutton, M. (2002). "Coheren X-ray Diffraction" *Third-Generation Hard X-Ray Synchrotron Radiation Sources: Source Properties, Optics, Experimental Techniques*, , edited by D.M. Mills, pp. 101–123. John Whiley & Sons, Inc.

Le Tacon, M., Bosak, A., Souliou, S. M., Dellea, G., Loew, T., Heid, R., Bohnen, K. P., Ghiringhelli, G., Krisch, M., & Keimer, B. (2013). "Inelastic X-ray scattering in YBa2Cu3O6.6 reveals giant phonon anomalies and elastic central peak due to charge-density-wave formation" *Nat Phys.* 10, 52–58. http://dx.doi.org/10.1038/nphys2805.

Le Tacon, M., Forrest, T. R., Rüegg, C., Bosak, A., Walters, A. C., Mittal, R., Rønnow, H. M., Zhigadlo, N. D., Katrych, S., Karpinski, J., et al. (2009). "Inelastic x-ray scattering study of superconducting SmFeAsO1-xFy single crystals: Evidence for strong momentum-dependent doping-induced renormalizations of optical phonons" *Phys. Rev. B.* 80, 220504. http://link.aps.org/doi/10.1103/PhysRevB.80.220504.

Le Tacon, M., Krisch, M., Bosak, A., Bos, J.-W. G., & Margadonna, S. (2008). "Phonon density of states in NdFeAsO1-xFx" *Phys. Rev. B.* 78, 140505. http://link.aps.org/doi/10.1103/PhysRevB.78.140505.

Tadano, T., Gohda, Y., & Tsuneyuki, T. (2014). "Anharmonic force constants extracted from first-principles molecular dynamics: applications to heat transfer simulations" *J. Phys. Cond. Mat.* 26, 225402. http://stacks.iop.org/0953-8984/26/i=22/a=225402.

Tagantsev, A. K., Vaideeswaran, K., Vakhrushev, S. B., Filimonov, A. V, Burkovsky, R. G., Shaganov, A., Andronikova, D., Rudskoy, A. I., Baron, A. Q. R., Uchiyama, H., et al. (2013). "The origin of antiferroelectricity in PbZrO3" *Nat Commun.* 4, Article Number 2229. http://dx.doi.org/10.1038/ncomms3229.

Tamtögl, A., Kraus, P., Mayrhofer-Reinhartshuber, M., Campi, D., Bernasconi, M., Benedek, G., & Ernst, W. E. (2013). "Surface and subsurface phonons of Bi(111) measured with helium atom scattering" *Phys. Rev. B.* 87, 35410. DOI: 10.1103/PhysRevB.87.035410.

Teitelbaum, S. W., Henighan, T., Huang, Y., Liu, H., Jiang, M. P., Zhu, D., Chollet, M., Sato, T., Murray, É. D., Fahy, S., et al. (2018). "Direct Measurement of Anharmonic Decay Channels of a Coherent Phonon" *Phys. Reveiw Lett.* 121, 125901. https://link.aps.org/doi/10.1103/PhysRevLett.121.125901.

Thomsen, C., Cardona, M., Gegenheimer, B., Liu, R., & Simon, A. (1988). "Raman study of the phonon anomaly in single-crystal YBa_{2}Cu_{3}O_{7- delta } in the presence of a magnetic field" *Phys. Rev. B.* 37, 9860.

Tian, Z., Li, M., Ren, Z., Ma, H., Alatas, A., Wilson, S. D., & Li, J. (2015). "Inelastic x-ray scattering measurements of phonon dispersion and lifetimes in PbTe 1− x Se x alloys" *J. Phys. Cond. Mat.* 27, 375403. http://stacks.iop.org/0953-8984/27/i=37/a=375403.

Toellner, T. S., Alatas, A., & Said, A. H. (2011). "Six-reflection meV-monochromator for synchrotron radiation" *J. Synch. Rad.* 18, 605–611. DOI: 10.1107/S0909049511017535.

Toellner, T. S., Hu, M. Y., Sturhahn, W., Alp, E. E., & Zhao, J. (2002). "Ultrahigh-resolution x-ray monochromators for elastic and inelastic x-ray scattering studies (abstract) (invited)" *Rev. Sci. Instrum.* 73,.

Toellner, T. S., Mooney, T., Shastri, S., & Alp, E. E. (1992). "High energy resolution, high angular acceptance crystal monochromator" *Opt. High-Brightness Synchrotron Beamlines.* SPIE Vol 1, 218–222.

Togo, A., Oba, F., & Tanaka, I. (2008). "First-principles calculations of the ferroelastic transition between rutile-type and CaCl2-SiO2 at high pressures" *Phys. Rev. B.* 78, 134106. http://link.aps.org/doi/10.1103/PhysRevB.78.134106.

Tornatzky, H., Gillen, R., Uchiyama, H., & Maultzsch, J. (2019). "Phonon dispersion in ${\mathrm{MoS}}_{2}$" *Phys. Rev. B.* 99, 144309. DOI: 10.1103/PhysRevB.99.144309.

Tóth, S., Wehinger, B., Rolfs, K., Birol, T., Stuhr, U., Takatsu, H., Kimura, K., Kimura, T., Rønnow, H. M., & Rüegg, C. (2016). "Electromagnon dispersion probed by inelastic X-ray scattering in LiCrO2" *Nat. Commun.* 7, 13547. http://dx.doi.org/10.1038/ncomms13547.

Trigo, M., Fuchs, M., Chen, J., Jiang, M. P., Cammarata, M., Fahy, S., Fritz, D. M., Gaffney, K., Ghimire, S., Higginbotham, A., et al. (2013). "Fourier-transform inelastic X-ray scattering from time- and momentum-dependent phonon-phonon correlations" *Nat Phys.* 9, 790–794. DOI: 10.1038/nphys2788http://www.nature.com/nphys/journal/v9/n12/abs/nphys2788.html#supplementary-information.

Tsutsui, S., Kobayashi, H., Ishikawa, D., Sutter, J. P., Baron, A. Q. R., Hasegawa, T., Ogita, N., Udagawa, M., Yoda, Y., Onodera, H., et al. (2008). "Direct Observation of Low Energy Sm phonon in SmRu4P12" *J. Phys. Soc. Japan.* 77, 033601.

Tsutsui, S., Uchiyama, H., Sutter, J. P., Baron, A. Q. R., Mizumaki, M., Kawamura, N., Uruga, T., Sugawara, H., Yamaura, J.-I., Ochiai, A., et al. (2012). "Atomic dynamics of low-lying rare-earth guest modes in heavy fermion filled skutterudites ROs4Sb12 (R = light rare-earth)" *Phys. Rev. B.* 86, 13. DOI: 10.1103/PhysRevB.86.195115.

Uchiyama, H., Baron, A. Q. R., Tsutsui, S., Tanaka, Y., Hu, W. Z., Yamamoto, A., Tajima, S., & Endoh, Y. (2004). "Softening of Cu-O Bond Stretching Phonons in Tetragonal HgBa[sub 2]CuO[sub 4 + delta]" *Phys. Rev. Lett.* 92, 197005. http://link.aps.org/abstract/PRL/v92/e197005.

Uchiyama, H., Oshima, Y., Patterson, R., Iwamoto, S., Shiomi, J., & Shimamura, K. (2018). "Phonon Lifetime Observation in Epitaxial ScN Film with Inelastic X-Ray Scattering Spectroscopy" *Phys. Rev. Lett.* 120, 235901. DOI: 10.1103/PhysRevLett.120.235901.

Uchiyama, H., Tsutsui, S., & Baron, A. Q. R. (2010). "Effects of anisotropic charge on transverse optical phonons in NiO: Inelastic x-ray scattering spectroscopy study" *Phys. Rev. B.* 81, 241103. DOI: 10.1103/PhysRevB.81.241103.

Umemoto, K., Hirose, K., Imada, S., Nakajima, Y., Komabayashi, T., Tsutsui, S., & Baron, A. Q. R. (2014). "Liquid iron-sulfur alloys at outer core conditions by first-principles calculations" *Geophys. Res. Lett.* 41, DOI: 10.1002/2014GL061233.

Upton, M. H., Forrest, T. R., Walters, A. C., Howard, C. A., Ellerby, M., Said, A. H., & McMorrow, D. F. (2010). "Phonons and superconductivity in YbC6 and related compounds" *Phys. Rev. B.* 82, 134515. http://link.aps.org/doi/10.1103/PhysRevB.82.134515.

Upton, M., Walters, A., Howard, C., Rahnejat, K., Ellerby, M., Hill, J., McMorrow, D., Alatas, A., Leu, B., & Ku, W. (2007). "Phonons in superconducting CaC6 studied via inelastic x-ray scattering" *Phys. Rev. B.* 76, 220501. DOI: 10.1103/PhysRevB.76.220501.

Usuda, K., Oyanagi, Y., Yokogawa, R., Uchiyama, H., Tsutsui, S., Yonenaga, I., & Ogura, A. (2020). "Phonon dispersion of bulk Ge-rich SiGe: inelastic X-ray scattering studies" *Jpn. J. Appl. Phys.* 59, 61003. DOI: 10.35848/1347-4065/ab8e07.

van Hove, L. (1954). "Correlations in Space and Time and Born Approximation Scattering in Systems of Interacting Particles" *Phys. Rev.* 95, 249–262.

Verbeni, R., Henriquet, C., Gambetti, D., Martel, K., Krisch, M., Monaco, G., & Sette, F. (2003). "Development of an Asymmetrically-cut Backscattering Monochromator for Very High Energy Resolution Inelastic X-ray Scattering" *ESRF Highlights.* Mehtods, 03. http://www.esrf.eu/UsersAndScience/Publications/Highlights/2003/Methods/Methods03.

Verbeni, R., Kocsis, M., Huotari, S., Krisch, M., Monaco, G., Sette, F., & Vanko, G. (2005). "Advances in crystal analyzers for inelastic X-ray scattering" *J. Phys. Chem. Solids.* 66, 2299–2305. DOI: http://dx.doi.org/10.1016/j.jpcs.2005.09.079.

Voneshen, D. J., Refson, K., Borissenko, E., Krisch, M., Bosak, A., Piovano, A., Cemal, E., Enderle, M., Gutmann, M. J., Hoesch, M., et al. (2013). "Suppression of thermal conductivity by rattling modes in thermoelectric sodium cobaltate" *Nat Mater.* **12**, 1028–1032. http://dx.doi.org/10.1038/nmat3739.

Waasmaier, D. & Kirfel, A. (1994). "New analytical scattering factor functions for free atoms and ions" *Acta Crystallogr.* **A51**, 416.

Wakabayashi, Y., Nakajima, D., Ishiguro, Y., Kimura, K., Kimura, T., Tsutsui, S., Baron, A. Q. R., Hayashi, K., Happo, N., Hosokawa, S., et al. (2016). "Chemical and orbital fluctuations in Ba3CuSb2 O9" *Phys. Rev. B.* **93**, DOI: 10.1103/PhysRevB.93.245117.

Wakiya, K., Onimaru, T., Tsutsui, S., Hasegawa, T., Matsumoto, K. T., Nagasawa, N., Baron, A. Q. R., Ogita, N., Udagawa, M., & Takabatake, T. (2016). "Low-energy optical phonon modes in the caged compound LaRu2Zn20" *Phys. Rev. B.* **93**, 64105. DOI: 10.1103/PhysRevB.93.064105.

Waller, O., Piekarz, P., Bosak, A., Jochym, P. T., Ibrahimkutty, S., Seiler, A., Krisch, M., Baumbach, T., Parlinski, K., & Stankov, S. (2016). "Lattice dynamics of neodymium: Influence of 4f electron correlations" *Phys. Rev. B.* **94**, 14303. http://link.aps.org/doi/10.1103/PhysRevB.94.014303.

Walters, A. C., Howard, C. A., Upton, M. H., Dean, M. P. M., Alatas, A., Leu, B. M., Ellerby, M., McMorrow, D. F., Hill, J. P., Calandra, M., et al. (2011). "Comparative study of the phonons in nonsuperconducting BaC6 and superconducting CaC6 using inelastic x-ray scattering" *Phys. Rev. B.* **84**, 14511. http://link.aps.org/doi/10.1103/PhysRevB.84.014511.

Weber, F., Hott, R., Heid, R., Bohnen, K.-P., Rosenkranz, S., Castellan, J.-P., Osborn, R., Said, A. H., Leu, B. M., & Reznik, D. (2013). "Optical phonons and the soft mode in 2H-NbSe2" *Phys. Rev. B.* **87**, 245111. DOI: 10.1103/PhysRevB.87.245111.

Weber, F., Kreyssig, A., Pintschovius, L., Heid, R., Reichardt, W., Reznik, D., Stockert, O., & Hradil, K. (2008). "Direct Observation of the Superconducting Gap in Phonon Spectra" *Phys. Rev. Lett.* **101**, 237002. http://link.aps.org/doi/10.1103/PhysRevLett.101.237002.

Weber, F., Rosenkranz, S., Castellan, J.-P., Osborn, R., Hott, R., Heid, R., Bohnen, K.-P., Egami, T., Said, A. H., & Reznik, D. (2011). "Extended Phonon Collapse and the Origin of the Charge-Density Wave in 2H-NbSe2" *Phys. Rev. Lett.* **107**, 107403. DOI: 10.1103/PhysRevLett.107.107403.

Wehinger, B., Bosak, A., & Jochym, P. T. (2016). "Soft phonon modes in rutile TiO2" *Phys. Rev. B.* **93**, 14303. http://link.aps.org/doi/10.1103/PhysRevB.93.014303.

Wehinger, B., Krisch, M., & Reichert, H. (2011). "High-frequency dynamics in the near-surface region studied by inelastic x-ray scattering: the case of liquid indium" *New J. Phys.* **13**, 023021. http://iopscience.iop.org/1367-2630/13/2/023021/fulltext/.

Wehinger, B., Mirone, A., Krisch, M., & Bosak, A. (2017). "Full Elasticity Tensor from Thermal Diffuse Scattering" *Phys. Rev. Lett.* **118**, 35502. http://link.aps.org/doi/10.1103/PhysRevLett.118.035502.

Winkler, B., Friedrich, A., Wilson, D. J., Haussühl, E., Krisch, M., Bosak, A., Refson, K., & Milman, V. (2008). "Dispersion Relation of an OH-Stretching Vibration from Inelastic X-Ray Scattering" *Phys. Rev. Lett.* **101**, 65501. http://link.aps.org/doi/10.1103/PhysRevLett.101.065501.

Wong, J., Krisch, M., Farber, D. L., Occelli, F., Schwartz, A. J., Chiang, T. C., Wall, M., Boro, C., & Xu, R. (2003). "Phonon dispersion of fcc delta-plutonium-gallium by inelastic X-ray scattering" *Science (80-. ).* **301**, 1078–1080. DOI: 10.1126/science.1087179.

Wong, J., Krisch, M., Farber, D. L., Occelli, F., Xu, R., Chiang, T. C., Clatterbuck, D., Schwartz, A. J., Wall, M., & Boro, C. (2005). "Crystal dynamics of delta fcc Pu-Ga alloy by high-resolution inelastic X-ray scattering" *Phys. Rev. B.* **72**, 64112–64115. DOI: 10.1103/PhysRevB.72.064115.

Xia, H., Patterson, R., Smyth, S., Feng, Y., Chung, S., Zhang, Y., Shrestha, S., Huang, S., Uchiyama, H., Tsutsui, S., et al. (2017). "Inelastic X-ray scattering measurements of III-V multiple quantum wells" *Appl. Phys. Lett.* **110**, DOI: 10.1063/1.4974478.

Xu, R. & Chiang, T. C. (2005). "Determination of phonon dispersion relations by X-ray thermal diffuse scattering" *Zeitschrift Fur Krist.* **220**, 1009–1016.

Yabashi, M., Tamasaku, K., Kikuta, S., & Ishikawa, T. (2001). "X-ray monochromator with an energy resolution of 8×10 –9 at 14.41 keV" *Rev. Sci. Instrum.* **72**, 4080.

Yavas, H., Ercan Alp, E., Sinn, H., Alatas, A., Said, A. H., Shvyd'ko, Y., Toellner, T., Khachatryan, R., Billinge, S. J. L., Zahid Hasan, M., et al. (2007). "Sapphire analyzers for high-resolution X-ray spectroscopy" *Nucl. Inst. Meth. Phys. Res. A.* **582**, 149–151. http://www.sciencedirect.com/science/article/B6TJM-4PDC17C-W/2/973612407121a98f86bb1c0a9bddbe36.

Yavaş, H., Sutter, J. P., Gog, T., Wille, H.-C., & Baron, A. Q. R. (2017). "New materials for high-energy-resolution X-ray optics" *MRS Bull.* **42**, 424–429. DOI: 10.1557/mrs.2017.94.

Yildirim, T., G¸lseren, O., Lynn, J. W., Brown, C. M., Udovic, T. J., Huang, Q., Rogado, N., Regan, K. A., Hayward, M. A., Slusky, J. S., et al. (2001). "Giant Anharmonicity and Nonlinear Electron-Phonon Coupling in MgB_{2}: A Combined First-Principles Calculation and Neutron Scattering Study" *Phys. Rev. Lett.* **87**, 37001. http://link.aps.org/abstract/PRL/v87/e037001.

Yin, Z. P., Kutepov, A., & Kotliar, G. (2013). "Correlation-Enhanced Electron-Phonon Coupling: Applications of GW and Screened Hybrid Functional to Bismuthates, Chloronitrides, and Other High-" *Phys. Rev. X.* **3**, 21011. http://link.aps.org/doi/10.1103/PhysRevX.3.021011.

Yoneda, A., Fukui, H., Gomi, H., Kamada, S., Xie, L., Hirao, N., Uchiyama, H., Tsutsui, S., & Baron, A. Q. R. (2017). "Single crystal elasticity of gold up to ~20 GPa: Bulk modulus anomaly and implication for a primary pressure scale" *Jpn. J. Appl. Phys.* **56**, 95801. http://stacks.iop.org/1347-4065/56/i=9/a=095801.

Yoneda, A., Fukui, H., Xu, F., Nakatsuka, A., Yoshiasa, A., Seto, Y., Ono, K., Tsutsui, S., Uchiyama, H., & Baron, A. Q. R. (2014). "Elastic anisotropy of experimental analogues of perovskite and post-perovskite help to interpret D" diversity" *Nat. Commun.* **5**, DOI: 10.1038/ncomms4453.

Zhang, T. T., Miao, H., Wang, Q., Lin, J. Q., Cao, Y., Fabbris, G., Said, A. H., Liu, X., Lei, H. C., Fang, Z., et al. (2019). "Phononic Helical Nodal Lines with $\mathcal{PT}$ Protection in ${\mathrm{MoB}}_{2}$" *Phys. Rev. Lett.* **123**, 245302. DOI: 10.1103/PhysRevLett.123.245302.

Zhang, Y., Yang, C., Alatas, A., Said, A. H., Salke, N. P., Hong, J., & Lin, J.-F. (2019). "Pressure effect on Kohn anomaly and electronic topological transition in single-crystal tantalum" *Phys. Rev. B.* **100**, 75145. DOI: 10.1103/PhysRevB.100.075145.

Zhernenkov, M., Bolmatov, D., Soloviov, D., Zhernenkov, K., Toperverg, B. P., Cunsolo, A., Bosak, A., & Cai, Y. Q. (2016). "Revealing the mechanism of passive transport in lipid bilayers via phonon-mediated nanometre-scale density fluctuations" *Nat. Commun.* **7**, 11575. http://dx.doi.org/10.1038/ncomms11575.

Zhu, D., Robert, A., Henighan, T., Lemke, H. T., Chollet, M., Glownia, J. M., Reis, D. A., & Trigo, M. (2015). "Phonon spectroscopy with sub-meV resolution by femtosecond x-ray diffuse scattering" *Phys. Rev. B.* **92**, 54303. http://link.aps.org/doi/10.1103/PhysRevB.92.054303.

Zocco, D. A., Krannich, S., Heid, R., Bohnen, K.-P., Wolf, T., Forrest, T., Bosak, A., & Weber, F. (2015). "Lattice dynamical properties of superconducting SrPt3P studied via inelastic x-ray scattering and density functional perturbation theory" *Phys. Rev. B.* **92**, 220504. DOI: 10.1103/PhysRevB.92.220504.

"Bilbao Crystallographic Server" http://www.cryst.ehu.es/.

"NSLS-II Conceptual Design Report" (2006). http://www.bnl.gov/nsls2/project/CDR/